\PassOptionsToPackage{pdfpagelabels=false}{hyperref}
\documentclass[useAMS,usenatbib]{mn2e}
\def \eg {e.g.}

\def \ie {i.e.}
\def \cf {cf.}
\def \lcdm {{\hbox{$\Lambda$CDM}}}
\def \omegam {{\hbox{$\Omega_m$}}}
\def \omegal {{\hbox{$\Omega_\Lambda$}}}
\def \hzero {{\hbox{$H_0$}}}
\def \arcmin {\hbox{$^\prime$}}
\def \arcsec {\hbox{$^{\prime\prime}$}}
\def \deg {\hbox{$^\circ$}}
\def \ktu {{\hbox{$kT_u$}}}
\def \ktd {{\hbox{$kT_d$}}}
\def \mach {{\hbox{$\mathcal{M}$}}}
\def \machkt {{\hbox{$\mathcal{M}_{\rm kT}$}}}
\def \machsb {{\hbox{$\mathcal{M}_{\rm SB}$}}}
\def \compr {{\hbox{$\mathcal{C}$}}}
\def \rat {{\hbox{$\mathcal{R}$}}}

\def \chisqdof {\hbox{$\chi^2 /\rm{d.o.f.}$}}
\def \cstatdof {\hbox{c-stat/d.o.f.}}
\def \msun {\hbox{${\rm M_\odot}$}}
\def \zsun {\hbox{${\rm Z_\odot}$}}

\def \mfive {\hbox{$M_{500}$}}

\newcommand{\kmsmpc }{\mbox{km s$^{-1}$ Mpc$^{-1}$}}

\newcommand{\kev }{\mbox{keV}}
\newcommand{\kevcmsq }{\mbox{keV cm$^{2}$}}

\newcommand{\acisi }{ACIS-I}
\newcommand{\aciss }{ACIS-S}

\newcommand{\obsid }{ObsID}
\newcommand{\faint }{FAINT}
\newcommand{\vfaint }{VFAINT}

\newcommand{\xspec }{\textsc{xspec}}
\newcommand{\contbin }{\textsc{contbin}}
\newcommand{\ciao }{\textsc{ciao}}
\newcommand{\caldb }{\textsc{caldb}}
\newcommand{\proffit }{\textsc{proffit}}
\newcommand{\xmm }{{\em XMM-Newton}}
\newcommand{\chandra }{{\em Chandra}}

\newcommand{\planck }{{\em Planck}}

\newcommand{\suzaku }{{\em Suzaku}}

\newcommand{\asca }{ASCA}

\newcommand{\rosat }{ROSAT}

\newcommand{\hstE }{Hubble Space Telescope}

\usepackage{xcolor}
\usepackage{stackengine}
\usepackage{subfig}
\usepackage{graphicx}
\usepackage{amssymb}
\usepackage{multirow,bigdelim}
\usepackage{cleveref}
\crefname{figure}{}{} % {<type>}{<singular>}{<plural>}
\crefname{equation}{}{} % {<type>}{<singular>}{<plural>}

\newcommand{\subfigimgwhitebig}[3][,]{%
  \setbox1=\hbox{\includegraphics[#1]{#3}}% Store image in box
  \leavevmode\rlap{\usebox1}% Print image
  \rlap{\hspace*{9.1cm}\raisebox{\dimexpr\ht1-1.8\height}{\textcolor{white}{#2}}}% Print label
  \phantom{\usebox1}% Insert appropriate spcing
}
\newcommand{\subfigimgwhiteggm}[3][,]{%
  \setbox1=\hbox{\includegraphics[#1]{#3}}% Store image in box
  \leavevmode\rlap{\usebox1}% Print image
  \rlap{\hspace*{3.9cm}\raisebox{\dimexpr\ht1-1.6\height}{\textcolor{white}{#2}}}% Print label
  \phantom{\usebox1}% Insert appropriate spcing
}
\newcommand{\subfigimgblack}[3][,]{%
  \setbox1=\hbox{\includegraphics[#1]{#3}}% Store image in box
  \leavevmode\rlap{\usebox1}% Print image
  \rlap{\hspace*{4.5cm}\raisebox{\dimexpr\ht1-2.5\baselineskip}{#2}}% Print label
  \phantom{\usebox1}% Insert appropriate spcing
}
\newcommand{\subfigimgsb}[3][,]{%
  \setbox1=\hbox{\includegraphics[#1]{#3}}% Store image in box
  \leavevmode\rlap{\usebox1}% Print image
  \rlap{\hspace*{4.8cm}\raisebox{\dimexpr\ht1-1.22\baselineskip}{#2}}% Print label
  \phantom{\usebox1}% Insert appropriate spcing
}
\title[Shocks and cold fronts in galaxy clusters]{Shocks and cold fronts in merging and massive galaxy clusters: new detections with \chandra}

\author[Botteon et al.]{A.~Botteon$^{1,2}$\thanks{E-mail: botteon@ira.inaf.it}, 
F.~Gastaldello$^{3}$ and G.~Brunetti$^{2}$ \\
$^{1}$Dipartimento di Fisica e Astronomia, Universit\`{a} di Bologna, via P.~Gobetti 93/2, I-40129 Bologna, Italy \\
$^{2}$INAF - IRA, via P.~Gobetti 101, I-40129 Bologna, Italy\\
$^{3}$INAF - IASF Milano, via E.~Bassini 15, I-20133 Milano, Italy\\}
\date{\today}

\date{Accepted XXX. Received YYY; in original form ZZZ}

\pubyear{2018}

\begin{document}
\label{firstpage}
\pagerange{\pageref{firstpage}--\pageref{lastpage}}
\maketitle

\begin{abstract}
A number of merging galaxy clusters shows the presence of shocks and cold fronts, \ie\ sharp discontinuities in surface brightness and temperature. The observation of these features requires an X-ray telescope with high spatial resolution like \chandra, and allows to study important aspects concerning the physics of the intra-cluster medium (ICM), such as its thermal conduction and viscosity, as well as to provide information on the physical conditions leading to the acceleration of cosmic rays and magnetic field amplification in the cluster environment. In this work we search for new discontinuities in 15 merging and massive clusters observed with \chandra\ by using different imaging and spectral techniques of X-ray observations. Our analysis led to the discovery of 22 edges: six shocks, eight cold fronts and eight with uncertain origin. All the six shocks detected have $\mach < 2$ derived from density and temperature jumps. This work contributed to increase the number of discontinuities detected in clusters and shows the potential of combining diverse approaches aimed to identify edges in the ICM. A radio follow-up of the shocks discovered in this paper will be useful to study the connection between weak shocks and radio relics.
\end{abstract}


\section{Introduction}

Galaxy clusters are the most massive collapsed objects in the Universe. Their total mass is dominated by the dark matter ($\sim80\%$) that shapes deep potential wells where the baryons ($\sim20\%$) virialize. The majority of the baryonic matter in clusters is in the form of intra-cluster medium (ICM), a hot ($kT \sim 2 - 10$ \kev) and tenuous ($n \sim 10^{-3} - 10^{-4}$ cm$^{-3}$) plasma emitting via thermal bremsstrahlung in the X-rays. In the past two decades, \chandra\ and \xmm\ established a dichotomy between cool-core (CC) and non cool-core (NCC) clusters \citep[\eg][]{molendi01}, depending whether their core region shows a drop in the temperature profile or not. The reason of this drop is a natural consequence of the strongly peaked X-ray emissivity of relaxed systems that leads to the gas cooling in this denser environment, counterbalanced by the active galactic nucleus (AGN) feedback \citep[\eg][for a review]{peterson06rev}. On the other hand, disturbed systems exhibit shallower X-ray emissivity, hence lower cooling rates. For this reason there is a connection between the properties of the cluster core to its dynamical state: relaxed (\ie\ in equilibrium) systems naturally end to form a CC while NCCs are typically found in unrelaxed objects \citep[\eg][]{leccardi10}, where the effects of energetic events such as mergers have tremendous impact on their core, leading either to its direct disruption \citep[\eg][]{russell12, rossetti13, wang16} or to its mixing with the surrounding hot gas \citep{zuhone10}. \\
\indent
In the hierarchical process of large-scale structures formation the cluster mass is assembled from aggregation and infall of smaller structures \citep[\eg][for a review]{kravtsov12rev}. Mergers between galaxy clusters are the most energetic phenomena in the Universe, with the total kinetic energy that is dissipated in a crossing time-scale ($\sim$ Gyr) during the collision in the range $10^{63-64}$ ergs. At this stage shock waves, cold fronts, hydrodynamic instabilities, turbulence, and non-thermal components are generated in the ICM, making merging clusters unique probes to study several aspects of plasma astrophysics. \\
\indent
The sub-arcsec resolution of \chandra\ in particular allowed detailed studies of previously unseen edges, \ie\ shocks and cold fronts \citep[see][for a review]{markevitch07rev}. Both are sharp surface brightness (SB) discontinuities that differ in the sign of the temperature jump across the front. Shocks mark pressure discontinuities where the gas is heated in the downstream (\ie\ post-shock) region, showing higher temperature values with respect to the upstream (\ie\ pre-shock) region. In cold fronts, instead, this jump is inverted and the pressure across the edge is almost continuous. \\
\indent
Shocks and cold fronts have been observed in several galaxy clusters that are clearly undergoing significant merging activity \citep[\eg][for some collections]{markevitch07rev, owers09sample, ghizzardi10, markevitch10arx}. The most remarkable example is probably the Bullet Cluster \citep{markevitch02bullet}, where an infalling subcluster (the ``Bullet'') creates a contact discontinuity between its dense and low-entropy core and the surrounding hot gas. Ahead of this cold front another drop in SB but with reversed temperature jump, \ie\ a shock front, is also detected. The observation of this kind of fronts requires that the collision occurs almost in the plane of the sky as projection effects can hide the SB and temperature jumps. Since shocks move quickly in cluster outskirts, in regions where the thermal brightness is fainter, they are more difficult to observe than cold fronts. \\
\indent
To be thorough, we mention that also morphologically relaxed clusters can exhibit SB discontinuities. However, in this case their origin is different: shocks can be associated with the outbursts of the central AGN \citep[\eg][]{forman05, mcnamara05, nulsen05} and cold fronts can be produced by sloshing motions of the cluster core that are likely induced by off-axis minor mergers \citep[\eg][]{ascasibar06, roediger11, roediger12}. \\
\indent
The observation of shocks and cold fronts allows to investigate relevant physical processes in the ICM. Shocks (and turbulence) are able to (re)-accelerate particles and amplify magnetic fields leading to the formation of cluster-scale diffuse radio emission known as radio relics (and radio halos; \eg\ \citealt{brunetti14rev}, for a review). In the presence of a strong shock it is also possible to investigate the electron-ion equilibration timescale in the ICM \citep{markevitch06}.  Cold fronts are complementary probes of the ICM microphysics \citep[see][for a recent review]{zuhone16rev}. The absence of Rayleigh-Taylor or Kelvin-Helmholtz instabilities in these sharp discontinuities indeed gives information on the suppression of transport mechanisms in the ICM (\eg\ \citealt{ettori00, vikhlinin01cold, vikhlinin01magnetic}; however, see \citealt{ichinohe17}). The cold fronts generated by the infalling of groups and galaxies in clusters \citep[\eg][for recent works]{eckert14, eckert17a2142, ichinohe15, degrandi16, su17ngc1404} enable the study of other physical processes, such as magnetic draping and ram pressure stripping, providing information on the plasma mixing in the ICM \citep[\eg][and references therein]{dursi08}. \\
\indent
Currently, the number of detected edges in galaxy clusters is modest for observational limitations. This is reflected in the handful of merger shocks that have been confirmed using both X-ray imaging and spectral analysis. In this work we aim to search in an objective way for new merger induced shocks and cold fronts in massive and NCC galaxy clusters. The reason is to look for elusive features that can be followed-up in the radio band. To do that in practice we analyzed 15 clusters that were essentially selected because of the existence of adequate X-ray data available in the \chandra\ archive. The \chandra\ satellite is the the best instrument capable to resolve these sharp edges thanks to its excellent spatial resolution. We applied different techniques for spatial and spectral analysis including the application of an edge detection algorithm on the cluster images, the extraction and fitting of SB profiles, the spectral modeling of the X-ray (astrophysical and instrumental) background and the production of maps of the ICM thermodynamical quantities. This analysis is designed to properly characterize sharp edges distinguishing shocks from cold fronts. \\
\indent
The paper is organized as follows. In Section~\ref{sec:sample} we present the cluster sample. In Section~\ref{sec:analysis} we outline the edge-detection procedure and provide details about the X-ray data reduction (see also Appendices~\ref{app:absorption} and \ref{app:nxb}). In Section~\ref{sec:search} we describe how shocks and cold fronts were characterized in the analysis and in Section~\ref{sec:results} we present our results. Finally, in Section~\ref{sec:conclusions} we summarize and discuss our work. \\
\indent
Throughout the paper, we assume a \lcdm\ cosmology with $\omegal = 0.7$, $\omegam = 0.3$, and $\hzero = 70$ \kmsmpc. Statistical errors are provided at the $1\sigma$ confidence level, unless stated otherwise.

\section{Cluster sample}\label{sec:sample}

\begin{table*}
 \centering
 \caption{The galaxy clusters analyzed in this work (\textit{top}) and the ones that have been excluded as the presence of a shock/cold front (or both) has been already claimed (\textit{bottom}). Reported values of \mfive\ and $K_0$ are taken from \citet{planck14xxix} and \citet{cavagnolo09}, respectively.}
 \label{tab:sample}
  \begin{tabular}{lccccccc} 
  \hline
  Cluster name & RA$_{\rm{J}2000}$ & DEC$_{\rm{J}2000}$ & \mfive\ & $z$ & $K_0$  & Shock & Cold front\\
  & (h,m,s) & (\deg,\arcmin,\arcsec) & ($10^{14}$ M$_\odot$) & & (keV cm$^2$) & (ref.) & (ref.) \\
  \hline
  A2813 & 00 43 24 & $-$20 37 17 & 9.16 & 0.292 & $268\pm44$ & $\ldots$ & $\ldots$ \\
  A370 & 02 39 50 & $-$01 35 08 & 7.63 & 0.375 & $322\pm91$ & \multicolumn{2}{c}{$^{\tilde{1}}$} \\
  A399 & 02 57 56 & +13 00 59 & 5.29 & 0.072 & $153\pm19$ & $\ldots$ & $^1$ \\
  A401 & 02 58 57 & +13 34 46 & 6.84 & 0.074 & $167\pm8$ & $\ldots$ & $^1$ \\
  MACS J0417.5-1154 & 04 17 35 & $-$11 54 34 & 11.7 & 0.440 & $27\pm7$ & $\ldots$ & $^1$ \\
  RXC J0528.9-3927 & 05 28 53 & $-$39 28 18 & 7.31 & 0.284  & $73\pm14$ & $\ldots$ & $^1$ \\
  MACS J0553.4-3342 & 05 53 27 & $-$33 42 53 & 9.39 & 0.407  & $\ldots$ & $^1$ & $^1$ \\
  AS592 & 06 38 46 & $-$53 58 45 & 6.71 & 0.222 & $59\pm14$ & $^1$ & $\ldots$ \\
  A1413 & 11 55 19 & +23 24 31 & 5.98 & 0.143 & $164\pm8$ & $\ldots$ & $\ldots$ \\ 
  A1689 & 13 11 29 & $-$01 20 17 & 8.86 & 0.183 & $78\pm8$ & $\ldots$ & $\ldots$ \\
  A1914 & 14 26 02 & +37 49 38 & 6.97 & 0.171 & $107\pm18$ & $^1$ & $^1$ \\
  A2104 & 15 40 07 & $-$03 18 29 & 5.91 & 0.153 & $161\pm42$ & $^1$ & $\ldots$ \\
  A2218 & 16 35 52 & +66 12 52 & 6.41 & 0.176 & $289\pm20$ & $^1$ & $\ldots$ \\
  Triangulum Australis & 16 38 20 & $-$64 30 59 & 7.91 & 0.051 & $\ldots$ & \multicolumn{2}{c}{$^{\tilde{1}}$} \\
  A3827 & 22 01 56 & $-$59 56 58 & 5.93 & 0.098 & $165\pm12$ & $\ldots$ & $\ldots$ \\
 \hline  
  A2744 & 00 14 19 & $-$30 23 22 & 9.56 & 0.308 & $438\pm59$ & $^2$ & $^3$ \\
  A115 & 00 55 60 & +26 22 41 & 7.20 & 0.197 & $\ldots$ & $^4$ & $\ldots$ \\
  El Gordo & 01 02 53 & $-$49 15 19 & 8.80 & 0.870 & $\ldots$ & $^5$ & $\ldots$ \\
  3C438 & 01 55 52 & +38 00 30 & 7.35 & 0.290 & $\ldots$ & $^6$ & $^6$ \\
  A520 & 04 54 19 & +02 56 49 & 7.06 & 0.199 & $325\pm29$ & $^7$ & $\ldots$ \\
  A521 & 04 54 09 & $-$10 14 19 & 6.90 & 0.253 & $260\pm36$ & $^8$ & $^8$ \\
  Toothbrush Cluster & 06 03 13 & +42 12 31 & 11.1 & 0.225 & $\ldots$ & $^{9,10}$ & $^{10}$ \\
  Bullet Cluster & 06 58 31 & $-$55 56 49 & 12.4 & 0.296 & $307\pm19$ & $^{11,12}$ & $^{11}$ \\
  MACS J0717.5+3745 & 07 17 31 & +37 45 30 & 11.2 & 0.546 & $220\pm96$ & $\ldots$ & $^{13}$ \\
  A665 & 08 30 45 & +65 52 55 & 8.23 & 0.182 & $135\pm23$ & $^{14}$ & $^{14}$ \\
  A3411 & 08 41 55 & $-$17 29 05 & 6.48 & 0.169 & $270\pm5$ & $^{15}$ & $\ldots$ \\
  A754 & 09 09 08 & $-$09 39 58 & 6.68 & 0.054 & $270\pm24$ & $^{16}$ & $^{17}$ \\
  MACS J1149.5+2223 & 11 49 35 & +22 24 11 & 8.55 & 0.544 & $281\pm39$ & $\ldots$ & $^{18}$ \\
  Coma Cluster & 12 59 49 & +27 58 50 & 5.29 & 0.023 & $\ldots$ & $^{19,20}$ & $\ldots$ \\
  A1758 & 13 32 32 & +50 30 37 & 7.99 & 0.279 & $231\pm37$ & $\ldots$ & $^{21}$ \\
  A2142 & 15 58 21 & +27 13 37 & 8.81 & 0.091 & $68\pm3$ & $\ldots$ & $^{22}$ \\
  A2219 & 16 40 21 & +46 42 21 & 11.0 & 0.226 & $412\pm43$ & $^{23}$ & $\ldots$ \\
  A2256 & 17 03 43 & +78 43 03 & 6.34 & 0.058 & $350\pm12$ & $^{24}$ & $^{25}$ \\
  A2255 & 17 12 31 & +64 05 33 & 5.18 & 0.081 & $529\pm28$ & $^{26}$ & $\ldots$ \\
  A2319 & 19 21 09 & +43 57 30 & 8.59 & 0.056 & $270\pm5$ & $\ldots$ & $^{27}$ \\
  A3667 & 20 12 30 & $-$56 49 55 & 5.77 & 0.056 & $160\pm15$ & $^{28,29}$ & $^{30}$ \\
  AC114 & 22 58 52 & $-$34 46 55 & 7.78 & 0.312 & $200\pm28$ & $\ldots$ & $^{31}$ \\
  \hline
  \multicolumn{8}{{p{.8\textwidth}}}{\textit{Notes.} References: $^1$ this work (if a tilde is superimposed the edge nature is uncertain); $^2$ \citet{eckert16}; $^3$ \citet{owers11}; $^4$ \citet{botteon16a115}; $^5$ \citet{botteon16gordo}; $^6$ \citet{emery17}; $^7$ \citet{markevitch05}; $^8$ \citet{bourdin13}; $^9$ \citet{ogrean13toothbrush}; $^{10}$ \citet{vanweeren16toothbrush}; $^{11}$ \citet{markevitch02bullet}; $^{12}$ \citet{shimwell15}; $^{13}$ \citet{vanweeren17macs0717}; $^{14}$ \citet{dasadia16a665}; $^{15}$ \citet{vanweeren17a3411}; $^{16}$ \citet{macario11}; $^{17}$ \citet{ghizzardi10}; $^{18}$ \citet{ogrean16} $^{19}$ \citet{akamatsu13coma}; $^{20}$ \citet{ogrean13coma}; $^{21}$ \citet{david04}; $^{22}$ \citet{markevitch00}; $^{23}$ \citet{canning17}; $^{24}$ \citet{trasatti15}; $^{25}$ \citet{sun02}; $^{26}$ \citet{akamatsu17a2255}; $^{27}$ \citet{ohara04}; $^{28}$ \citet{finoguenov10}; $^{29}$ \citet{storm17arx}; $^{30}$ \citet{vikhlinin01cold}; $^{31}$ \citet{defilippis04}.}
  \end{tabular}
\end{table*}

\begin{table*}
 \centering
 \caption{Summary of the \chandra\ observations analyzed in this work. The net exposure time is after the flare filtering. The averaged values of $N_{\rm H_{I}}$ \citep{kalberla05} and $N_{\rm H,tot}$ \citep{willingale13} measured in the direction of the clusters are also reported; these are compared in Fig.~\ref{fig:nh-vs-nh}.}
 \label{tab:chandra_obs}
  \begin{tabular}{lccccccc} 
  \hline
  Cluster name & & Observation & Detector & Exposure & Total exposure & $N_{\rm H_{I}}$ & $N_{\rm H,tot}$ \\
       & & ID & (ACIS) & (ks) & (net ks) & $10^{20}$ cm$^{-2}$ & $10^{20}$ cm$^{-2}$ \\
  \hline
  
  \multirow{2}*{A2813} & \ldelim\{{2}{1mm} & 9409, 16278, 16366 & I, I, S & 20, 8, 37 & 
  \multirow{2}*{114} & \multirow{2}*{1.83} & \multirow{2}*{1.93} \\
                      & & 16491, 16513 & S, S & 37, 30 \\
                       
  A370 & & 515$^\dagger$, 7715 & S$^\ast$, I & 90, 7 & 64 & 3.01 & 3.32 \\
  
  A399 & & 3230 & I & 50 & 42 & 10.6 & 17.1 \\
  
  \multirow{2}*{A401} & \ldelim\{{2}{1mm} & 518$^\dagger$, 2309, 10416, 10417 & I$^\ast$, I$^\ast$, I, I & 18, 12, 5, 5 & \multirow{2}*{176} & \multirow{2}*{9.88} & \multirow{2}*{15.2} \\ 
                      & & 10418, 10419, 14024 & I, I, I & 5, 5, 140 \\
                      
  MACS J0417.5-1154 & & 3270, 11759, 12010 & I, I, I & 12, 54, 26 & 87 & 3.31 & 3.87 \\

  RXC J0528.9-3927 & & 4994, 15177, 15658 & I, I, I & 25, 17, 73 & 96 & 2.12 & 2.26 \\
  
  MACS J0553.4-3342 & & 5813, 12244 & I, I & 10, 75 & 77 & 3.32 & 3.79 \\
  
  \multirow{2}*{AS592} & \ldelim\{{2}{1mm} & 9420, 15176 & I, I & 20, 20 & \multirow{2}*{98} & 
  \multirow{2}*{6.07}  & \multirow{2}*{8.30}\\
                     & & 16572, 16598 & I, I & 46, 24 \\
                       
  \multirow{2}*{A1413} & \ldelim\{{2}{1mm} & 537, 1661, 5002 & I, I, I & 10, 10, 40 & \multirow{2}*{128} & 
  \multirow{2}*{1.84} & \multirow{2}*{1.97} \\
                     & & 5003, 7696 & I, I & 75, 5 \\
  
  \multirow{2}*{A1689} & \ldelim\{{2}{1mm} & 540, 1663, 5004 & I$^\ast$, I$^\ast$, I & 10, 10, 20 & \multirow{2}*{185} & 
  \multirow{2}*{1.83} & \multirow{2}*{1.98}\\
                     & & 6930, 7289, 7701 & I, I, I & 80, 80, 5 \\
  
  A1914 & & 542$^\dagger$, 3593 & I, I & 10, 20 & 23 & 1.06 & 1.10 \\
  
  %A1995 & & 906, 7021, 7713 & S$^\ast$, I, I & 57, 50, 7 & 109 & 1.19 \\
    
  A2104 & & 895 & S$^\ast$ & 50 & 48 & 8.37 & 14.5 \\
    
  \multirow{2}*{A2218} & \ldelim\{{2}{1mm} & 553$^\dagger$, 1454$^\dagger$ & I$^\ast$, I$^\ast$ & 7, 13 & \multirow{2}*{47} & 
  \multirow{2}*{2.60}  & \multirow{2}*{2.83} \\
                     & & 1666, 7698  & I, I & 50, 5 \\
  
  Triangulum Australis & & 17481 & I & 50 & 49 & 11.5 & 17.0 \\
  
  A3827 & & 3290 & S & 50 & 45 & 2.65 & 2.96 \\  
  \hline
  \multicolumn{8}{{p{.95\textwidth}}}{\textit{Notes.} \obsid s marked with $^\dagger$ were excluded in the spectral analysis as the focal plane temperature was warmer than the standard $-119.7$ \deg C observations and there is not Charge Transfer Inefficiency correction available to apply to this data with subsequent uncertainties in the spectral analysis of these datasets. All the observations were taken in \vfaint\ mode except the ones marked by $^\ast$ that were instead taken in \faint\ mode.}
  \end{tabular}
\end{table*}

We selected a number of galaxy clusters where it is likely to detect merger-induced discontinuities searching for (i) massive systems in a dynamical disturbed state and (ii) with an adequate X-ray count statistics, based on current observations available in the \chandra\ archive. In particular the following.

\begin{enumerate}
 \item Using the \planck\ Sunyaev-Zel'dovich (SZ) catalog (PSZ1; \citealt{planck14xxix}) we selected clusters with mass\footnote{\mfive\ is the mass within the radius that encloses a mean overdensity of 500 with respect to the critical density at the cluster redshift.}, as inferred from the SZ signal, $\mfive > 5 \times 10^{14}$ \msun. Searching for diffuse radio emission connected with shocks (radio relics and edges of radio halos) is a natural follow-up of our study, hence this high mass threshold has been set mainly because non-thermal emission is more easily detectable in massive merging systems \citep[\eg][]{cassano13, degasperin14, cuciti15}. As a second step, we select only dynamically active systems excluding all the CC clusters. To do that we used the Archive of \chandra\ Cluster Entropy Profile Tables (ACCEPT; \citealt{cavagnolo09}) and the recent compilation by \citet{giacintucci17} to look for the so-called core entropy value $K_0$ \citep[see Eq.~4 in][]{cavagnolo09}, which is a good proxy to identify NCC systems \citep[\eg][]{mccarthy07}: clusters with $K_0 < 30 - 50$ \kevcmsq\ exhibit all the properties of a CC hence they were excluded in our analysis. 
 \item Detecting shocks and cold fronts requires adequate X-ray count statistics as in particular the latter discontinuities are found in cluster outskirts, where the X-ray brightness is faint. For this reason, among the systems found in the \chandra\ data archive\footnote{http://cda.harvard.edu/chaser/} satisfying (i), we excluded clusters with $\lesssim 4-5 \times 10^4$ counts in the \chandra\ broad-band $0.5-7.0$ \kev\ with the exposure available at the time of writing. We did that by converting the \rosat\ flux in the $0.1-2.4$ \kev\ band reported in the main X-ray galaxy cluster catalogs [Brightest Cluster Sample (BCS), \citealt{ebeling98}; extended Brightest Cluster Sample (eBCS), \citealt{ebeling00}; Northern ROSAT All-Sky (NORAS), \citealt{bohringer00}; \rosat-ESO Flux Limited X-ray (REFLEX), \citealt{bohringer04}; MAssive Cluster Survey (MACS), \citealt{ebeling07, ebeling10}] into a \chandra\ count rate using the PIMMS software\footnote{http://heasarc.gsfc.nasa.gov/Tools/w3pimms.html} and assuming a thermal emission model. Clusters without a reported \rosat\ flux in the catalogs were individually checked by measuring the counts in circle enclosing the cluster SB profile when it is below the level of the background and thus rejected adopting the same count threshold.
\end{enumerate}

We ended up with 37 massive and NCC cluster candidates for our study (Tab.~\ref{tab:sample}). In 22 of these systems (bottom of Tab.~\ref{tab:sample}) shocks/cold fronts (or both) have been already discovered  and consequently we focused on the analysis of the remaining 15 clusters (top of Tab.~\ref{tab:sample}). We anticipate that the results on the detection of shocks and cold fronts in these clusters are summarized in Section~\ref{sec:summary}.

\section{Methods and data analysis}\label{sec:analysis}

To firmly claim the presence of a shock or a cold front in the ICM, both imaging and spectral analysis are required. Our aim is to search for SB and temperature discontinuities in the most objective way as possible, without being too much biased by prior constraints due to guesses of the merger geometry or presence of features at other wavelengths (\eg\ a radio relic). To do so did the following.

\begin{enumerate}
 \item Applied an edge detection filter to pinpoint possible edges in the clusters that were also searched visually in the X-ray images for a comparison. 
 \item Selected the most clear features three times above the root mean square noise level of the filtered images following a coherent arc-shaped structure extending for $>100$ kpc in length.
 \item Investigated deeper the pre-selected edges with the extraction and fitting of SB profiles.
 \item Performed the spectral analysis in dedicated spectral regions to confirm the nature of the jumps.
\end{enumerate}

\noindent
In addition, we produced maps of the ICM thermodynamical quantities to help in the interpretation of the features found with the above-mentioned procedure. \\
\indent
In the following sections we describe into details the X-ray data analysis performed in this work.

\subsection{X-ray data preparation}

In Tab.~\ref{tab:chandra_obs} we report all the \chandra\ Advanced CCD Imaging Spectrometer I-array (\acisi) and Advanced CCD Imaging Spectrometer S-array (\aciss) observations of our cluster sample. Data were reprocessed with \ciao\ v4.9 and \chandra\ \caldb\ v4.7.3 starting from \texttt{level=1} event file. Observation periods affected by soft proton flares were excluded using the \texttt{deflare} task after the inspection of the light curves extracted in the $0.5-7.0$ \kev\ band. For \acisi, these where extracted from the front-illuminated S2 chip that was kept on during the observation or in one front-illuminated \acisi\ chip, avoiding the cluster diffuse emission, if S2 was turned off. In \aciss\ observations the target is imaged in the back-illuminated S3 chip hence light curves were extracted in S1, also back-illuminated\footnote{In the \aciss\ \obsid\ 515 the light curve was extracted in S2 as S1 was turned off.}. \\
Cluster images were created in the $0.5-2.0$ \kev\ band and combined with the corresponding monochromatic exposure maps (given the restricted energy range) in order to produce exposure-corrected images binned to have a pixel size of 0.984 arcsec. The datasets of clusters observed multiple times (11 out of 15) were merged with \texttt{merge\_obs} before this step. \\
The \texttt{mkpsfmap} script was used to create and match point spread function (PSF) map at $1.5$ \kev\ with the corresponding exposure map for every \obsid. For cluster with multiple \obsid s we created a single exposure-corrected PSF map with minimum size. Thus, point sources were detected with the \texttt{wavdetect} task, confirmed by eye and excised in the further analysis.

\subsection{Edge detection filter}

In practice, the visual inspection of X-ray images allows to identify the candidate discontinuities \citep{markevitch07rev}. We complement this approach with the visual inspection of filtered images. \citet{sanders16ggm} presented a Gaussian gradient magnitude (GGM) filter that aims to highlight the SB gradients in an image, similarly to the Sobel filter (but assuming Gaussian derivatives); in fact, it has been shown that these GGM images are particularly useful to identify candidate sharp edges, such as shocks and cold fronts \citep[\eg][]{walker16}. The choice of the Gaussian width $\sigma$ in which the gradient is computed depends on the physical scale of interest, magnitude of the jump and data quality: edges become more visible with increasing jump size and count rate; this requires images filtered on multiple scales to best identify candidate discontinuities \citep[\eg][]{sanders16ggm, sanders16centaurus}. In this respect, we applied the GGM filter adopting $\sigma = 1, 2, 4$, and $8$ pixels (a pixel corresponds to 0.984 arcsec) to the exposure-corrected images of the clusters in our sample. We noticed that the use of small length filters (1 and 2 pixels) is generally ineffective in detecting discontinuities in cluster outskirts due to the low counts in these peripheral regions \citep[see also][]{sanders16ggm}. Instead Gaussian widths of $\sigma = 4$ and $8$ pixels better highlight the SB gradients without saturating too much the ICM emission (as it would result with the application of filters with scales $\sigma = 16$ and $32$ pixels). For this reason, here we will report GGM filtered images with these two scales.

\subsection{Surface brightness profiles}

After looking at X-ray and GGM images, we extracted and fitted SB profiles of the candidate discontinuities on the $0.5-2.0$ \kev\ exposure-corrected images of the clusters using \proffit\ v1.4 \citep{eckert11}. A background image was produced by matching (with \texttt{reproject\_event}) the background templates to the corresponding event files for every \obsid. This was normalized by counts in the $9.5-12.0$ \kev\ band and subtracted in the SB analysis. Corrections were typically within 10\% except for the S3 chip in \faint\ mode (\obsid s 515 and 895) where the correction was $\sim45\%$. For clusters observed multiple times, all the \obsid s were used in the fits. In the profiles, data were grouped to reach a minimum signal-to-noise ratio threshold per bin of 7.

\subsection{Spectra}

\begin{figure}
 \centering
 \includegraphics[width=\hsize]{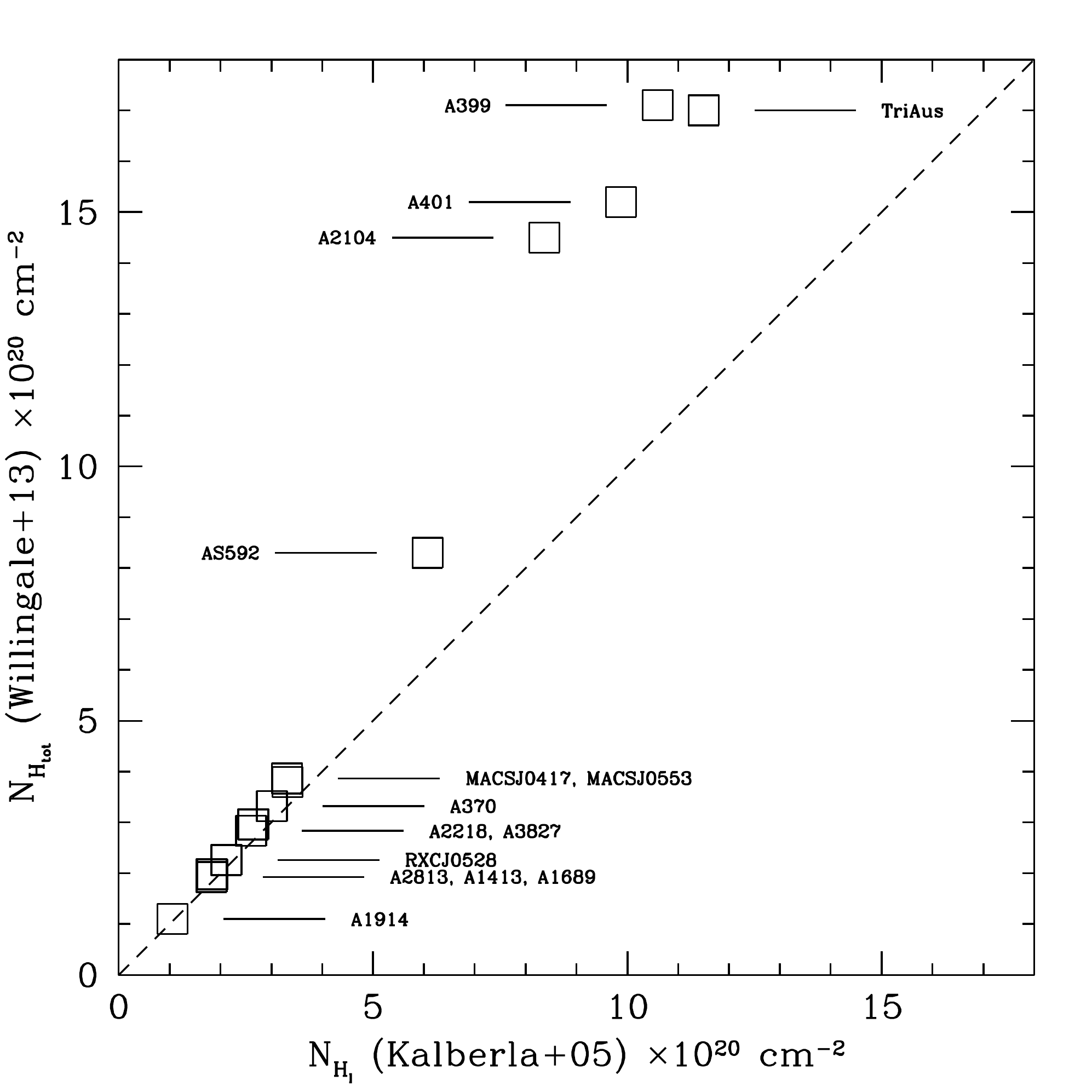}
 \caption{Comparison between the H$_{\rm I}$ density column from \citet{kalberla05} and the total (H$_{\rm I}$+H$_{\rm 2}$) density column from \citet{willingale13}. The dashed line indicates the linear correlation as a reference.}
 \label{fig:nh-vs-nh}
\end{figure}

The scientific scope of our work requires a careful treatment of the background of X-ray spectra, as in particular shock fronts are typically observed in the cluster outskirts, where the source counts are low. We modeled the background by extracting spectra in source free regions at the edge of the field-of-view. This was not possible for \acisi\ observations of nearby objects and for clusters observed with \aciss\ as the ICM emission covers all the chip area. In this respect we used observations within 3\deg\ to the target pointing (\ie\ \obsid\ 15068 for A399 and A401, \obsid\ 3142 for A2104, \obsid\ 2365 for Triangulum Australis and \obsid\ 17881 for A3827) to model the components due to the cosmic X-ray background (CXB) and to the Galactic local foreground. The former is due to the superposition of the unresolved emission from distant point sources and can be modeled as a power-law with photon index $\Gamma_{\rm cxb} = 1.42$ \citep[\eg][]{lumb02}. The latter can be decomposed in two-temperature thermal emission components \citep{kuntz00} due to the Galactic Halo (GH) emission and the Local Hot Bubble (LHB), with temperature $kT_{\rm gh} = 0.25$ \kev\ and $kT_{\rm lhb} = 0.14$ \kev\ and solar metallicity. Galactic absorption for GH and CXB was taken into account using the averaged values measured in the direction of the clusters from the Leiden/Argentine/Bonn (LAB) Survey of Galactic H$_{\rm I}$ \citep{kalberla05}. However, it has to be noticed that the total hydrogen column density is formally $N_{\rm H,tot} = N_{\rm H_{I}} + 2N_{\rm H_{2}}$, where $N_{\rm H_{2}}$ accounts for molecular hydrogen whose contribution seems to be neglected for low-density columns. In Tab.~\ref{tab:chandra_obs} we reported the values of $N_{\rm H_{I}}$ \citep{kalberla05} and $N_{\rm H,tot}$ \citep{willingale13} in the direction of the clusters in our sample, while in Fig.~\ref{fig:nh-vs-nh} we compared them. In Appendix~\ref{app:absorption} we discuss the five clusters (A399, A401, AS592, A2104 and Triangulum Australis) that do not lay on the linear correlation of Fig.~\ref{fig:nh-vs-nh}. \\
\indent
Additionally to the astrophysical CXB, GH and LHB emission, an instrumental non-X-ray background (NXB) component due to the interaction of high-energy particles with the satellite and its electronics was considered. Overall, the background model we used can be summarized as 

\begin{equation}\label{eq:bgk_x}
 apec_{\rm lhb} + phabs * (apec_{\rm gh} + powerlaw_{\rm cxb}) + bkg_{\rm nxb}
\end{equation}

\noindent
where the $bkg_{\rm nxb}$ was modeled with

\begin{equation}\label{eq:bkg_model}
 \begin{array}{ll}
 expdec + power + \sum gaussian, & \mbox{for\quad \acisi}  \\
\\
 expdec + bknpower + \sum gaussian, & \mbox{for \quad \aciss}  \\
 \end{array}
\end{equation}

\noindent
where a number of Gaussian fluorescence emission lines were superimposed on to the continua. For more details on the NXB modeling the reader is referred to Appendix~\ref{app:nxb}. \\
\indent
The ICM emission was described with a thermal model taking into account the Galactic absorption in the direction of the clusters (\cf\ Tab.~\ref{tab:chandra_obs} and Appendix~\ref{app:absorption}) 

\begin{equation}\label{eq:source_icm}
 phabs*apec_{\rm icm}\:,
\end{equation}

\noindent
the metallicity of the ICM was set to $0.3$ \zsun\ \citep[\eg][]{werner13}. \\
\indent
Spectra were simultaneously fitted (using all the \obsid s available for each cluster, unless stated otherwise) in the $0.5-11.0$ \kev\ energy band for \acisi\ and in the $0.7-10.0$ \kev\ band for \aciss, using the package \xspec\ v12.9.0o with \citet{anders89} abundances table. Since the counts in cluster outskirts are poor, Cash statistics \citep{cash79} was adopted.

\subsubsection{Contour binning maps}

We used \contbin\ v1.4 \citep{sanders06contbin} to produce projected maps of temperature, pressure and entropy for all the clusters of our sample. The clusters were divided in regions varying the geometric constraint value \citep[see][for details]{sanders06contbin} according to the morphology of each individual object to better follow the SB contour of the ICM. We required $\sim 2500$ background-subtracted counts per bin in the $0.5-2.0$ \kev\ band. Spectra were extracted and fitted as described in the previous section. \\
\indent
While the temperature is a direct result of the spectral fitting, pressure and entropy require the passage through the normalization value of the thermal model, \ie

\begin{equation}\label{eq:norm-xspec}
 \mathcal{N} = \frac{10^{-14}}{4\pi[D_{A}(1+z)]^2} \int n_e n_H\:dV\quad({\rm cm^{-5}})
\end{equation}

\noindent
where $D_A$ is the angular size distance to the source (cm) whereas $n_e$ and $n_H$ are the electron and hydrogen density (cm$^{-3}$), respectively. The projected emission measure is

\begin{equation}\label{eq:pseudo-em}
 \textrm{EM} = \mathcal{N}/A\quad({\rm cm^{-5}\,arcsec^{-2}})
\end{equation}

\noindent
with $A$ the area of each bin, and it is proportional to the square of the electron density integrated along the line of sight. Using Eq.~\ref{eq:pseudo-em} we can compute the pseudo-pressure

\begin{equation}\label{eq:pseudo-pressure}
 P=kT(\textrm{EM})^{1/2}\quad({\rm keV\,cm^{-5/2}\,arcsec^{-1}})
\end{equation}

\noindent
and pseudo-entropy 

\begin{equation}\label{eq:pseudo-entropy}
 K=kT(\textrm{EM})^{-1/3}\quad({\rm keV\,cm^{5/3}\,arcsec^{-2/3}})
\end{equation}

\noindent
values for each spectral bin. The prefix pseudo- underlines that these quantities are projected along the line of sight \citep[\eg][]{mazzotta04}.

\section{Characterization of the edges}\label{sec:search}

The inspection of the cluster X-ray and GGM filtered images provide the first indication of putative discontinuities in the ICM. These need to be characterized with standard imaging and spectral analysis techniques to be firmly claimed as edges. \\
\indent
The SB profiles of the candidate shocks and cold fronts were modeled assuming that the underlying density profile follows a broken power-law \citep[\eg][and references therein]{markevitch07rev}. In the case of spherical symmetry, the downstream and upstream (subscripts $d$ and $u$) densities differ by a factor $\compr \equiv n_d/n_u$ at the distance of the jump $r_j$

\begin{equation}\label{eq:bknpow}
 \begin{array}{ll}
 n_d (r) = \compr n_0 \left( \frac{r}{r_j} \right)^{a_1}, & \mbox{if} \quad r \leq r_j \\
\\
 n_u (r) = n_0 \left( \frac{r}{r_j} \right)^{a_2}, & \mbox{if} \quad r > r_j
 \end{array}
\end{equation}

\noindent
where $a_1$ and $a_2$ are the power-law indices, $n_0$ is a normalization factor and $r$ denotes the radius from the center of the sector. In the fitting procedure all these quantities were free to vary. We stress that the values of \compr\ reported throughout the paper have been deprojected along the line of sight under the spherical assumption by \proffit\ \citep{eckert11}. \\
\indent
A careful choice of the sector where the SB profile is extracted is needed to properly describe a sharp edge due to a shock or a cold front. In this respect, the GGM filtered images give a good starting point to delineate that region. During the analysis, we adopted different apertures, radial ranges and positions for the extracting sectors, then we used the ones that maximize the jump with the best-fitting statistics. Errors reported for \compr\ however do not account for the systematics due to the sector choice. \\
\indent 
Spectral fitting is necessary to discriminate the nature of a discontinuity as the temperature ratio $\rat \equiv T_{d}/T_{u}$ is $>1$ in the case of a shock and $<1$ in the case of a cold  front \citep[\eg][]{markevitch02bullet}. The temperature map can already provide indication about the sign of the jump. However, once that the edge position is well identified by the SB profile analysis, we can use a sector with the same aperture and center of that maximizing the SB jump to extract spectra in dedicated sectors covering the downstream and upstream regions. In this way we can carry out a self-consistent analysis and avoid possible contamination due to large spectral bins that might contain plasma at different temperature unrelated to the shock/cold front. \\
\indent
In the case of a shock, the Mach number \mach\ can be determined by using the Rankine-Hugoniot jump conditions \citep[\eg][]{landau59} for the density

\begin{equation}\label{eq:mach-from-dens}
 \compr \equiv \frac{n_d}{n_u} = \frac{4\mach_{\rm{SB}}^2}{\mach_{\rm{SB}}^2 + 3}
\end{equation}

\noindent
and temperature

\begin{equation}\label{eq:mach-from-temp}
 \rat \equiv \frac{T_d}{T_u} = \frac{5\mach_{\rm kT}^4 + 14\mach_{\rm kT}^2 -3}{16\mach_{\rm kT}^2}
\end{equation}

\noindent
here reported for the case of monatomic gas.

\section{Results}\label{sec:results}

We find 29 arc-shaped features three times above the root mean square noise level in the GGM filtered images, 22 of them were found to trace edges in the SB profiles. In Fig.~\cref{fig:a2813,fig:a370,fig:a399,fig:a401,fig:macsj0417,fig:rxcj0528,fig:macsj0553,fig:as592,fig:a1413,fig:a1689,fig:a1914,fig:a2104,fig:a2218,fig:triangulum,fig:a3827} we show a \chandra\ image in the $0.5-2.0$ \kev\ energy band, the products of the GGM filters, the maps of the ICM thermodynamical quantities and the SB profiles for each cluster of the sample. The \cstatdof\ and the temperature fractional error for each spectral region are reported in Appendix~\ref{app:errors}. The edges are highlighted in the \chandra\ images in white for shocks and in green for cold fronts. Discontinuities for whose spectral analysis does not firmly allow this distinction are reported in yellow. The temperature values obtained by fitting spectra in dedicated upstream and downstream regions are reported in shaded boxes (whose lengths cover the radial extent of the spectral region) in the panels showing the SB profiles. If the jump was detected also in temperature, the box is colored in red for the hot gas and in blue for the cold gas; conversely, if the upstream and downstream temperatures are consistent (within $1\sigma$), the box is displayed in yellow. As a general approach, in the case of weak discontinuities we also compare results with the best fit obtained with a single power-law model. \\
\indent
In the following we discuss the individual cases. In particular, in Sections~\ref{sec:detection} and \ref{sec:non-detection} we report the clusters with and without detected edges, respectively. The results of our detections are summarized in Section~\ref{sec:summary} and in Tab.~\ref{tab:results}. In Appendix~\ref{app:null} we show the seven arc-like features selected by the GGM filtered images that do not present a discontinuity in the SB profile fitting.

\subsection{Detections}\label{sec:detection}

\begin{figure*}
 \centering
 \begin{tabular}{cc}
  \multirow{2}{*}{\subfloat{\subfigimgwhitebig[width=.6\textwidth]{\quad  a)}{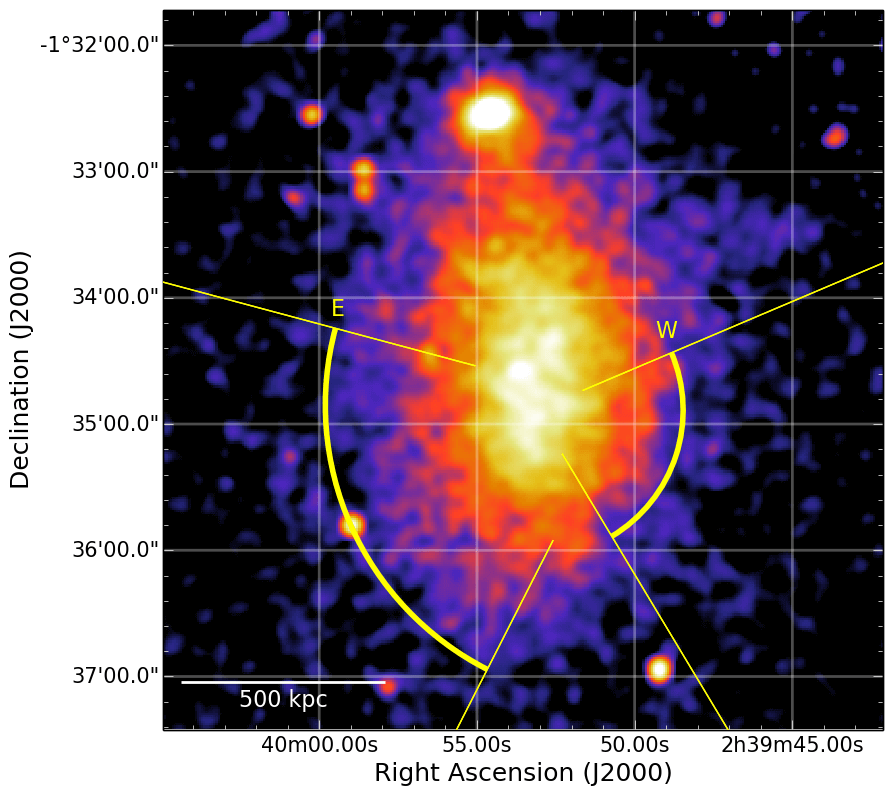}}} & \\
  & \vspace{0.15cm}\hspace{-0.3cm}\subfloat{\subfigimgwhiteggm[width=.28\textwidth]{\quad b)}{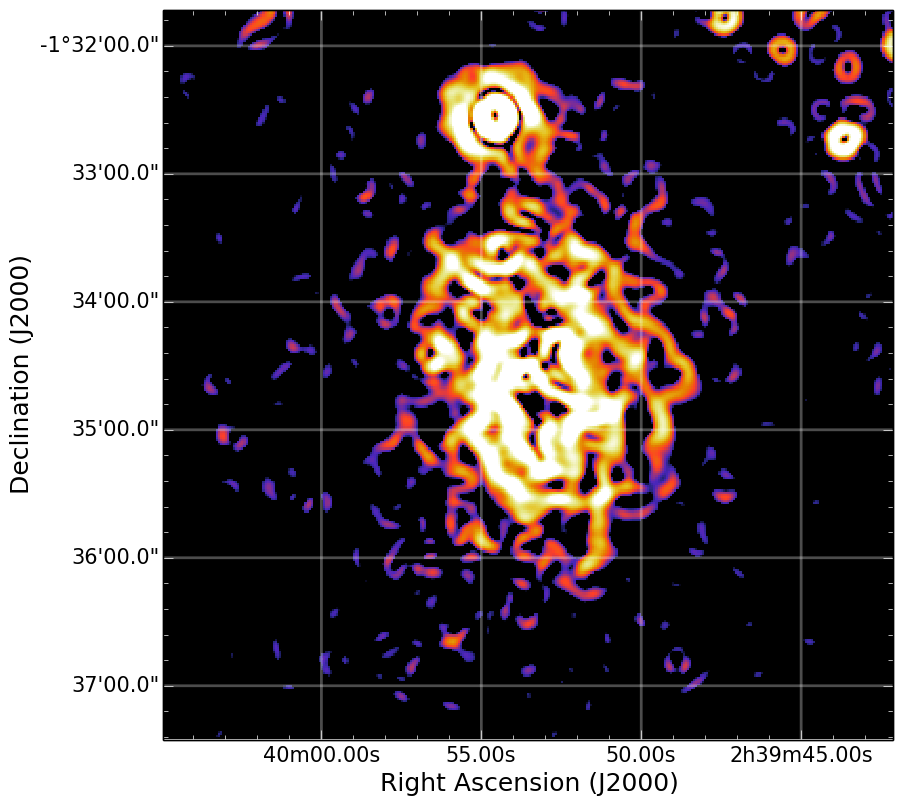}} \\
  & \hspace{-0.3cm}\subfloat{\subfigimgwhiteggm[width=.28\textwidth]{\quad c)}{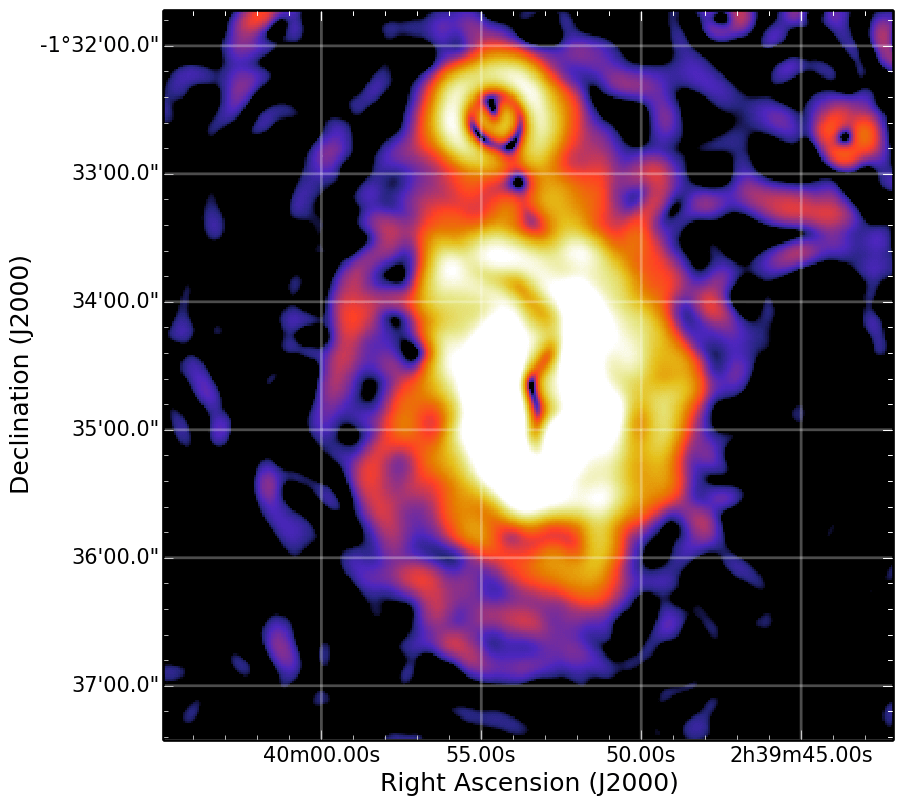}}
 \end{tabular}
 \subfloat{\subfigimgsb[width=.3\textwidth,trim={0cm 0cm 4cm 0cm},clip]{d)}{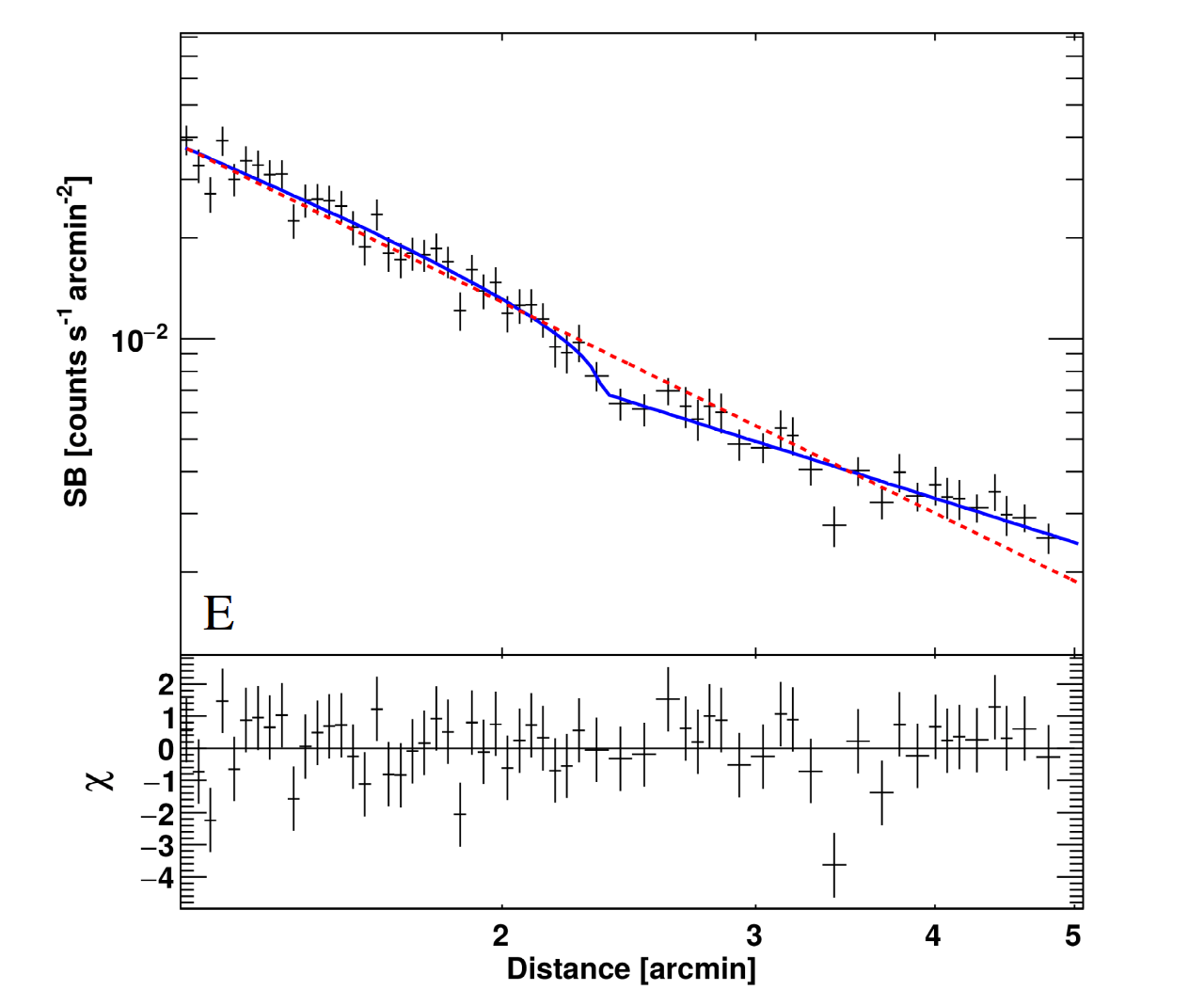}}
 \subfloat{\subfigimgsb[width=.3\textwidth,trim={0cm 0cm 4cm 0cm},clip]{e)}{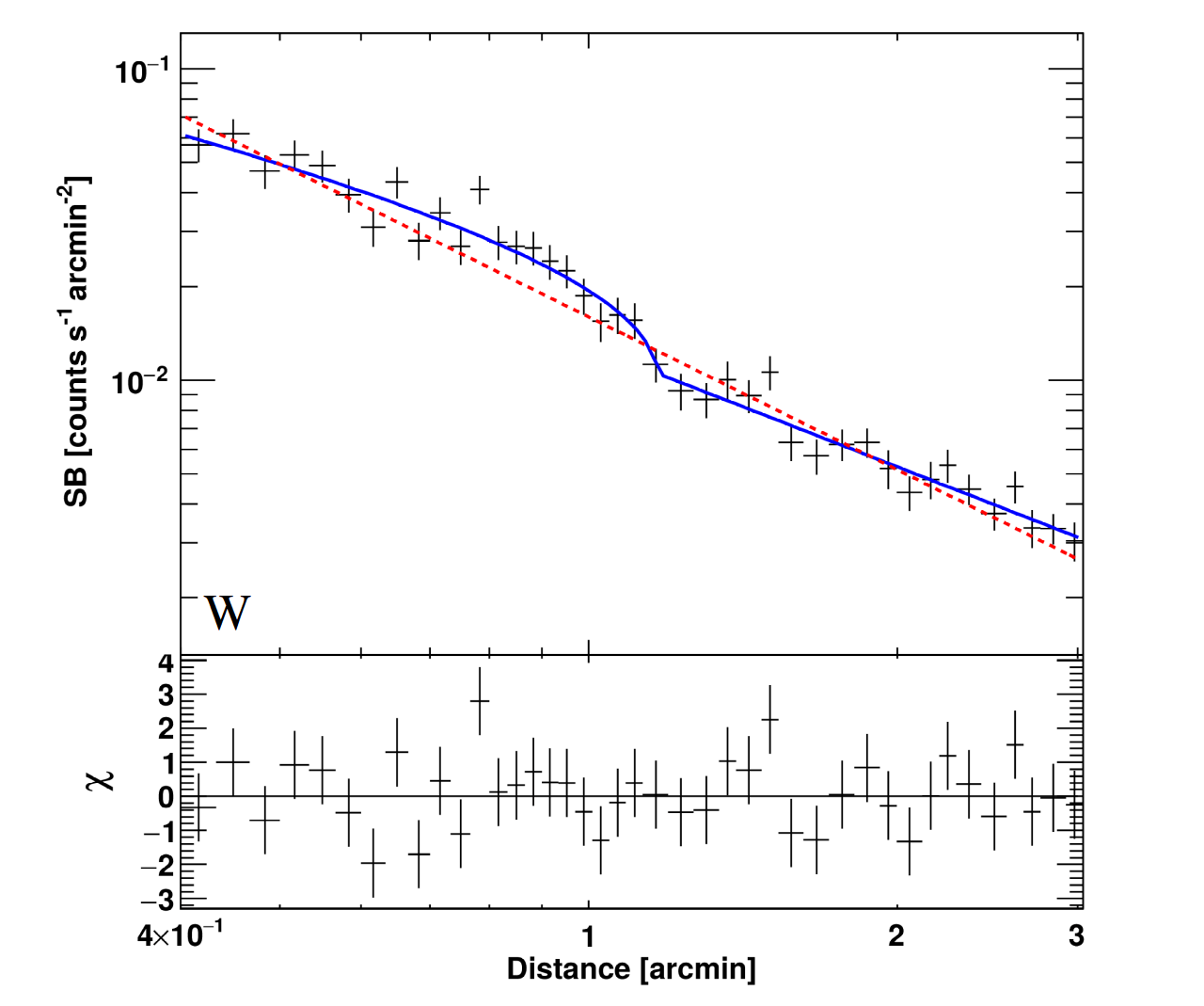}}
 \caption{A370. \chandra\ $0.5-2.0$ \kev\ image (\textit{a}), GGM filtered images on scales of 4 (\textit{b}) and 8 (\textit{c}) pixels and best-fitting broken power-law (solid blue) and single power-law (dashed red) models (residuals on the bottom are referred to the former) of the extracted SB profiles (\textit{d,e}). The sectors where the SB profiles were fitted and the positions of the relative edges are marked in the \chandra\ image in yellow.}
 \label{fig:a370}
\end{figure*}

\subparagraph{A370.} This represents the most distant object in Abell catalog \citep{abell89}, at a redshift of $z=0.375$. It is famous to be one of the first galaxy clusters where a gravitational lens was observed \citep{soucail87, kneib93}. The X-ray emission is elongated in the N-S direction (Fig~\ref{fig:a370}a); the bright source to the north is a nearby ($z=0.044$) elliptical galaxy not associated with the cluster. \\
A370 was observed two times with \chandra. The longer observation (\obsid\ 515) was performed in an early epoch after \chandra\ launch in which an accurate modeling of the ACIS background is not possible, making the spectral analysis of this dataset unfeasible (see notes in Tab.~\ref{tab:chandra_obs} for more details). The other observation of A370 (\obsid\ 7715) is instead very short. For this reason we did only a spatial analysis for this target. \\
The GGM images in Fig.~\ref{fig:a370}b,c suggest the presence of a rapid SB variation both in the W and E direction. The SB profiles taken across these directions were precisely modeled in our fits in Fig.~\ref{fig:a370}d,e, revealing jumps with similar entity ($\compr\ \sim 1.5$). The inability of performing spectral analysis in this cluster leaves their origin unknown. 
An additional SB gradient suggested by the GGM images toward the S direction did not reveal the presence of an edge with the SB profile fitting (Fig.~\ref{fig:a370_noedge}).

\begin{figure*}
 \centering
 \begin{tabular}{cc}
  \multirow{2}{*}{\subfloat{\subfigimgwhitebig[width=.6\textwidth]{\quad  a)}{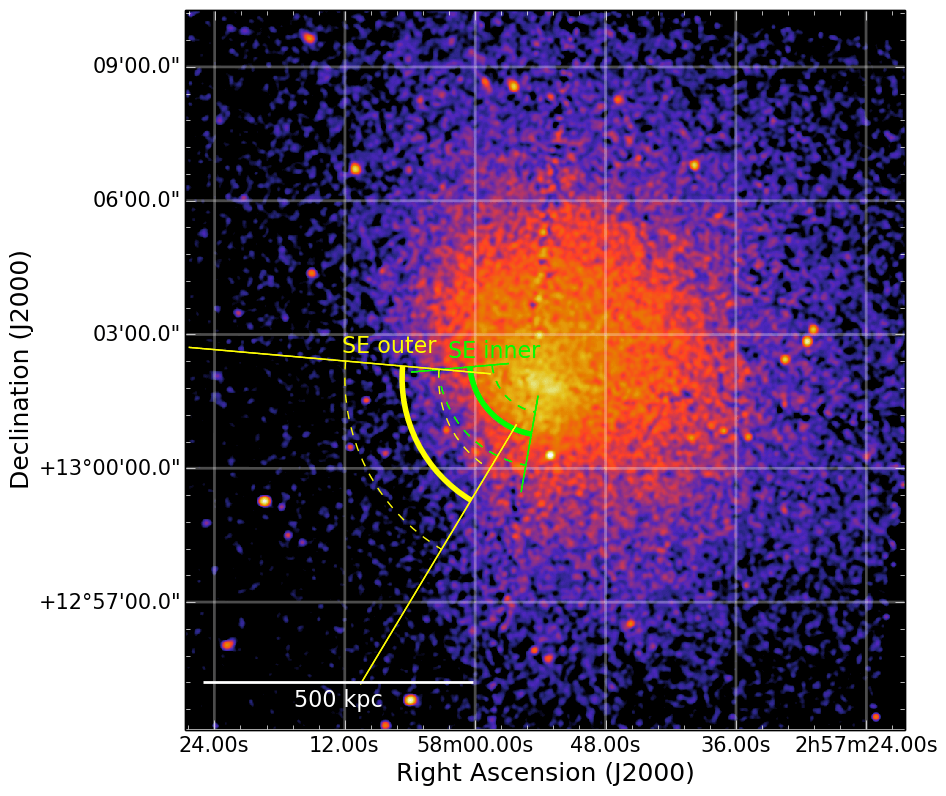}}} & \\
  & \vspace{0.15cm}\hspace{-0.3cm}\subfloat{\subfigimgwhiteggm[width=.28\textwidth]{\quad b)}{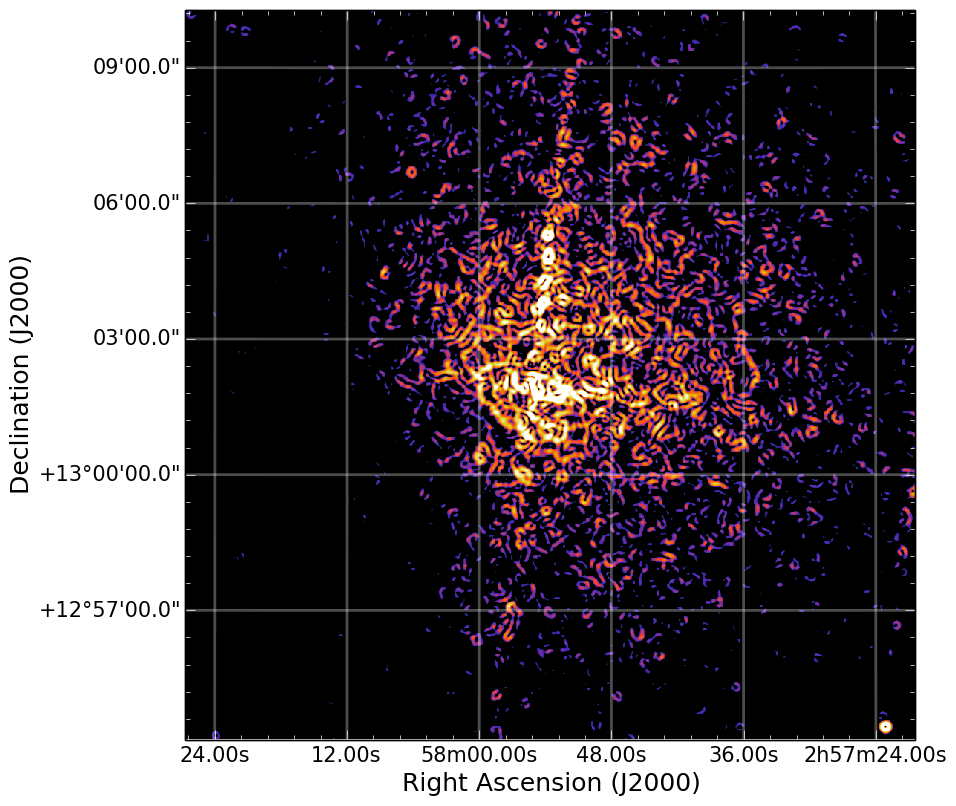}} \\
  & \hspace{-0.3cm}\subfloat{\subfigimgwhiteggm[width=.28\textwidth]{\quad c)}{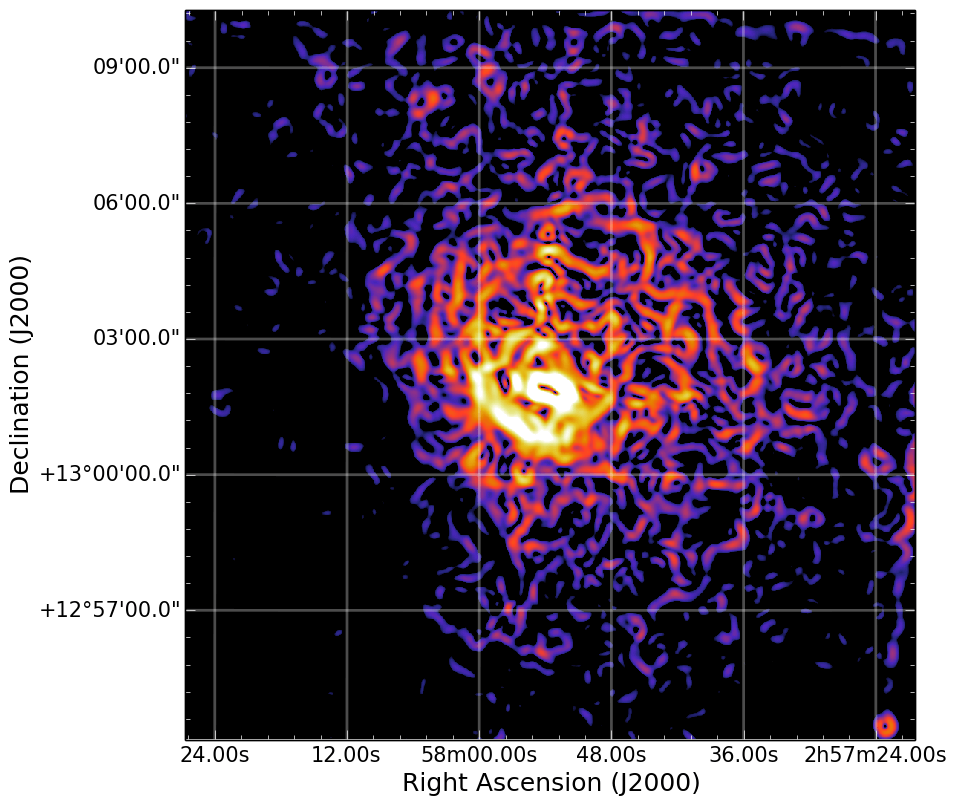}}
 \end{tabular}
 \subfloat{\subfigimgblack[width=.3\textwidth]{\enspace d)}{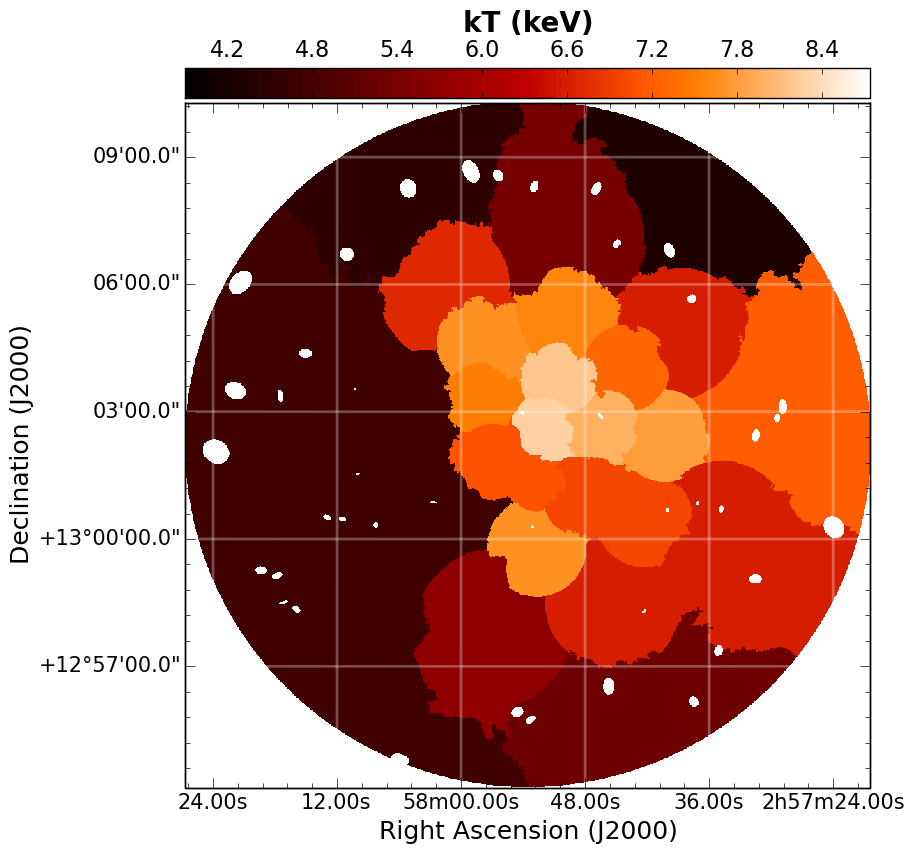}}
 \subfloat{\subfigimgblack[width=.3\textwidth]{\enspace e)}{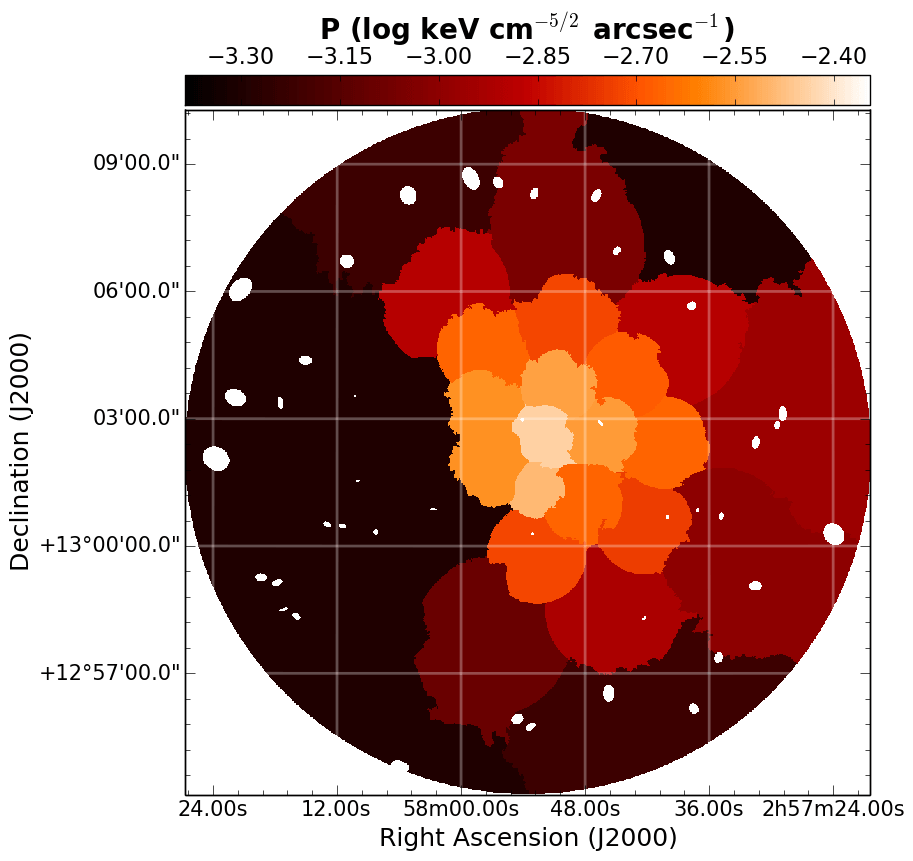}}
 \subfloat{\subfigimgblack[width=.3\textwidth]{\enspace f)}{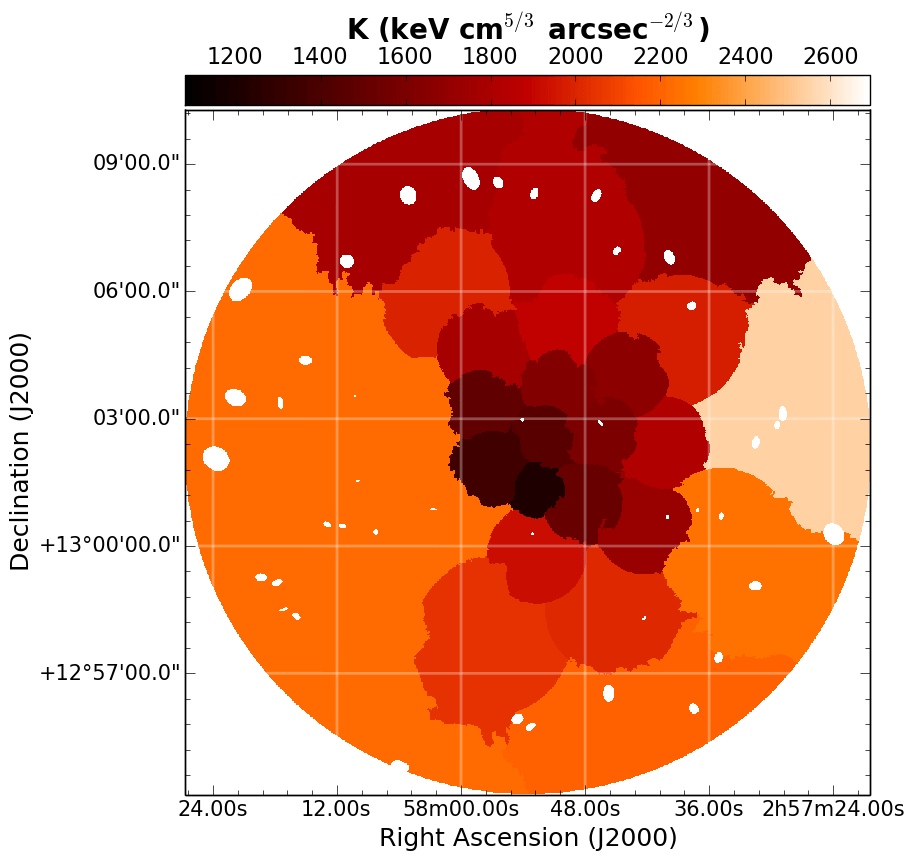}}\\
 \subfloat{\subfigimgsb[width=.3\textwidth,trim={0cm 0cm 4cm 0cm},clip]{g)}{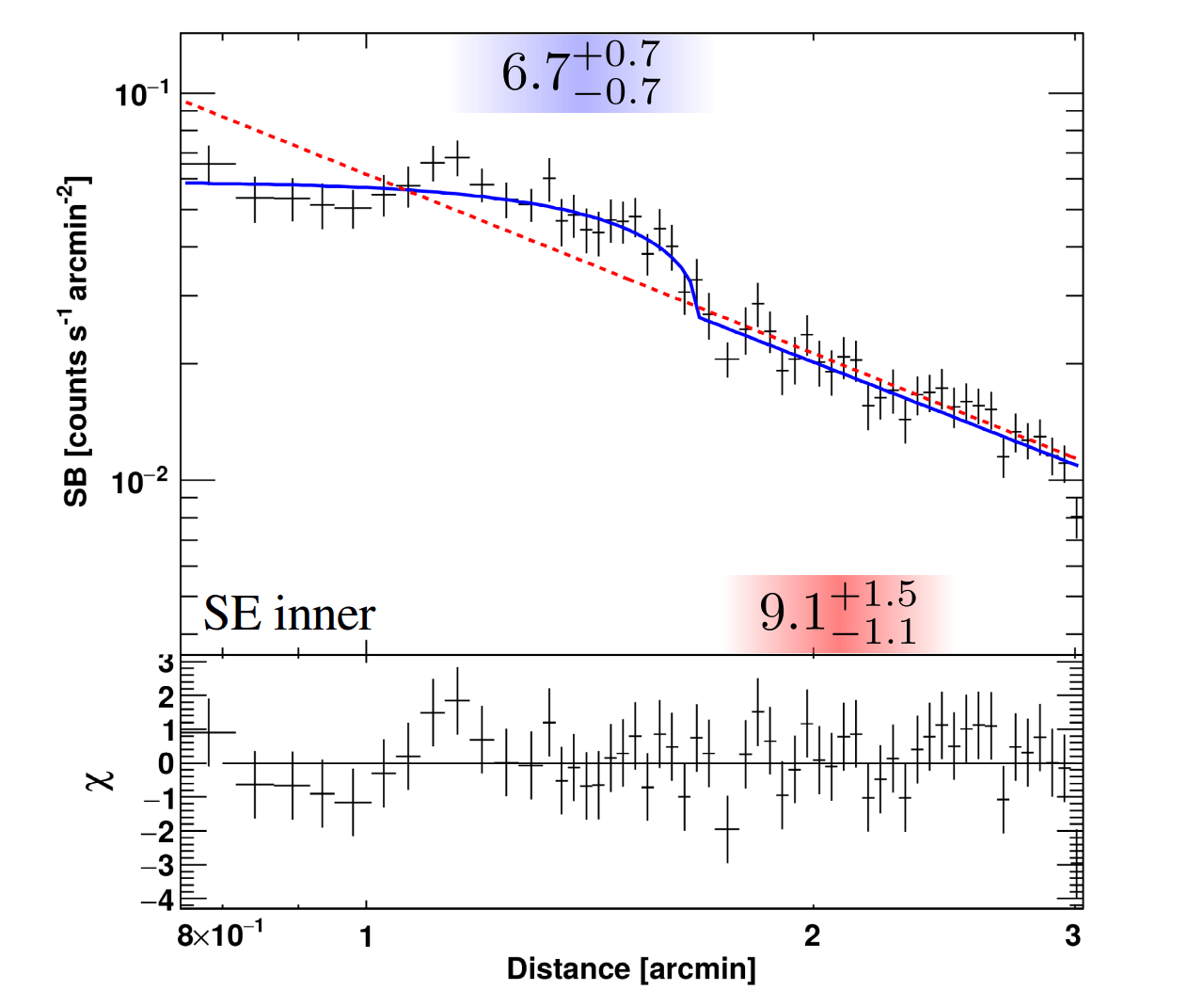}}
 \subfloat{\subfigimgsb[width=.3\textwidth,trim={0cm 0cm 4cm 0cm},clip]{h)}{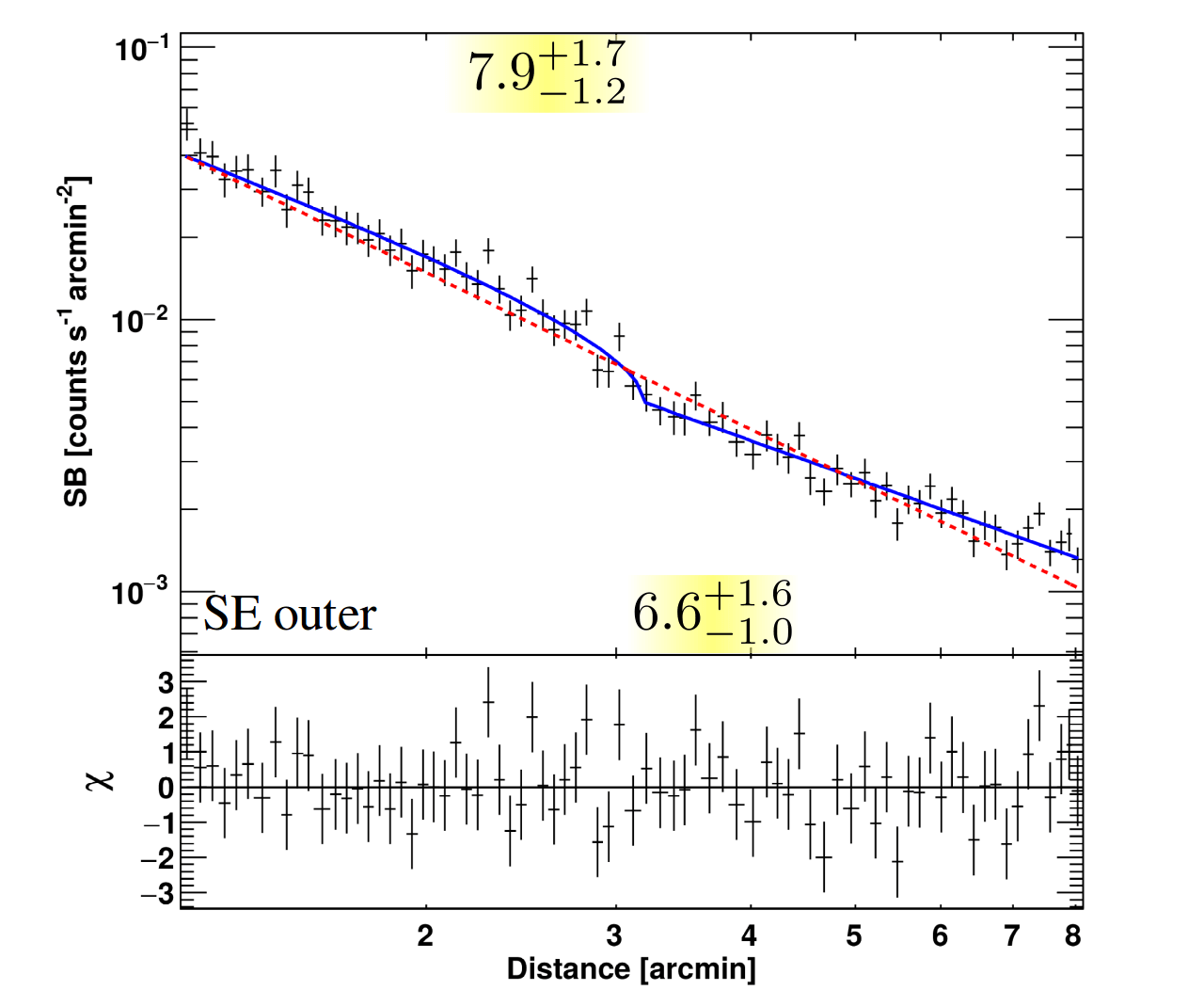}}
 \caption{A399. \chandra\ $0.5-2.0$ \kev\ image (\textit{a}), GGM filtered images on scales of 4 (\textit{b}) and 8 (\textit{c}) pixels, projected maps of temperature (\textit{d}), pressure (\textit{e}), entropy (\textit{f}) and best-fitting broken power-law (solid blue) and single power-law (dashed red) models (residuals on the bottom are referred to the former) of the extracted SB profiles (\textit{g,h}). The goodness of fits is reported in Fig.~\ref{fig:a399_errors}. The sectors where the SB profiles were fitted and the positions of the relative edges are marked in the \chandra\ image in green (cold front) and yellow. The dashed arcs show the radial limits used for measuring the temperature downstream and upstream the front, which values (in \kev) are reported in the shaded boxes in the SB profiles. Note that in the GGM filtered images the straight and perpendicular features are artifacts due to chip gaps.}
 \label{fig:a399}
\end{figure*}

\begin{figure*}
 \centering
 \begin{tabular}{cc}
  \multirow{2}{*}{\subfloat{\subfigimgwhitebig[width=.6\textwidth]{\quad  a)}{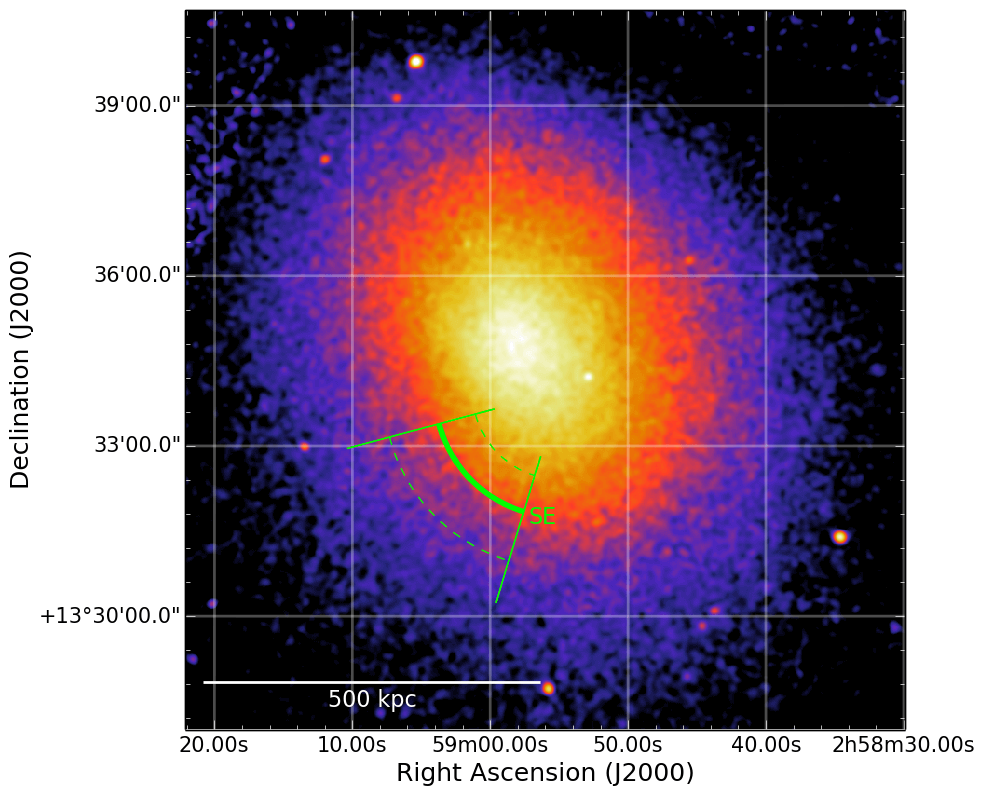}}} & \\
  & \vspace{0.15cm}\hspace{-0.3cm}\subfloat{\subfigimgwhiteggm[width=.28\textwidth]{\quad b)}{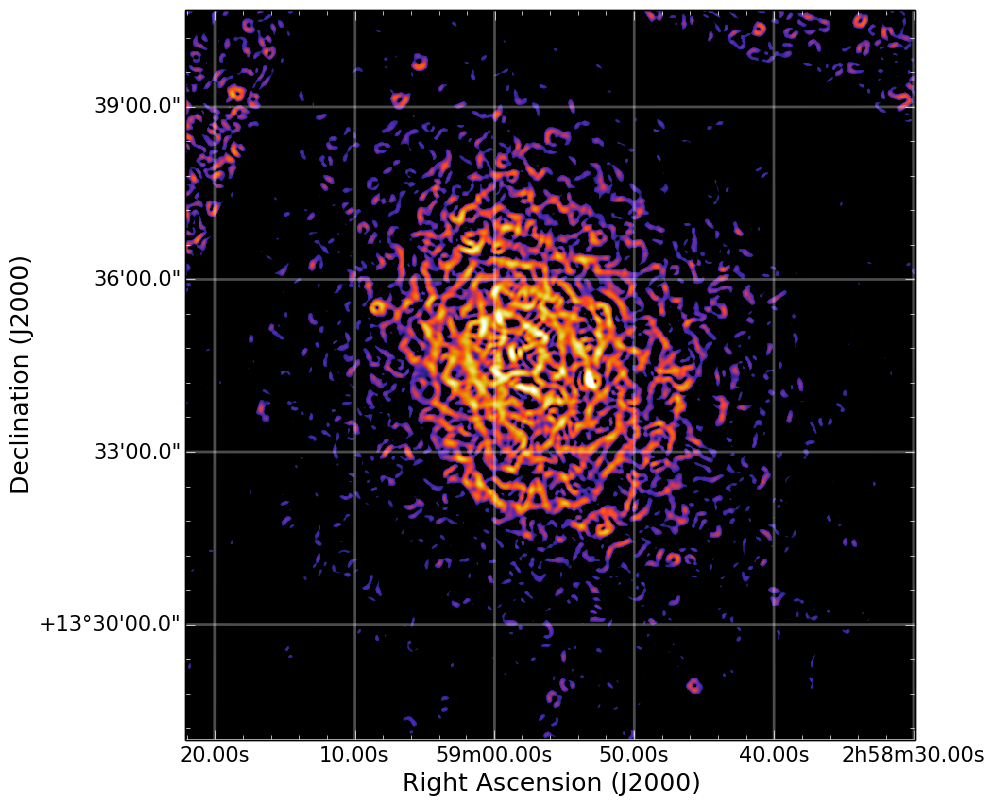}} \\
  & \hspace{-0.3cm}\subfloat{\subfigimgwhiteggm[width=.28\textwidth]{\quad c)}{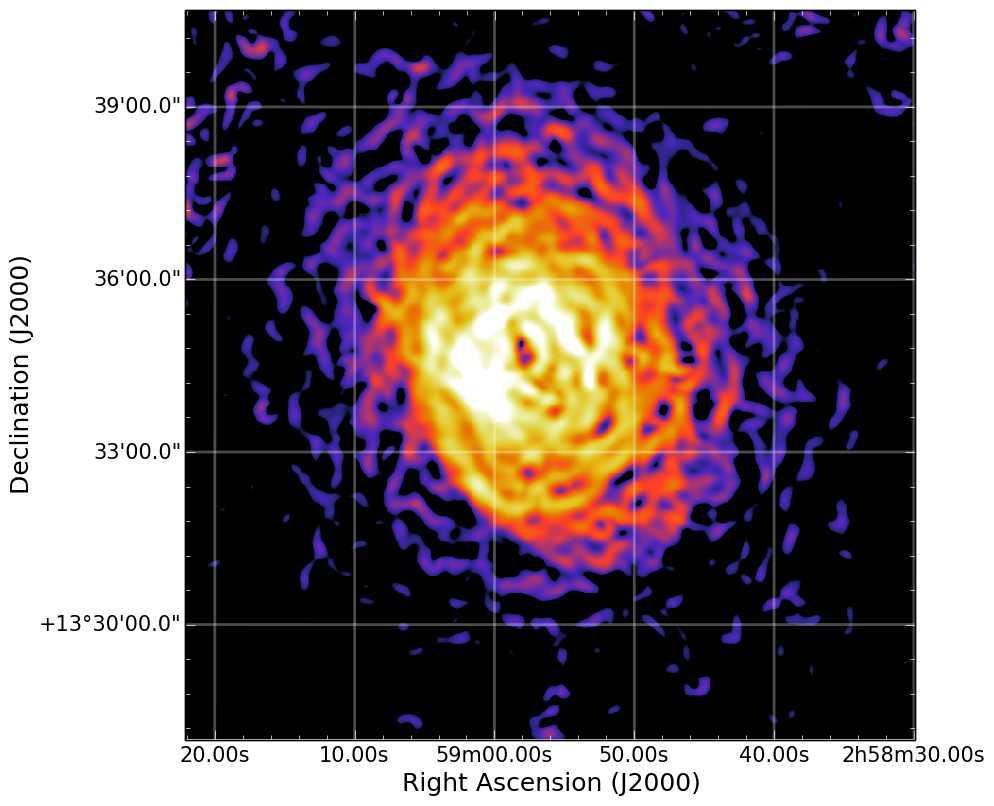}}
 \end{tabular}
 \subfloat{\subfigimgblack[width=.3\textwidth]{d)}{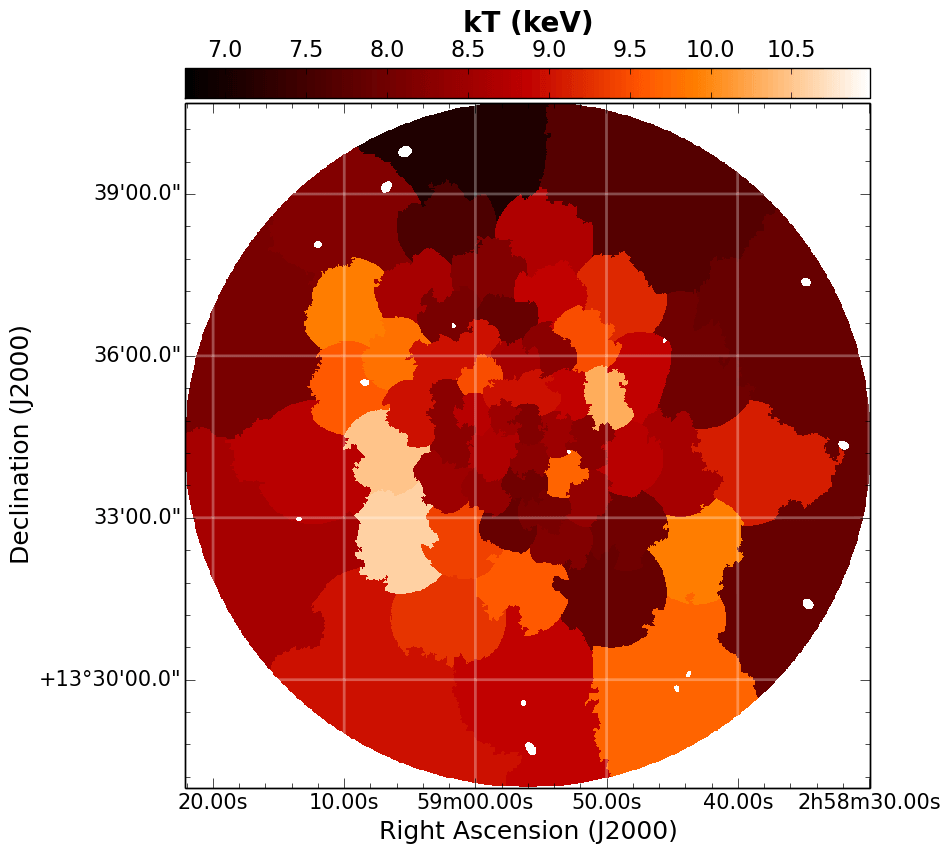}}
 \subfloat{\subfigimgblack[width=.3\textwidth]{e)}{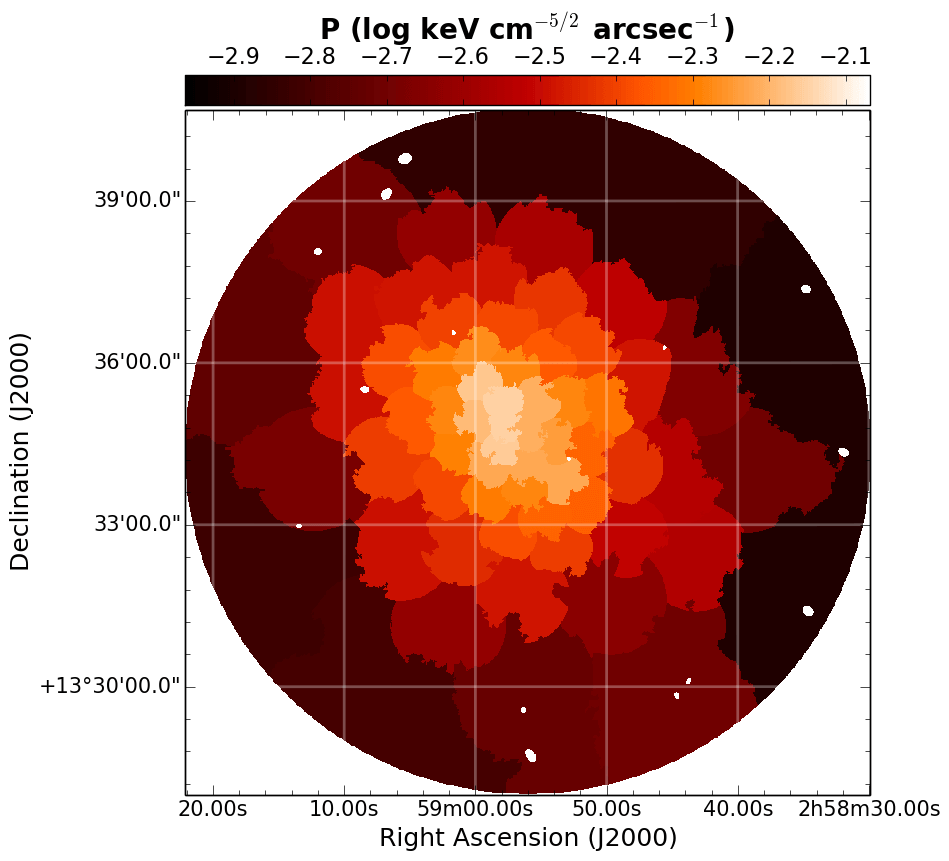}}
 \subfloat{\subfigimgblack[width=.3\textwidth]{f)}{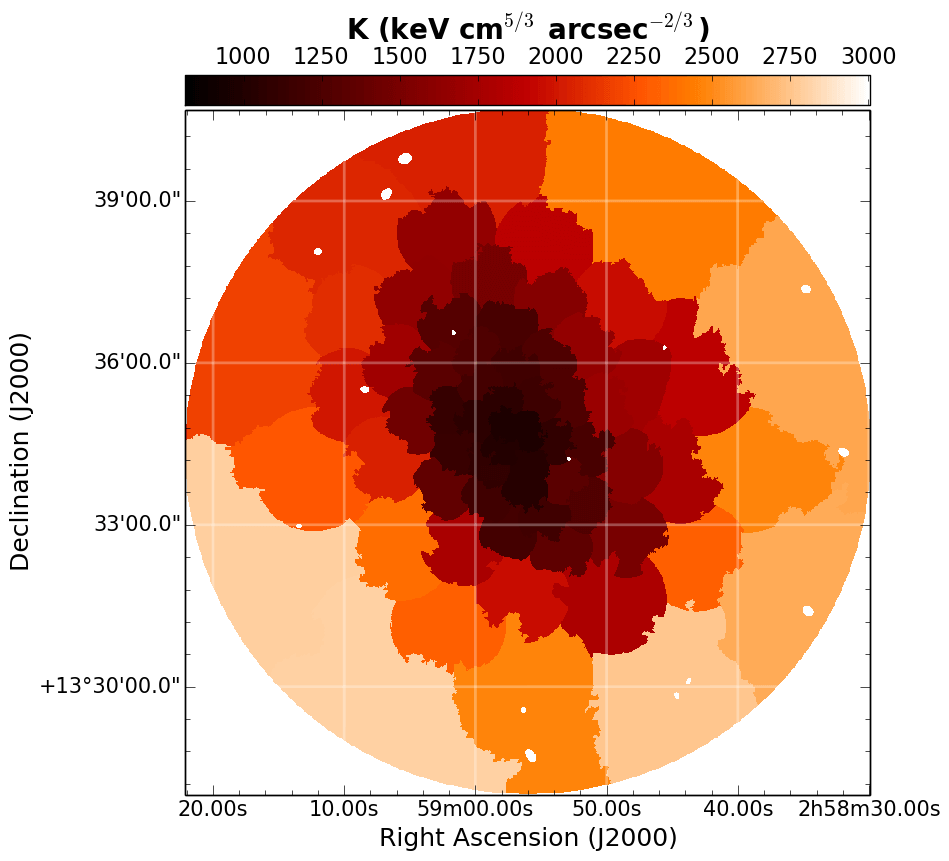}}\\
 \subfloat{\subfigimgsb[width=.3\textwidth,trim={0cm 0cm 4cm 0cm},clip]{g)}{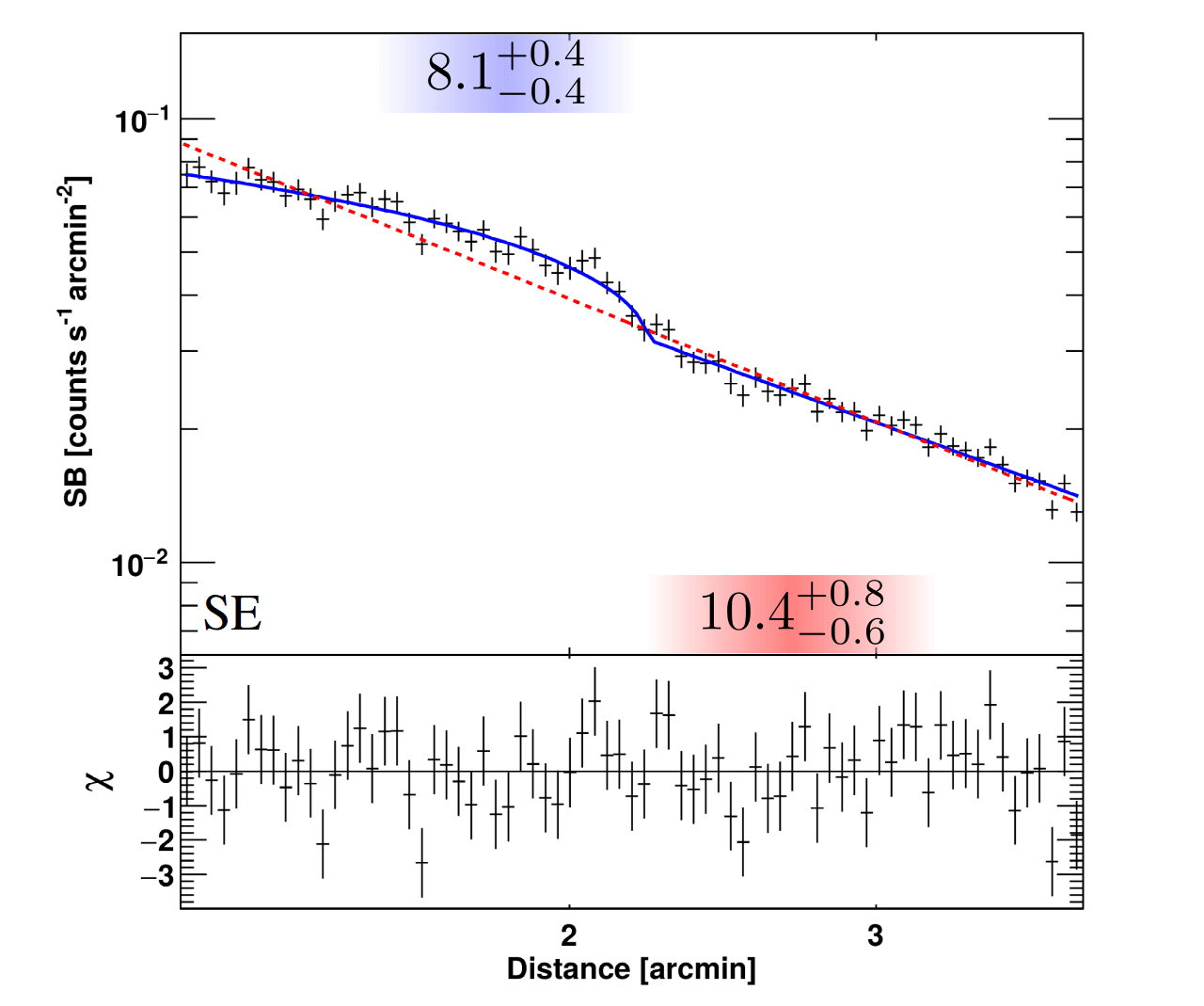}}
 \caption{A401. The same as for Fig~\ref{fig:a399}. The goodness of fits is reported in Fig.~\ref{fig:a401_errors}. The position of the edge is marked in the \chandra\ image in green (cold front).}
 \label{fig:a401}
\end{figure*}

\begin{figure*}
 \centering
 \begin{tabular}{cc}
  \multirow{2}{*}{\subfloat{\subfigimgwhitebig[width=.6\textwidth]{\quad  a)}{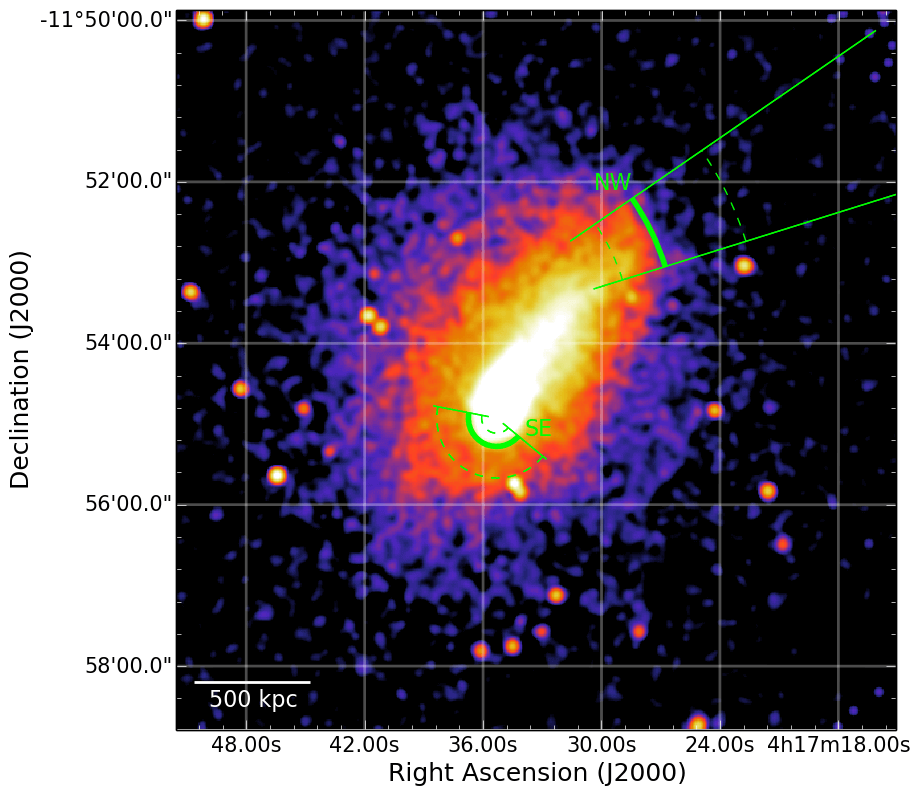}}} & \\
  & \vspace{0.15cm}\hspace{-0.3cm}\subfloat{\subfigimgwhiteggm[width=.28\textwidth]{\quad b)}{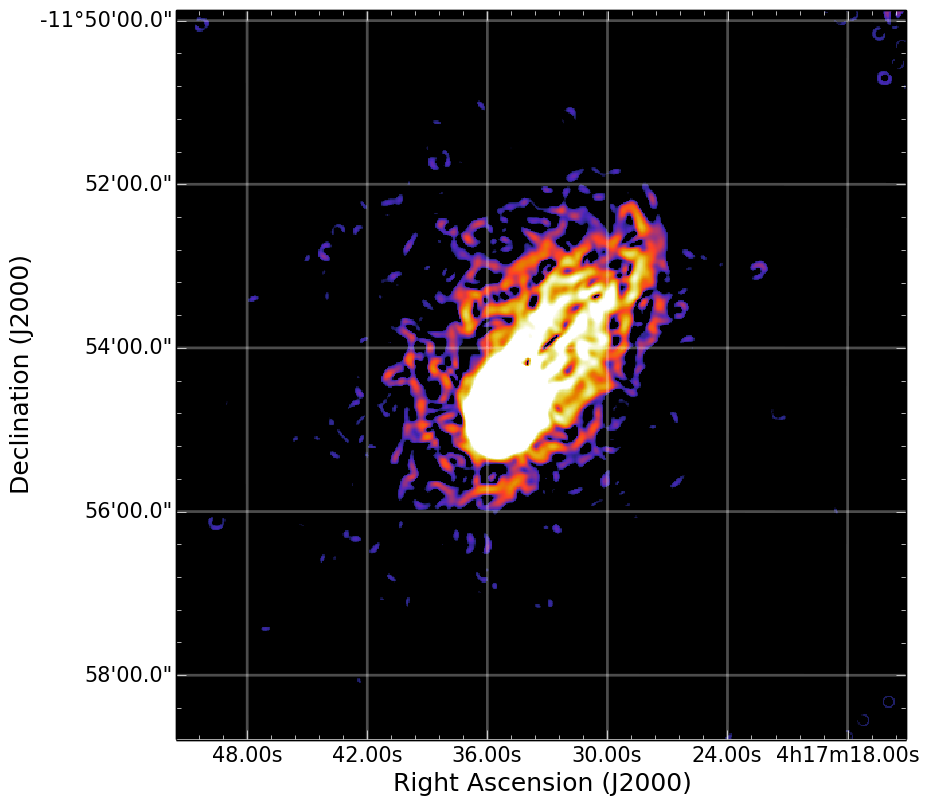}} \\
  & \hspace{-0.3cm}\subfloat{\subfigimgwhiteggm[width=.28\textwidth]{\quad c)}{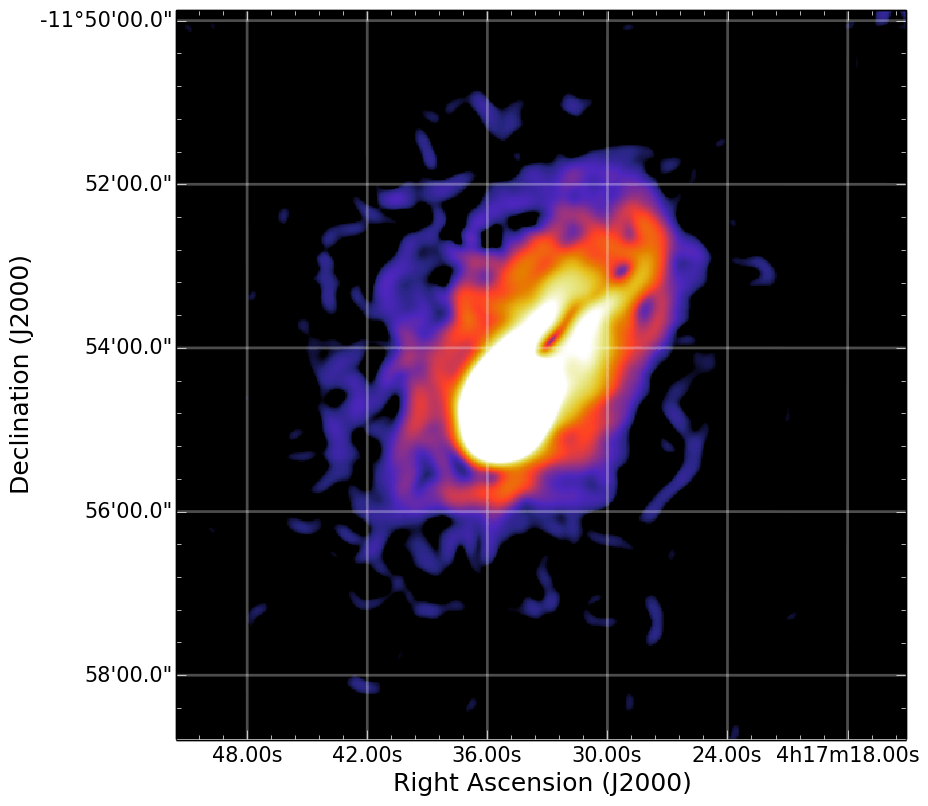}}
 \end{tabular}
 \subfloat{\subfigimgblack[width=.3\textwidth]{\quad d)}{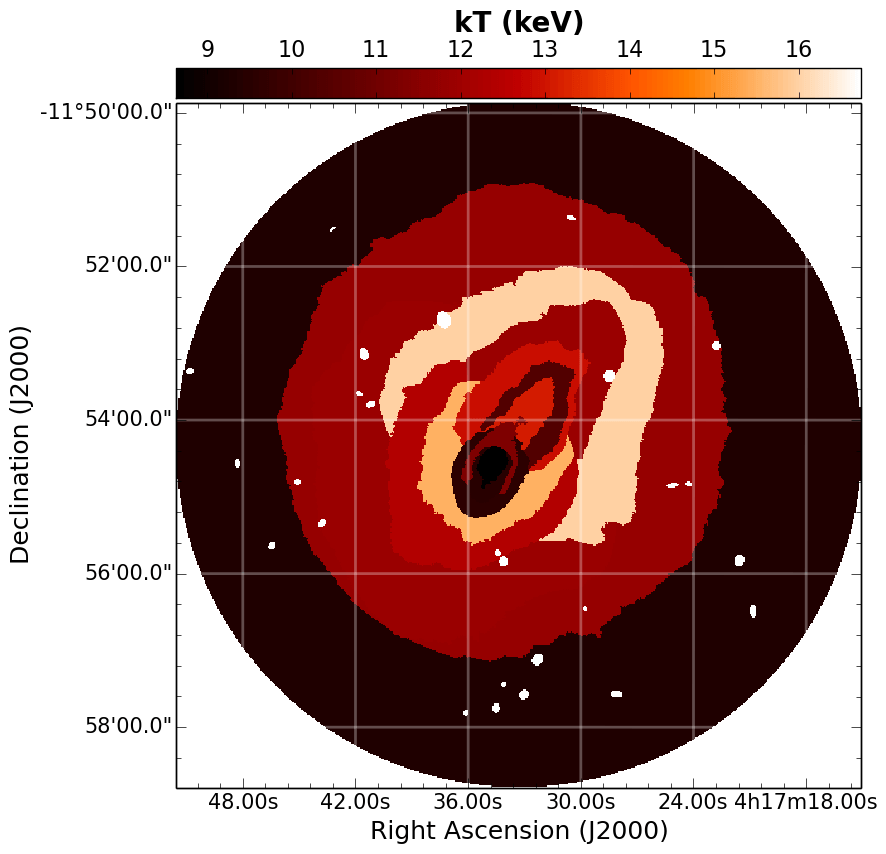}}
 \subfloat{\subfigimgblack[width=.3\textwidth]{\quad e)}{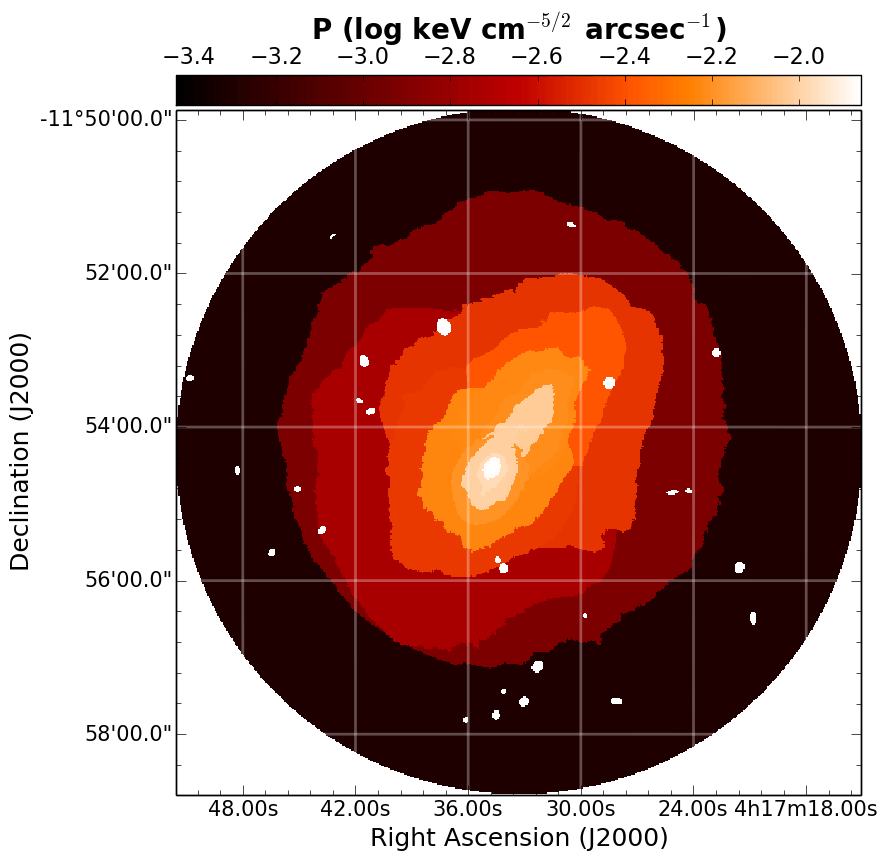}}
 \subfloat{\subfigimgblack[width=.3\textwidth]{\quad f)}{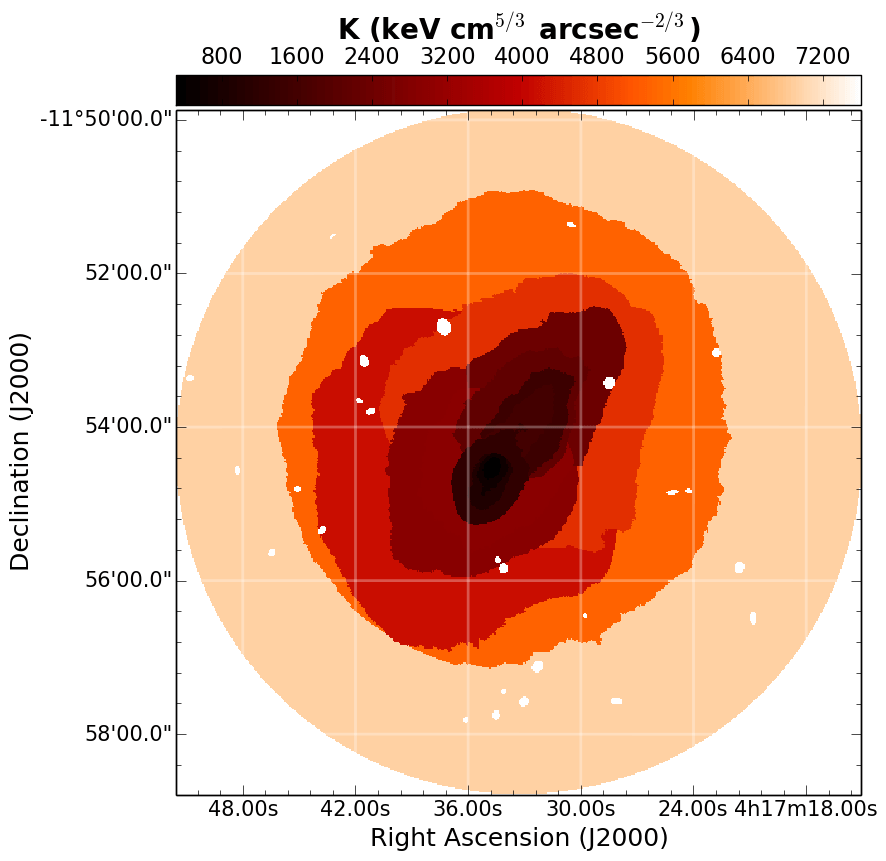}}\\
 \subfloat{\subfigimgsb[width=.3\textwidth,trim={0cm 0cm 4cm 0cm},clip]{g)}{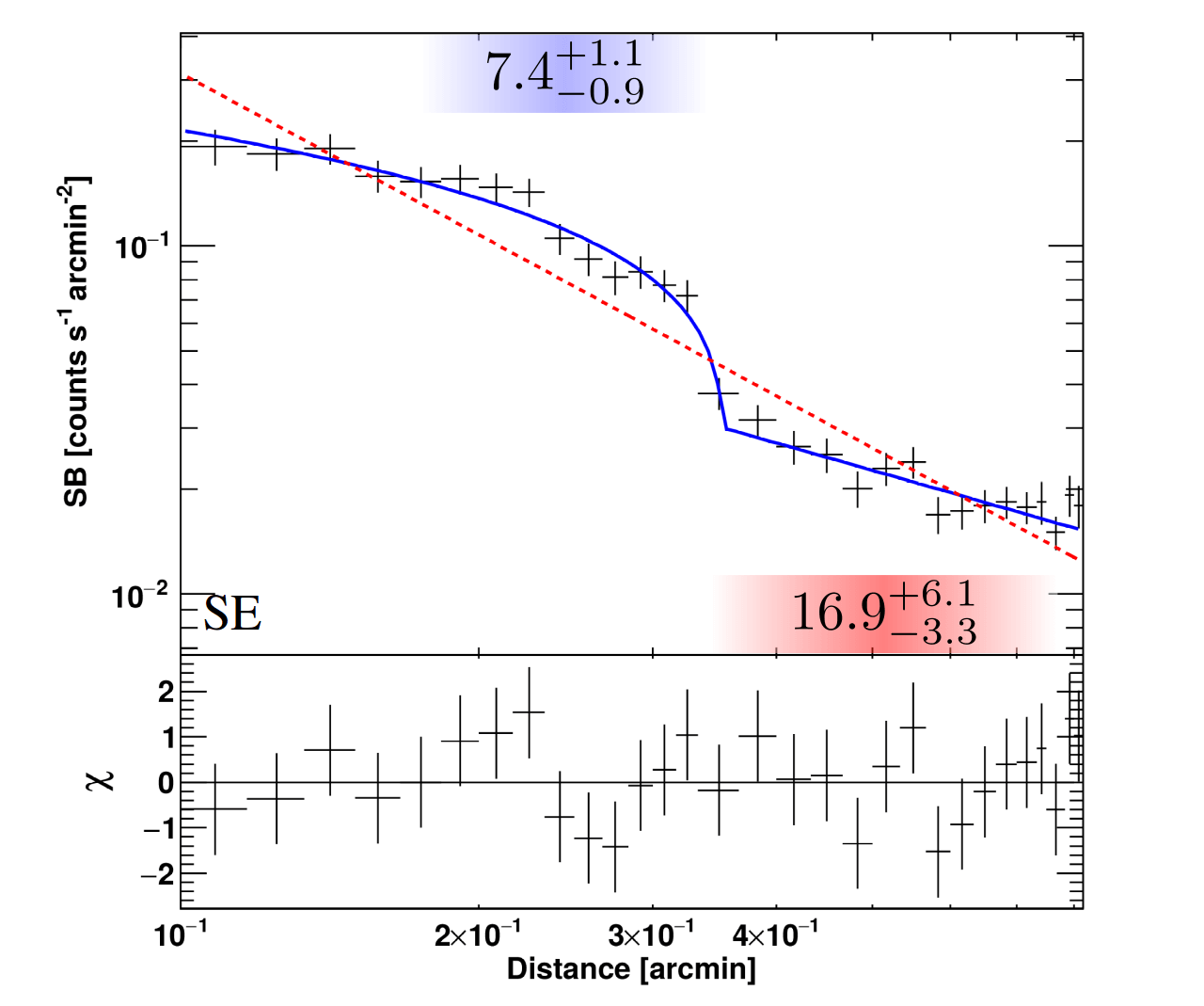}}
 \subfloat{\subfigimgsb[width=.3\textwidth,trim={0cm 0cm 4cm 0cm},clip]{h)}{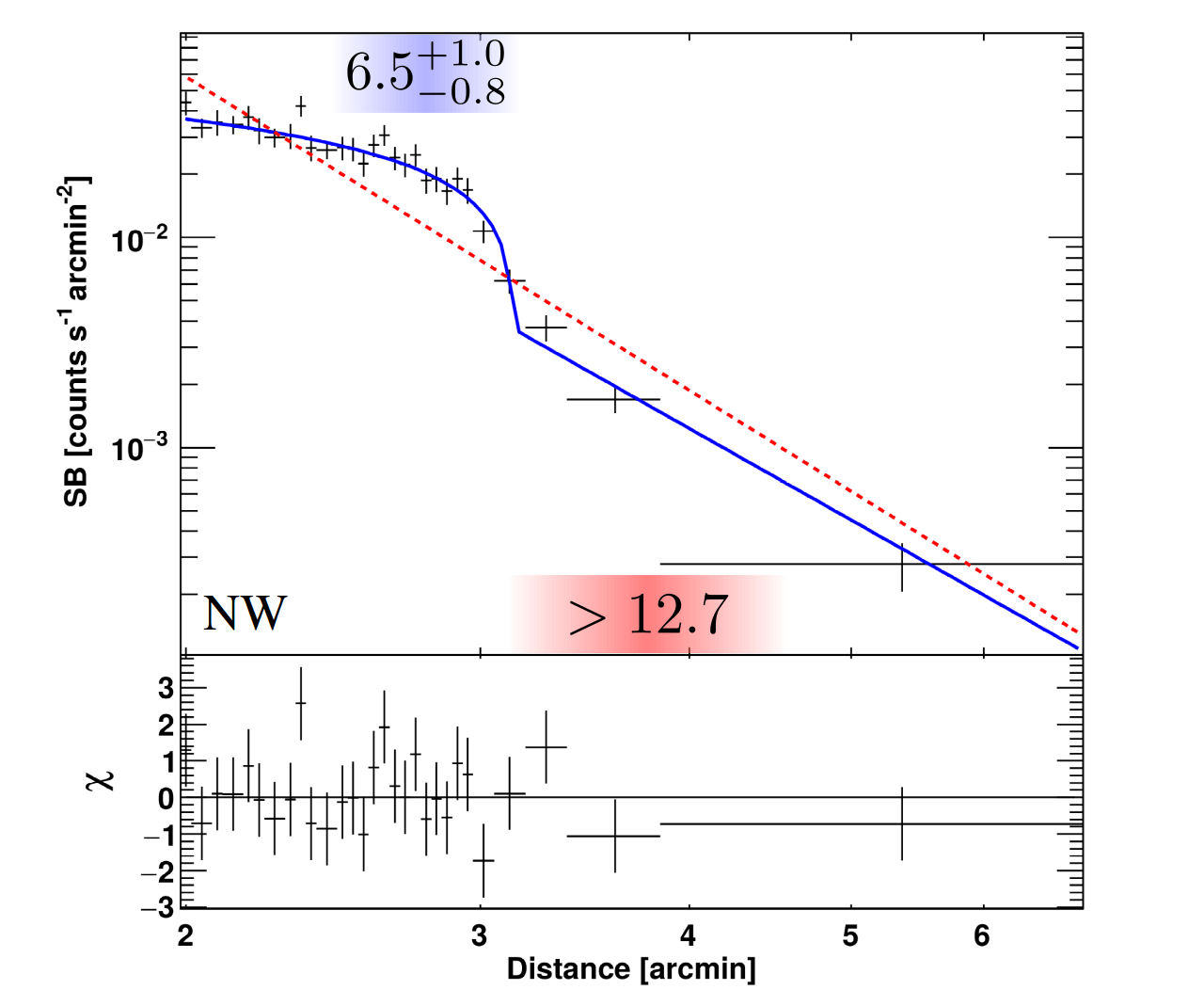}}
 \caption{MACSJ0417. The same as for Fig~\ref{fig:a399}. The goodness of fits is reported in Fig.~\ref{fig:macsj0417_errors}. The positions of the edges are marked in the \chandra\ image in green (cold fronts).}
 \label{fig:macsj0417}
\end{figure*}

\begin{figure*}
 \centering
 \begin{tabular}{cc}
  \multirow{2}{*}{\subfloat{\subfigimgwhitebig[width=.6\textwidth]{\quad  a)}{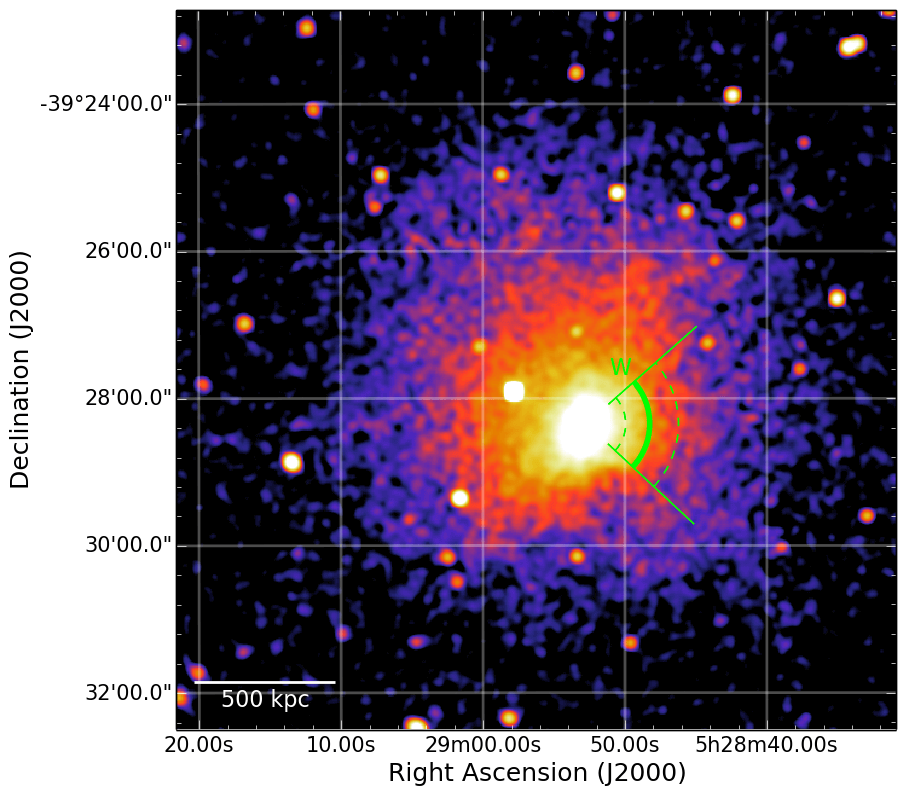}}} & \\
  & \vspace{0.15cm}\hspace{-0.3cm}\subfloat{\subfigimgwhiteggm[width=.28\textwidth]{\quad b)}{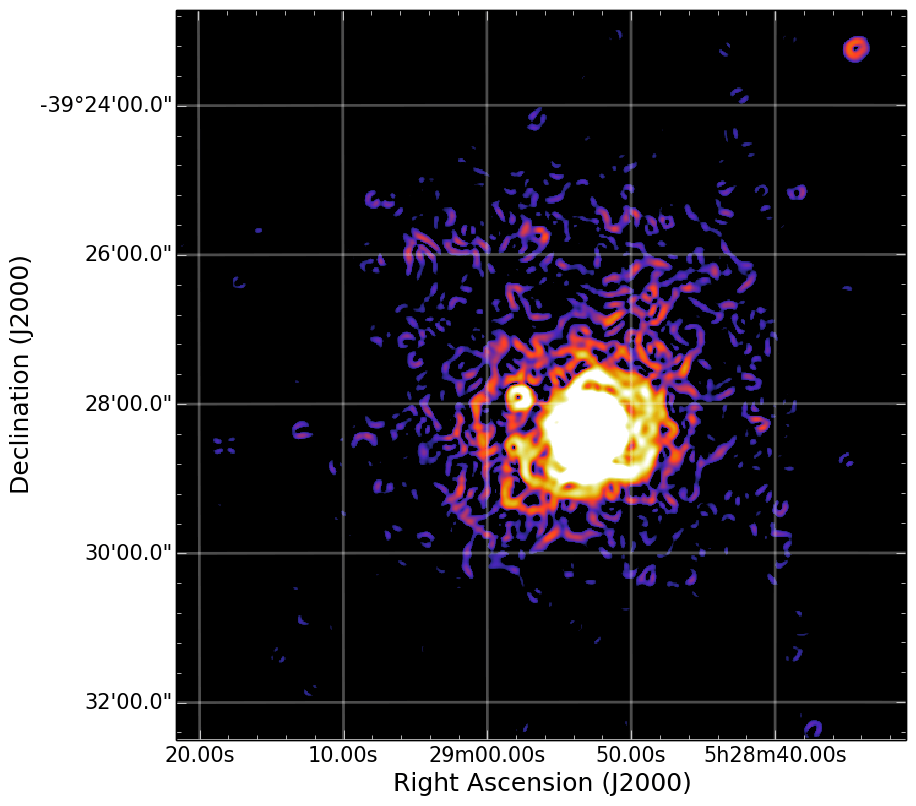}} \\
  & \hspace{-0.3cm}\subfloat{\subfigimgwhiteggm[width=.28\textwidth]{\quad c)}{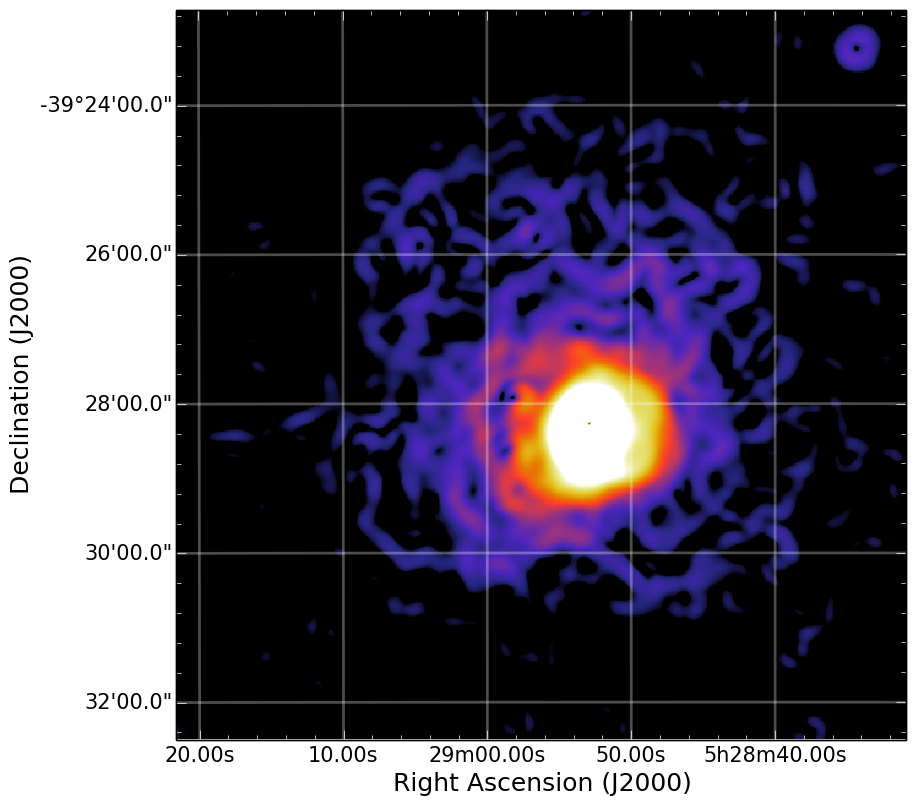}}
 \end{tabular}
 \subfloat{\subfigimgblack[width=.3\textwidth]{\quad d)}{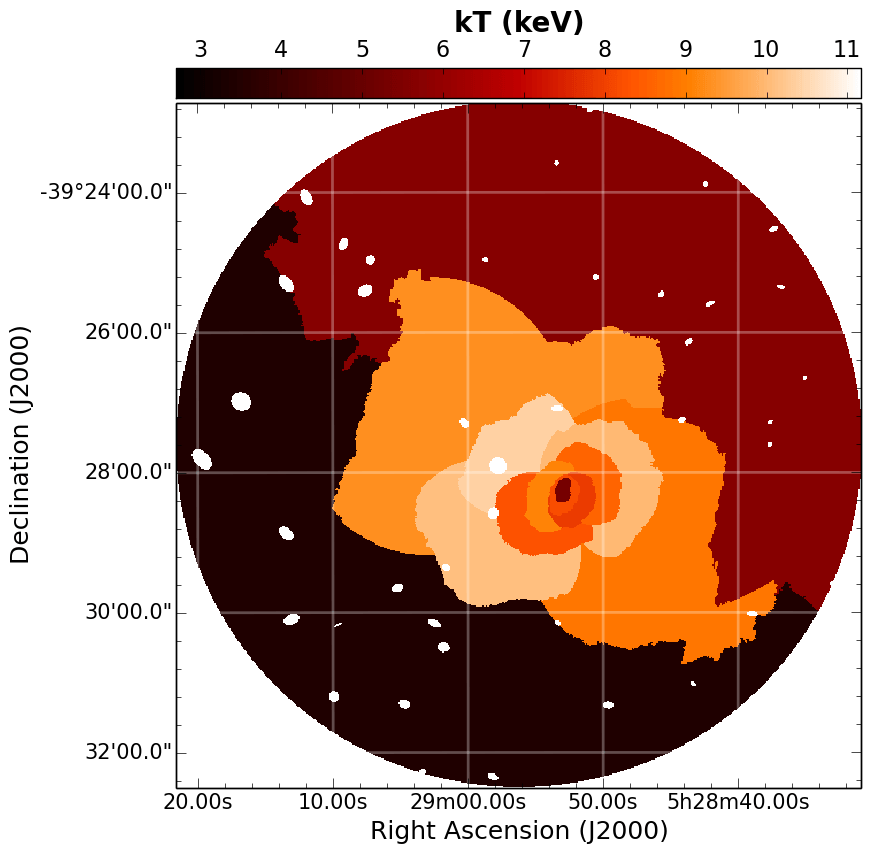}}
 \subfloat{\subfigimgblack[width=.3\textwidth]{\quad e)}{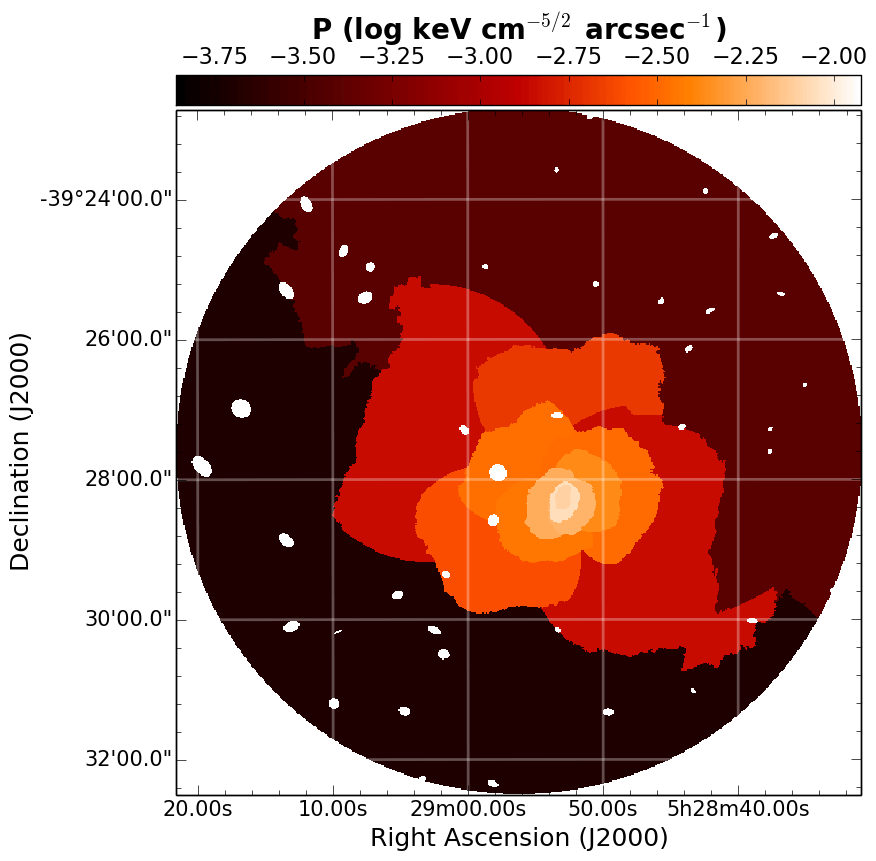}}
 \subfloat{\subfigimgblack[width=.3\textwidth]{\quad f)}{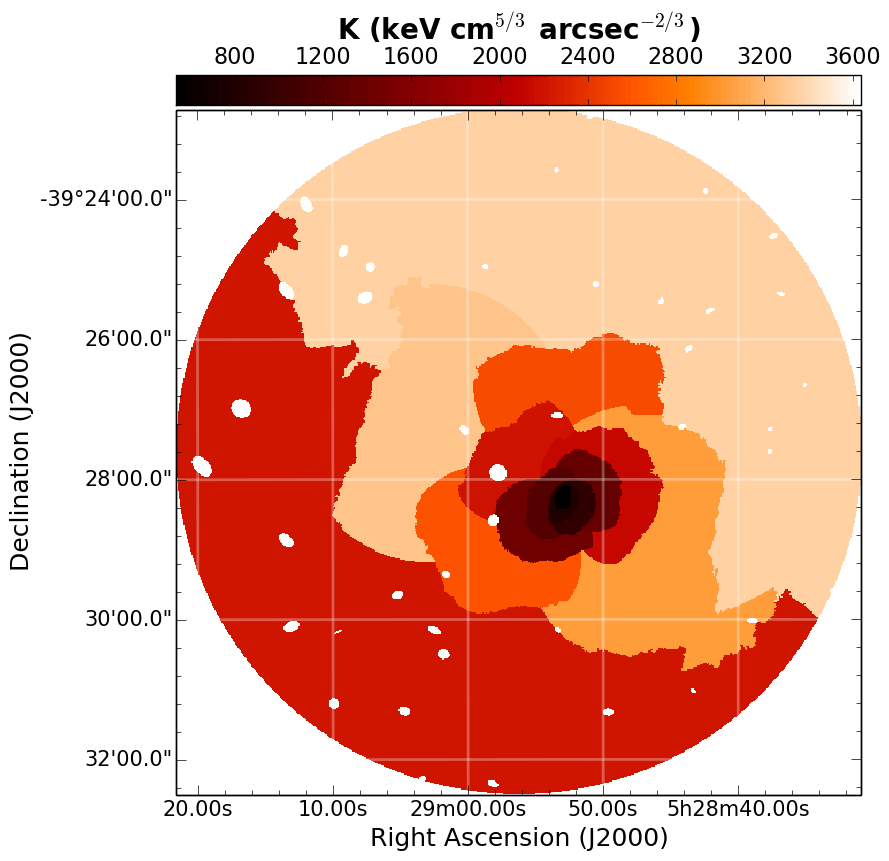}}\\
 \subfloat{\subfigimgsb[width=.3\textwidth,trim={0cm 0cm 4cm 0cm},clip]{g)}{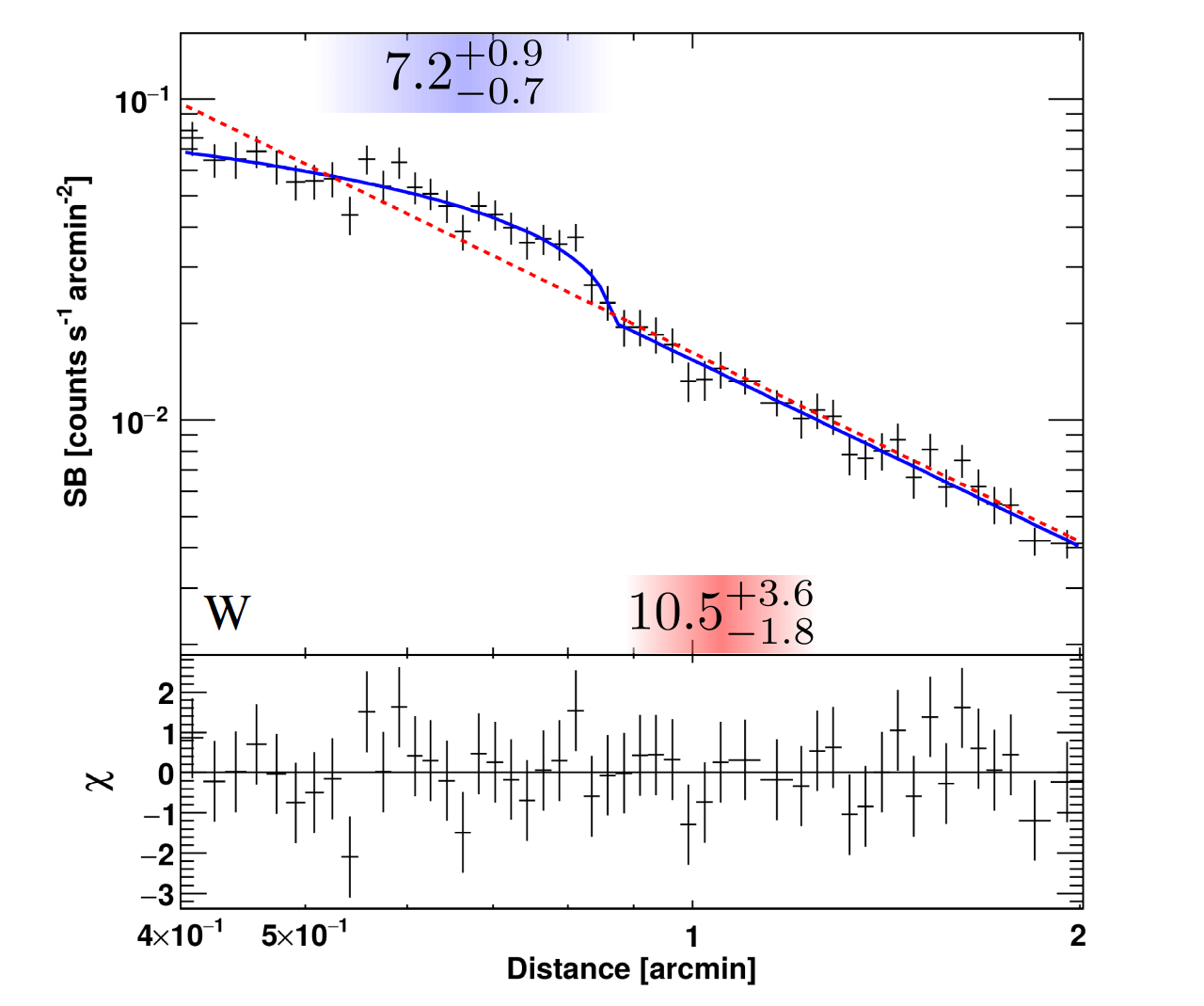}}
 \caption{RXCJ0528. The same as for Fig~\ref{fig:a399}. The goodness of fits is reported in Fig.~\ref{fig:rxcj0528_errors}. The position of the edge is marked in the \chandra\ image in green (cold front).}
 \label{fig:rxcj0528}
\end{figure*}

\begin{figure*}
 \centering
 \begin{tabular}{cc}
  \multirow{2}{*}{\subfloat{\subfigimgwhitebig[width=.6\textwidth]{\quad  a)}{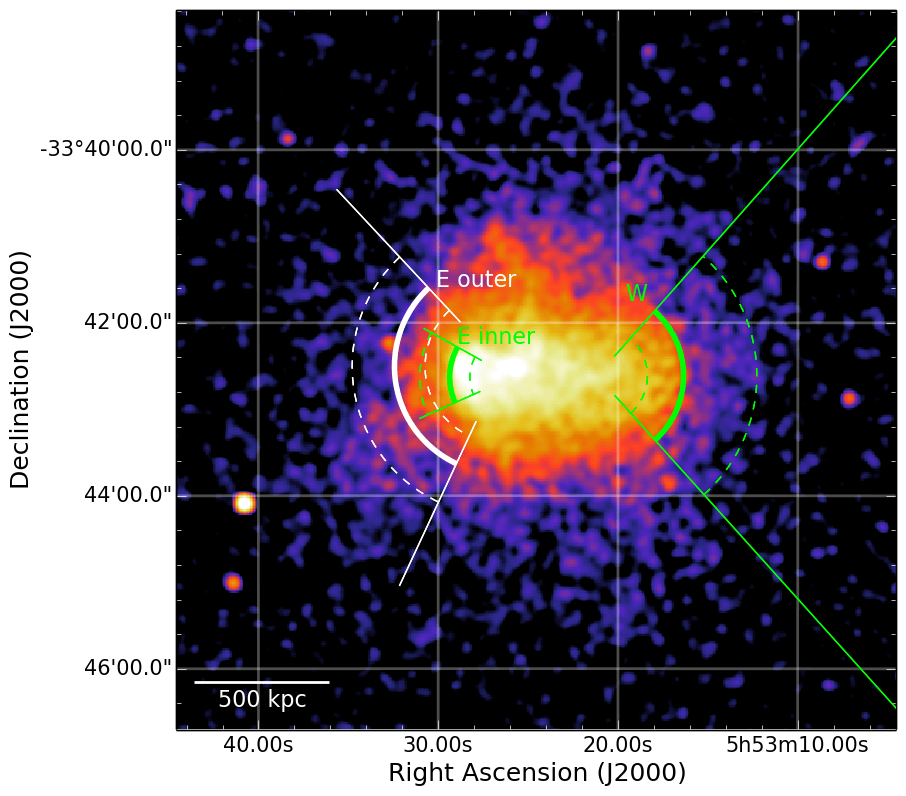}}} & \\
  & \vspace{0.15cm}\hspace{-0.3cm}\subfloat{\subfigimgwhiteggm[width=.28\textwidth]{\quad b)}{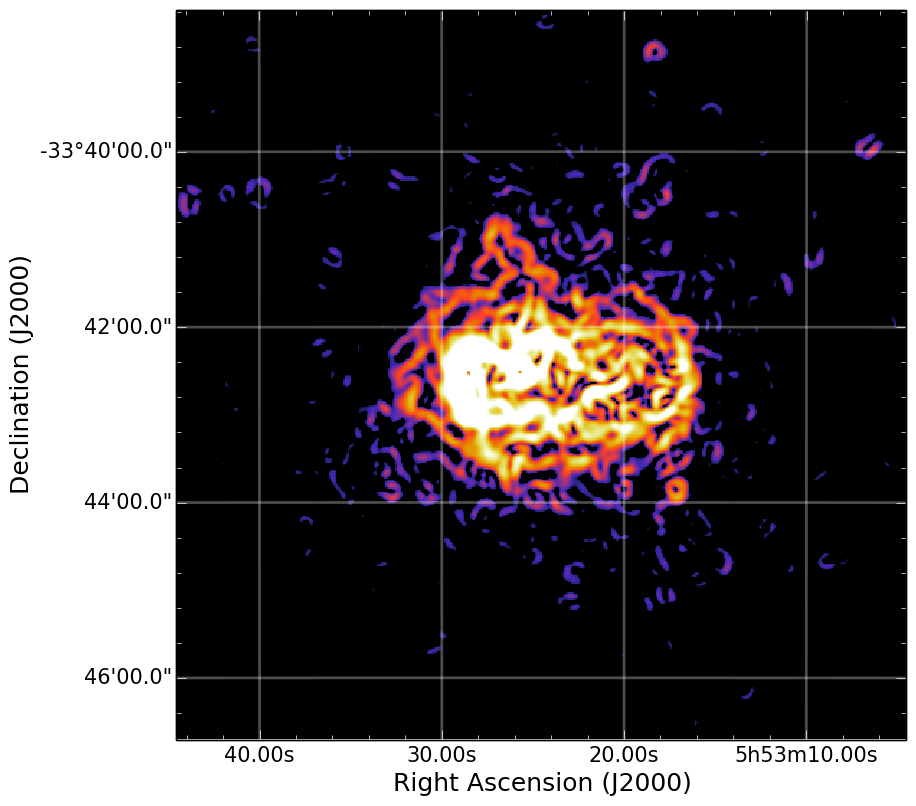}} \\
  & \hspace{-0.3cm}\subfloat{\subfigimgwhiteggm[width=.28\textwidth]{\quad c)}{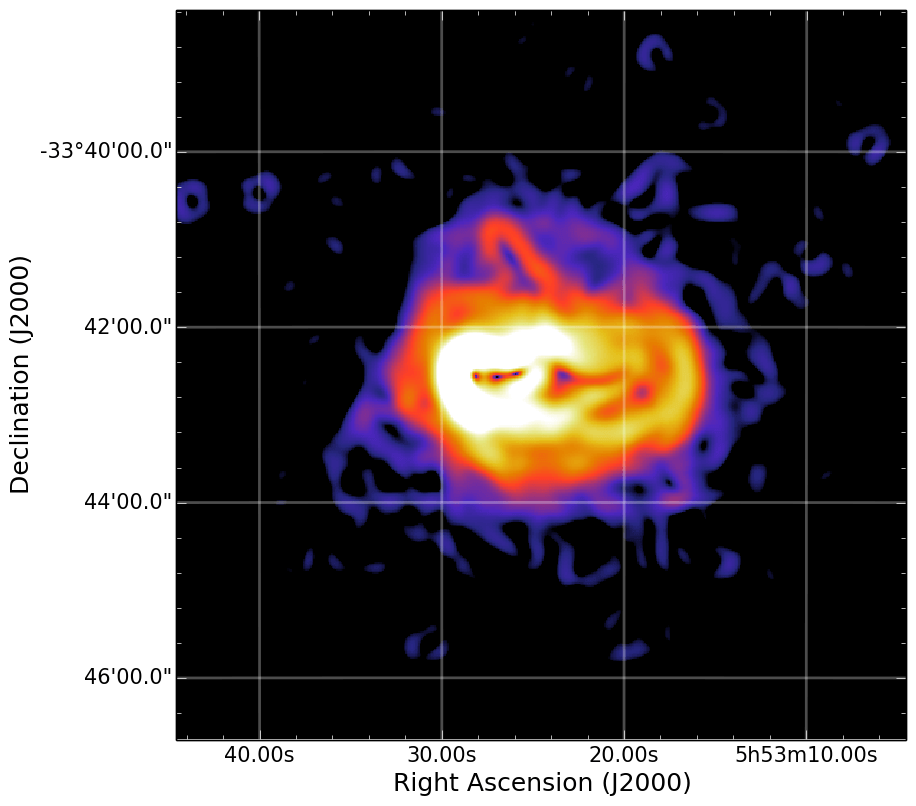}}
 \end{tabular}
 \subfloat{\subfigimgblack[width=.3\textwidth]{\quad d)}{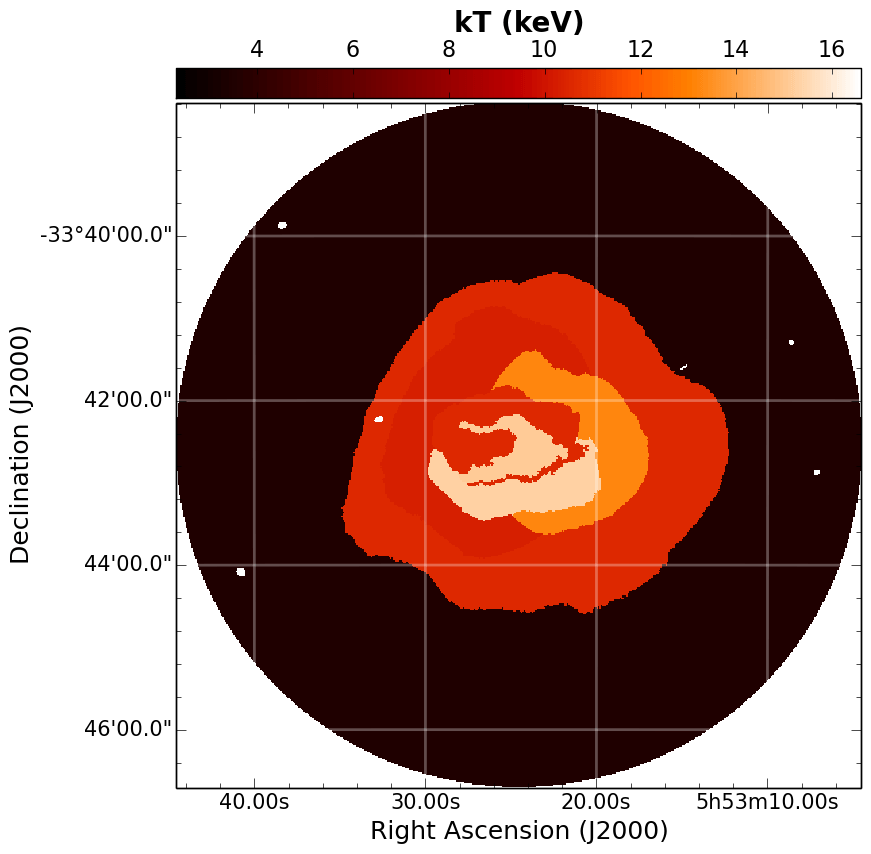}}
 \subfloat{\subfigimgblack[width=.3\textwidth]{\quad e)}{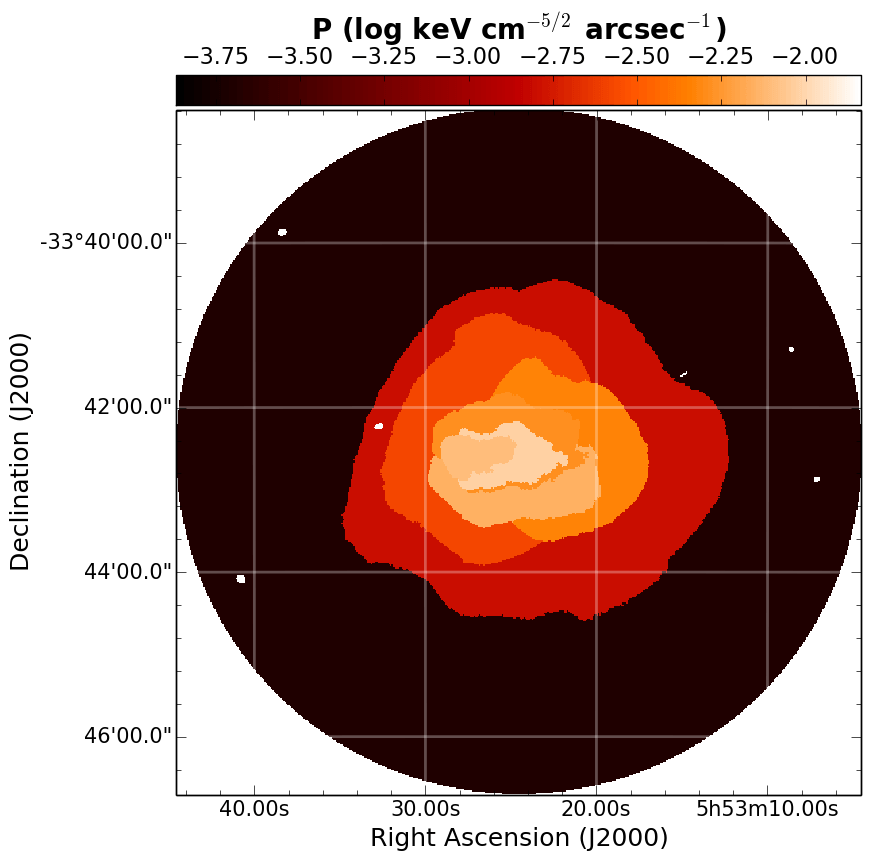}}
 \subfloat{\subfigimgblack[width=.3\textwidth]{\quad f)}{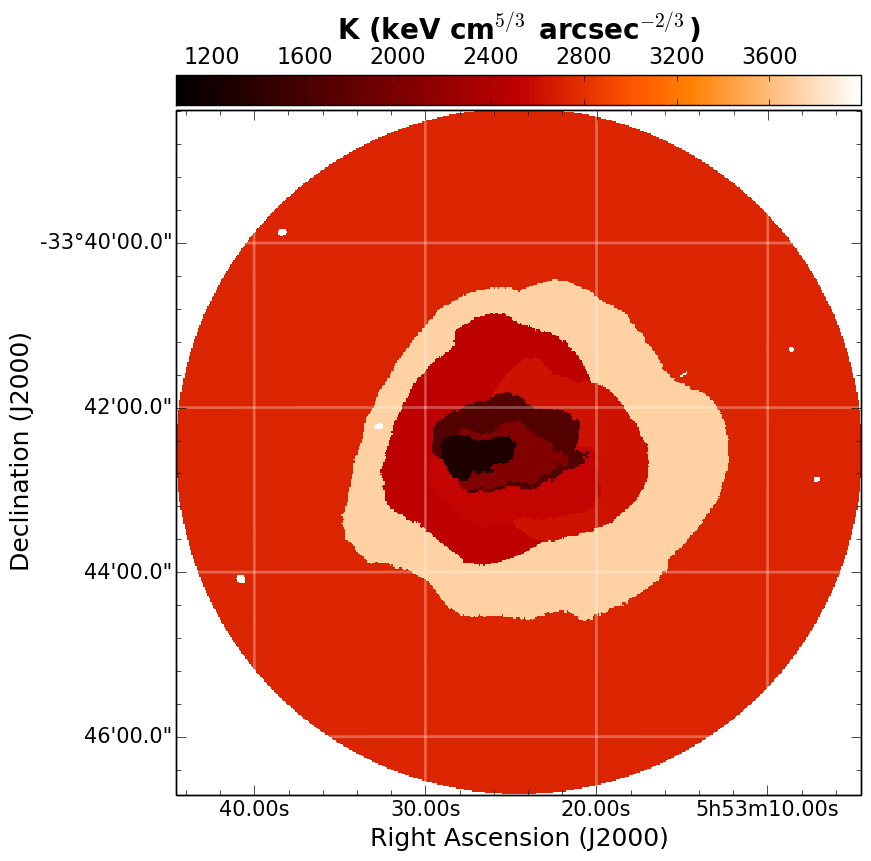}}\\
 \subfloat{\subfigimgsb[width=.3\textwidth,trim={0cm 0cm 4cm 0cm},clip]{g)}{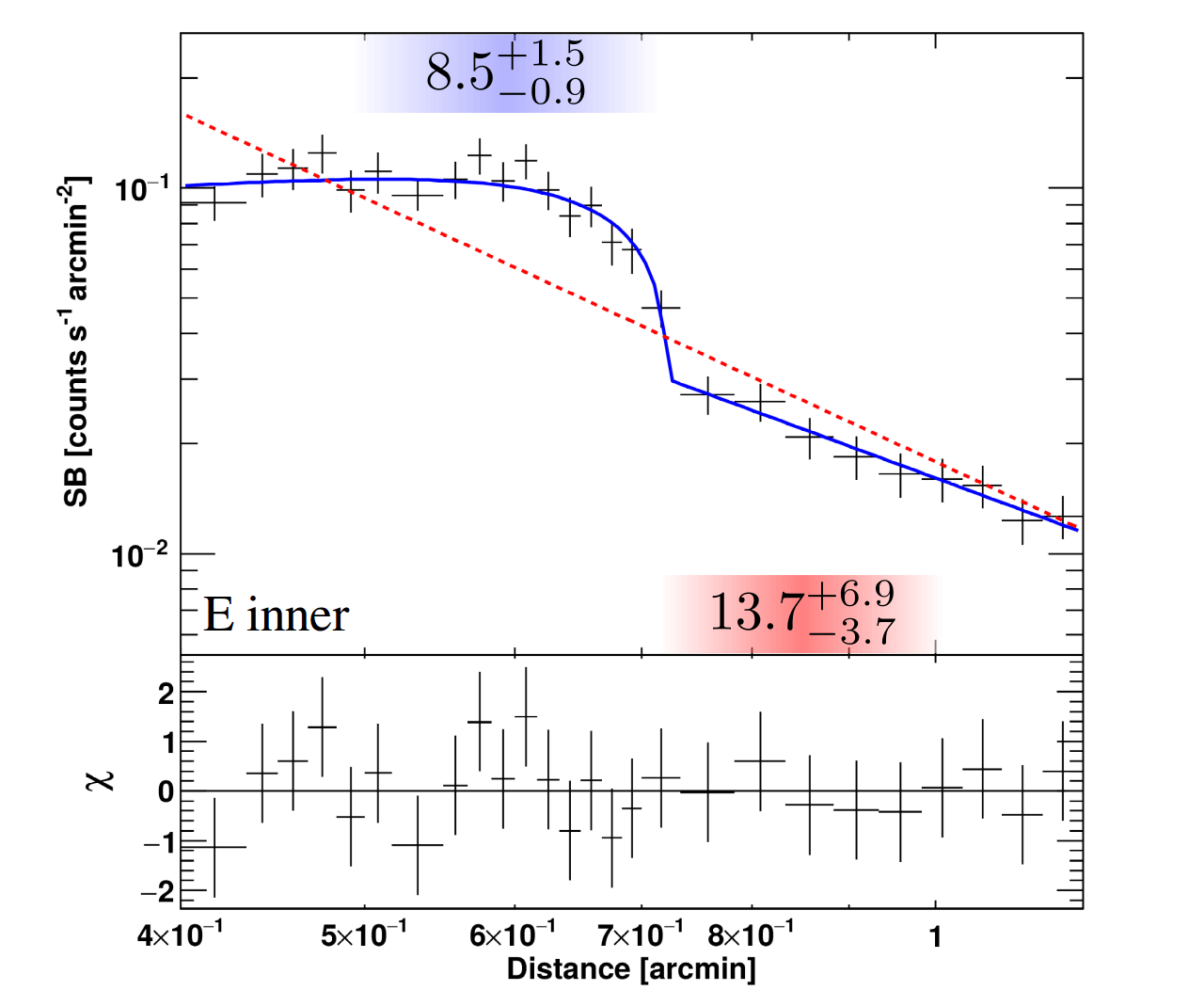}}
 \subfloat{\subfigimgsb[width=.3\textwidth,trim={0cm 0cm 4cm 0cm},clip]{h)}{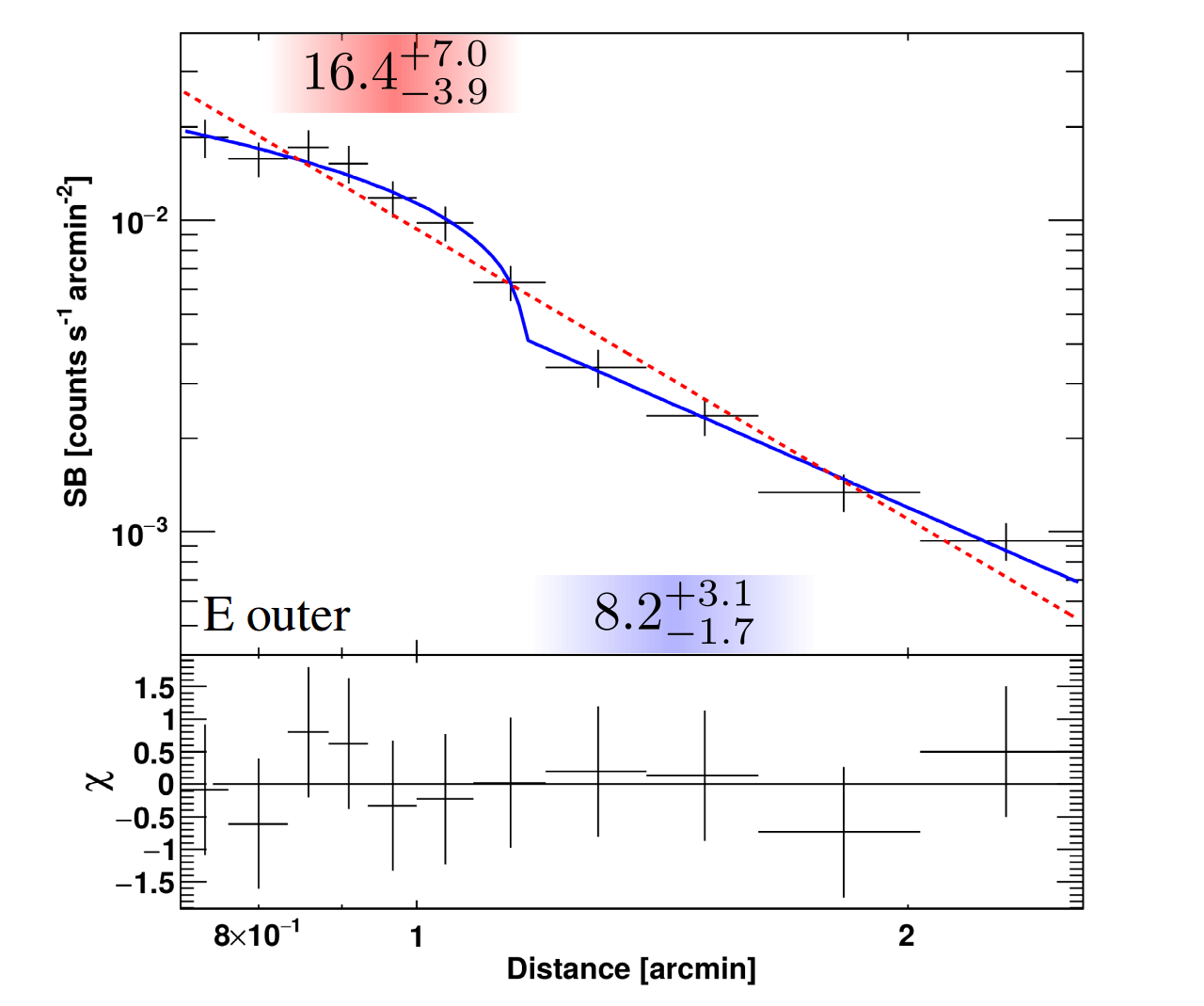}}
 \subfloat{\subfigimgsb[width=.3\textwidth,trim={0cm 0cm 4cm 0cm},clip]{i)}{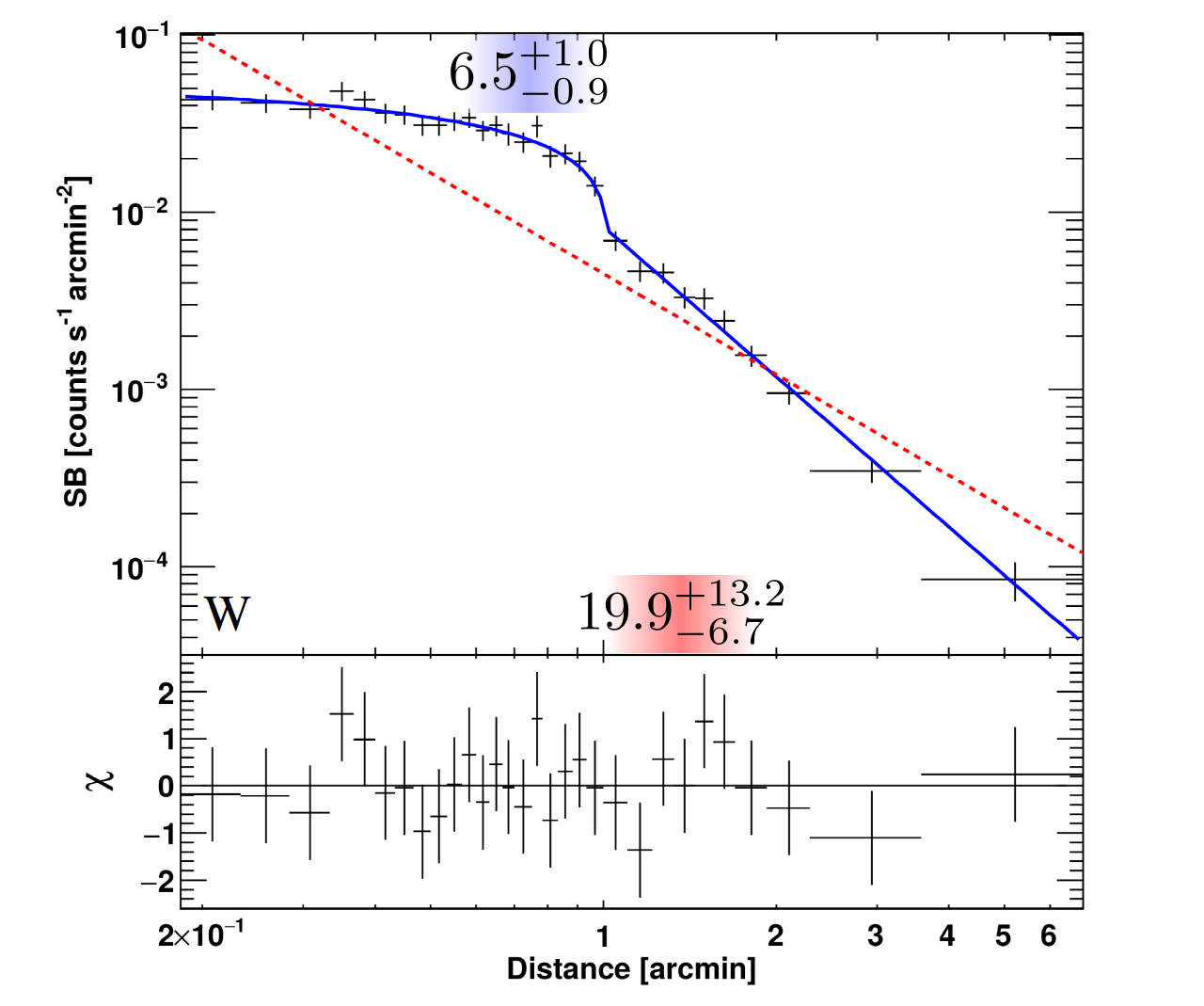}}
 \caption{MACS J0553. The same as for Fig~\ref{fig:a399}. The goodness of fits is reported in Fig.~\ref{fig:macsj0553_errors}. The positions of the edges are marked in the \chandra\ image in green (cold fronts) and white (shock).}
 \label{fig:macsj0553}
\end{figure*}

\begin{figure*}
 \centering
 \begin{tabular}{cc}
  \multirow{2}{*}{\subfloat{\subfigimgwhitebig[width=.6\textwidth]{\quad  a)}{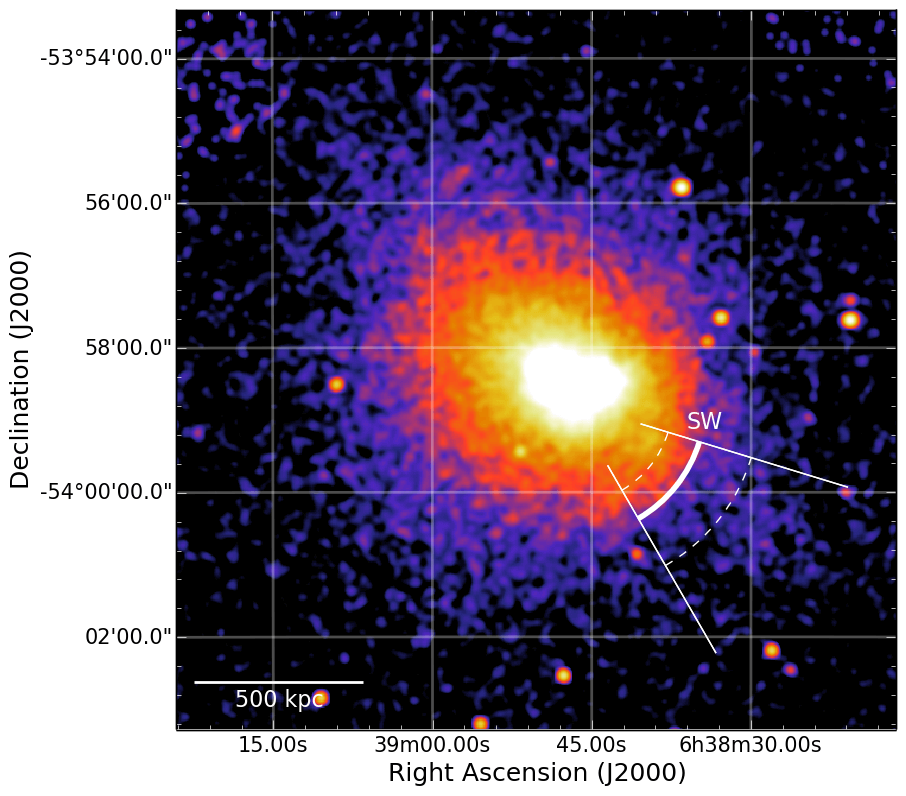}}} & \\
  & \vspace{0.15cm}\hspace{-0.3cm}\subfloat{\subfigimgwhiteggm[width=.28\textwidth]{\quad b)}{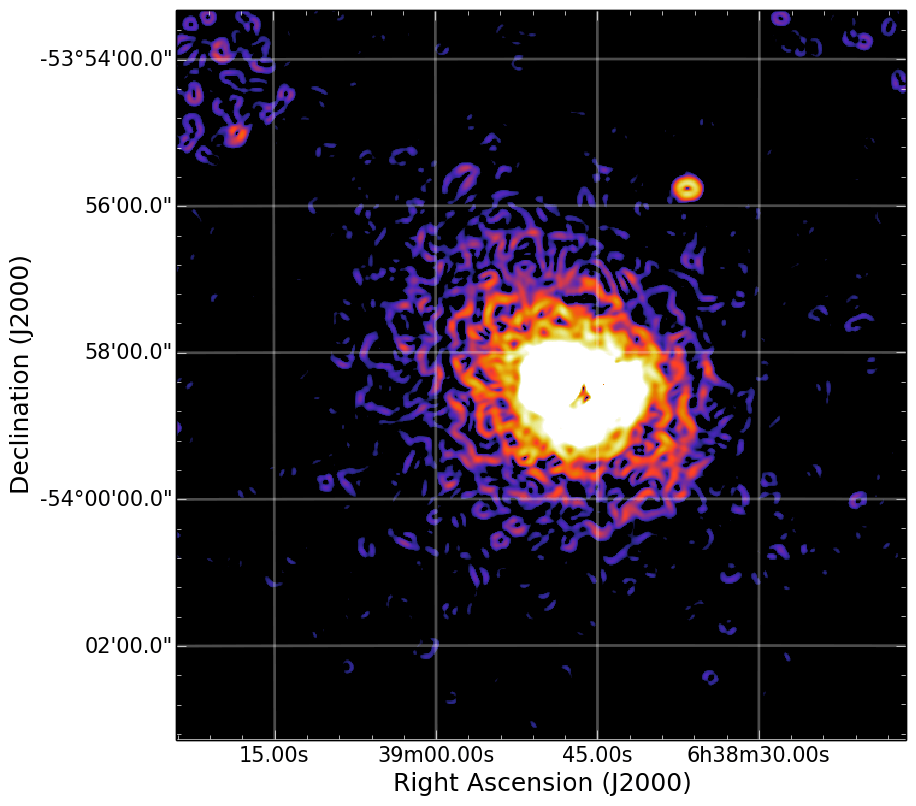}} \\
  & \hspace{-0.3cm}\subfloat{\subfigimgwhiteggm[width=.28\textwidth]{\quad c)}{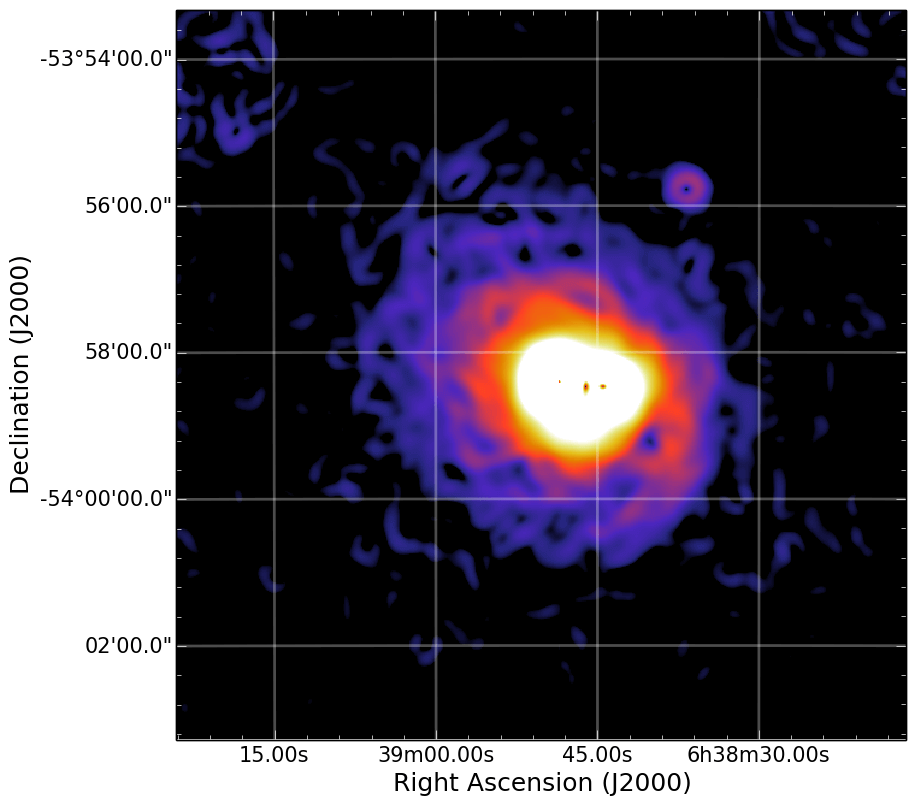}}
 \end{tabular}
 \subfloat{\subfigimgblack[width=.3\textwidth]{\quad d)}{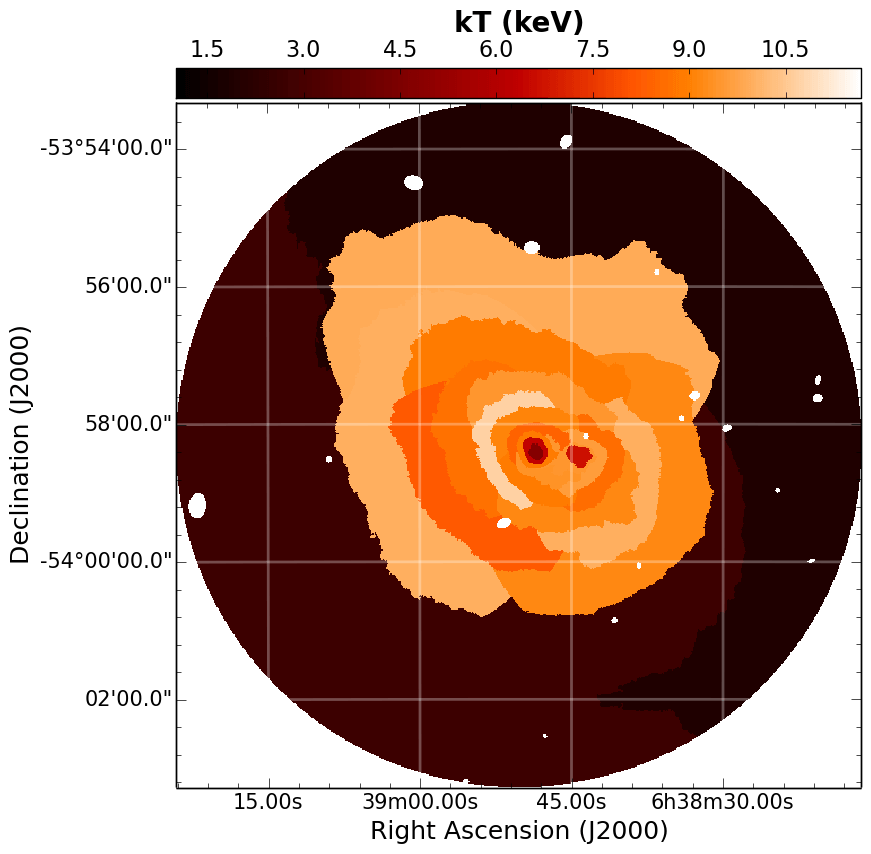}}
 \subfloat{\subfigimgblack[width=.3\textwidth]{\quad e)}{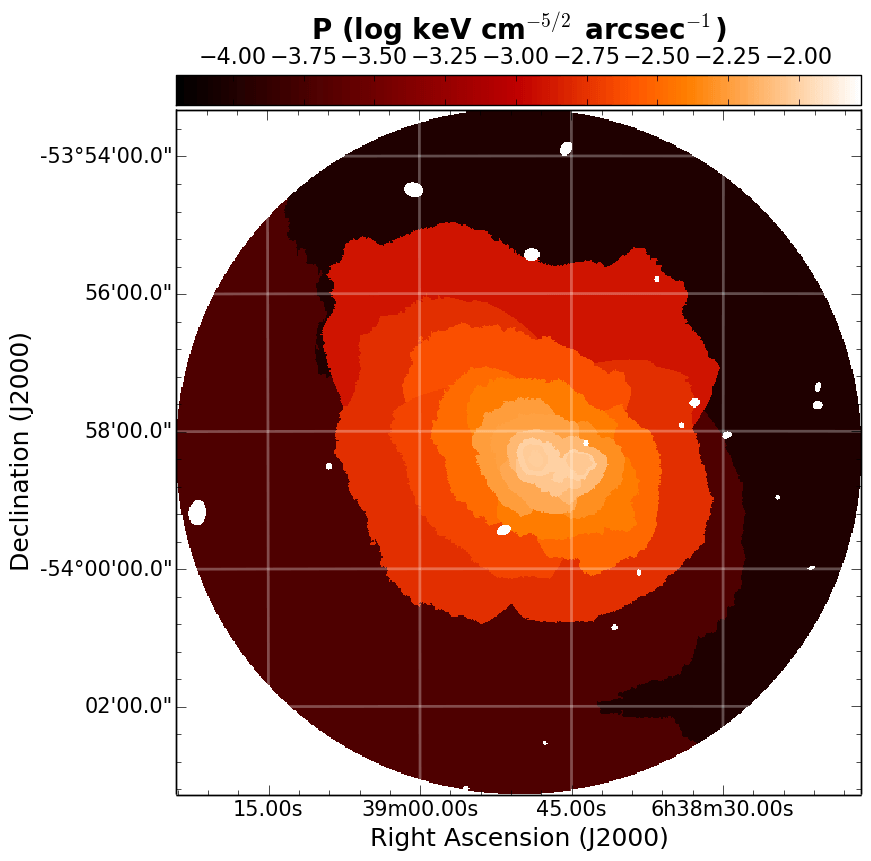}}
 \subfloat{\subfigimgblack[width=.3\textwidth]{\quad f)}{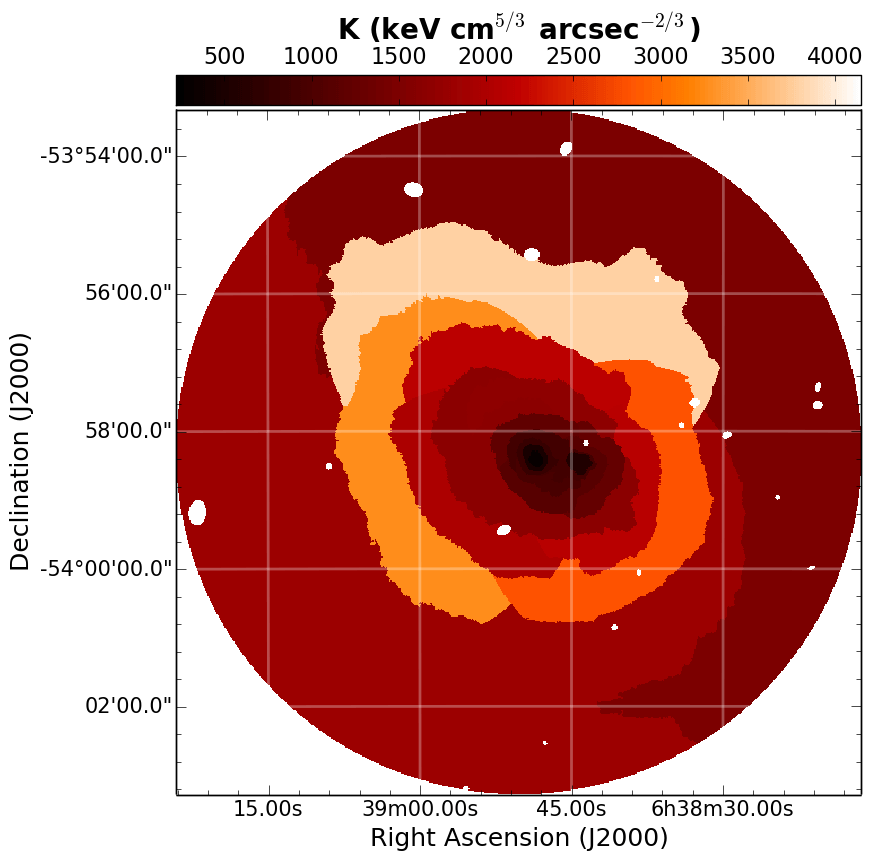}}\\
 \subfloat{\subfigimgsb[width=.3\textwidth,trim={0cm 0cm 4cm 0cm},clip]{g)}{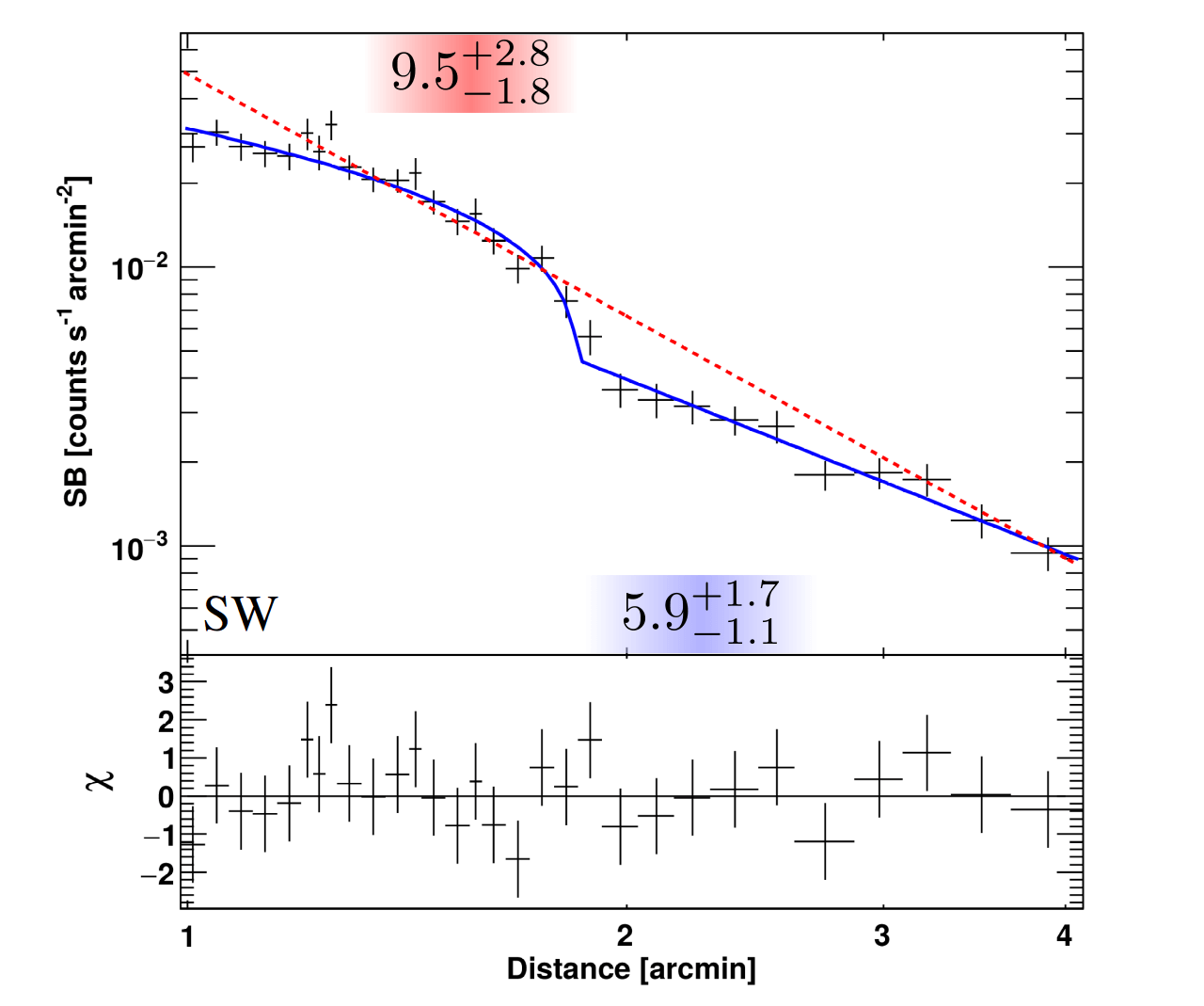}}
 \caption{AS592. The same as for Fig~\ref{fig:a399}. The goodness of fits is reported in Fig.~\ref{fig:as592_errors}. The position of the edge is marked in the \chandra\ image in white (shock).}
 \label{fig:as592}
\end{figure*}

\begin{figure*}
 \centering
 \begin{tabular}{cc}
  \multirow{2}{*}{\subfloat{\subfigimgwhitebig[width=.6\textwidth]{\quad  a)}{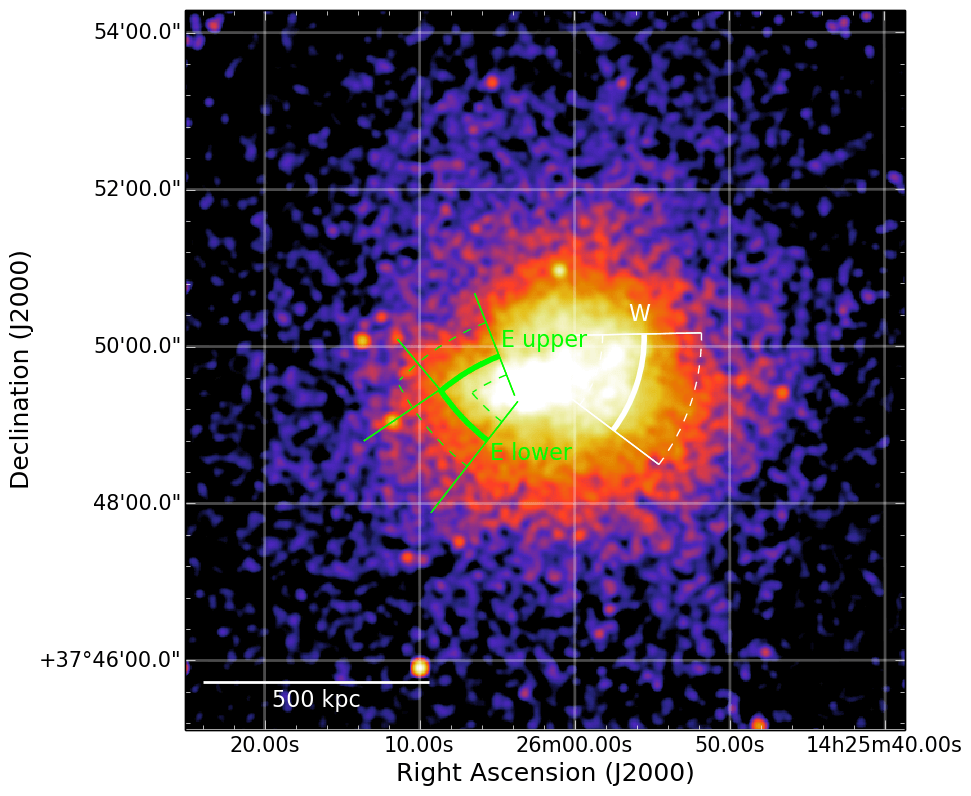}}} & \\
  & \vspace{0.15cm}\hspace{-0.3cm}\subfloat{\subfigimgwhiteggm[width=.28\textwidth]{\quad b)}{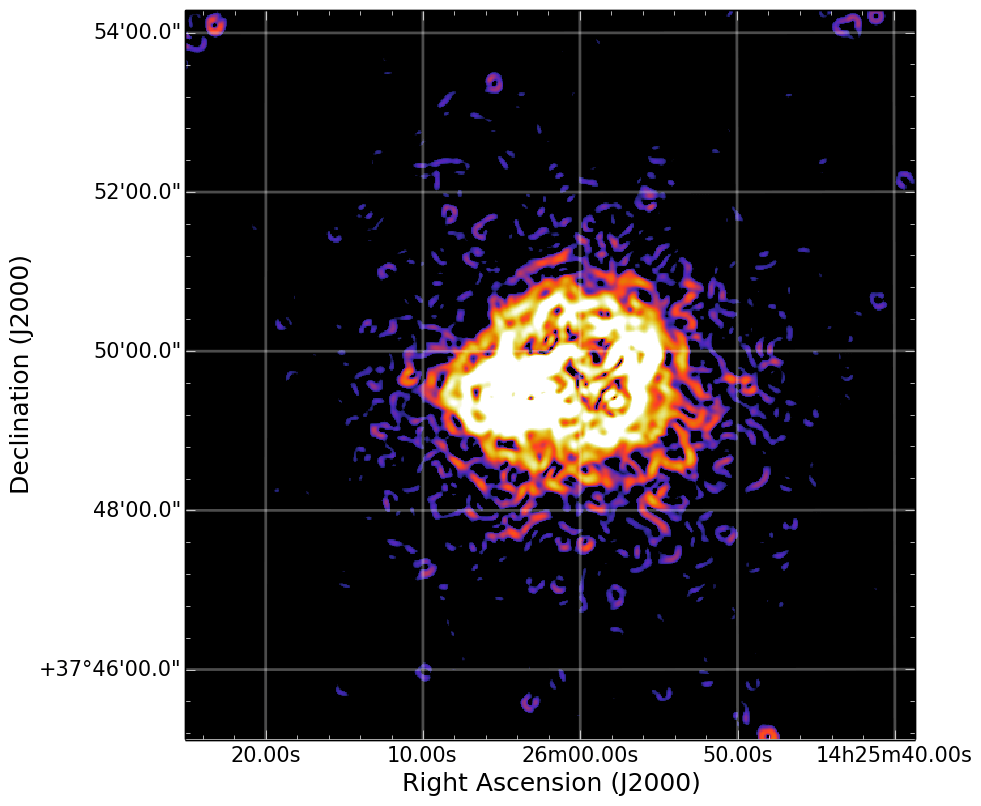}} \\
  & \hspace{-0.3cm}\subfloat{\subfigimgwhiteggm[width=.28\textwidth]{\quad c)}{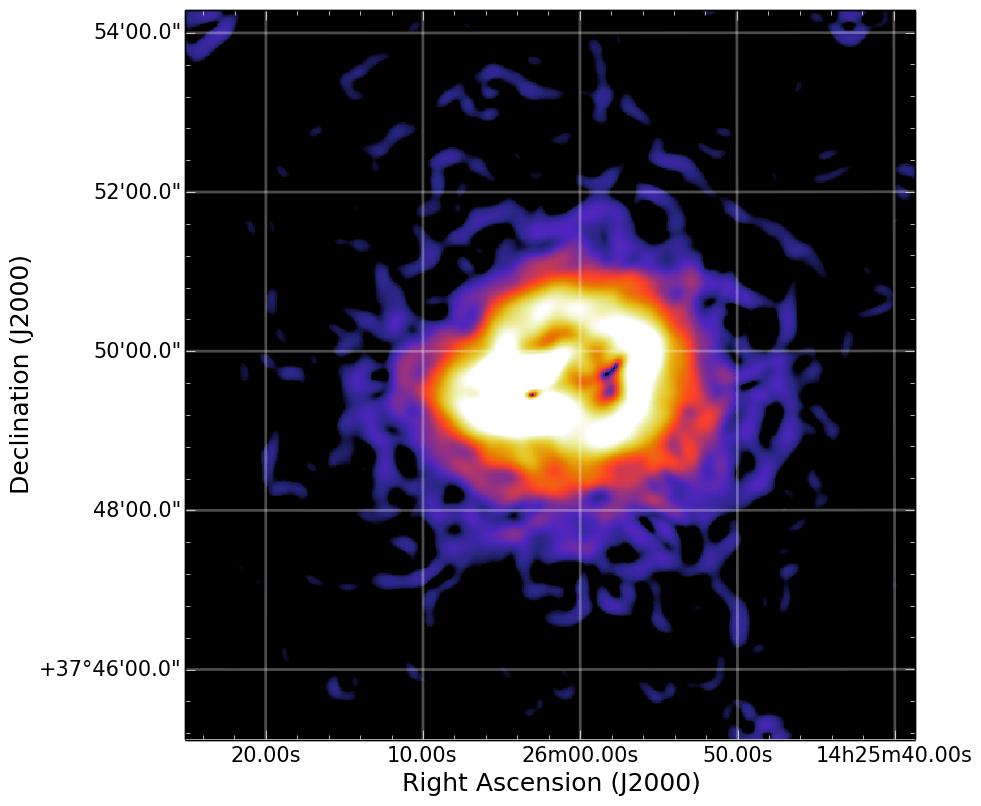}}
 \end{tabular}
 \subfloat{\subfigimgblack[width=.3\textwidth]{\enspace d)}{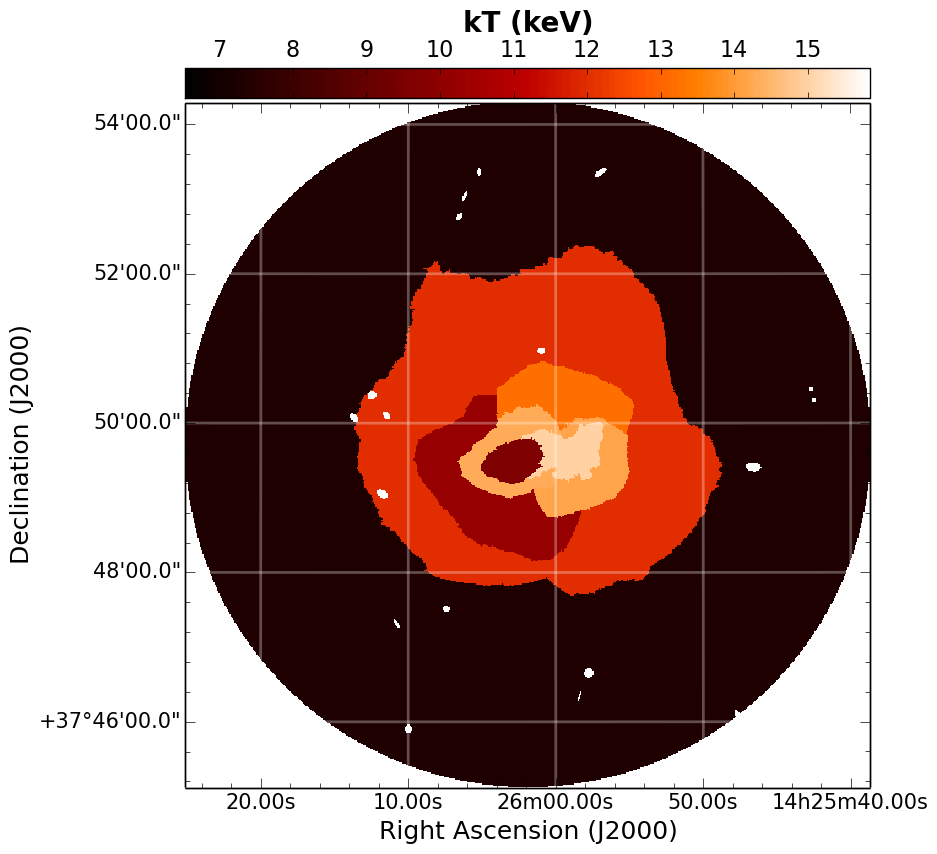}}
 \subfloat{\subfigimgblack[width=.3\textwidth]{\enspace e)}{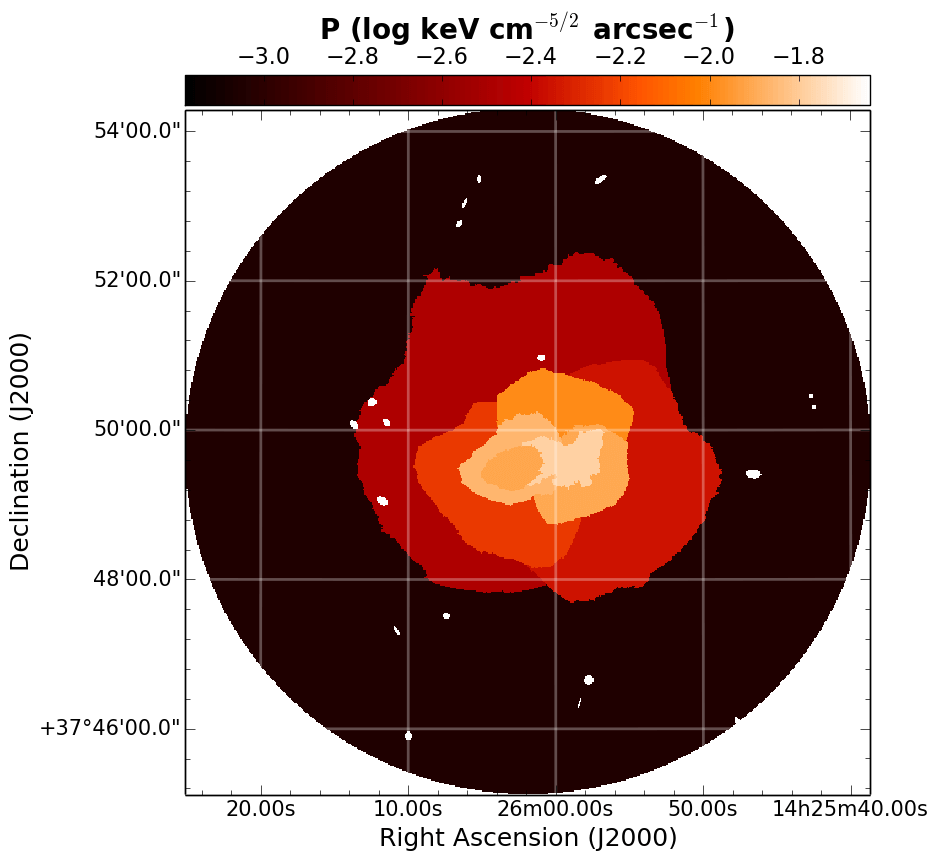}}
 \subfloat{\subfigimgblack[width=.3\textwidth]{\enspace f)}{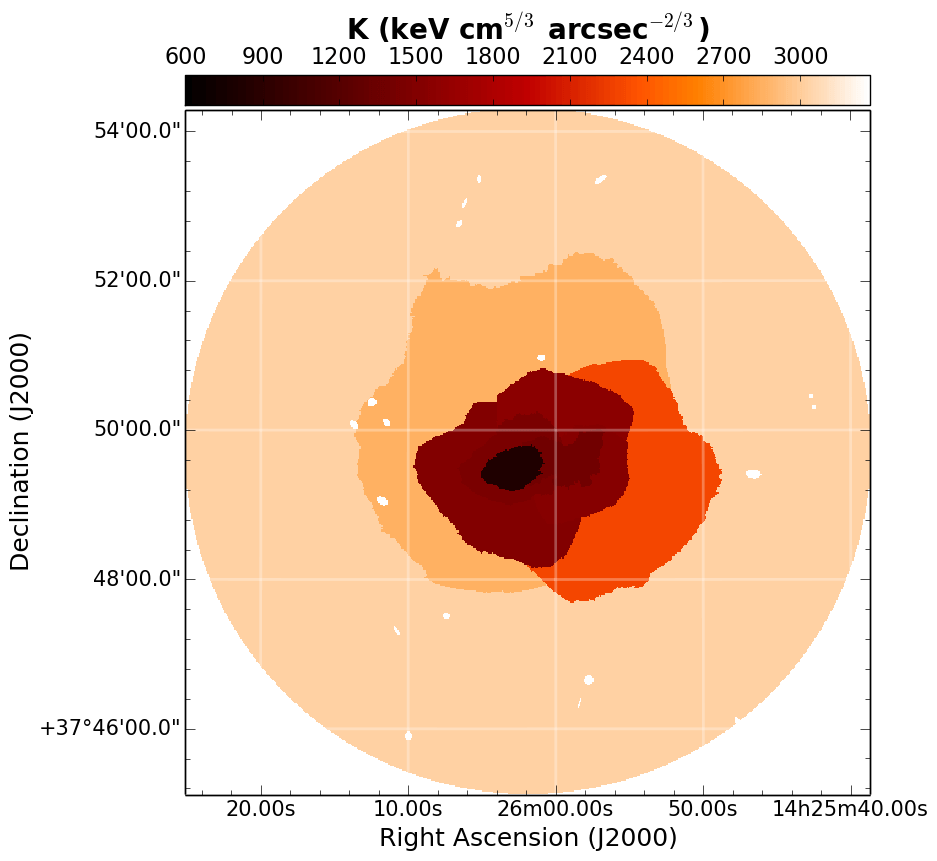}}\\
 \subfloat{\subfigimgsb[width=.3\textwidth,trim={0cm 0cm 4cm 0cm},clip]{g)}{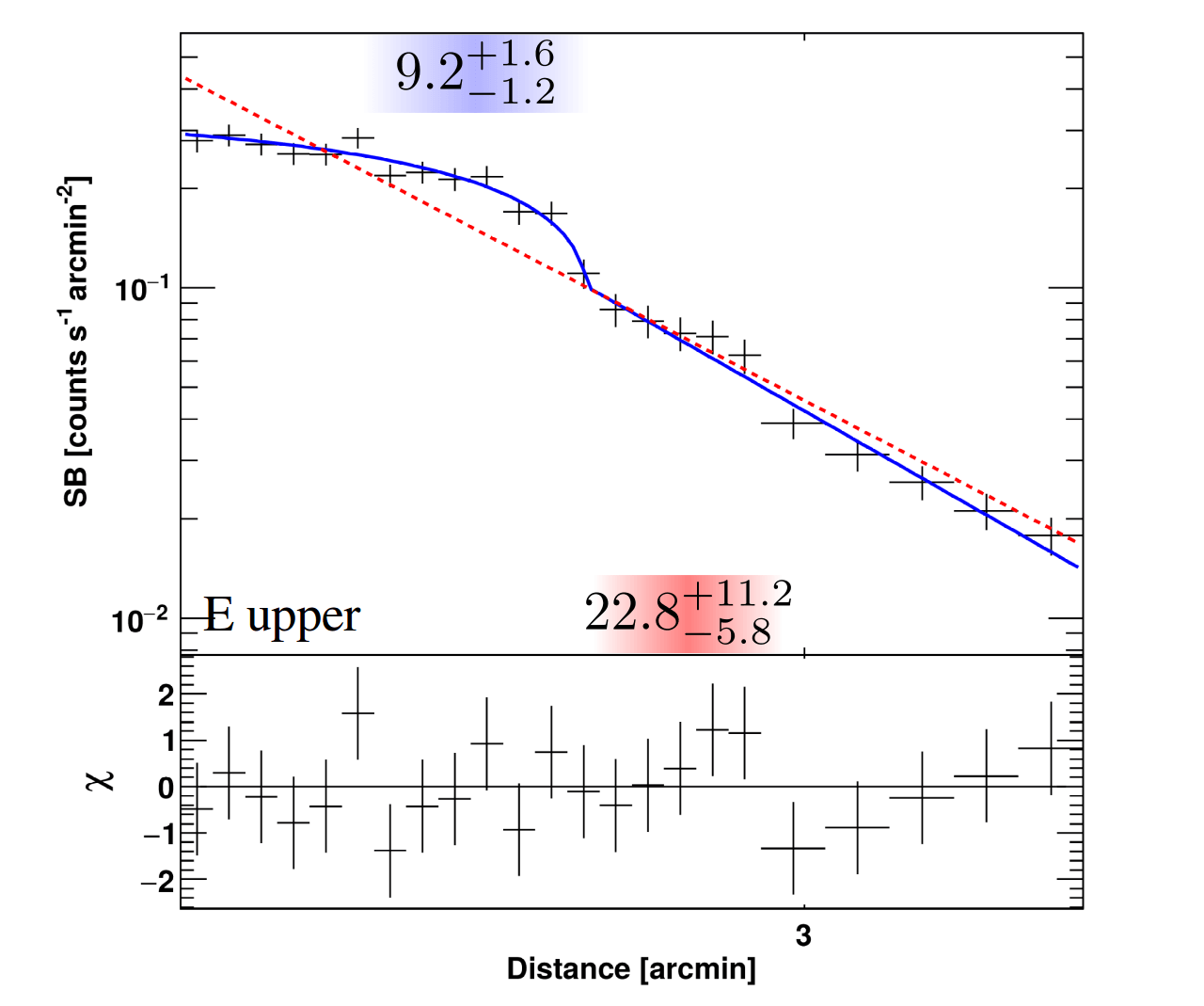}}
 \subfloat{\subfigimgsb[width=.3\textwidth,trim={0cm 0cm 4cm 0cm},clip]{h)}{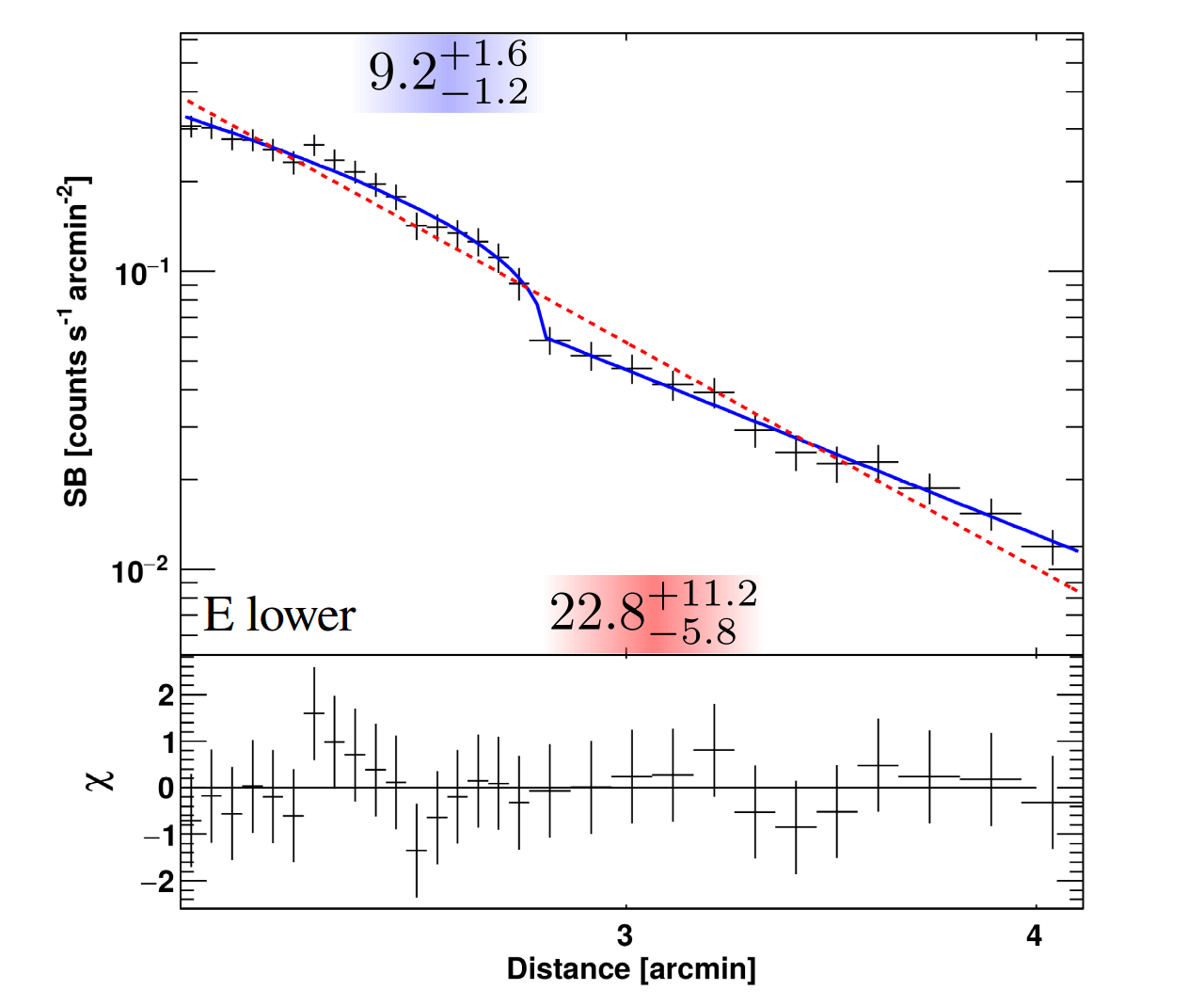}}
 \subfloat{\subfigimgsb[width=.3\textwidth,trim={0cm 0cm 4cm 0cm},clip]{i)}{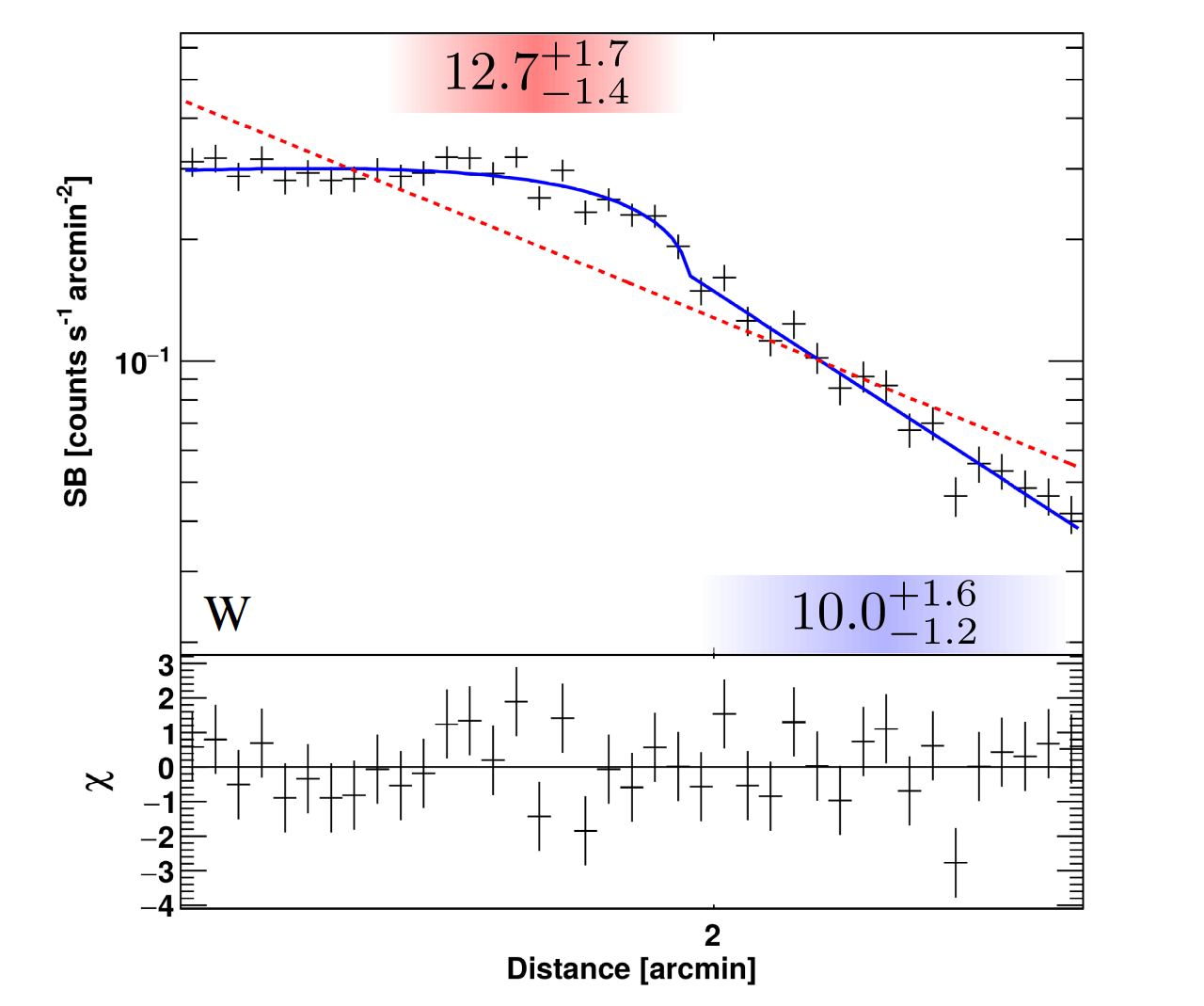}}
 \caption{A1914. The same as for Fig~\ref{fig:a399}. The goodness of fits is reported in Fig.~\ref{fig:a1914_errors}. The positions of the edges are marked in the \chandra\ image in green (cold front) and white (shock).}
 \label{fig:a1914}
\end{figure*}

\begin{figure*}
 \centering
 \begin{tabular}{cc}
  \multirow{2}{*}{\subfloat{\subfigimgwhitebig[width=.6\textwidth]{\quad  a)}{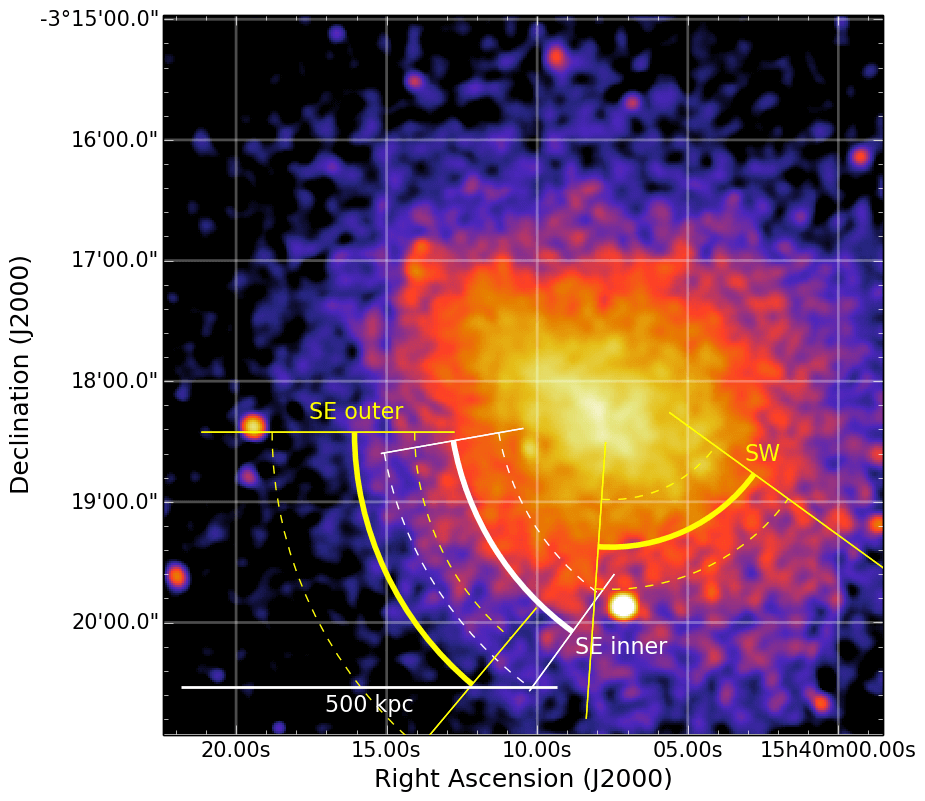}}} & \\
  & \vspace{0.15cm}\hspace{-0.3cm}\subfloat{\subfigimgwhiteggm[width=.28\textwidth]{\quad b)}{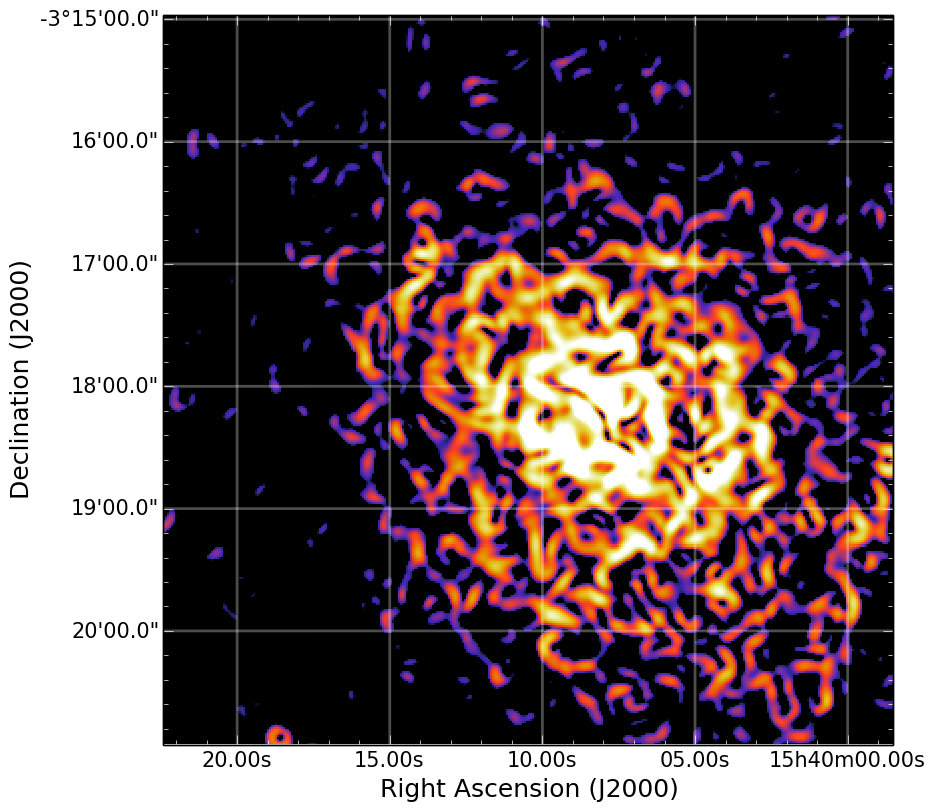}} \\
  & \hspace{-0.3cm}\subfloat{\subfigimgwhiteggm[width=.28\textwidth]{\quad c)}{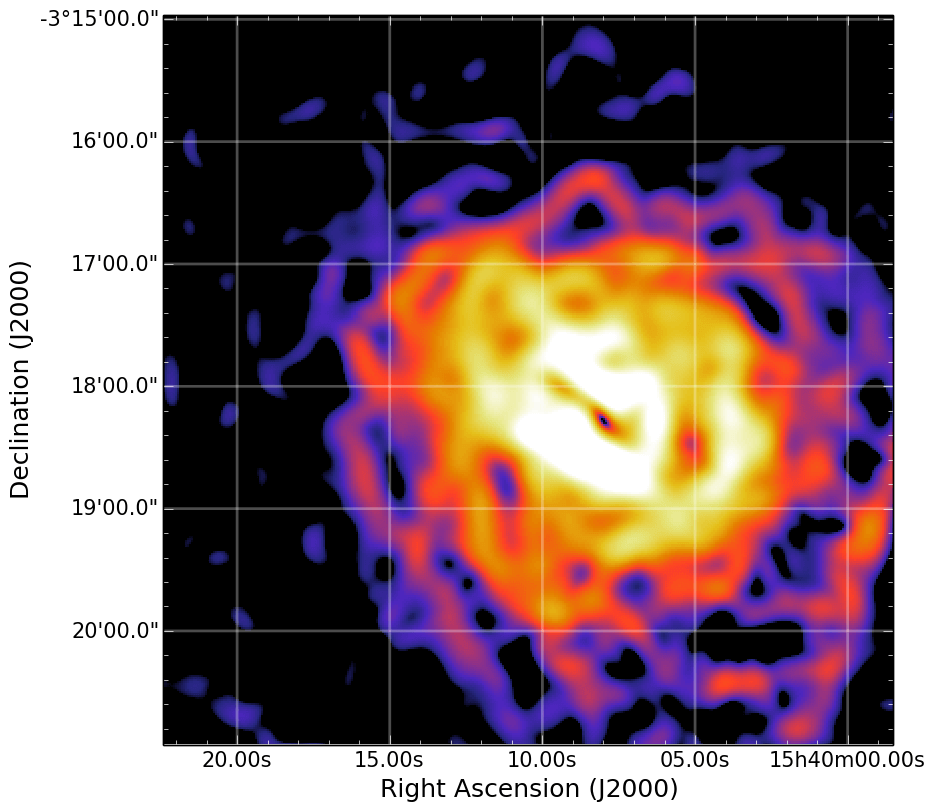}}
 \end{tabular}
 \subfloat{\subfigimgblack[width=.3\textwidth]{\enspace d)}{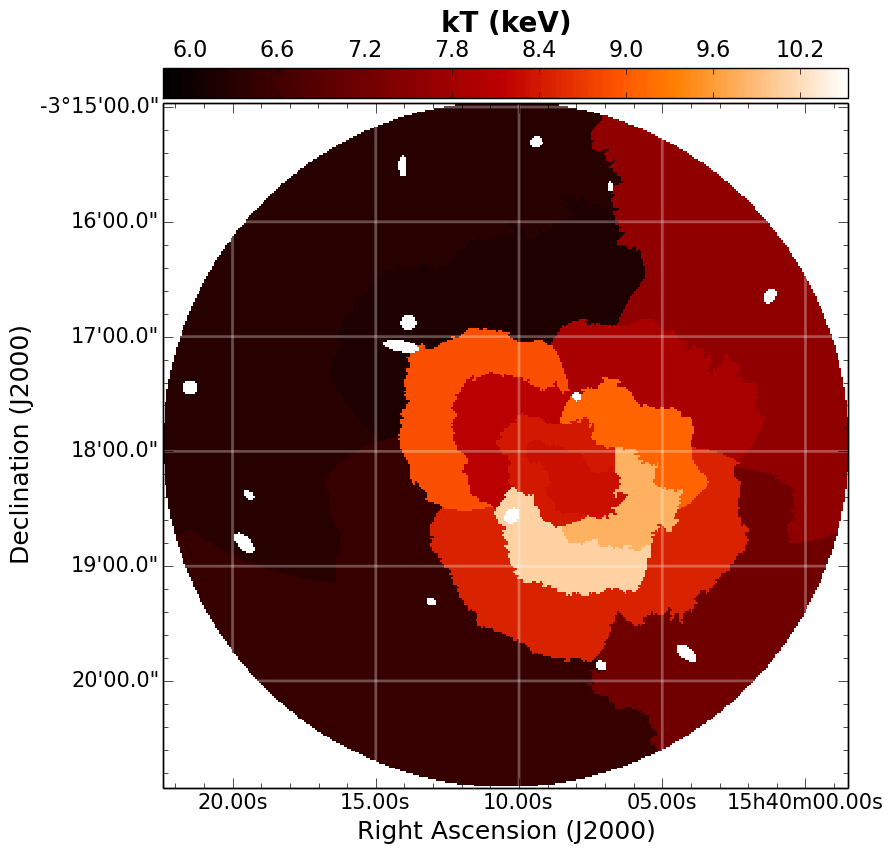}}
 \subfloat{\subfigimgblack[width=.3\textwidth]{\enspace e)}{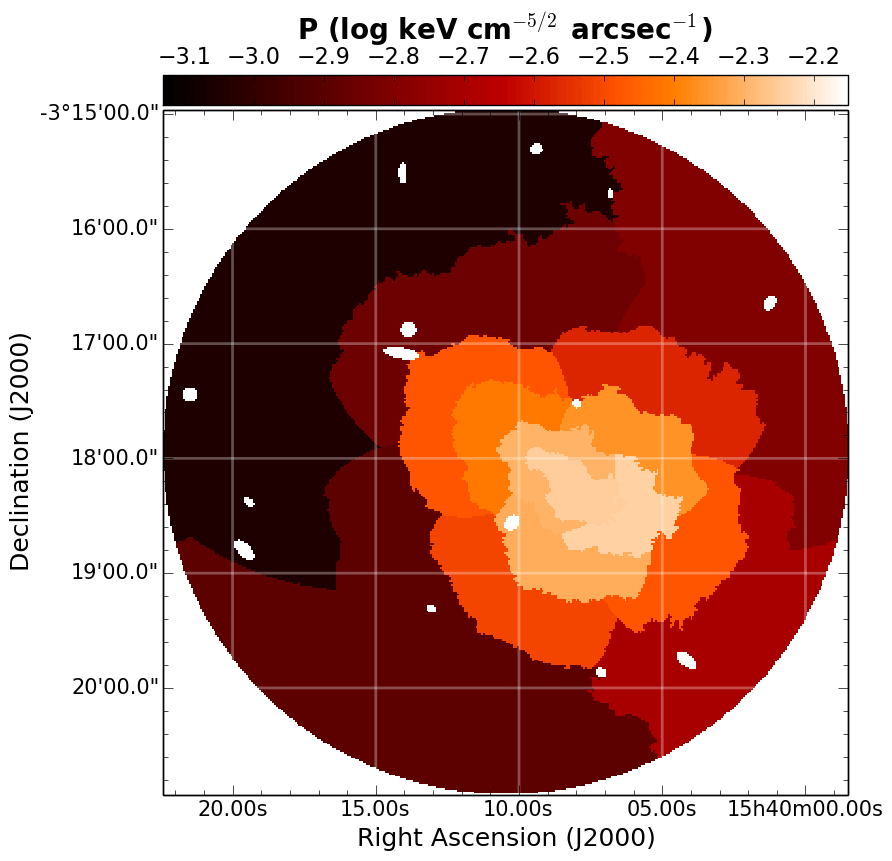}}
 \subfloat{\subfigimgblack[width=.3\textwidth]{\enspace f)}{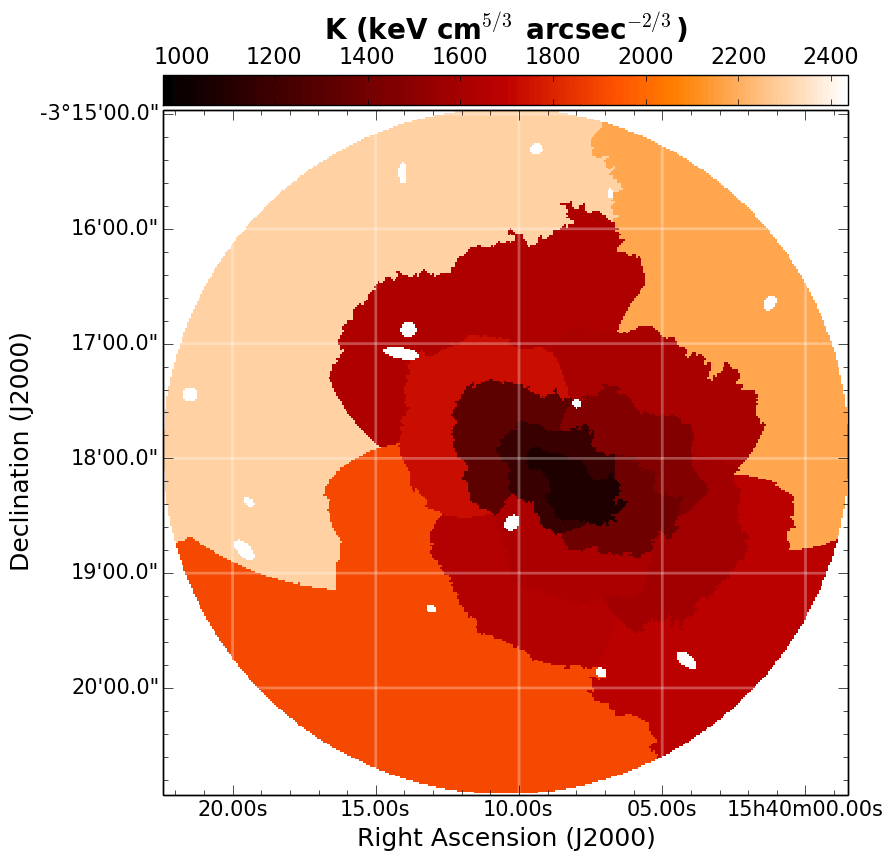}}\\
 \subfloat{\subfigimgsb[width=.3\textwidth,trim={0cm 0cm 4cm 0cm},clip]{g)}{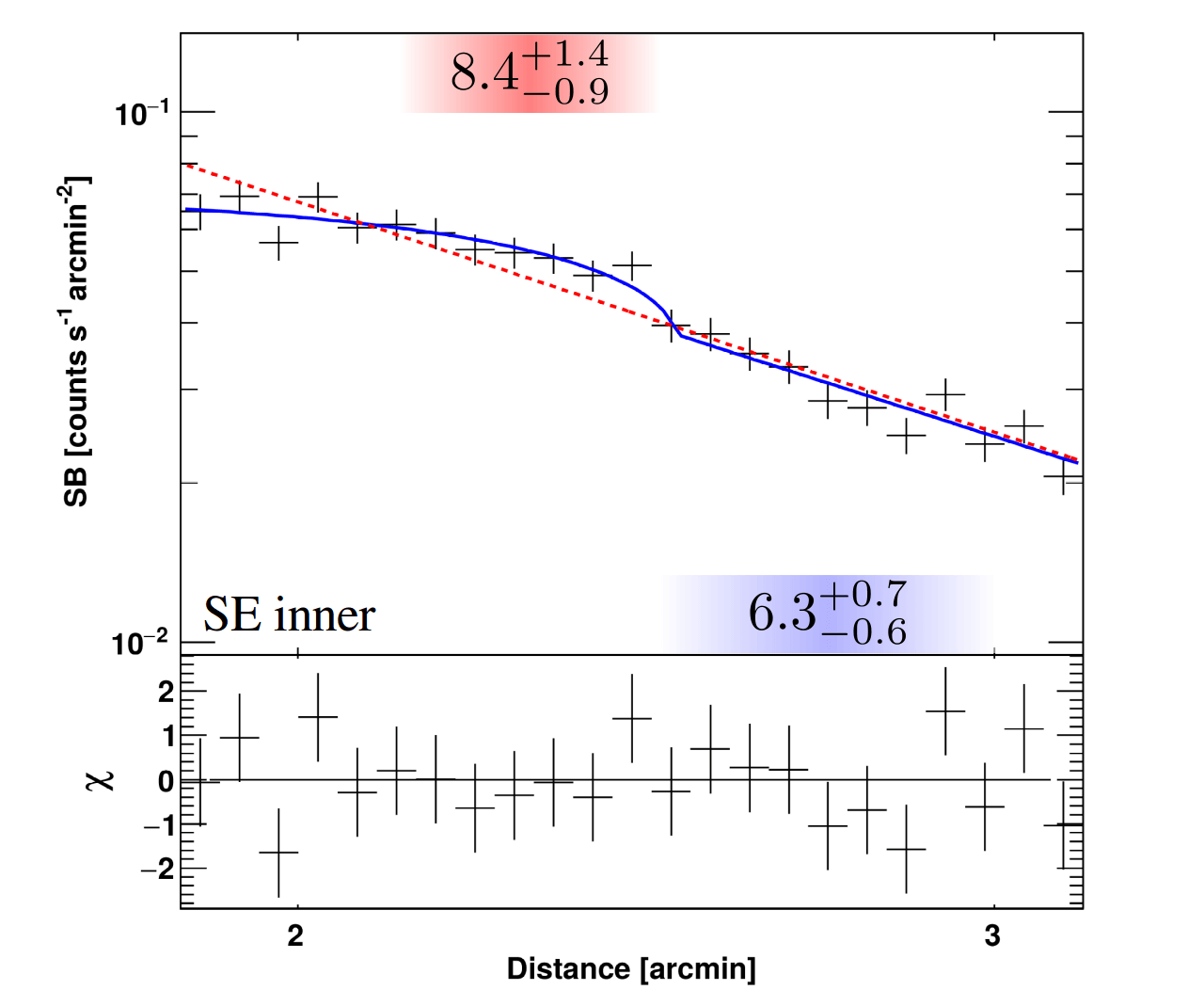}}
 \subfloat{\subfigimgsb[width=.3\textwidth,trim={0cm 0cm 4cm 0cm},clip]{h)}{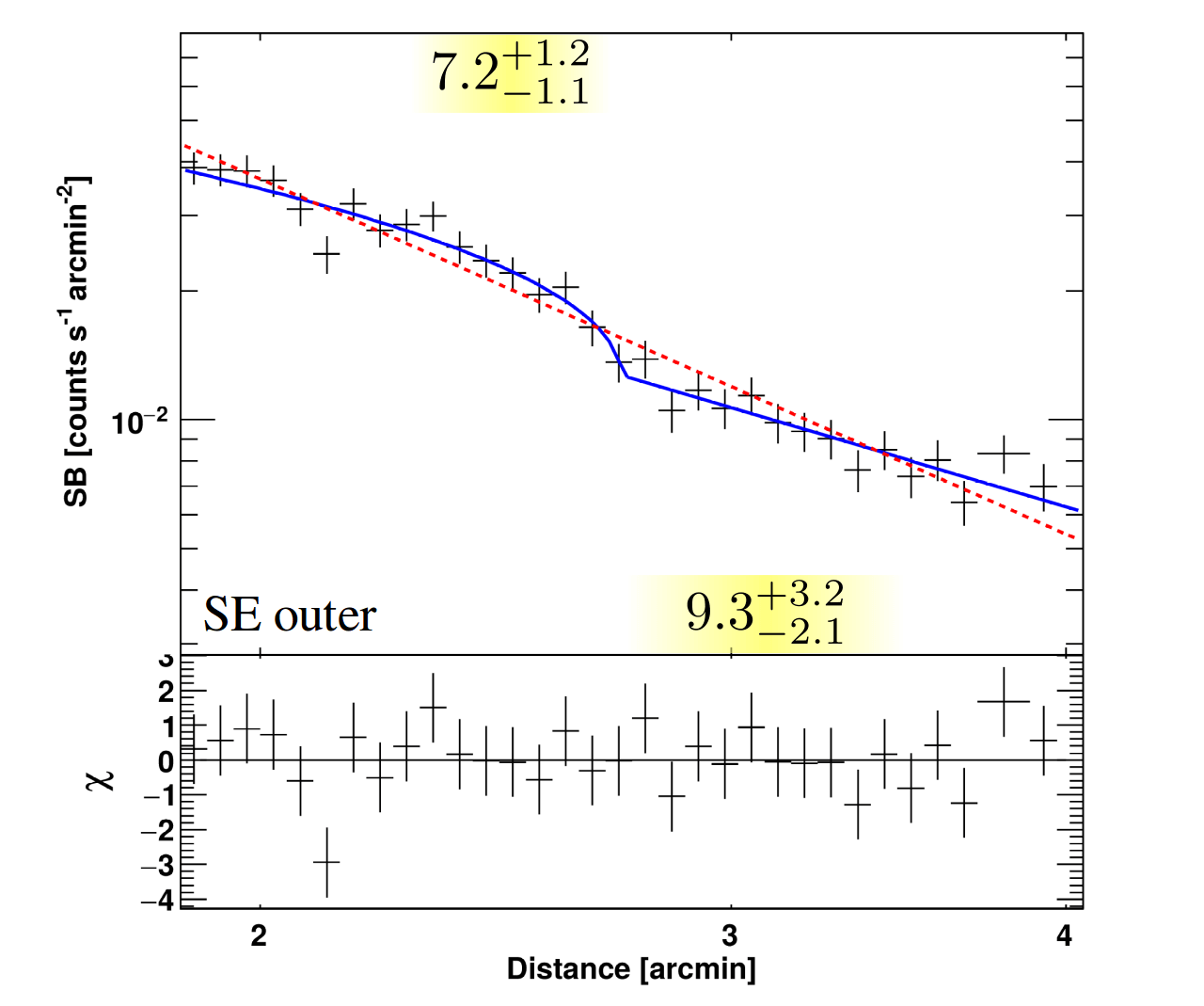}}
 \subfloat{\subfigimgsb[width=.3\textwidth,trim={0cm 0cm 4cm 0cm},clip]{i)}{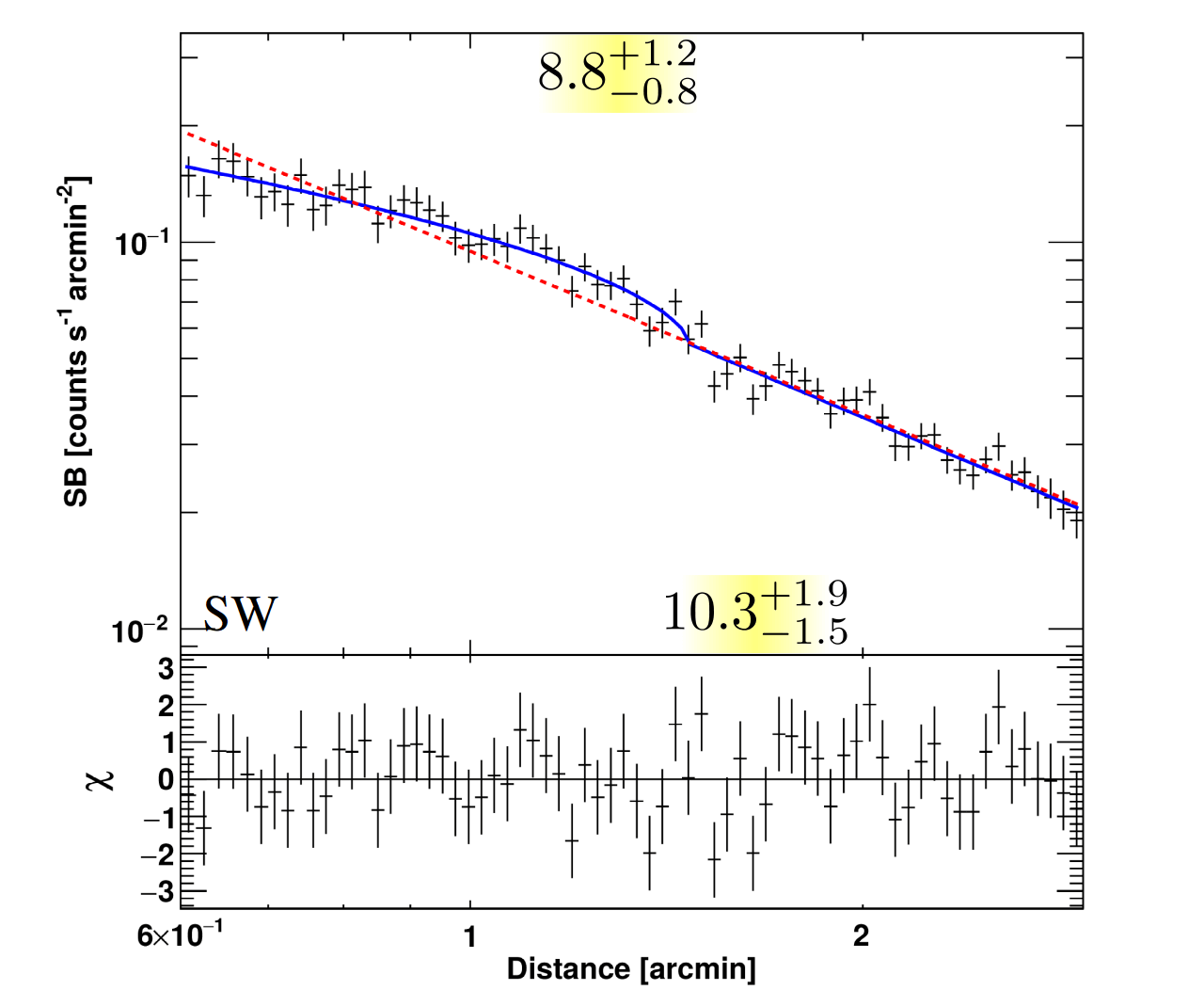}}
 \caption{A2104. The same as for Fig~\ref{fig:a399}. The goodness of fits is reported in Fig.~\ref{fig:a2104_errors}. The positions of the edges are marked in the \chandra\ image in white (shock) and in yellow.}
 \label{fig:a2104}
\end{figure*}

\begin{figure*}
 \centering
 \begin{tabular}{cc}
  \multirow{2}{*}{\subfloat{\subfigimgwhitebig[width=.6\textwidth]{\quad  a)}{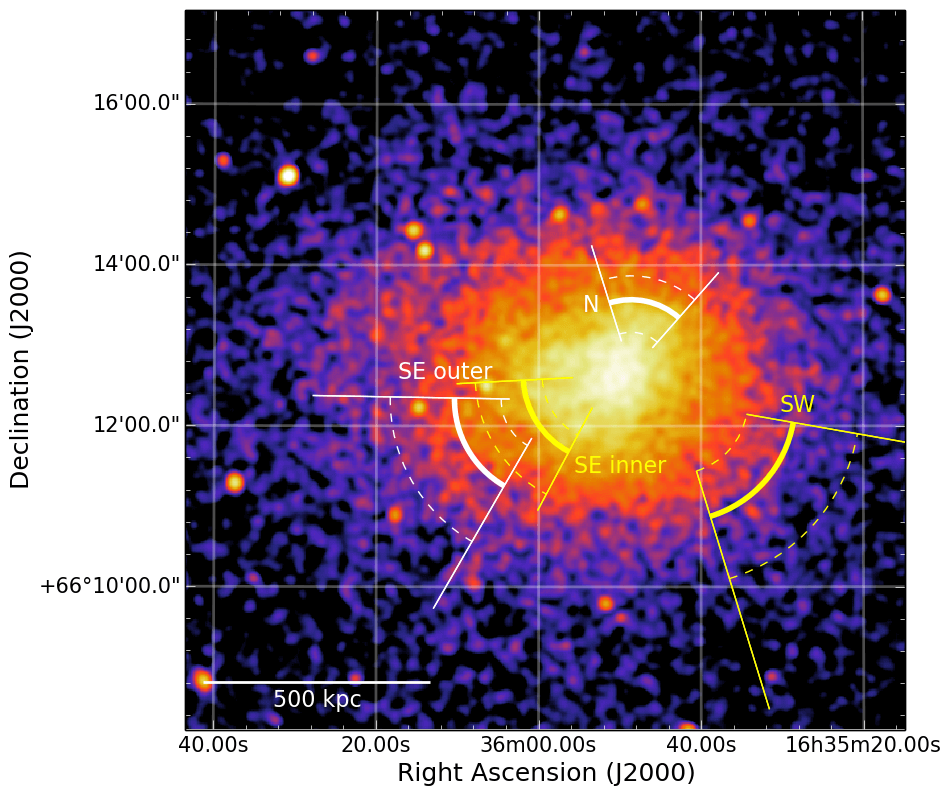}}} & \\
  & \vspace{0.15cm}\hspace{-0.3cm}\subfloat{\subfigimgwhiteggm[width=.28\textwidth]{\quad b)}{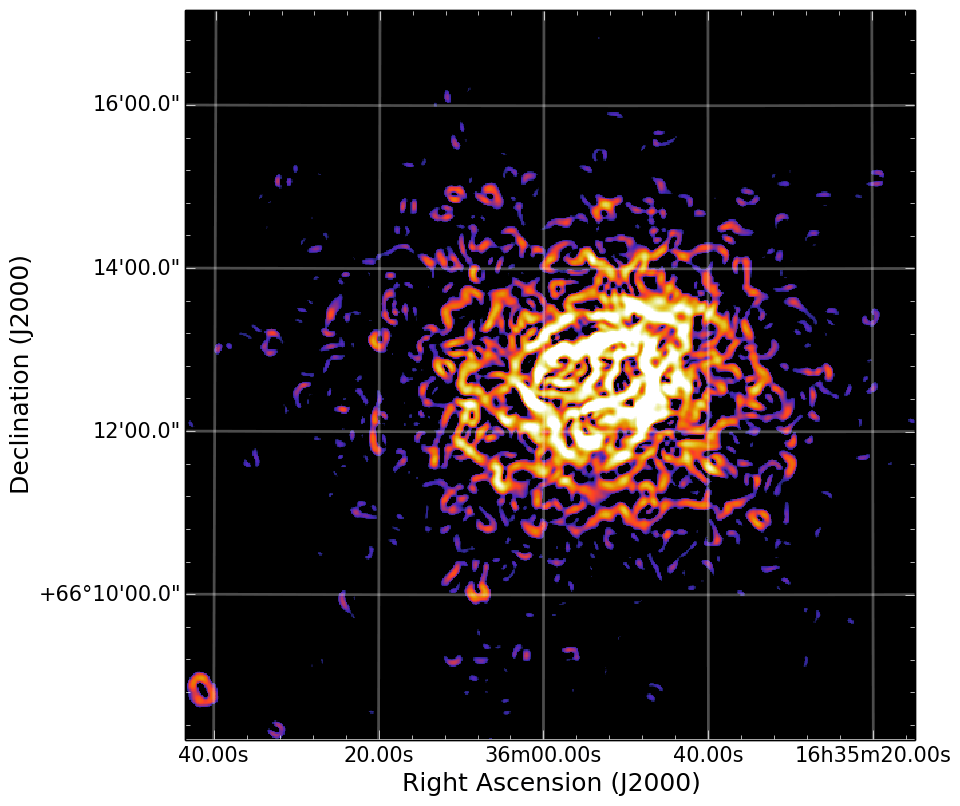}} \\
  & \hspace{-0.3cm}\subfloat{\subfigimgwhiteggm[width=.28\textwidth]{\quad c)}{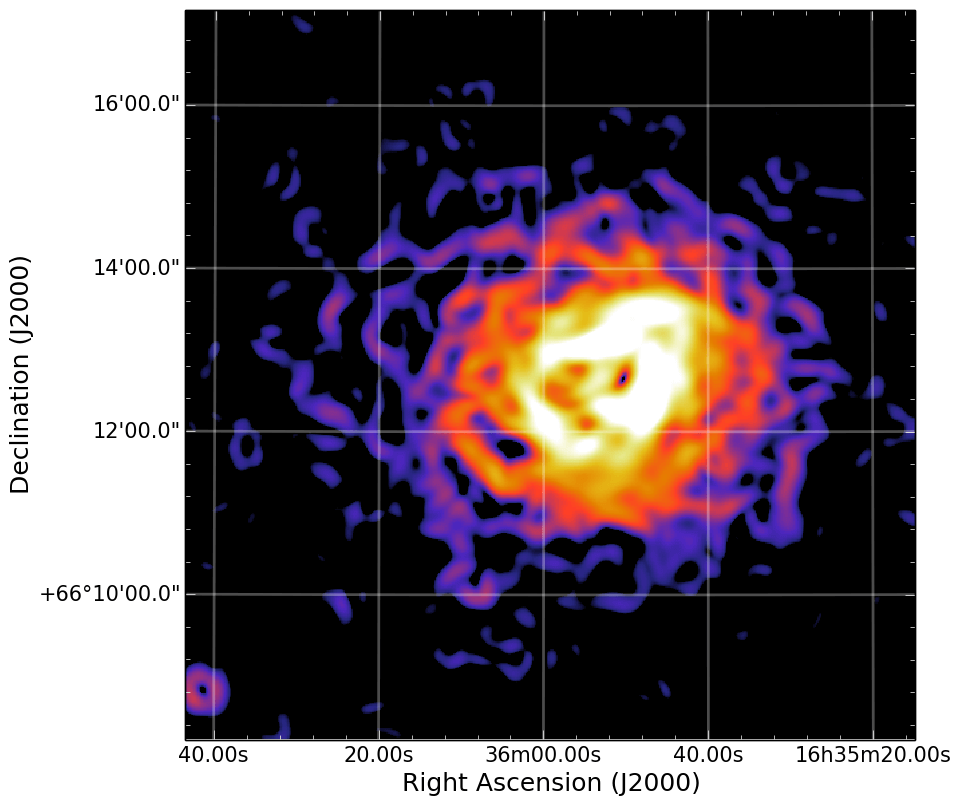}}
 \end{tabular}
 \subfloat{\subfigimgblack[width=.3\textwidth]{\enspace d)}{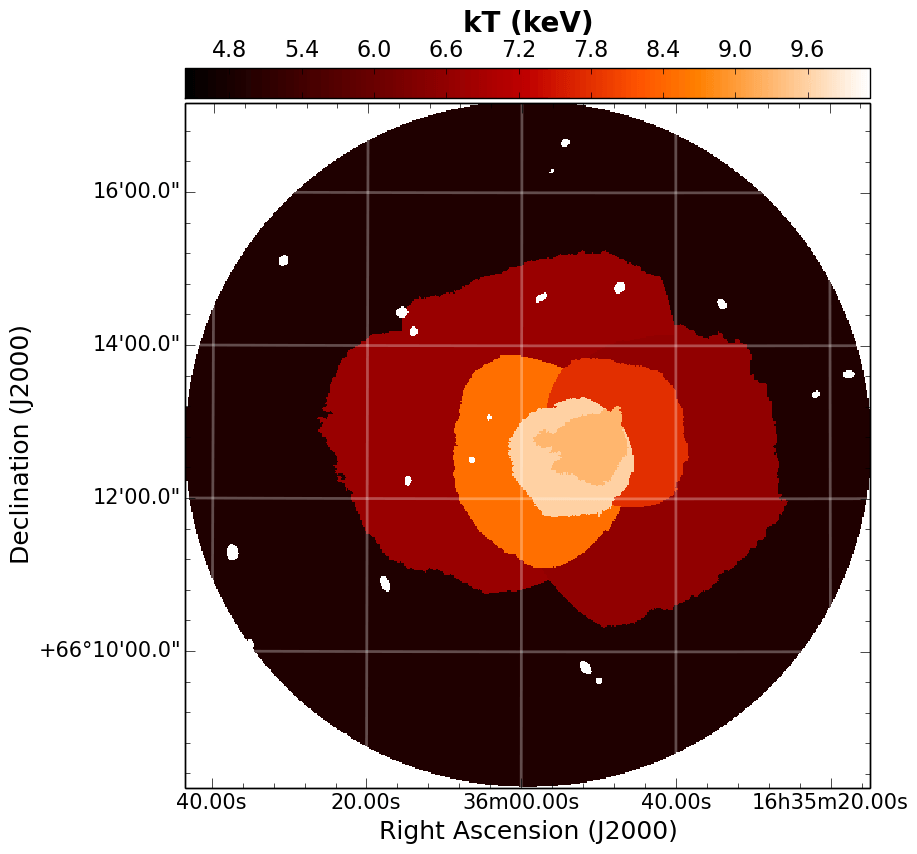}}
 \subfloat{\subfigimgblack[width=.3\textwidth]{\enspace e)}{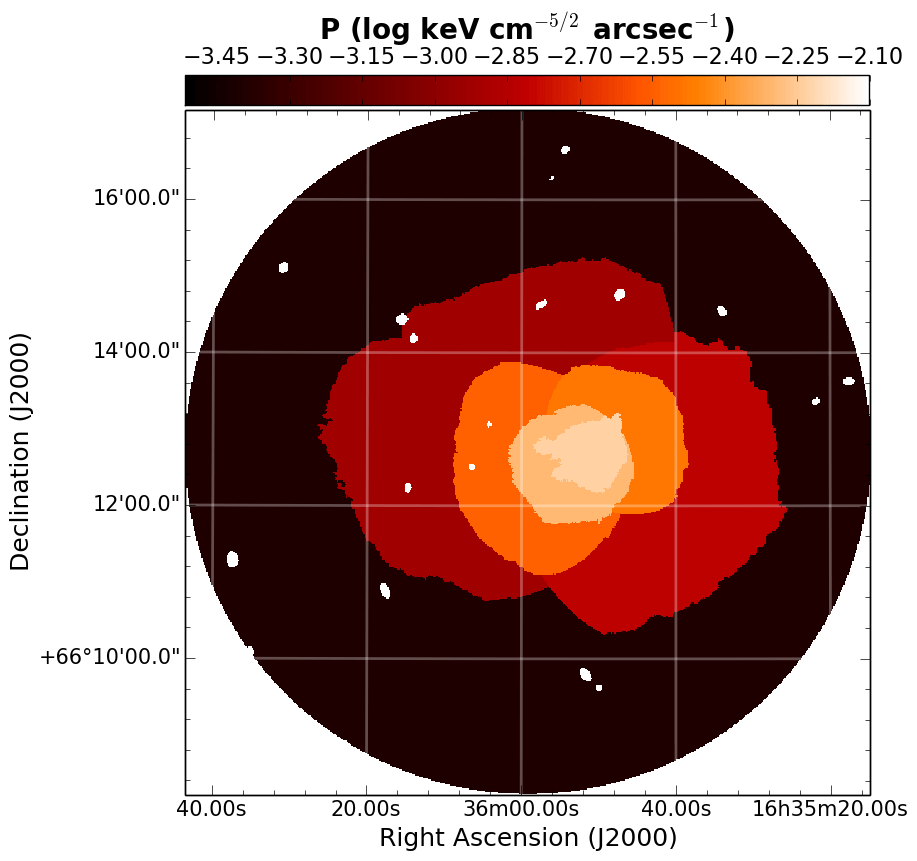}}
 \subfloat{\subfigimgblack[width=.3\textwidth]{\enspace f)}{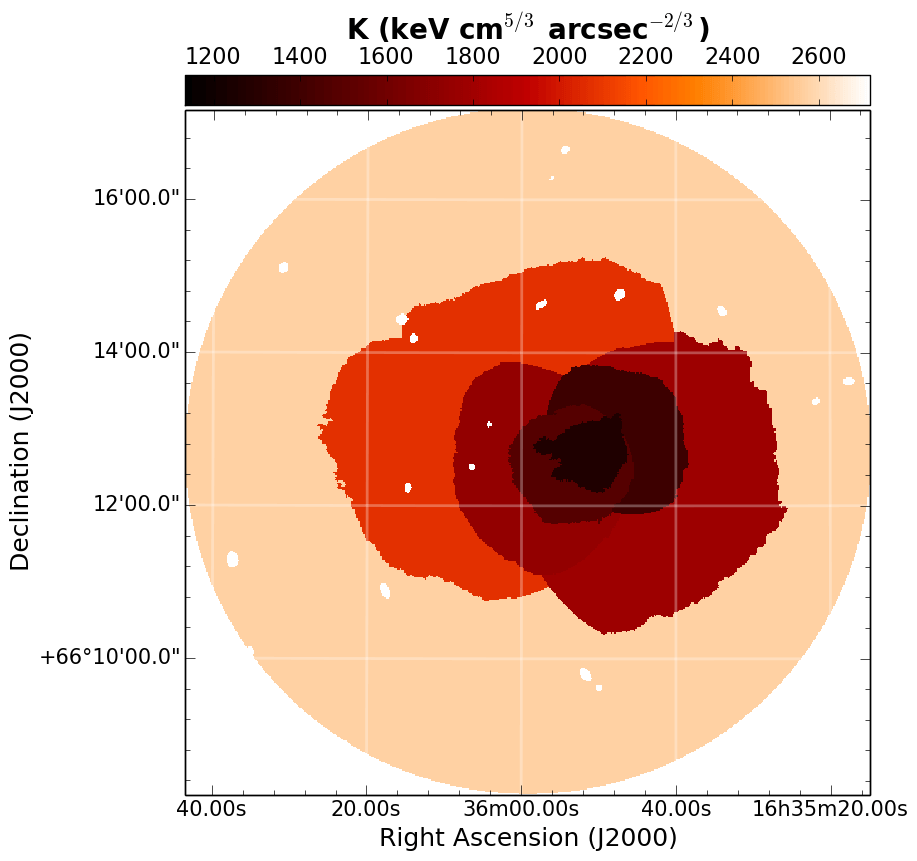}}\\
 \subfloat{\subfigimgsb[width=.3\textwidth,trim={0cm 0cm 4cm 0cm},clip]{g)}{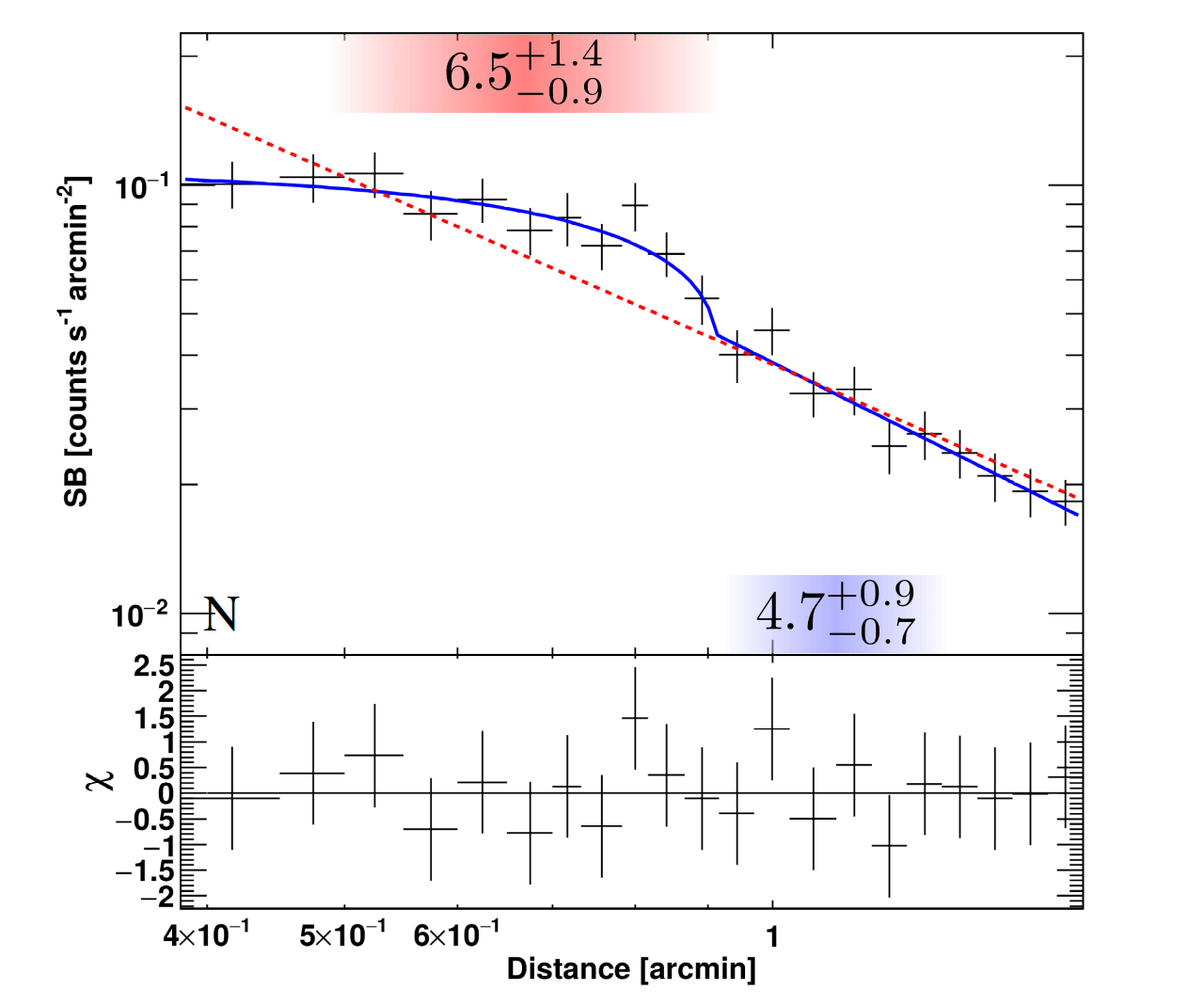}}
 \subfloat{\subfigimgsb[width=.3\textwidth,trim={0cm 0cm 4cm 0cm},clip]{h)}{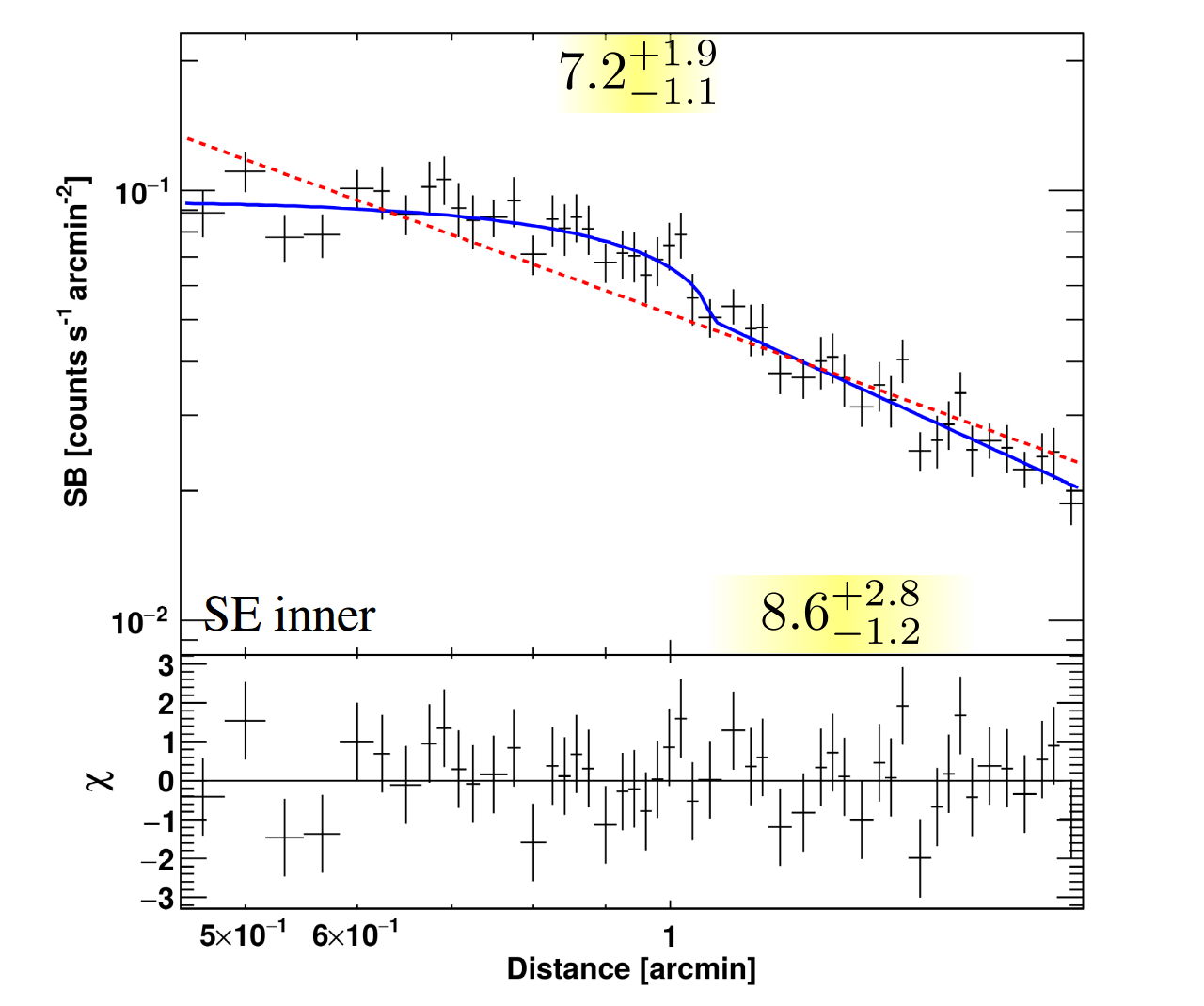}}
 \subfloat{\subfigimgsb[width=.3\textwidth,trim={0cm 0cm 4cm 0cm},clip]{i)}{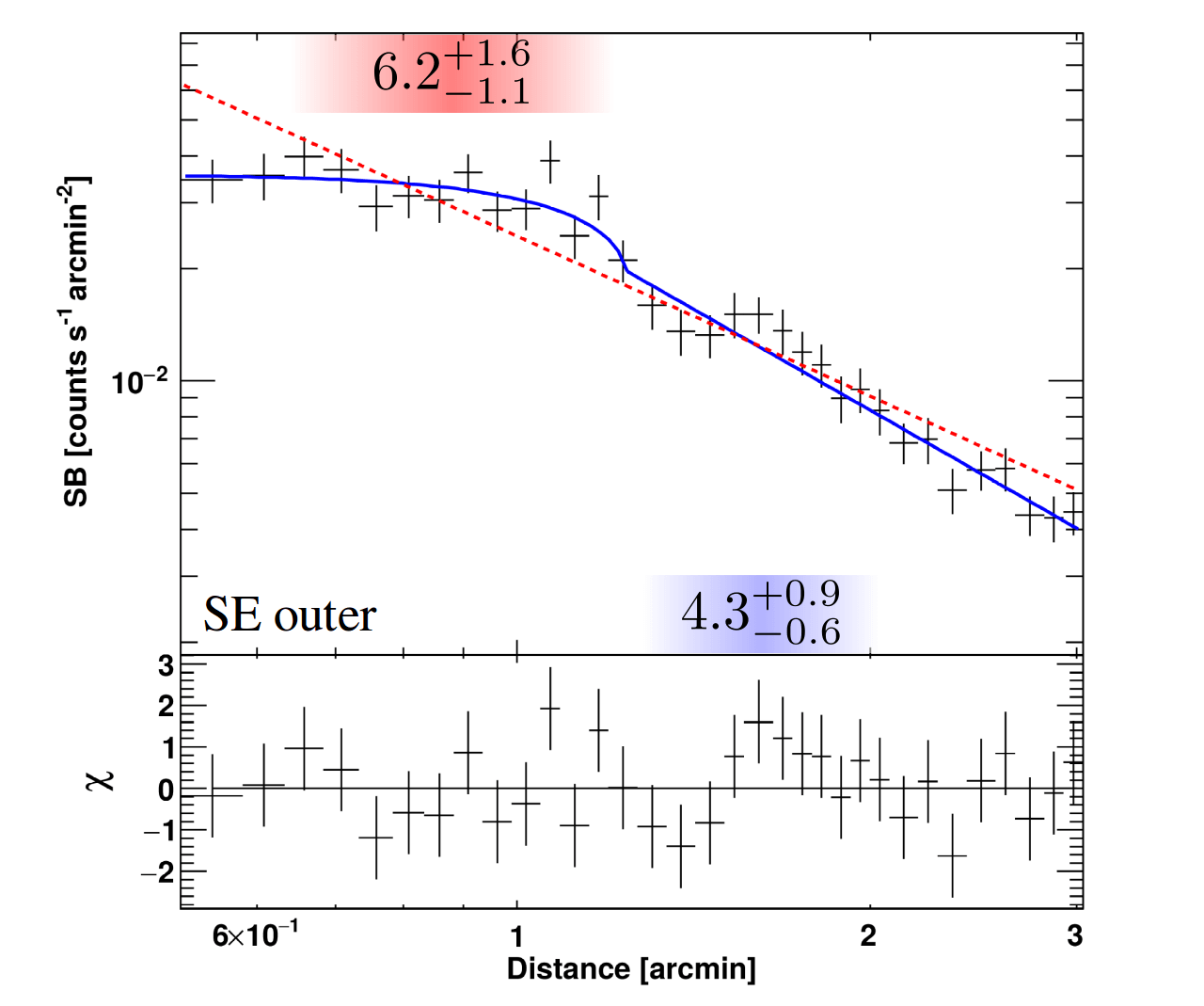}}\\
 \subfloat{\subfigimgsb[width=.3\textwidth,trim={0cm 0cm 4cm 0cm},clip]{j)}{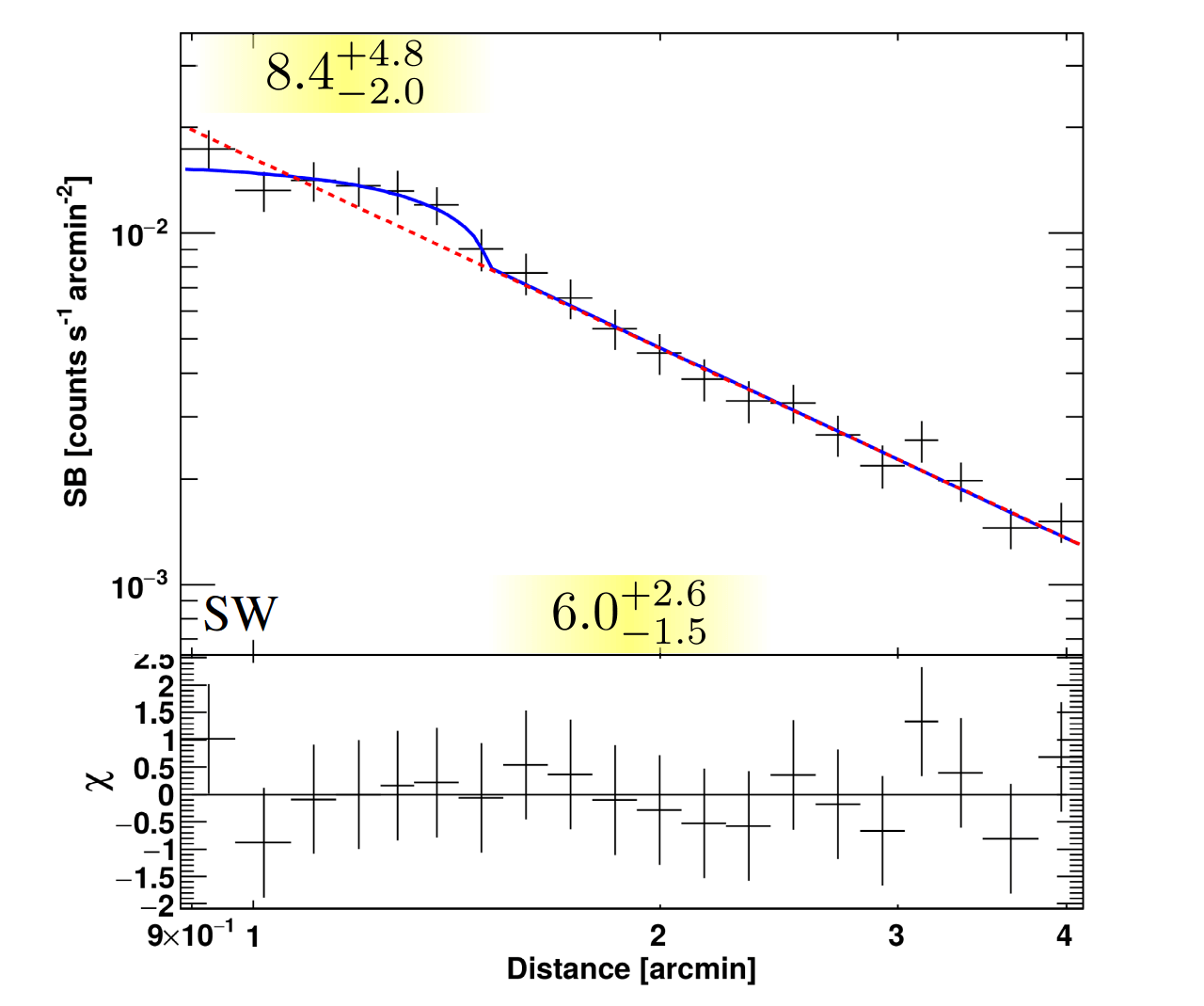}}
 \caption{A2218. The same as for Fig~\ref{fig:a399}. The goodness of fits is reported in Fig.~\ref{fig:a2218_errors}. The positions of the edges are marked in the \chandra\ image in green (cold front), white (shocks) and yellow.}
 \label{fig:a2218}
\end{figure*}

\begin{figure*}
 \centering
 \begin{tabular}{cc}
  \multirow{2}{*}{\subfloat{\subfigimgwhitebig[width=.6\textwidth]{\quad  a)}{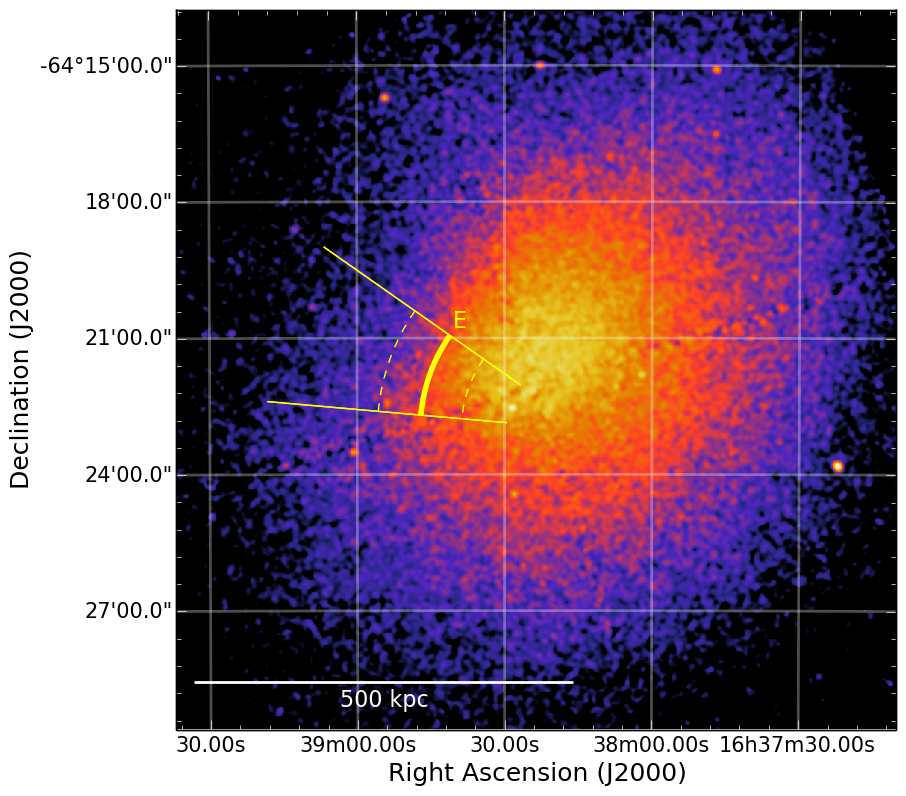}}} & \\
  & \vspace{0.15cm}\hspace{-0.3cm}\subfloat{\subfigimgwhiteggm[width=.28\textwidth]{\quad b)}{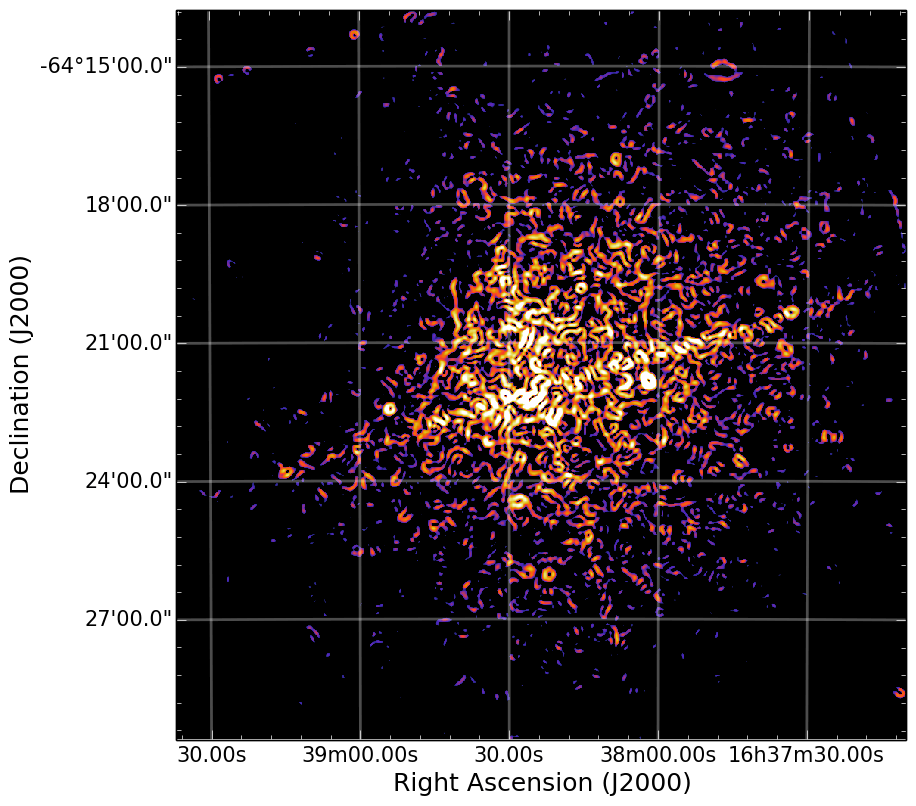}} \\
  & \hspace{-0.3cm}\subfloat{\subfigimgwhiteggm[width=.28\textwidth]{\quad c)}{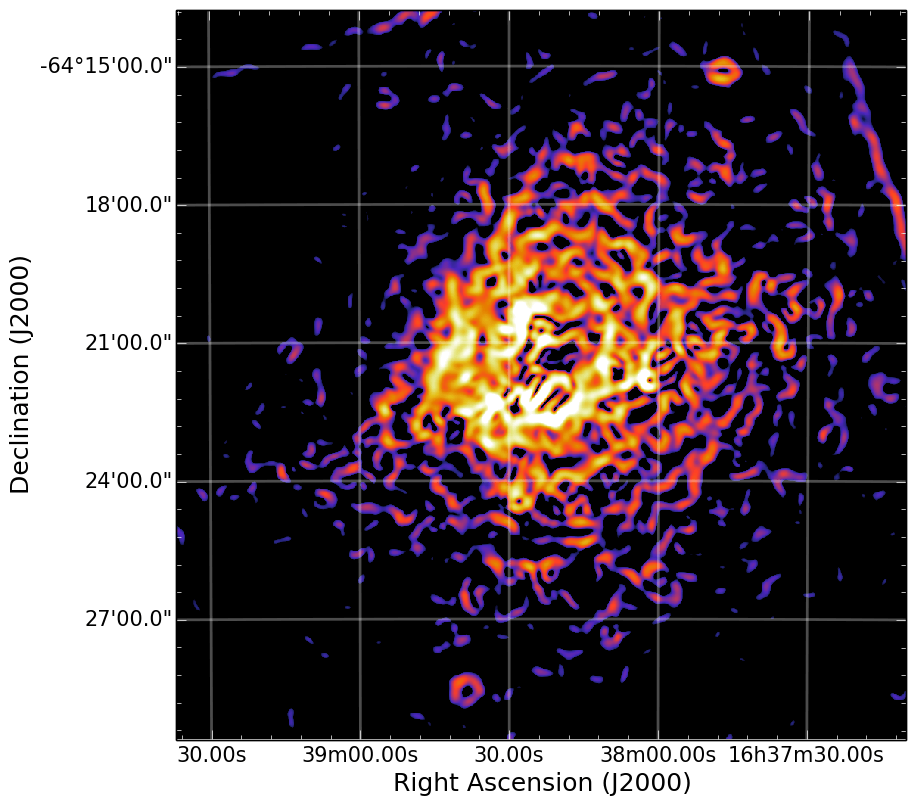}}
 \end{tabular}
 \subfloat{\subfigimgblack[width=.3\textwidth]{\quad d)}{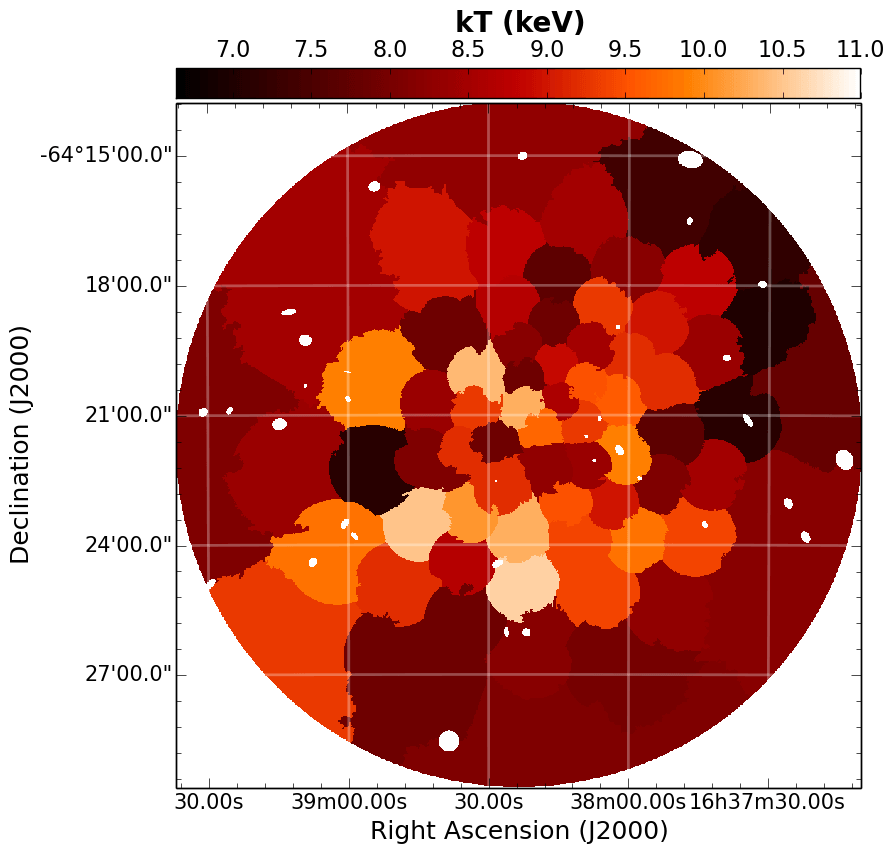}}
 \subfloat{\subfigimgblack[width=.3\textwidth]{\quad e)}{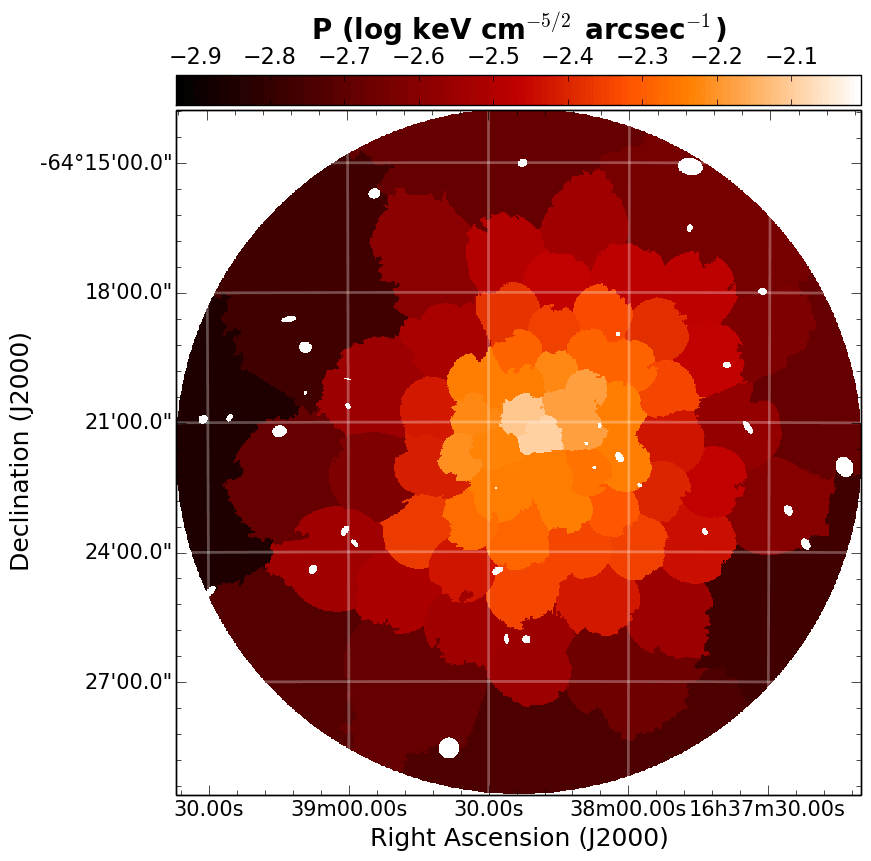}}
 \subfloat{\subfigimgblack[width=.3\textwidth]{\quad f)}{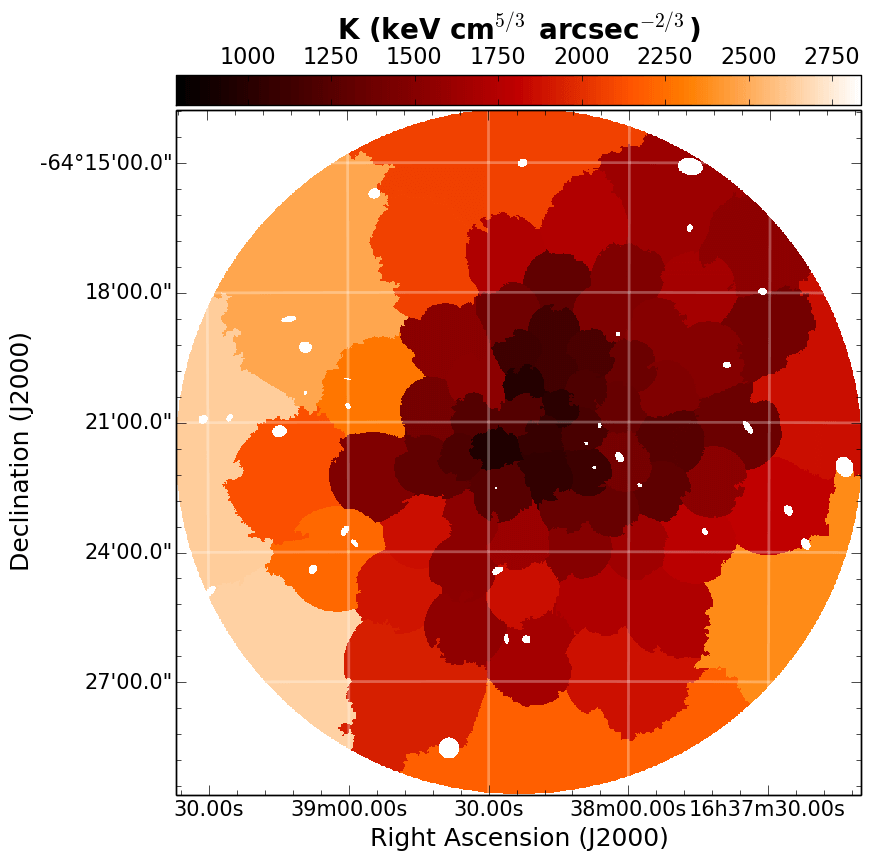}}\\
 \subfloat{\subfigimgsb[width=.3\textwidth,trim={0cm 0cm 4cm 0cm},clip]{g)}{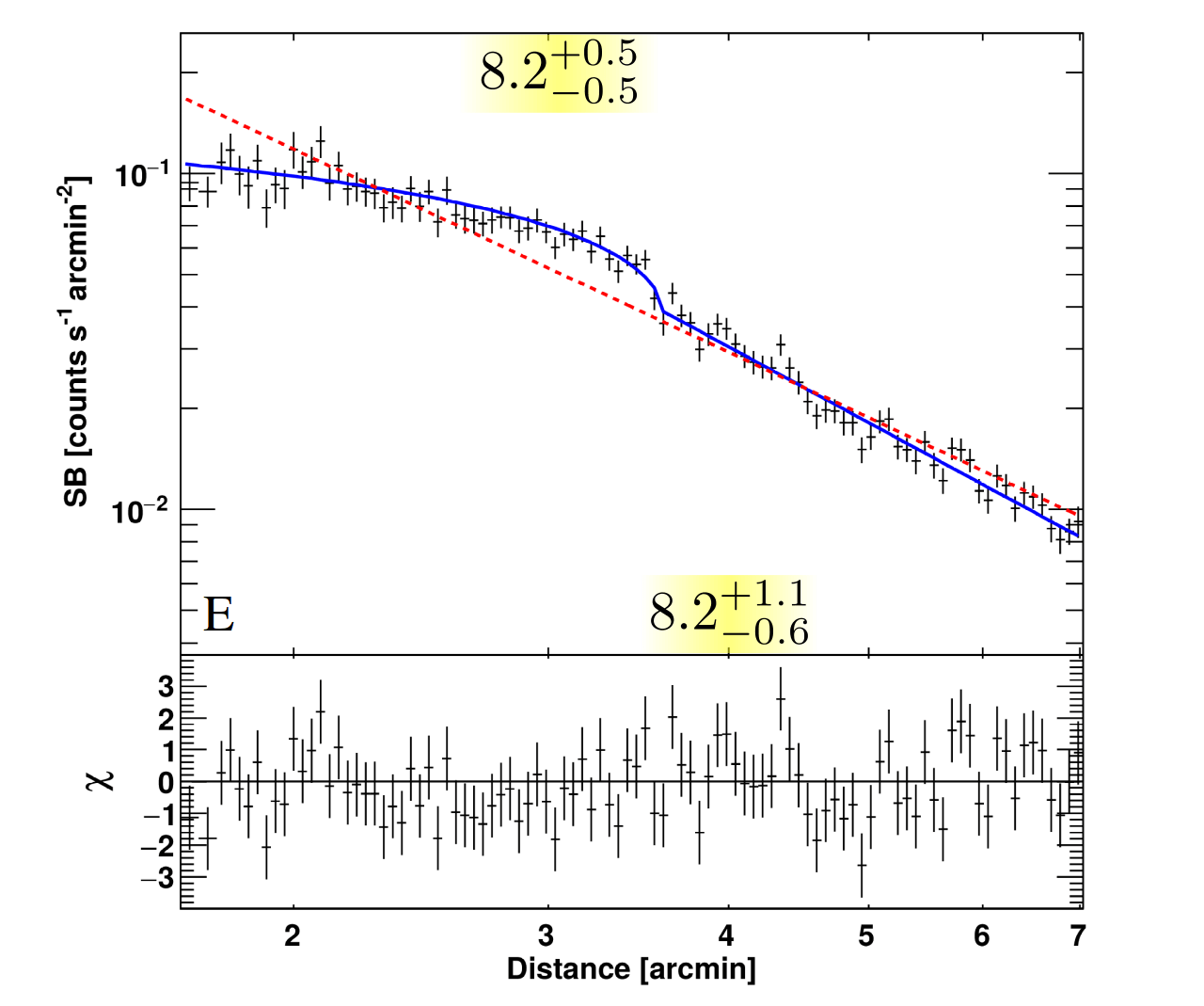}}
 \caption{Triangulum Australis. The same as for Fig~\ref{fig:a399}. The goodness of fits is reported in Fig.~\ref{fig:triangulum_errors}. The position of the edge is marked in the \chandra\ image in  yellow. Note that in the GGM filtered images the straight and perpendicular features are artifacts due to chip gaps.}
 \label{fig:triangulum}
\end{figure*}

\subparagraph{A399 and A401.} These two objects constitute a close system ($z=0.072$ and $z=0.074$, respectively) of two interacting galaxy clusters \citep[\eg][]{fujita96}. Their X-ray morphology is disturbed (Fig.~\ref{fig:a399}a and \ref{fig:a401}a) and the ICM temperature distribution irregular \citep{bourdin08}, revealing the unrelaxed state of the clusters. Recently, \citet{akamatsu17filament} claimed the presence of an accretion shock between the two using \suzaku\ data. This cluster pair hosts two radio halos \citep{murgia10}. The boundary of the halo in A399 is coincident with an X-ray edge, as already suggested by \xmm\ observations \citep{sakelliou04}. \\
Only one \chandra\ observation is available for A399, whereas several observations were performed on A401. Despite this, we only used \obsid\ 14024 (which constitutes the 74\% of the total observing time) to produce the maps shown in Fig.~\ref{fig:a401}d,e,f as the remainder \obsid s are snapshots that partially cover the cluster emission. This is also the only case where we required $\sim5000$ counts in each spectral bin given the combination of high brightness and long exposure on A401. \\
The temperature maps in Fig.~\ref{fig:a399}d and \ref{fig:a401}d indicate an overall hot ICM and the presence of some hot substructures, in agreement with previous studies \citep{sakelliou04, bourdin08}. \\
The GGM images of A399 reveal a SB gradient toward the SE direction. The SB profile across this region and its temperature jump reported in Fig.~\ref{fig:a399}g show that this ``inner'' edge is a cold front with $\rat = 0.74^{+0.14}_{-0.12}$ and $\compr = 1.72^{+0.13}_{-0.12}$. Ahead of that, the X-ray SB rapidly fades away, as well as the radio emission of the halo \citep{murgia10}. The ``outer'' SB profile in this direction indeed shows another discontinuity with $\compr = 1.45^{+0.10}_{-0.10}$ (Fig.~\ref{fig:a399}h). The broken-power law model provides a better description of the data ($\chisqdof = 68.6/72$) compared to a single power-law fit ($\chisqdof = 122.6/74$), corresponding to a null-hypothesis probability of $8 \times 10^{-10}$ ($6.1\sigma$ level) with the F-test. In this case, however, the temperatures across the edge are consistent, not allowing us to firmly claim the nature of the SB jump. We mention that the presence of a shock would be in agreement with the fact that cold fronts sometimes follow shocks \citep[\eg][]{markevitch02bullet} and that shocks might (re)accelerate cosmic rays producing the synchrotron emission at the boundary of some radio halos \citep[\eg][]{shimwell14}. \\
A401 has a more elliptical X-ray morphology and an average temperature higher than A399. The hottest part of the ICM is found on the E direction. Indeed, the GGM image with $\sigma=8$ pixels in Fig.~\ref{fig:a401}c highlights a kind of spiral structure in SB on this side of the cluster, with maximum contrast toward the SE. The SB profile in this sector is well described by a broken power-law with compression factor $\compr = 1.39^{+0.04}_{-0.04}$ (Fig.~\ref{fig:a401}g). The higher temperature in the upstream region ($\ktu = 10.4^{+0.8}_{-0.6}$ \kev\ against $\ktd = 8.1^{+0.4}_{-0.4}$ \kev) confirms that this is a cold front. This could be part of a bigger spiral-shaped structure generated by a sloshing motion.

\subparagraph{MACS J0417.5-1154.} It is the most massive ($\mfive = 1.2\times10^{15}$ \msun) and most distant ($z=0.440$) cluster of our sample. Its extremely elongated X-ray morphology (Fig.~\ref{fig:macsj0417}a) suggests that this cluster is undergoing a high speed merger \citep{ebeling10, mann12}. Despite this, the value of $K_0 = 27\pm7$ \kevcmsq\ indicates that its compact core has not been disrupted yet, acting as a ``bullet'' in the ICM \citep[\eg][for a similar case]{markevitch02bullet}. Radio observations show the presence of a giant radio halo that remarkably follows the ICM thermal emission \citep{dwarakanath11, parekh17six}. \\
The most striking feature of MACS J0417.5-1154 is certainly its prominent cold front in the SE generated by an infalling cold and low-entropy structure, as highlighted by our maps in Fig.~\ref{fig:macsj0417}d,e,f. The SB across this region abruptly drops ($\compr = 2.44^{+0.31}_{-0.25}$) in the upstream region (Fig.~\ref{fig:macsj0417}g), for which spectral analysis provided a clear jump in temperature of $\rat = 0.44^{+0.17}_{-0.10}$, leading us to confirm the cold front nature of the discontinuity. The high-temperature value of $\ktu = 16.9^{+6.1}_{-3.3}$ \kev\ found upstream is an indication of a shock-heated region; a shock is indeed expected in front of the CC similarly to other clusters observed in an analogous state \citep[\eg][]{markevitch02bullet, russell12, botteon16gordo} and is also suggested by our temperature and pseudo-pressure maps. Nonetheless, we were not able to characterize the SB jump of this potential feature. On the opposite side, GGM images pinpoint another edge toward the NW direction, representing again a huge jump ($\compr = 2.50^{+0.29}_{-0.25}$) in the SB profile (Fig.~\ref{fig:macsj0417}h). The spectral analysis in a dedicated region upstream of this feature allowed us only to set a lower limit of $\ktu > 12.7$ \kev, suggesting the presence of a hot plasma, in agreement with our temperature map and the one reported in \citet{parekh17six}, where the pressure is almost continuous (Fig.~\ref{fig:macsj0417}e), as expected for a cold front.

\subparagraph{RXC J0528.9-3927.} No dedicated studies exist on this cluster located at $z=0.284$. The ICM emission is peaked on the cluster core, the coldest region in the cluster \citep{finoguenov05}, and fades in the outskirts where the emission is faint and diffuse (Fig.~\ref{fig:rxcj0528}a). \\
Our maps of the ICM thermodynamical quantities in Fig.~\ref{fig:rxcj0528}d,e,f are rather affected by large spectral bins due to the low counts of the cluster. The X-ray emission is peaked on the central low entropy region, which is surrounded by hot gas. An edge on the W is suggested both from the GGM images and from the above mentioned maps. The SB profile in Fig.~\ref{fig:rxcj0528}g is well fitted with a broken power-law with $\compr = 1.51^{+0.10}_{-0.09}$ and the dedicated spectral analysis confirms the value reported in the temperature map ($\ktu = 10.5^{+3.6}_{-1.8}$ \kev\ and $\ktd = 7.2^{+0.9}_{-0.7}$ \kev), indicating the presence of a cold front. Two more SB gradients pinpointed in the GGM images to the E and W directions did not provide evidence for any edge with the SB profile fitting (Fig.~\ref{fig:rxcj0528_noedge}).

\subparagraph{MACS J0553.4-3342.} It is a distant cluster ($z=0.407$) in a disturbed dynamical state, as shown from both optical and X-ray observations \citep{ebeling10, mann12}. The X-ray morphology (Fig.~\ref{fig:macsj0553}a) suggests that a binary head-on merger is occurring approximately in the plane of the sky \citep{mann12}. No value of the central entropy $K_0$ is reported in \citet{cavagnolo09} nor in \citet{giacintucci17}. A radio halo that follows the ICM emission has been detected in this system \citep{bonafede12}. At the time of this writing, two more papers on MACS J0553.4-3342, both containing a joint analysis of \hstE\ and \chandra\ observations, were published \citep{ebeling17, pandge17}. \\
The maps of the ICM thermodynamical quantities shown in Fig.~\ref{fig:macsj0553}d,e,f further support the scenario of an head-on merger in the E-W direction for MACS J0553.4-3342 in which a low-entropy structure is moving toward E, where GGM images highlight a steep SB gradient. This is confirmed by the SB profile fit (Fig.~\ref{fig:macsj0553}g) that leads to a compression factor of $\compr = 2.49^{+0.32}_{-0.26}$, while the temperature jump found by spectral analysis of $\rat = 0.62^{+0.33}_{-0.18}$ indicates that this discontinuity is a cold front \citep[see also][]{ebeling17, pandge17}. The high value of $\ktu=13.7^{+6.9}_{-3.7}$ \kev\ suggests a shock-heated region to the E of the cold front; indeed the ``outer'' SB profile of Fig.~\ref{fig:macsj0553}h indicates the presence of an edge in the cluster outskirts. We used for the characterization of the SB profile a sector of aperture $133\deg-193\deg$ (where the angles are measured in an anticlockwise direction from W) whereas we used a wider sector ($133\deg-245\deg$) as depicted in Fig.~\ref{fig:macsj0553}a to extract the spectra in order to ensure a better determination of the downstream and upstream temperatures, which ratio of $\rat = 2.00^{+1.14}_{-0.63}$ confirms the presence of a shock with Mach number $\machsb = 1.58^{+0.30}_{-0.22}$ and $\machkt = 1.94^{+0.77}_{-0.56}$, respectively derived from the SB and temperature jumps. This edge is spatially connected with the boundary of the radio halo found by \citet{bonafede12}. On the opposite side of the cluster, another roundish SB gradient is suggested from the inspection of the GGM images (Fig.~\ref{fig:macsj0553}b,c). The W edge is well described by our fit (Fig.~\ref{fig:macsj0553}i) that leads to $\compr = 1.70^{+0.12}_{-0.11}$, while spectral analysis provides $\rat = 0.33^{+0.22}_{-0.12}$, consistent with the presence of another cold front. Even though the upstream temperature is poorly constrained, the spectral fit suggests high temperature values, also noticed in \citet{ebeling17}, possibly indicating another shock-heated region ahead of this cold front; however, the presence of a possible discontinuity associated with this shock can not be claimed with current data. The symmetry of the edges strongly supports the scenario of a head-on merger in the plane of the sky. However, the serious challenges to this simple interpretation described in \citet{ebeling17} in term of the relative positions of the brightest central galaxies, X-ray peaks, and dark matter distributions need to be reconsidered in view of the presence and morphology of the extended X-ray tail discussed in \citet{pandge17} and clearly highlighted by the GGM image (see Fig.~\ref{fig:macsj0553}c).

\subparagraph{AS592.} Known also with the alternative name RXC J0638.7-5358, this cluster located at $z=0.222$ is one of those listed in the supplementary table of southern objects of \citet{abell89}. The ICM has an overall high temperature \citep{menanteau10, mantz10scaling} and is clearly unrelaxed (Fig.~\ref{fig:as592}a), despite the fact that AS592 has one of the lowest $K_0$ value of our sample (\cf\ Tab.~\ref{tab:sample}). \\
The maps in Fig.~\ref{fig:as592}d,e,f highlight the presence of two low entropy and low temperature CCs surrounded by an overall hot ICM. In the SW, a feature in SB is suggested from the GGM image with $\sigma=8$ pixels. The analysis of the X-ray profile and spectra across it result in a SB discontinuity with compression factor $\compr = 1.99^{+0.17}_{-0.15}$ and temperature ratio $\rat = 1.61^{+0.66}_{-0.43}$ (Fig.~\ref{fig:as592}g), leading us to claim the presence of a shock front with Mach number derived from the SB jump of $\machsb = 1.72^{+0.15}_{-0.12}$, in agreement with that derived by the temperature jump $\machkt = 1.61^{+0.54}_{-0.42}$. The SB variation indicated by the GGM images toward the NE direction did not show the presence of a discontinuity with the SB profile fitting (Fig.~\ref{fig:as592_noedge}).

\subparagraph{A1914.} It is a system at $z=0.171$ in a complex merger state \citep[\eg][]{barrena13}, geometry of which is still not understood well \citep{mann12}. In particular, the irregular mass distribution inferred from weak lensing data \citep{okabe08} is puzzling if compared to near-spherical X-ray emission of the ICM on larger scales (Fig.~\ref{fig:a1914}a). Previous \chandra\ studies highlighted the presence of a heated ICM with temperature peak in the cluster center \citep{govoni04chandra, baldi07}. At low frequency, a bright steep spectrum source 4C~38.39 \citep{roland85} and a radio halo \citep{kempner01} are detected. \\
Among the two \chandra\ observations on A1914 retrieved from the archive we had to discard \obsid\ 542 since it took place in an early epoch of the \chandra\ mission, as described above for the case of A370 (see also notes in Tab.~\ref{tab:chandra_obs}). We mention that in the \chandra\ archive other four datasets (\obsid s 12197, 12892, 12893, 12894) can be found for A1914. However these are 5~ks snapshots pointed in four peripheral regions of the cluster that are not useful for our edge research; for this reason, they were excluded in our analysis. \\
Our maps of the ICM thermodynamical quantities in Fig.~\ref{fig:a1914}d,e,f indicate the presence of a bright low-entropy region close to the cluster center with a lower temperature with respect to an overall hot ICM. The adjacent spectral bin toward the E suggests the presence of high-temperature gas while GGM images indicate a rapid SB variation. This feature is quite sharpened, recalling the shape of a tip, and can not be described under a spherical assumption. For this reason two different, almost perpendicular, sectors were chosen to extract the SB profiles to the E, one in an ``upper'' (toward the NE) and one a ``lower'' (toward the SE) direction of the tip. Their fits in Fig.~\ref{fig:a1914}g,h both indicate a similar drop in SB ($\compr \sim 1.5$). Spectra were instead fitted in joint regions downstream and upstream of the two SB sectors, leading to a single value for \ktu\ and \ktd. The temperature jump is consistent with a cold front ($\rat = 0.40^{+0.21}_{-0.12}$). Although the large uncertainties, spectral analysis provides indication of a high upstream temperature, likely suggesting the presence of a shock-heated region. This scenario is similar to the Bullet Cluster \citep{markevitch02bullet} and to the above-mentioned MACS J0417.5-1154. A shock moving into the outskirts can not be claimed with the current data but it is already suggested in Fig.~\ref{fig:a1914}g,h by the hint of a slope change in the upstream power-law in correspondence of the outer edge of the region that we used to extract the upstream spectrum. Another SB feature to the W direction is highlighted by the GGM images and confirmed by the profile shown in Fig.~\ref{fig:a1914}i. Its compression factor of $\compr = 1.33^{+0.08}_{-0.07}$ and temperature ratio achieved from spectral analysis of $\rat = 1.27^{+0.26}_{-0.21}$ allow us to claim the presence of a weak shock with Mach number consistently derived from the SB and temperature jumps, \ie\ $\machsb = 1.22^{+0.06}_{-0.05}$ and $\machkt = 1.28^{+0.26}_{-0.21}$ respectively. This underlines the striking similarly of A1914 with other head-on mergers where a counter-shock (\ie\ a shock in the opposite direction of the infalling subcluster) has been detected, such as the Bullet cluster \citep{shimwell15} and El Gordo \citep{botteon16gordo}, for which it also shares a similar double tail X-ray morphology. 

\subparagraph{A2104.} This is a rich cluster at $z=0.153$. Few studies exist in the literature on A2104. \citet{pierre94} first revealed with \rosat\ that this system is very luminous in the X-rays and has a hot ICM. This result was confirmed more recently with \chandra\ \citep{gu09}, which also probed a slight elongation of the ICM in the NE-SW direction (Fig.~\ref{fig:a2104}a), and a temperature profile declining toward the cluster center \citep{baldi07}. \\
The maps of the ICM thermodynamical quantities (Fig.~\ref{fig:a2104}d,e,f) and GGM filtered images (Fig.~\ref{fig:a2104}b,c) of A2104 confirm an overall high temperature of the system as well as some SB contrasts in the ICM. We extract SB profiles across two sectors toward the SE and one toward the SW directions. The most evident density jump ($\compr = 1.54^{+0.16}_{-0.14}$) is detected for the SE ``outer'' sector shown in Fig.~\ref{fig:a2104}h, while the others show only the hint of a discontinuity (Fig.~\ref{fig:a2104}g,i). However, the fit statistics of the broken power-law and single power-law models indicate that the jump model is in better agreement with the data in both the cases, being respectively $\chisqdof=17.2/16$ and $\chisqdof=37.4/18$ for the SE ``inner'' sector ($3.1\sigma$ significance, F-test analysis) whereas it is $\chisqdof=64.5/63$ and $\chisqdof=122.5/65$ for the SW sector ($6.0\sigma$ significance, F-test analysis). Spectral analysis allowed us only to find a clear temperature jump for the SE ``inner'' edge, leaving the nature of the other two SB jumps more ambiguous. The temperature ratio across the SE ``inner'' sector is $\rat = 1.33^{+0.27}_{-0.19}$, leading us to claim a shock with Mach number $\machkt = 1.34^{+0.26}_{-0.20}$, comparable to the one computed from the upper limit on the compression factor ($\compr < 1.47$) of the SB jump, \ie\ $\machsb < 1.32$. 

\subparagraph{A2218.} Located at $z=0.176$, this cluster is one of the most spectacular gravitational lens known \citep{kneib96}. The system is in a dynamically unrelaxed state, as revealed by its irregular X-ray emission (Fig.~\ref{fig:a2218}a; \citealt{machacek02}) and by the substructures observed in optical \citep{girardi97}. Detailed spectral analysis already provided indication of a hot ICM in the cluster center \citep{govoni04chandra, pratt05, baldi07}. A small and faint radio halo has also been detected in this system \citep{giovannini00}. \\
Four \chandra\ observations exist on A2218. Unfortunately, two of these (\obsid s 553 and 1454) can not be used for the spectral analysis because, as mentioned above for A370 and A1914, they are early \chandra\ observations for which the ACIS background modeling is not possible (see notes in Tab.~\ref{tab:chandra_obs} for more details), hence we only used the remainder two \obsid s to produce the maps shown in Fig.~\ref{fig:a2218}. \\
The low counts on A2218 result in maps of the ICM thermodynamical quantities with large bins, as shown in Fig.~\ref{fig:a2218}d,e,f. The ICM temperature is peaked toward the cluster center, in agreement with previous studies \citep{pratt05, baldi07}. The analysis of GGM images highlights the presence of rapid SB variations in more than one direction. The SB profile toward the N shows the greatest of these jumps, corresponding to $\compr = 1.47^{+0.21}_{-0.18}$ (Fig.~\ref{fig:a2218}g). From the spectral analysis we achieve a temperature ratio $\rat = 1.38^{+0.40}_{-0.28}$ across the edge, indicating the presence of a shock with consistent Mach number derived from the SB jump, \ie\ $\machsb = 1.32^{+0.15}_{-0.13}$, and from the temperature jump, \ie\ $\machkt = 1.39^{+0.37}_{-0.29}$. The presence of a shock in this cluster region is consistent with the temperature map variations reported in \citet{govoni04chandra}. In the SE direction, there is indication of two discontinuities from the SB profile analysis (Fig.~\ref{fig:a2218}h,i): spectra suggest that the ``inner'' discontinuity is possibly a cold front (however the temperature jump is not clearly detected, \ie\ $\rat = 0.84^{+0.35}_{-0.17}$) while the ``outer'' one is consistent with a shock ($\rat = 1.44^{+0.48}_{-0.33}$) and might be connected with the SE edge of the radio halo. The shock Mach numbers derived from SB and temperature jump are $\machsb = 1.17^{+0.10}_{-0.09}$ and $\machkt = 1.45^{+0.43}_{-0.33}$, respectively. The SB profile taken in the SW region shows the hint of a kink (Fig.~\ref{fig:a2218}j); in this case the broken power-law model ($\chisqdof = 7.0/15$) yields to an improvement compared to a single power-law fit ($\chisqdof = 15.0/17$), which according to the F-test corresponds to a null-hypothesis probability of $3 \times 10^{-3}$ ($3.0\sigma$ level). Spectral analysis leaves the nature of this feature uncertain.

\subparagraph{Triangulum Australis.} It is the closest ($z=0.051$) cluster of our sample. Despite its proximity, it has been overlooked in the literature due to its low Galactic latitude. \citet{markevitch96triangulum} performed the most detailed X-ray analysis to date on this object using \asca\ and \rosat\ and revealed an overall hot temperature ($\sim10$ \kev) in its elongated ICM (Fig.~\ref{fig:triangulum}a). Neither \xmm\ nor \chandra\ dedicated studies have been published on this system. Its $K_0$ value is reported neither in \citet{cavagnolo09} nor in \citet{giacintucci17}, nonetheless its core was excluded to have low entropy by \citet{rossetti10}. Recently, a diffuse radio emission classified as a halo has been detected \citep{scaife15, bernardi16}. \\
Three observations of Triangulum Australis can be found in the \chandra\ data archive. However, the oldest two (\obsid s 1227 and 1281) are calibration observations from the commissioning phase and took place less than two weeks after \chandra\ first light, when the calibration products had very large uncertainties. For this reason, we only used \obsid\ 17481 in our analysis. \\
From the maps of the ICM thermodynamical quantities in Fig.~\ref{fig:triangulum}d,e,f, one can infer the complex dynamical state of Triangulum Australis. The GGM filtered on the larger scale gives a hint of a straight structure in SB in the E direction, and it is described by our broken power-law fit ($\compr \sim 1.3$) in Fig.~\ref{fig:triangulum}g. However, no temperature jump is detected across the edge, giving no clue about the origin of this SB feature. We mention that this region was also highlighted by \citet{markevitch96triangulum} with \asca\ and \rosat\ as a direct proof of recent or ongoing heating of the ICM in this cluster.

\subsubsection{Summary of the detected edges}\label{sec:summary}

\begin{table*}
 \centering
 \caption{Properties of the jumps detected. Upper and lower bound errors on $\rat$ and $\mathcal{P}$ were computed adding separately the negative error bounds and the positive error bounds in quadrature. Mach numbers from SB and temperature jumps are reported for shocks (S), for discontinuities whose nature is still uncertain (U) only the Mach derived from the SB is displayed while for spectroscopically confirmed cold fronts (CF) the Mach number determination is not applicable (n.a.).}
 \label{tab:results}
  \begin{tabular}{lcccccccc} 
  \hline
  Cluster name & & Position & \compr\ & $\rat$ & $\mathcal{P}$ & \machsb\ & \machkt\ & Nature \\
  \hline
  \multirow{2}*{A370} & \ldelim\{{2}{1mm} & E & $1.48^{+0.11}_{-0.10}$ & $\ldots$ & $\ldots$ & $1.33^{+0.08}_{-0.07}$ & $\ldots$ & U \\
  & & W & $1.56^{+0.13}_{-0.12}$ & $\ldots$ & $\ldots$ & $1.38^{+0.10}_{-0.09}$ & $\ldots$ & U \\
  \multirow{2}*{A399} & \ldelim\{{2}{1mm} & SE inner & $1.72^{+0.13}_{-0.12}$ & $0.74^{+0.14}_{-0.12}$ & $1.27^{+0.26}_{-0.22}$ & n.a. & n.a. & CF \\
  & & SE outer & $1.45^{+0.10}_{-0.10}$ & $1.20^{+0.39}_{-0.26}$ & $1.74^{+0.58}_{-0.40}$ & $1.31^{+0.07}_{-0.07}$ & $\ldots$ & U \\
  A401 & & SE & $1.39^{+0.04}_{-0.04}$ & $0.78^{+0.07}_{-0.06}$ & $1.08^{+0.10}_{-0.09}$ & n.a. & n.a. & CF \\
  \multirow{2}*{MACS J0417.5-1154} & \ldelim\{{2}{1mm} & NW & $2.50^{+0.29}_{-0.25}$ & $<0.59$ & $<1.64$ & n.a. & n.a. & CF \\ 
  & & SE & $2.44^{+0.31}_{-0.25}$ & $0.44^{+0.17}_{-0.10}$ & $1.07^{+0.44}_{-0.27}$ & n.a. & n.a. & CF \\
  RXC J0528.9-3927 & & E & $1.51^{+0.10}_{-0.09}$ & $0.73^{+0.25}_{-0.14}$ & $1.10^{+0.38}_{-0.22}$ & n.a. & n.a. & CF \\
  \multirow{3}*{MACS J0553.4-3342} & \ldelim\{{3}{1mm} & E inner & $2.49^{+0.32}_{-0.26}$ & $0.62^{+0.33}_{-0.18}$ & $1.54^{+0.85}_{-0.48}$ & n.a. & n.a. & CF \\
  & & E outer & $1.82^{+0.35}_{-0.29}$ & $2.00^{+1.14}_{-0.63}$ & $3.64^{+2.19}_{-1.28}$ & $1.58^{+0.30}_{-0.22}$ & $1.94^{+0.77}_{-0.56}$ & S \\
  & & W & $1.70^{+0.12}_{-0.11}$ & $0.33^{+0.22}_{-0.12}$ & $0.56^{+0.38}_{-0.21}$ & n.a. & n.a. & CF \\
  AS592 & & SW & $1.99^{+0.17}_{-0.15}$ & $1.61^{+0.66}_{-0.43}$ & $3.20^{+1.34}_{-0.89}$ & $1.72^{+0.15}_{-0.12}$ & $1.61^{+0.54}_{-0.42}$ & S \\
  \multirow{3}*{A1914} & \ldelim\{{3}{1mm} & E upper & $1.48^{+0.11}_{-0.12}$ & \multirow{2}*{$0.40^{+0.21}_{-0.12}$} & $0.59^{+0.31}_{-0.18}$ & \multirow{2}*{n.a.} & \multirow{2}*{n.a.} & \multirow{2}*{CF} \\
  & & E lower & $1.64^{+0.13}_{-0.12}$ & & $0.66^{+0.35}_{-0.20}$ & & \\
  & & W & $1.33^{+0.08}_{-0.07}$ & $1.27^{+0.26}_{-0.21}$ & $1.69^{+0.36}_{-0.29}$ & $1.22^{+0.06}_{-0.05}$ & $1.28^{+0.26}_{-0.21}$ & S \\ 
  \multirow{3}*{A2104} & \ldelim\{{3}{1mm} & SE inner & $<1.47$ & $1.33^{+0.27}_{-0.19}$ & $<2.36$ & $<1.32$ & $1.34^{+0.26}_{-0.20}$ & S \\
  & & SE outer & $1.54^{+0.16}_{-0.14}$ & $0.77^{+0.30}_{-0.21}$ & $1.19^{+0.48}_{-0.34}$ & $1.37^{+0.12}_{-0.10}$ & $\ldots$ & U \\
  & & SW & $1.27^{+0.07}_{-0.06}$ & $0.85^{+0.20}_{-0.15}$ & $1.08^{+0.26}_{-0.20}$ & $1.18^{+0.05}_{-0.04}$ & $\ldots$ & U \\
  \multirow{4}*{A2218} & \ldelim\{{4}{1mm} & N & $1.47^{+0.21}_{-0.18}$ & $1.38^{+0.40}_{-0.28}$ & $2.03^{+0.66}_{-0.48}$ & $1.32^{+0.15}_{-0.13}$ & $1.39^{+0.37}_{-0.29}$ & S \\
  & & SE inner & $1.38^{+0.14}_{-0.11}$ & $0.84^{+0.35}_{-0.17}$ & $1.16^{+0.50}_{-0.25}$ & $1.26^{+0.10}_{-0.08}$ & $\ldots$ & U \\
  & & SE outer & $1.26^{+0.14}_{-0.14}$ & $1.44^{+0.48}_{-0.33}$ & $1.81^{+0.64}_{-0.46}$ & $1.17^{+0.10}_{-0.09}$ & $1.45^{+0.43}_{-0.33}$ & S \\
  & & SW & $1.41^{+0.23}_{-0.21}$ & $1.41^{+0.83}_{-0.49}$ & $1.99^{+1.21}_{-0.75}$ & $1.28^{+0.17}_{-0.14}$ & $\ldots$ & U \\
  Triangulum Australis & & E & $1.34^{+0.04}_{-0.04}$ & $1.00^{+0.15}_{-0.10}$ & $1.34^{+0.20}_{-0.14}$ & $1.23^{+0.03}_{-0.03}$ & $\ldots$ & U \\
  \hline
  \end{tabular}
\end{table*}

\begin{figure}
 \centering
 \includegraphics[width=\hsize,trim={0.1cm 13.8cm 0.5cm 0},clip]{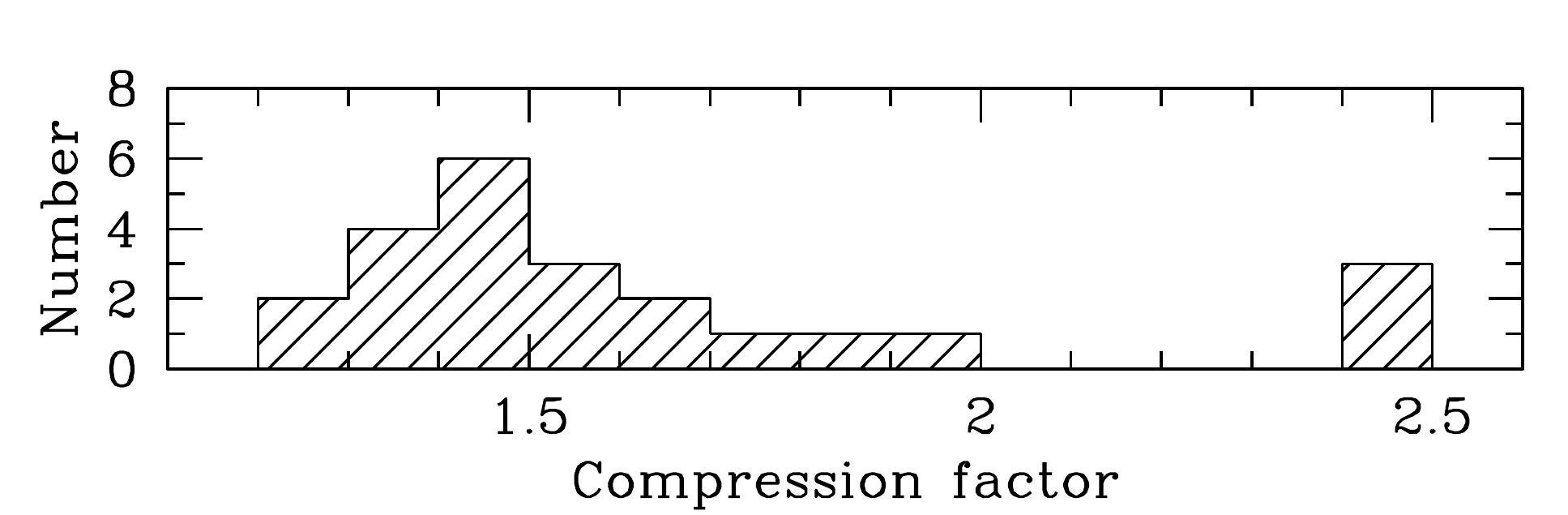}
 \includegraphics[width=\hsize,trim={0.1cm 13.8cm 0.5cm 0},clip]{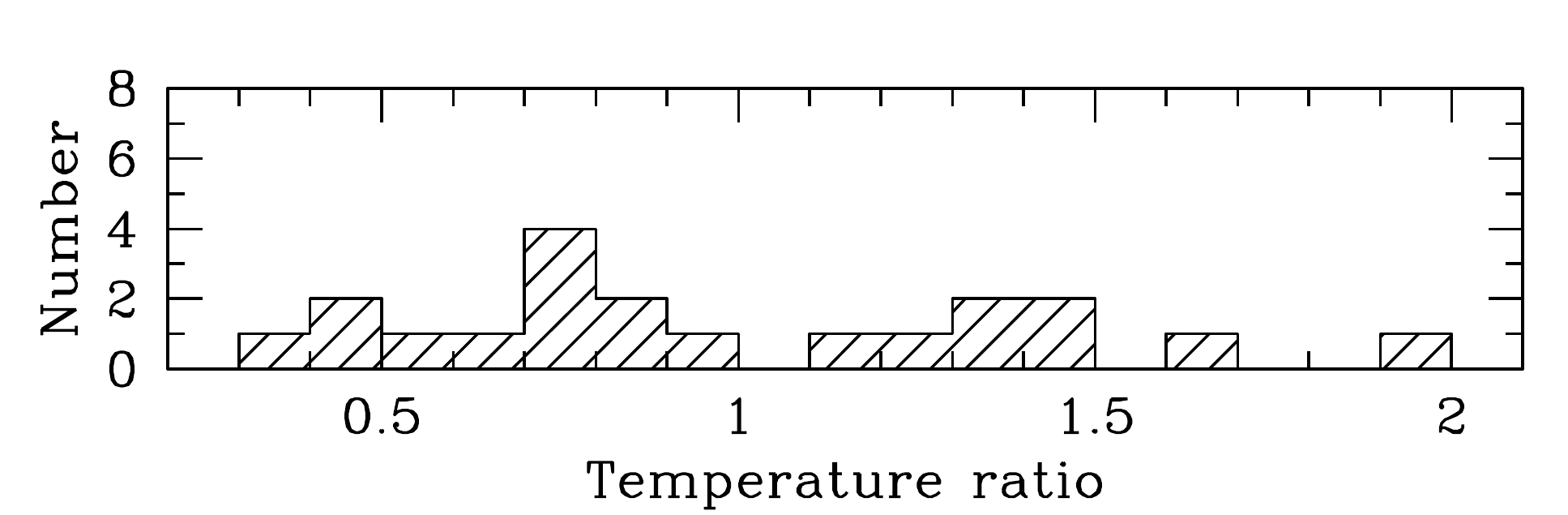}
 \caption{Distribution of the central values of \compr\ (\textit{top}) and $\rat$ (\textit{bottom}) reported in Tab.~\ref{tab:results}.}
 \label{fig:histogram}
\end{figure}

Overall, we found six shocks, eight cold fronts and other eight discontinuities with uncertain origin due to the poorly constrained temperature jump. The properties of the detected edges are summarized in Tab.~\ref{tab:results}, while the distributions of \compr\ and $\rat$ are displayed in Fig.~\ref{fig:histogram}. Although we are not carrying out a statistical analysis of shocks and cold fronts in galaxy clusters, we notice that the majority of the reported jumps are associated with weak discontinuities with $\compr < 1.7$ and $0.5<\rat<1.5$. This may indicate that the GGM filters allow to pick up also small SB jumps that are usually lost in a visual inspection of unsmoothed cluster images. \\
\indent
We mention that in the case of a shock the SB and temperature jumps allow to give two independent constraints on the Mach number (Eq.~\ref{eq:mach-from-dens}, \ref{eq:mach-from-temp}). However, so far, only few shocks reported in the literature have Mach number consistently derived from both the jumps (\eg\ A520, \citealt{markevitch05}; A665, \citealt{dasadia16a665}; A115, \citealt{botteon16a115}). Instead, in our analysis there is a general agreement between these two quantities, further supporting the robustness of the results. \\
\indent
One could argue that the nature of the weakest discontinuities claimed is constrained at slightly more than $1\sigma$ level from the temperature ratio. This is a consequence of the small temperature jump implied by these fronts and the large errors associated with the spectral analysis (despite the careful background treatment performed). However, we can check the presence of pressure discontinuities at these edges by combining the density and temperature jumps achieved from SB and spectral analysis. The values of $\mathcal{P} \equiv P_d / P_u = \compr \times \rat$ computed for all the discontinuities are reported in Tab.~\ref{tab:results} and show at higher confidence levels the presence of a pressure discontinuity in the shocks and the absence of a pressure jump in the cold fronts, strengthening our claims. Although this procedure combines a deprojected density jump with a temperature evaluated along the line of sight, we verified that given the uncertainties on the temperature determination and the errors introduced by a deprojection analysis, the projected and deprojected values of temperature and pressure ratios are statistically consistent even in the cases of the innermost edges (\ie\, those more affected by projection effects). \\
\indent
With the present work, we increased the number of known shocks and cold fronts in galaxy clusters. The detected shocks have all $\mach < 2$ likely due to the combination of the fact that shocks crossing the central Mpc regions of galaxy clusters are weak \citep[\eg][and references therein]{vazza12why} and that fast moving shocks would be present for a short time in the ICM. \\
\indent
The distinction between shock and cold fronts for the eight discontinuities with uncertain origin can tentatively be inferred from the current values of \rat\ and $\mathcal{P}$ reported in Tab.~\ref{tab:results}. In this respect, deeper observations of these edges will definitely shed light about their nature.

%\begin{figure}
% \centering
% \includegraphics[width=\hsize,trim={0.1cm 10.5cm 0.5cm 0},clip]{figure/machsb-isto.pdf}
% \includegraphics[width=\hsize,trim={0.1cm 10.5cm 0.5cm 0},clip]{figure/machkt-isto.pdf}
% \includegraphics[width=\hsize,trim={0.1cm 10.5cm 0.5cm 0},clip]{figure/cf-isto.pdf}
% \caption{Contribution of our work (\textit{red}) on the total number (\textit{black}) of shocks and cold fronts reported in Tab.~\ref{tab:sample}. The central values of Mach numbers and compression factors found in literature were used to create the histograms.}
% \label{fig:edges-isto}
%\end{figure}

\subsection{Non-detections}\label{sec:non-detection}

\begin{figure*}
 \centering
 \begin{tabular}{cc}
  \multirow{2}{*}{\subfloat{\subfigimgwhitebig[width=.6\textwidth]{\quad  a)}{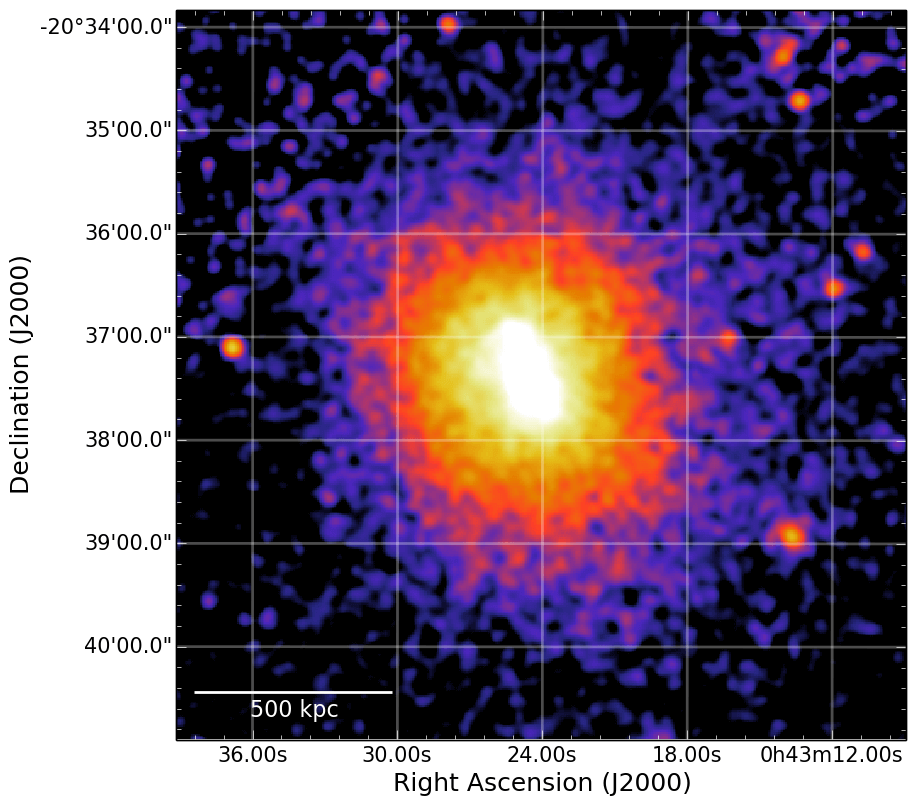}}} & \\
  & \vspace{0.15cm}\hspace{-0.3cm}\subfloat{\subfigimgwhiteggm[width=.28\textwidth]{\quad b)}{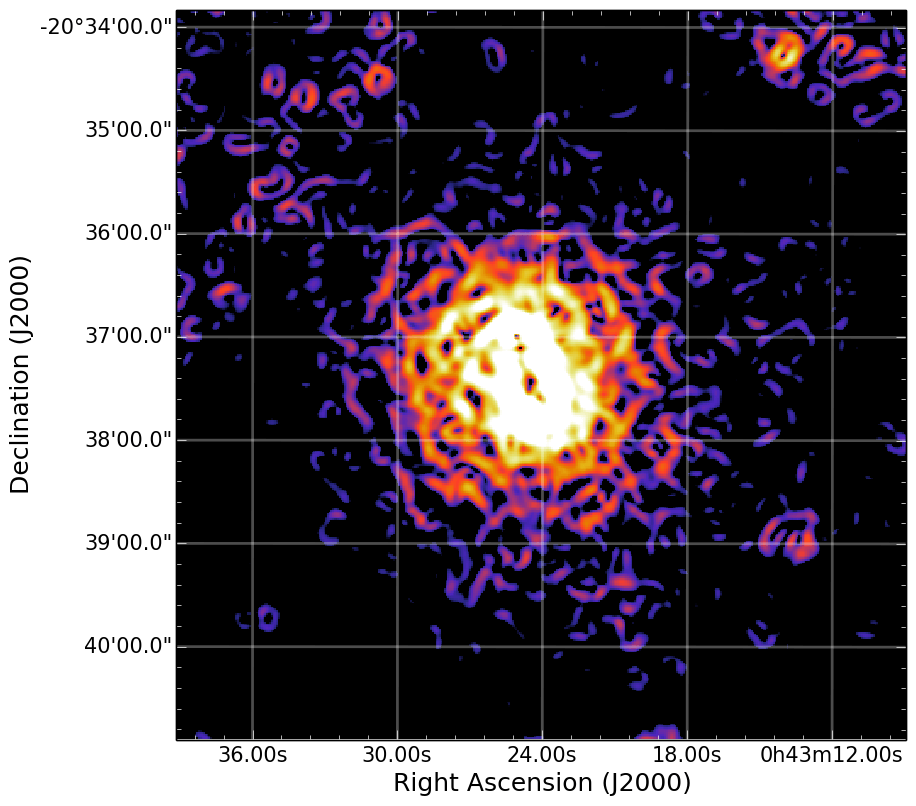}} \\
  & \hspace{-0.3cm}\subfloat{\subfigimgwhiteggm[width=.28\textwidth]{\quad c)}{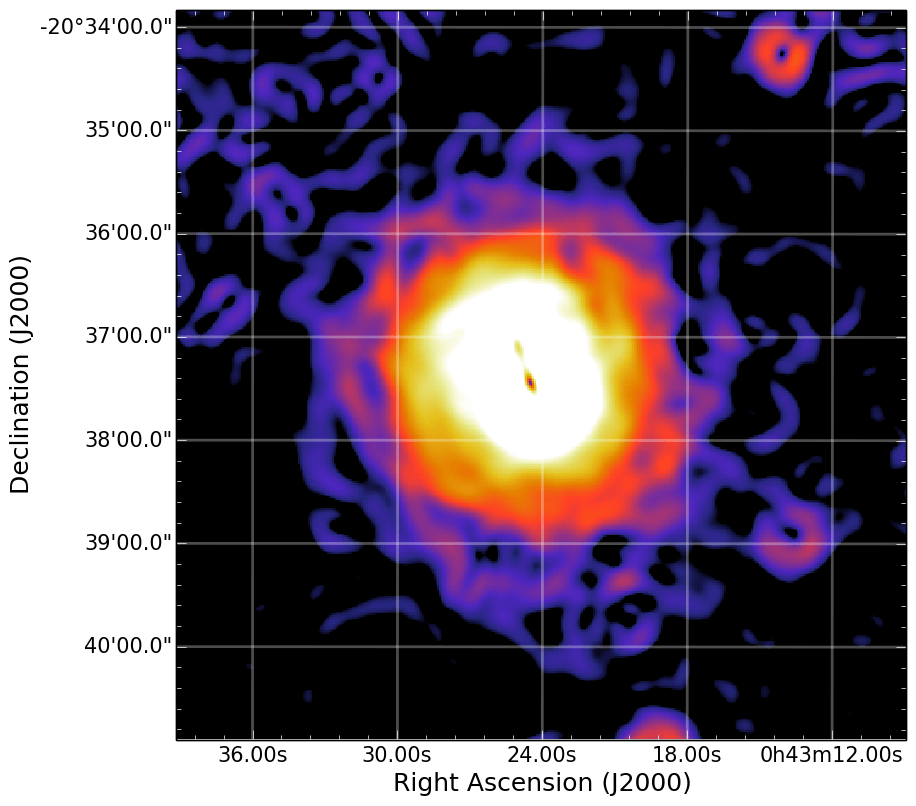}}
 \end{tabular}
 \subfloat{\subfigimgblack[width=.3\textwidth]{\quad d)}{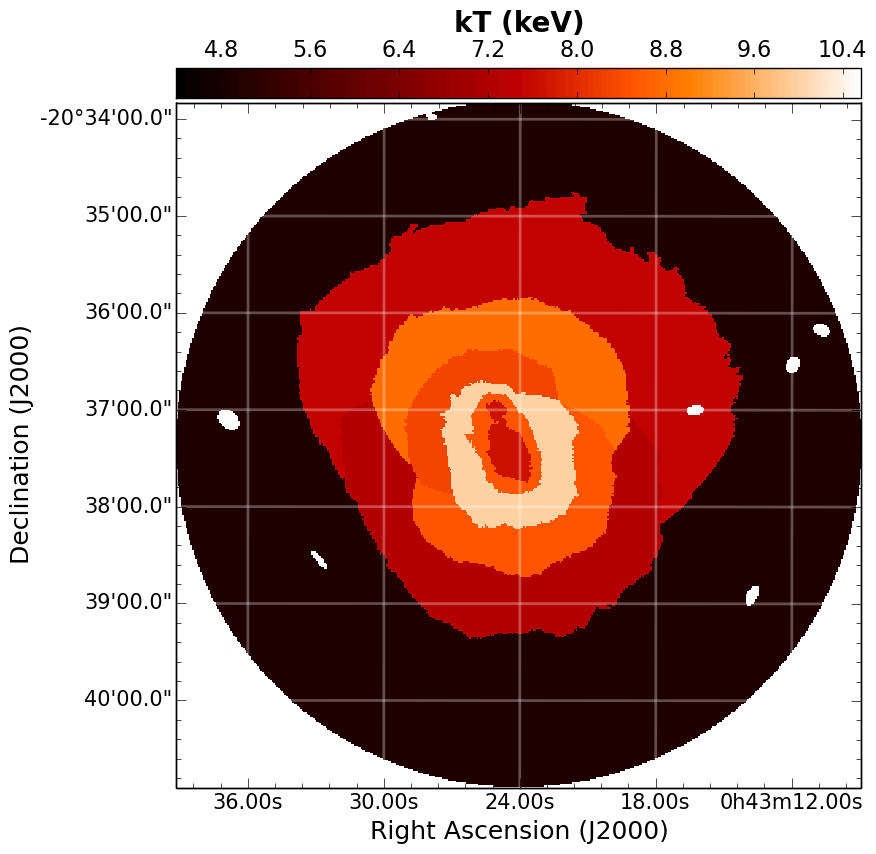}}
 \subfloat{\subfigimgblack[width=.3\textwidth]{\quad e)}{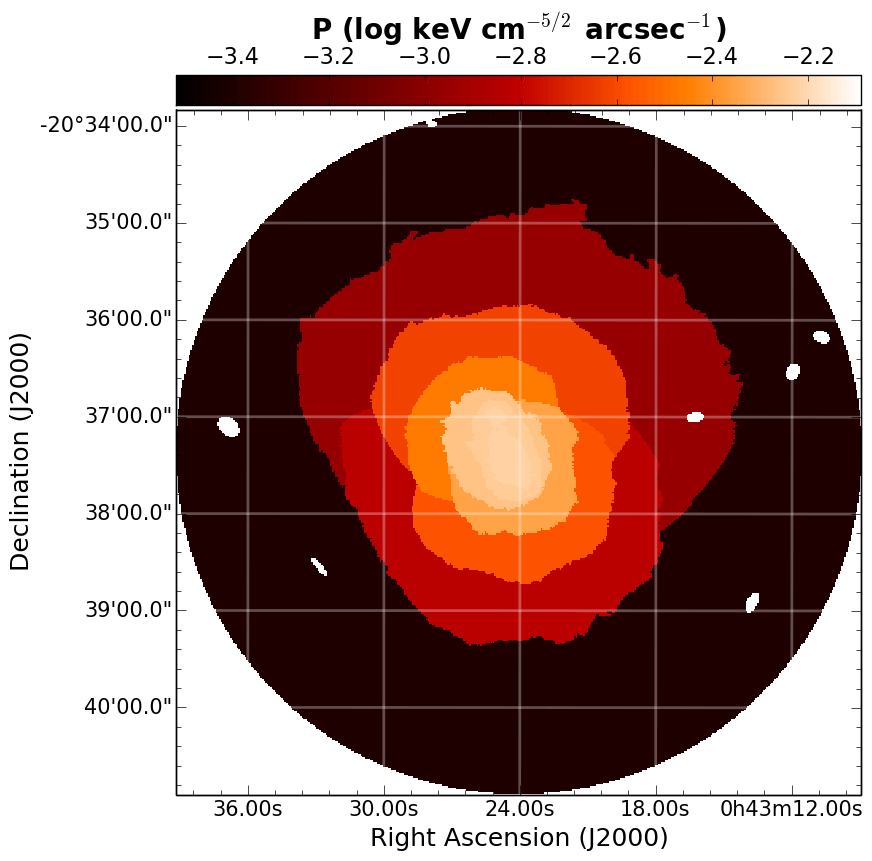}}
 \subfloat{\subfigimgblack[width=.3\textwidth]{\quad f)}{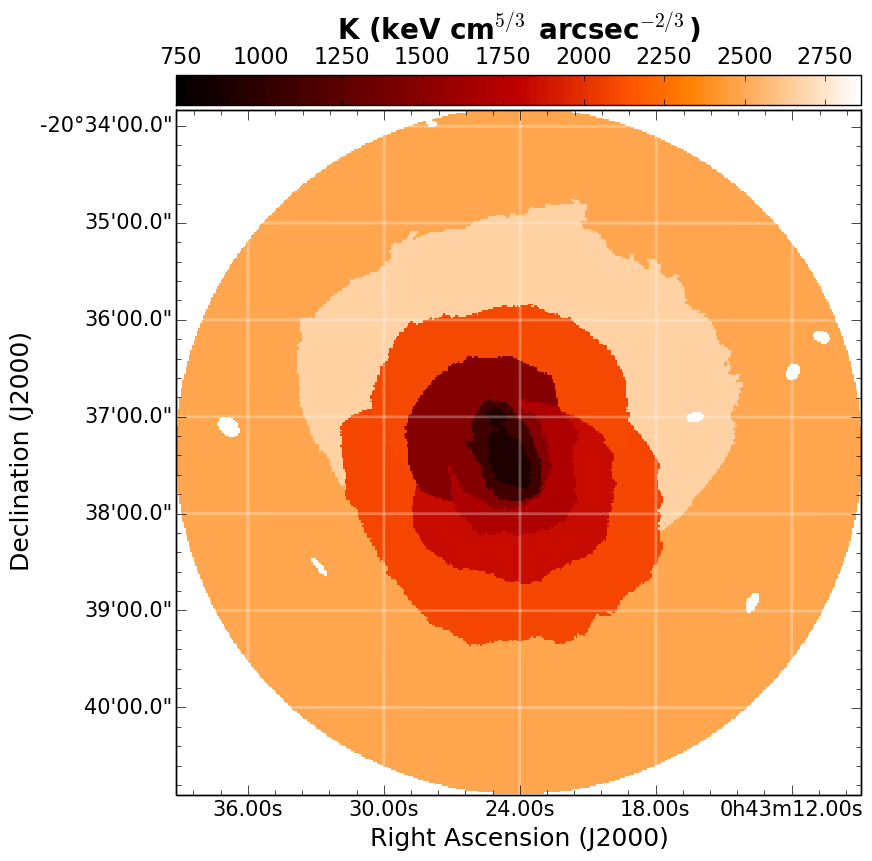}}
 \caption{A2813. The same as for Fig~\ref{fig:a399}. The goodness of fits is reported in Fig.~\ref{fig:a2813_errors}.}
 \label{fig:a2813}
\end{figure*}

\begin{figure*}
 \centering
 \begin{tabular}{cc}
  \multirow{2}{*}{\subfloat{\subfigimgwhitebig[width=.6\textwidth]{\quad  a)}{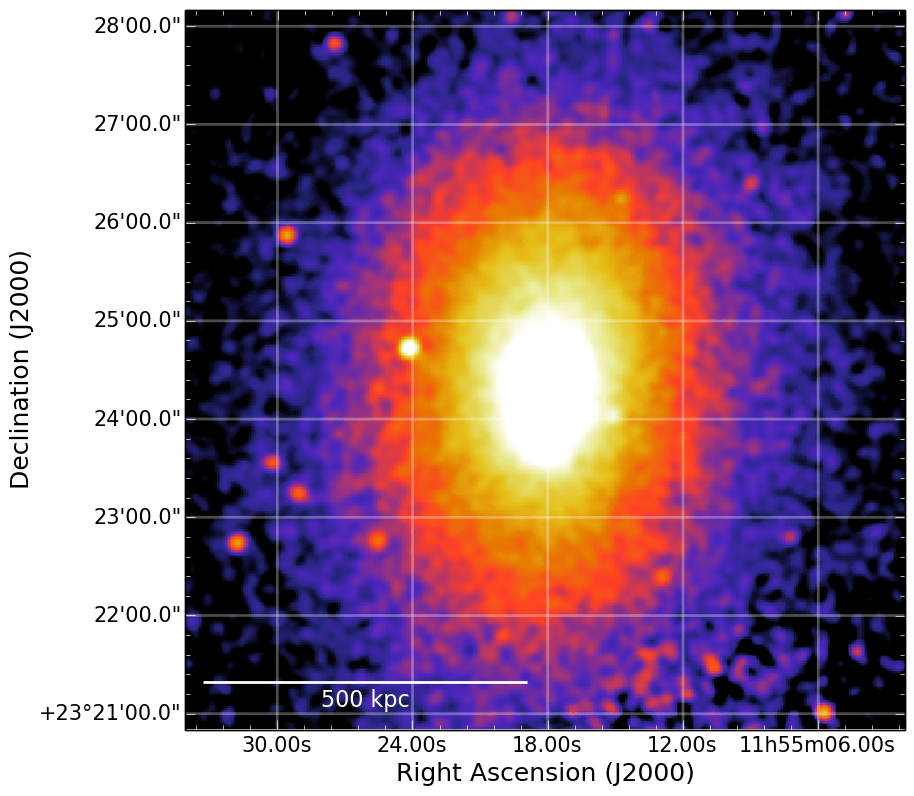}}} & \\
  & \vspace{0.15cm}\hspace{-0.3cm}\subfloat{\subfigimgwhiteggm[width=.28\textwidth]{\quad b)}{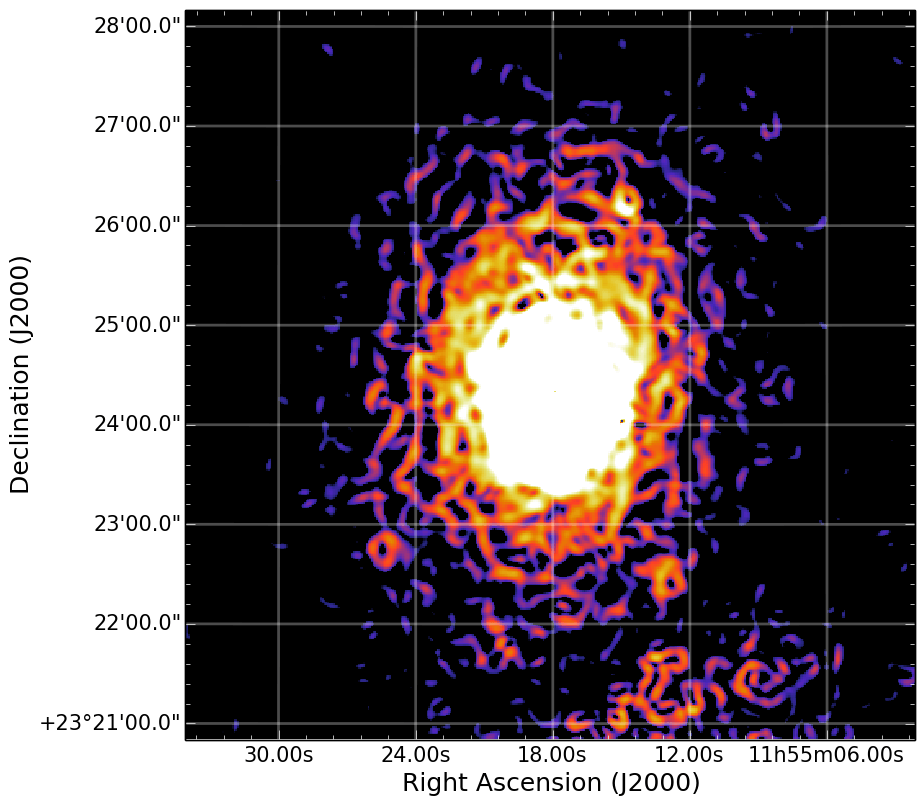}} \\
  & \hspace{-0.3cm}\subfloat{\subfigimgwhiteggm[width=.28\textwidth]{\quad c)}{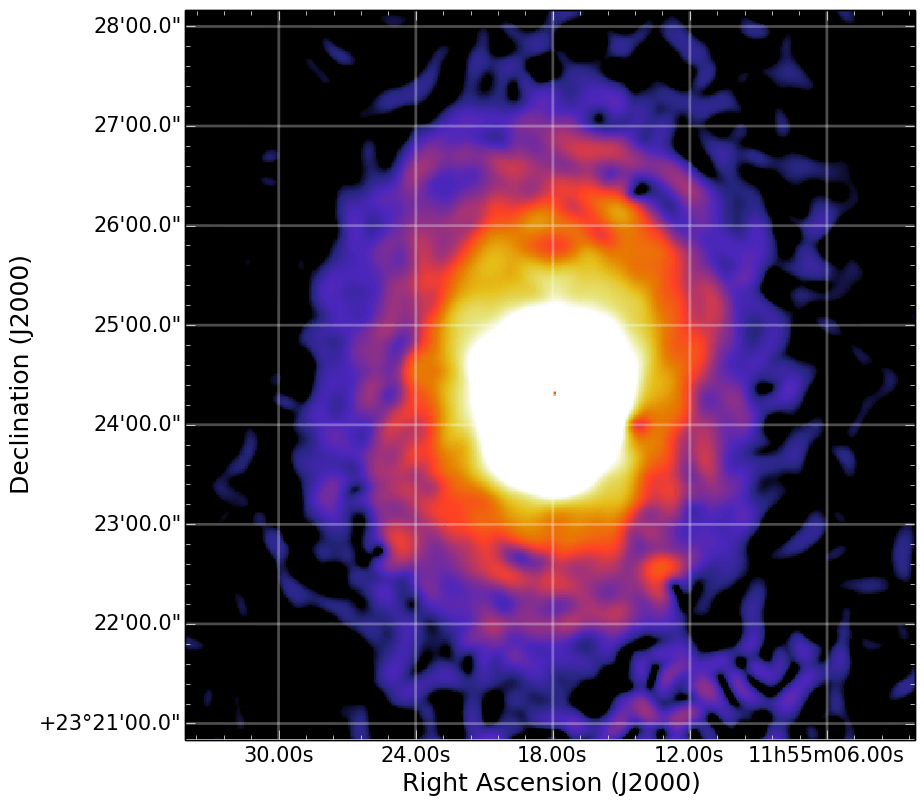}}
 \end{tabular}
 \subfloat{\subfigimgblack[width=.3\textwidth]{\quad d)}{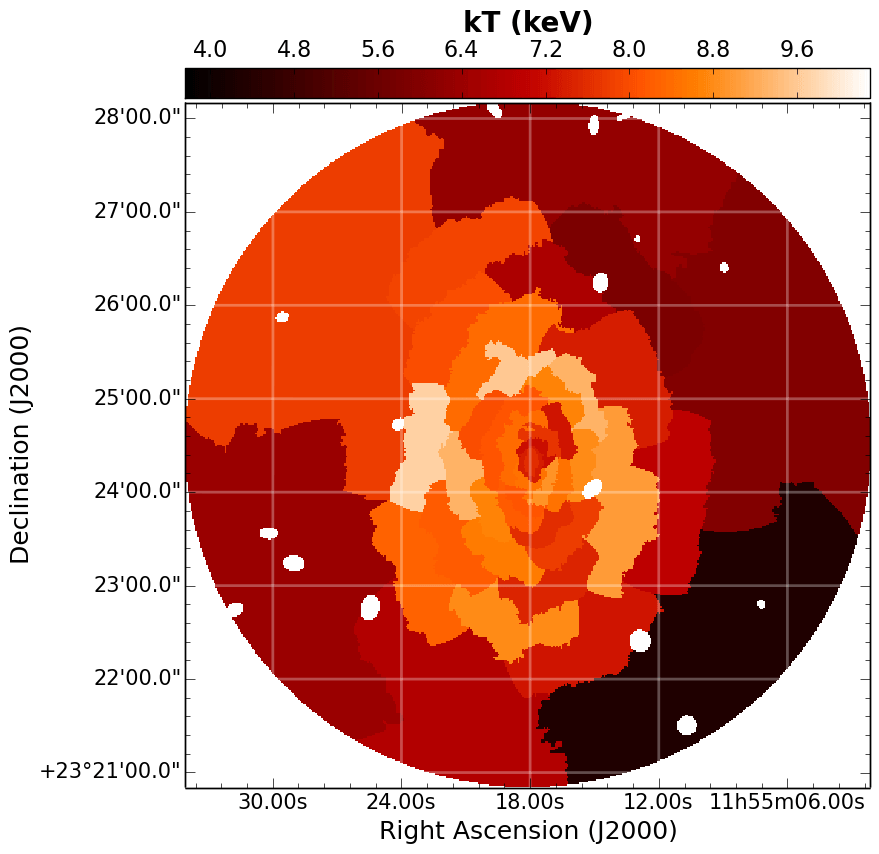}}
 \subfloat{\subfigimgblack[width=.3\textwidth]{\quad e)}{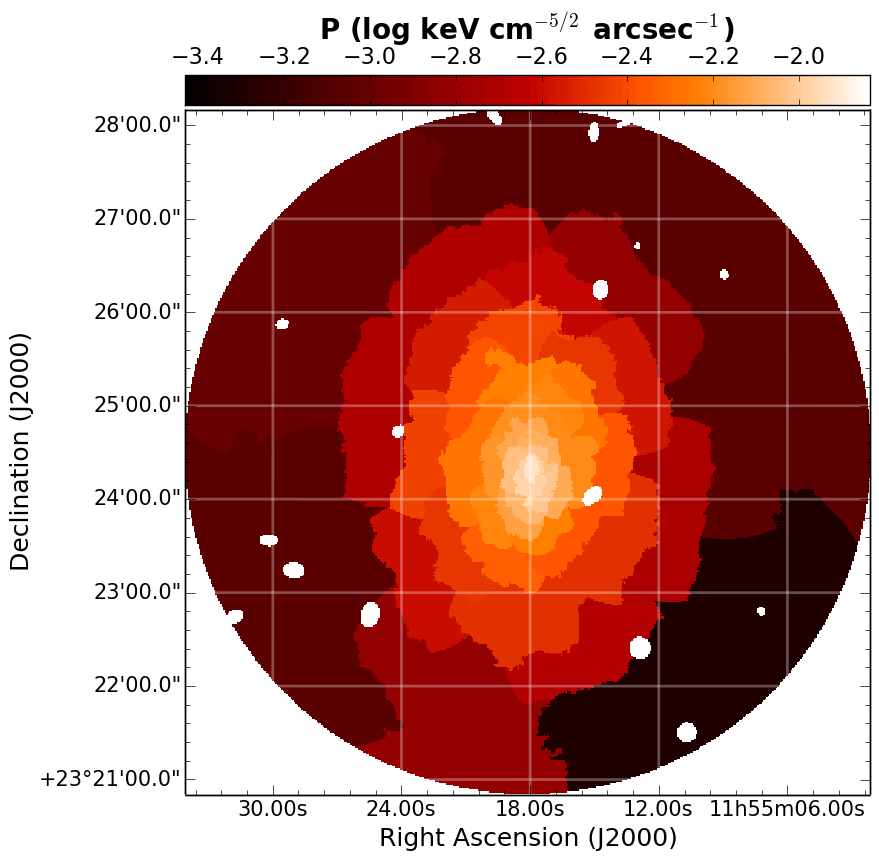}}
 \subfloat{\subfigimgblack[width=.3\textwidth]{\quad f)}{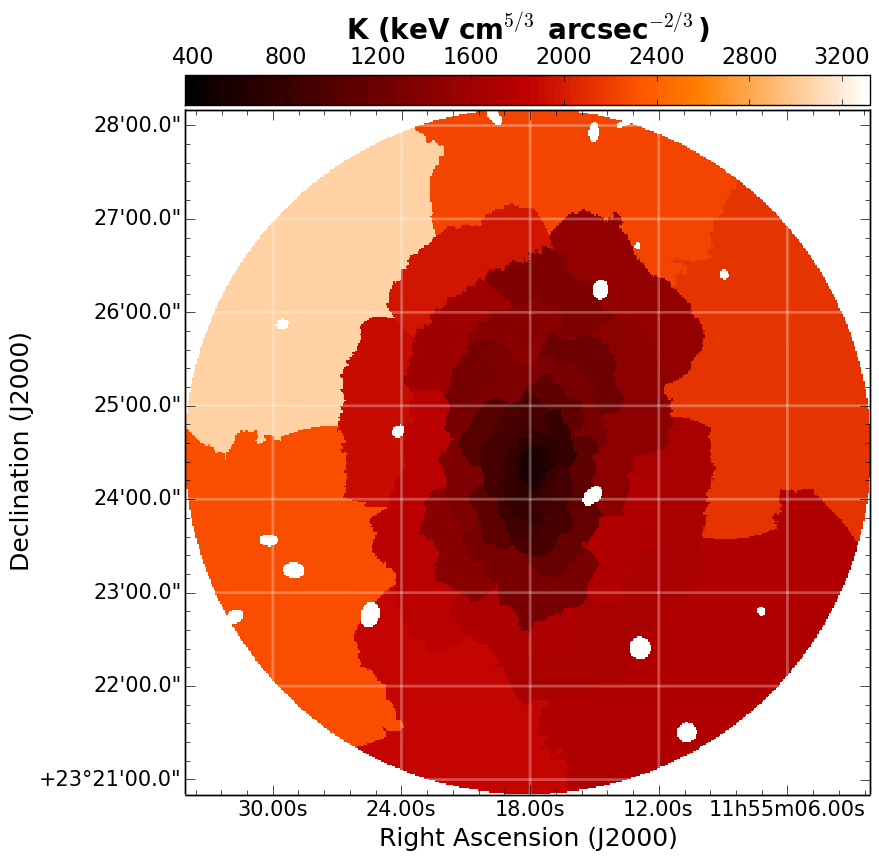}}
 \caption{A1413. The same as for Fig~\ref{fig:a399}. The goodness of fits is reported in Fig.~\ref{fig:a1413_errors}.}
 \label{fig:a1413}
\end{figure*}

\begin{figure*}
 \centering
 \begin{tabular}{cc}
  \multirow{2}{*}{\subfloat{\subfigimgwhitebig[width=.6\textwidth]{\quad  a)}{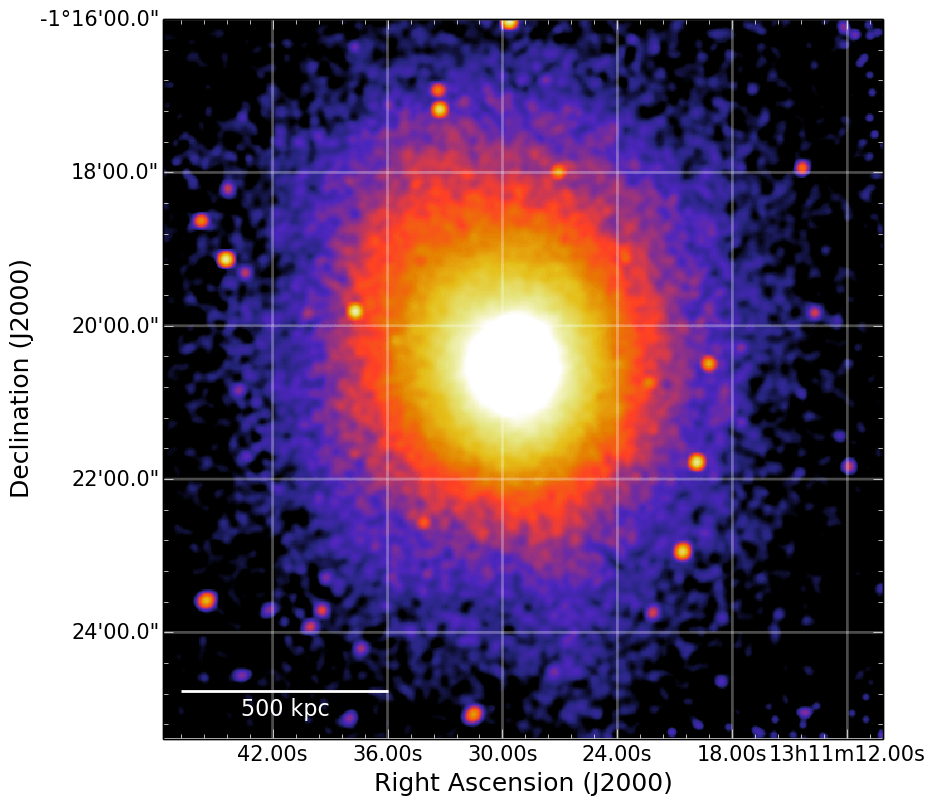}}} & \\
  & \vspace{0.15cm}\hspace{-0.3cm}\subfloat{\subfigimgwhiteggm[width=.28\textwidth]{\quad b)}{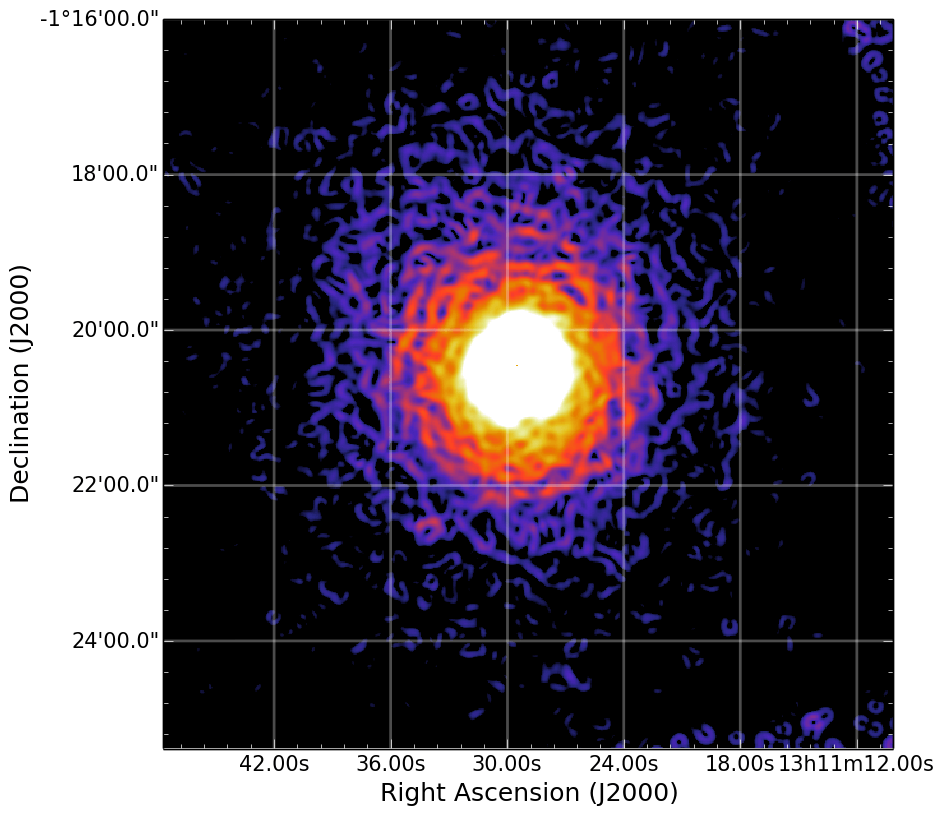}} \\
  & \hspace{-0.3cm}\subfloat{\subfigimgwhiteggm[width=.28\textwidth]{\quad c)}{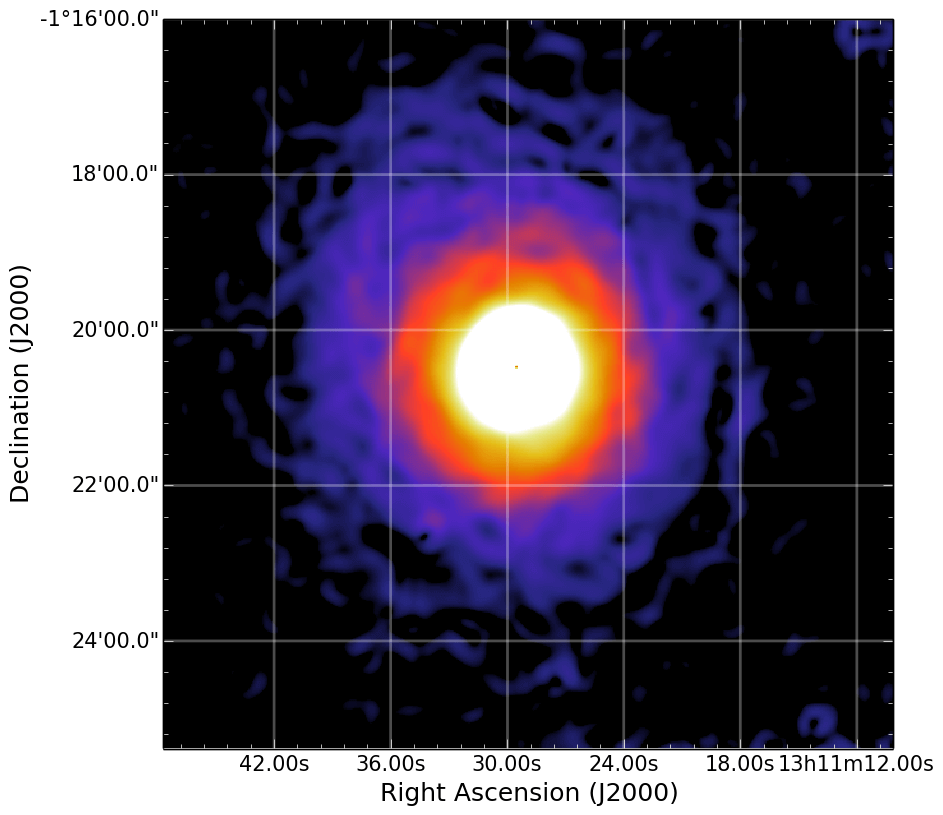}}
 \end{tabular}
 \subfloat{\subfigimgblack[width=.3\textwidth]{\enspace d)}{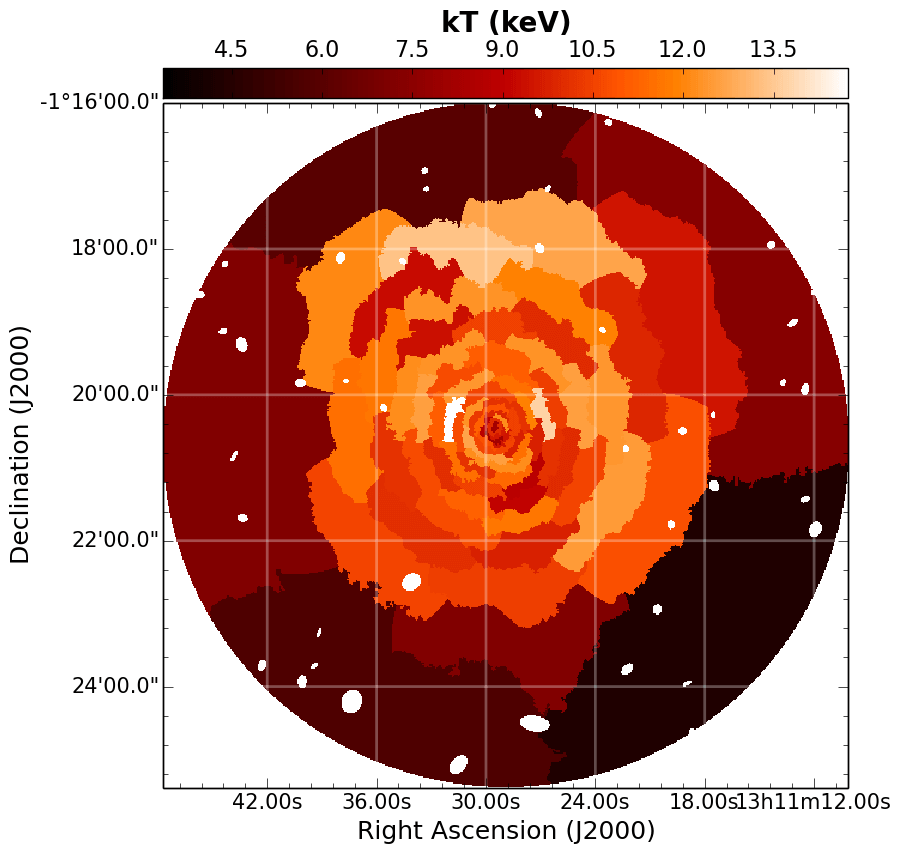}}
 \subfloat{\subfigimgblack[width=.3\textwidth]{\enspace e)}{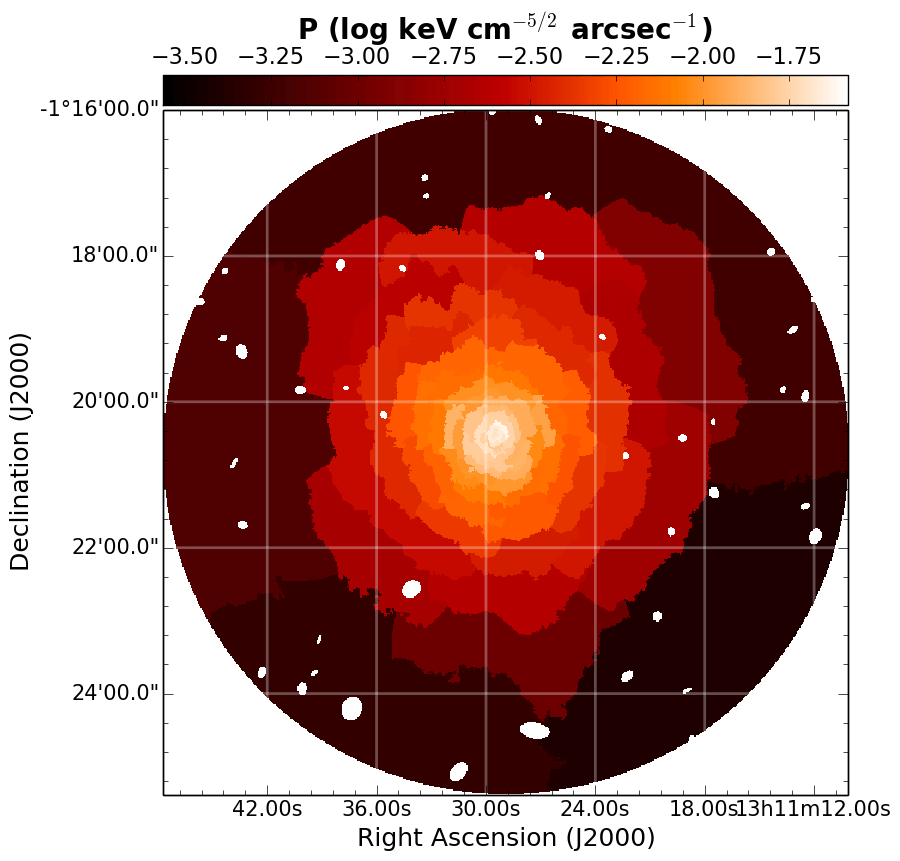}}
 \subfloat{\subfigimgblack[width=.3\textwidth]{\enspace f)}{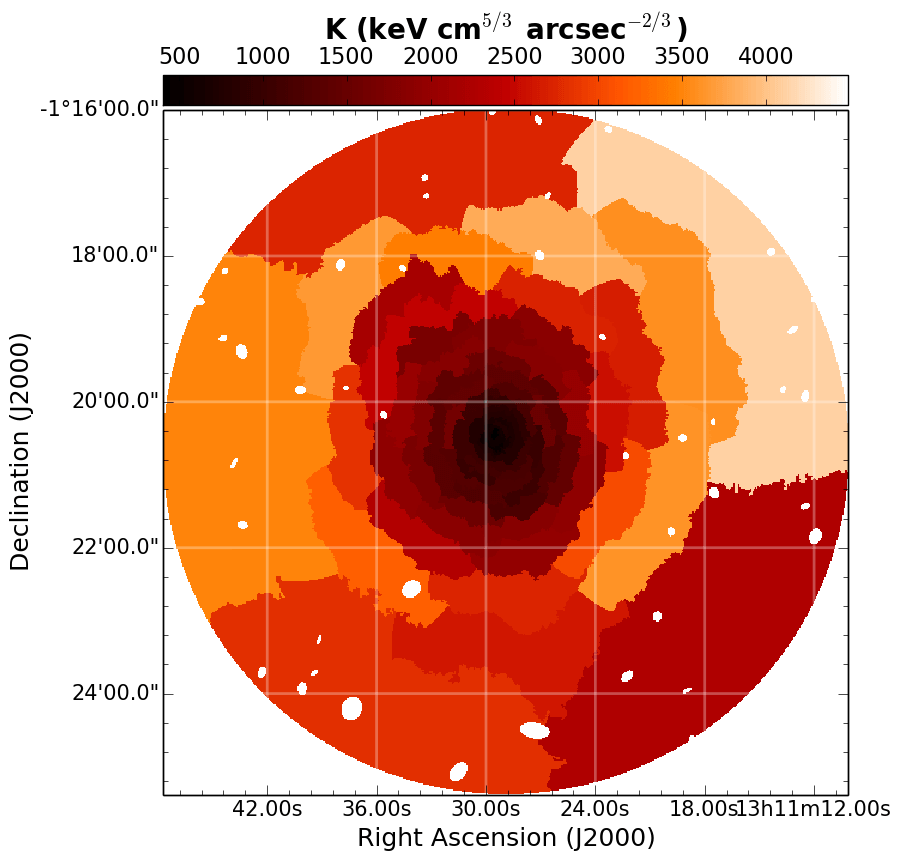}}
 \caption{A1689. The same as for Fig~\ref{fig:a399}. The goodness of fits is reported in Fig.~\ref{fig:a1689_errors}.}
 \label{fig:a1689}
\end{figure*}

\begin{figure*}
 \centering
 \begin{tabular}{cc}
  \multirow{2}{*}{\subfloat{\subfigimgwhitebig[width=.6\textwidth]{\quad  a)}{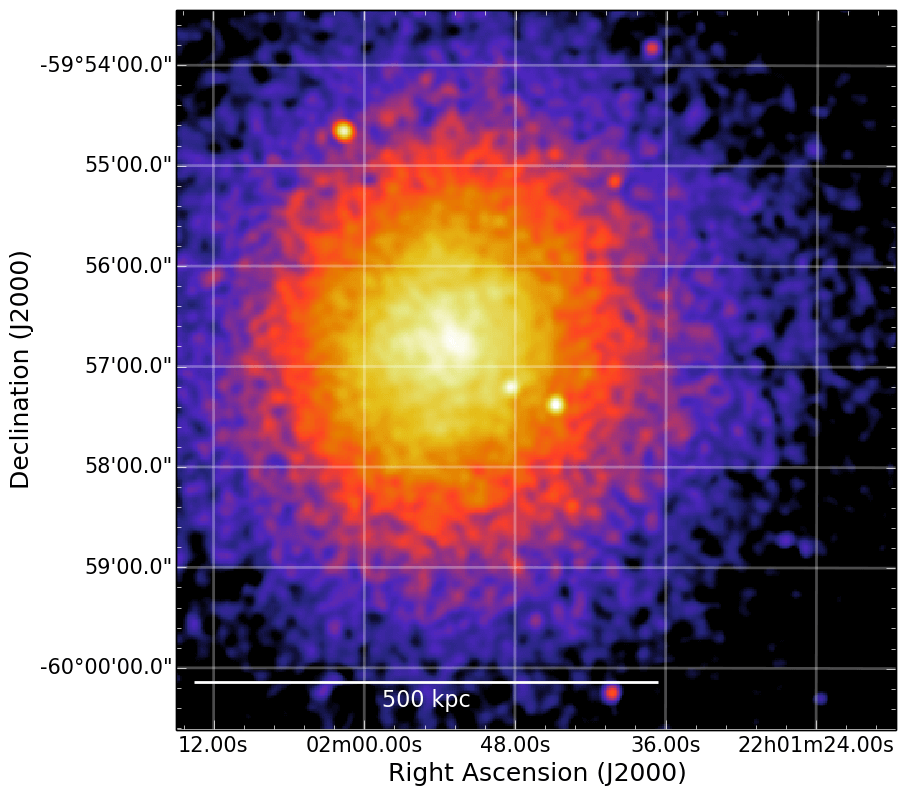}}} & \\
  & \vspace{0.15cm}\hspace{-0.3cm}\subfloat{\subfigimgwhiteggm[width=.28\textwidth]{\quad b)}{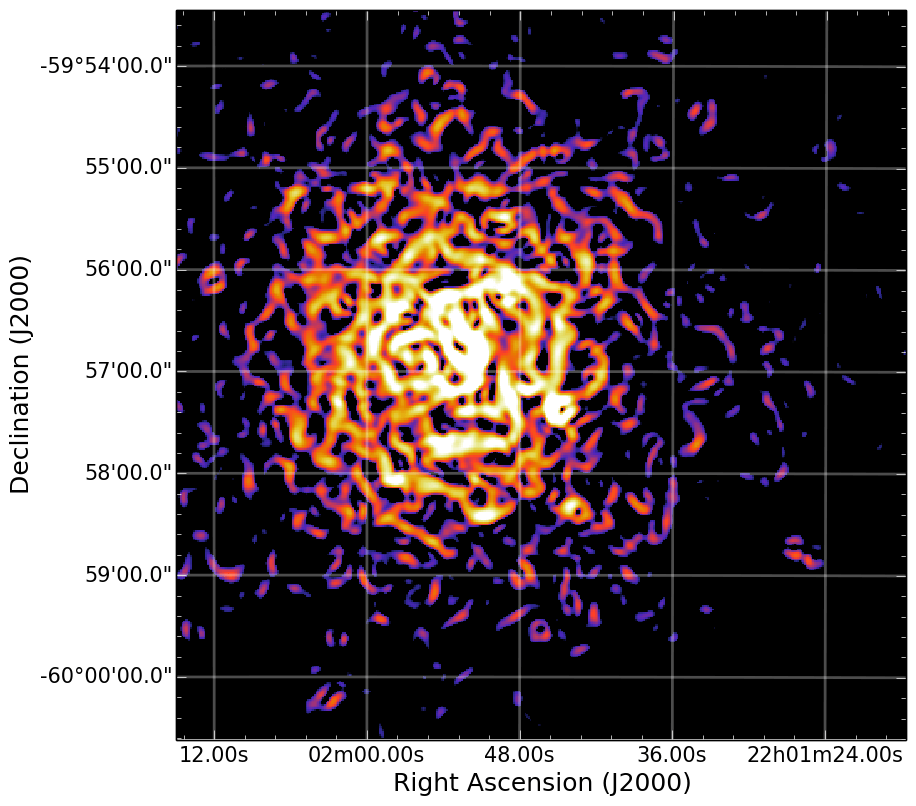}} \\
  & \hspace{-0.3cm}\subfloat{\subfigimgwhiteggm[width=.28\textwidth]{\quad c)}{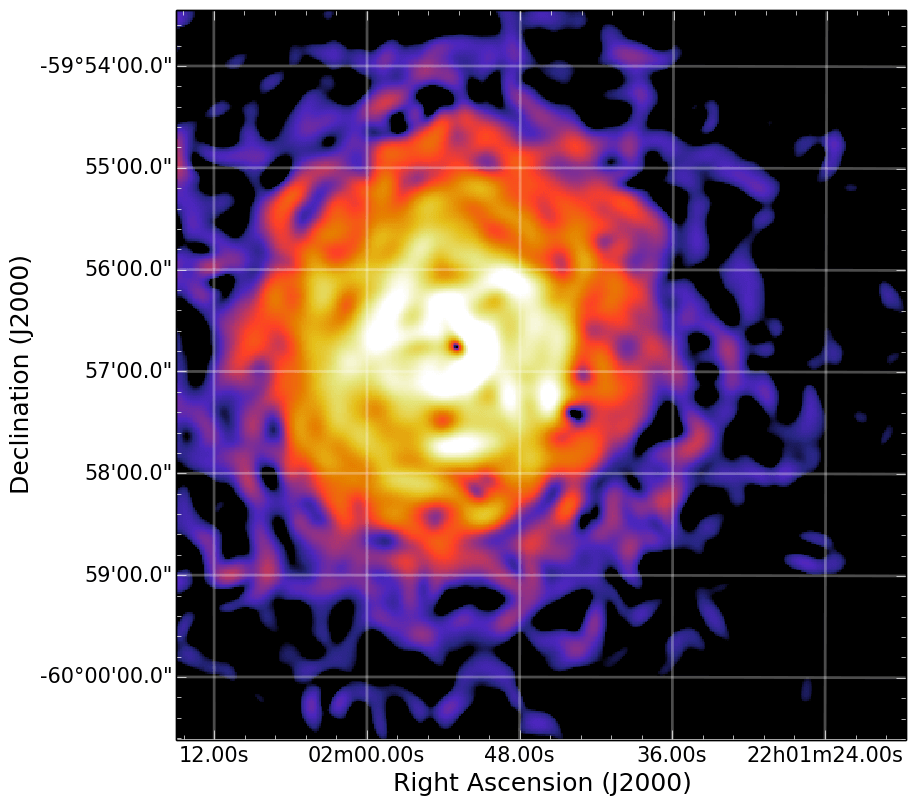}}
 \end{tabular}
 \subfloat{\subfigimgblack[width=.3\textwidth]{\quad d)}{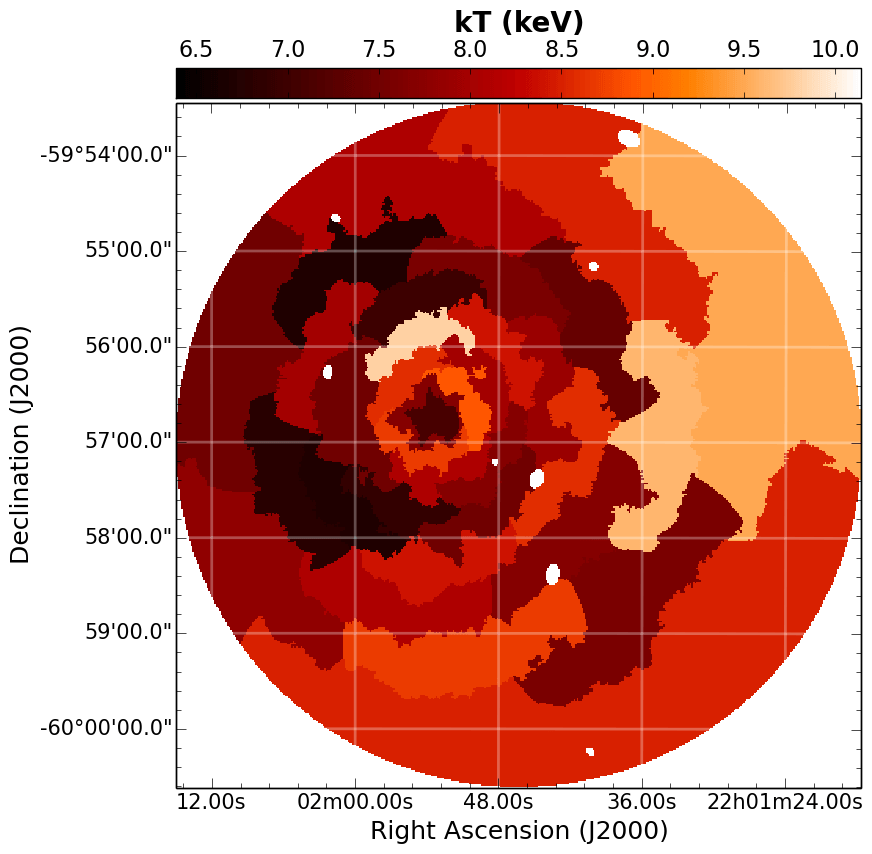}}
 \subfloat{\subfigimgblack[width=.3\textwidth]{\quad e)}{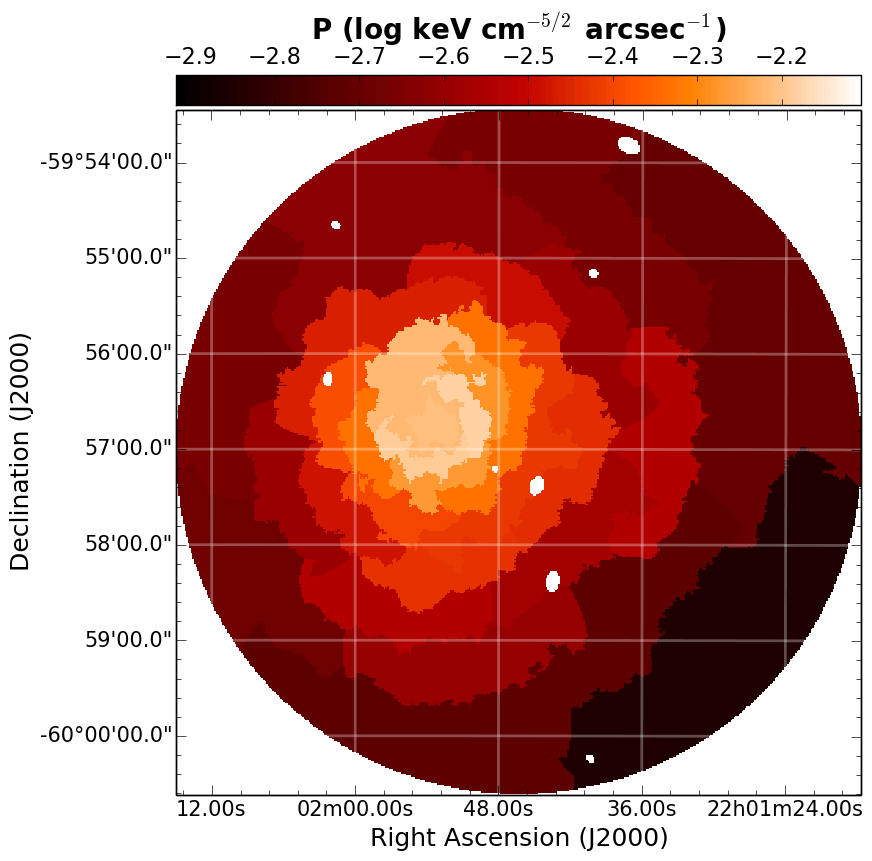}}
 \subfloat{\subfigimgblack[width=.3\textwidth]{\quad f)}{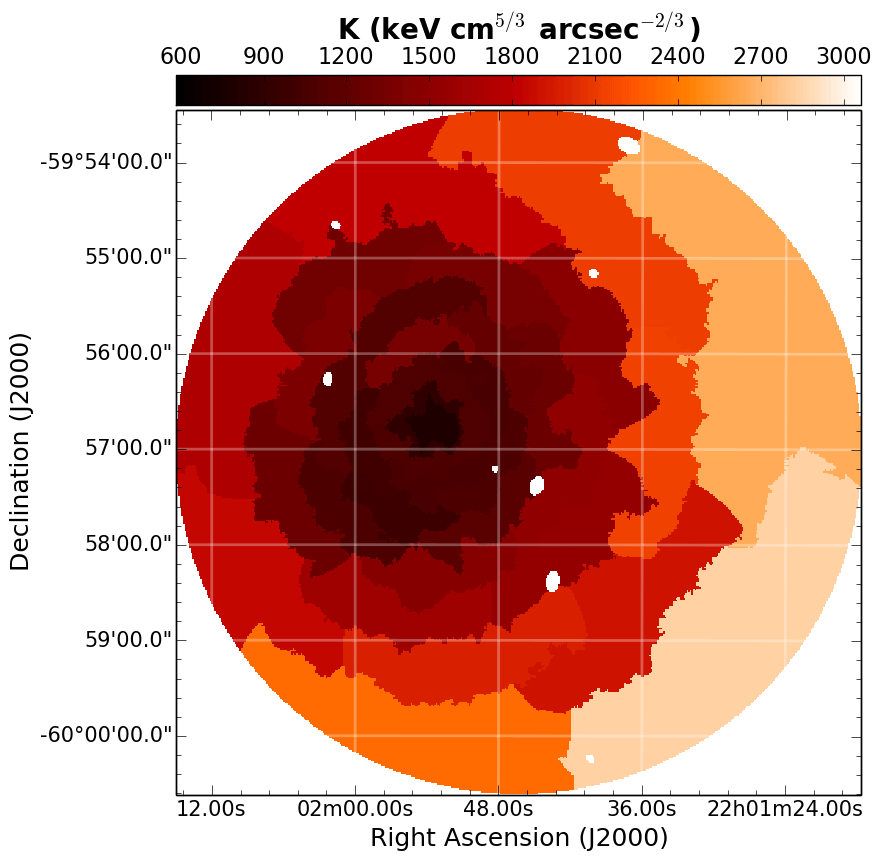}}
 \caption{A3827. The same as for Fig~\ref{fig:a399}. The goodness of fits is reported in Fig.~\ref{fig:a3827_errors}.}
 \label{fig:a3827}
\end{figure*}

Our analysis did not allow us to detect any edge in the following objects: A2813 ($z=0.292$), A1413 ($z=0.143$), A1689 ($z=0.183$) and A3827 ($z=0.098$). All these systems seem to have a more regular X-ray morphology (Fig.~\cref{fig:a2813,fig:a1413,fig:a1689,fig:a3827}) with respect to the other clusters of the sample.

\subparagraph{A2813.} This cluster has a roundish ICM morphology (Fig.~\ref{fig:a2813}a), nonetheless its value of $K_0=268\pm44$ \kevcmsq\ is among the highest in our sample (\cf\ Tab.~\ref{tab:sample}). The core is slightly elongated in the NE-SW direction and has a temperature $\sim7.7$ \kev, consistently with the \xmm\ value reported by \citet{finoguenov05}. The maps shown in Fig.~\ref{fig:a2813} were produced using all the \obsid s listed in Tab.~\ref{tab:chandra_obs}. We mention that the original target of the \aciss\ datasets (\obsid s 16366, 16491, 16513) is XMMUJ0044.0-2033; however, A2813 is found to entirely lay on an \acisi\ chip that was kept on during the observations. These data provide the largest amount ($\sim80\%$) of the total exposure time on A2813 and were used in our analysis although the unavoidable degradation of the instrument spatial resolution due to the \acisi\ chip being off-axis with this observing configuration. 

\subparagraph{A1413.} It has a border line value of $K_0$ (\cf\ Tab.~\ref{tab:sample}) from the threshold set in this work. The distribution of cluster gas is somewhat elliptical, elongated in the N-S direction (Fig.~\ref{fig:a1413}a). Our analysis and previous \chandra\ temperature profiles \citep{vikhlinin05, baldi07} are in contrast with \xmm\ that does not provide evidence of a CC \citep{pratt02}. This discrepancy is probably due to the poorer PSF of the latter instrument. A radio mini-halo covering the CC region is also found by \citet{govoni09}. The region in the NW direction with a possible discontinuity suggested by the GGM filtered images did not show the evidence for an edge with the SB profile fitting (Fig.~\ref{fig:a1413_noedge}).

\subparagraph{A1689.} It represents a massive galaxy cluster deeply studied in the optical band because its weak and strong gravitational lensing \citep[\eg][]{broadhurst05, limousin07}. The X-ray emission is quasi-spherical and centrally peaked (Fig.~\ref{fig:a1689}a), features that apparently indicate a CC. Nevertheless, optical \citep{girardi97} and \xmm\ observations \citep{andersson04} both suggest that the system is undergoing to a head-on merger seen along the line of sight due either to the presence of optical substructures or to the asymmetric temperature of the ICM, hotter in the N. Our results confirm the presence of asymmetry in temperature distribution (Fig.~\ref{fig:a1689}d). The fact that a radio halo is also detected \citep{vacca11} fits with the dynamically unrelaxed nature of the system. 

\subparagraph{A3827.} It constitutes another cluster studied in detail mainly for its optical properties. Its central galaxy is one of the most massive known found in a cluster center and exhibits strong lensing features \citep{carrasco10}. Gravitational lensing also indicates a separation between the stars and the center of mass of the dark matter in the central galaxies \citep{massey15}, making A3827 a good candidate to investigate the dark matter self-interactions \citep{kahlhoefer15}. On the X-ray side, the cluster emission is roughly spherical (Fig.~\ref{fig:a3827}a), with an irregular temperature distribution (Fig.~\ref{fig:a3827}d) and a mean value of $\sim7$ \kev\ \citep{leccardi08}. Two regions to the E and W directions suggested by the GGM images did not show any discontinuity with the SB profile fitting (Fig.~\ref{fig:a3827_noedge}).

\section{Conclusions}\label{sec:conclusions}

Shocks and cold fronts produced in a collision between galaxy clusters give information on the dynamics of the merger and can be used to probe the microphysics of the ICM. Nonetheless their detection is challenged by the low number of X-ray counts in cluster outskirts and by possible projection effects that can hide these sharp edges. For this reason only a few of them have been successfully detected both in SB and in temperature jumps. \\
\indent
In this work we explored a combination of different analysis approaches of X-ray observations to firmly detect and characterize edges in NCC massive galaxy clusters. Starting from GGM filtered images on different scales and the maps of the ICM thermodynamical quantities of the cluster, one can pinpoint ICM regions displaying significant SB and/or temperature variations. These can be thus investigated with the fitting of SB profiles, whose extracting sectors have to be accurately chosen in order to properly describe the putative shock or cold front. Once that the edge is well located, spectral analysis on dedicated upstream and downstream regions can also be performed in an optimized way. The discontinuity is firmly detected if the jump is observed both in images and in spectra. \\
\indent
In this paper we selected 37 massive NCC clusters with adequate X-ray data in the \chandra\ archive to search for new discontinuities driven in the ICM by merger activity. In particular we looked at 15 of these systems for which no claim of edges was published. We were able to characterize at least one SB jump in 11 out of these 15 clusters of the sample. The performed SB analysis relies on the spherical assumption. Among the detected edges, we also constrained the temperature jump for 14 discontinuities, six shocks and eight cold fronts, while for eight edges the classification is still uncertain. As a further check, we also computed the pressure ratios across the edges and verified the presence of the pressure discontinuity in the shocks and the absence of a pressure jump in the cold fronts. \\
\indent
Our work provides a significant contribution to the search for shocks and cold fronts in merging galaxy clusters demonstrating the strength of combining diverse techniques aimed to identify edges in the ICM. Indeed, many shocks and cold fronts reported in the literature have been discovered  because either they were evident in the unsmoothed cluster images or there were priors suggesting their existence (\eg\ merger geometry and/or presence of a radio relic). The usage of edge detection algorithms (as the GGM filter) in particular helps in highlighting also small SB gradients to investigate with SB profile and spectral fitting. Among the small jumps detected we found low Mach numbers ($\mach < 2$) shocks; this is a possible consequence of the fact that the central regions of the ICM are crossed by weak shocks while the strongest ones quickly fades in the cluster outskirts, making their observation more difficult (see also discussion in \citealt{vazza12why} on the occurrence of radio relics in clusters as a function of radius). \\
\indent
Many shocks in the literature were found thanks to the presence of previously observed radio relics (or edges of radio halos). As a consequence, the radio follow-up of the shocks detected in this paper will be useful to study the connection between weak shocks and non-thermal phenomena in the ICM.

\section*{Acknowledgments}

We thank the anonymous referee for useful comments. AB and GB acknowledge partial support from PRIN-INAF 2014 grant. AB thanks V.~Cuciti and A.~Ignesti for probing the legibility of the images. This research has made use of the SZ-Cluster Database operated by the Integrated Data and Operation Center (IDOC) at the Institut d'Astrophysique Spatiale (IAS) under contract with CNES and CNRS. The scientific results reported in this paper are based on observations made by the \chandra\ X-ray Observatory. This research made use of the NASA/IPAC Extragalactic Database (NED), operated by the Jet Propulsion Laboratory (California Institute of Technology), under contract with the National Aeronautics and Space Administration. This research made use of APLpy, an open-source plotting package for Python hosted at http://aplpy.github.com.

\bibliographystyle{mn2e}
\bibliography{library.bib}

\appendix

\section{Galactic absorption}\label{app:absorption}

From Fig.~\ref{fig:nh-vs-nh} it seems that for five clusters of our sample (A399, A401, AS592, A2104 and Triangulum Australis) the molecular component of the hydrogen density column can not be neglected. Here we want to compare the density column values as derived from the LAB survey of H$_{\rm I}$ \citep{kalberla05}, the \citet{willingale13} work (where the molecular hydrogen density column was derived using a function depending on the product between $N_{\rm H_{I}}$ and the dust extinction), the fits in \citet{cavagnolo09} and ours, obtained by fitting \chandra\ spectra extracted in central regions of the above-mentioned clusters and keeping the density column parameter free to vary. Values are compared in Tab.~\ref{tab:nh_table} and the results of our fits can be summarized as follows.

\begin{itemize}
 \item[-] \textit{A399 and A401} are in line with the values reported from \citet{kalberla05} and \citet{cavagnolo09}, indicating lower values with respect to \citet{willingale13}. % The higher density columns found by \citet{willingale13} could be introduced by a bias in their relation; the low infrared emission of this system appears to exclude a significant presence of H$_{2}$.
 \item[-] \textit{AS592} is in line with \citet{willingale13} while \citet{kalberla05} and \citet{cavagnolo09} suggest lower column densities. However, the discrepancy is $\lesssim25\%$.
 \item[-] \textit{A2104} appears to be in a direction with a higher density column with the respect to the one expected from H$_{\rm I}$ \citep{kalberla05}, in agreement with \citet{willingale13} and \citet{cavagnolo09}.
 \item[-] \textit{Triangulum Australis} is known lay in a region with high absorption, our density column is more in line with \citet{willingale13} than with \citet{kalberla05}.
\end{itemize}

\noindent
We carried out the analysis that led to the results presented in Section~\ref{sec:results} adopting the density column values achieved in our fits (Tab.~\ref{tab:nh_table}) for these five clusters.

\begin{table}
 \centering
 \caption{Density columns reported from \citet{kalberla05} (K05), \citet{willingale13} (W13) and \citet{cavagnolo09} (C09) compared with the results of our fits. Values are reported in units of $10^{20}$ cm$^{-2}$.}
 \label{tab:nh_table}
  \begin{tabular}{lrrrr} 
  \hline
  Cluster name & K05 & W13 & C09 & Fit \\
  \hline
  
  A399 & 10.6 & 17.1 & 11.5 & $9.8\pm1.1$ \\
  
  A401 & 9.88 & 15.2 & 12.5 & $10.8\pm0.4$ \\
                      
  AS592 & 6.07 & 8.30 & 6.41 & $8.0\pm0.5$ \\
  
  A2104 & 8.37 & 14.5 & 14.9 & $15.8\pm0.7$ \\
    
  Triangulum Australis & 11.5 & 17.0 & $\ldots$ & $18.4\pm0.8$ \\

  \hline
  \end{tabular}
\end{table}

\section{NXB modeling}\label{app:nxb}

\begin{figure}
 \centering
 \includegraphics[width=.73\hsize,angle=-90]{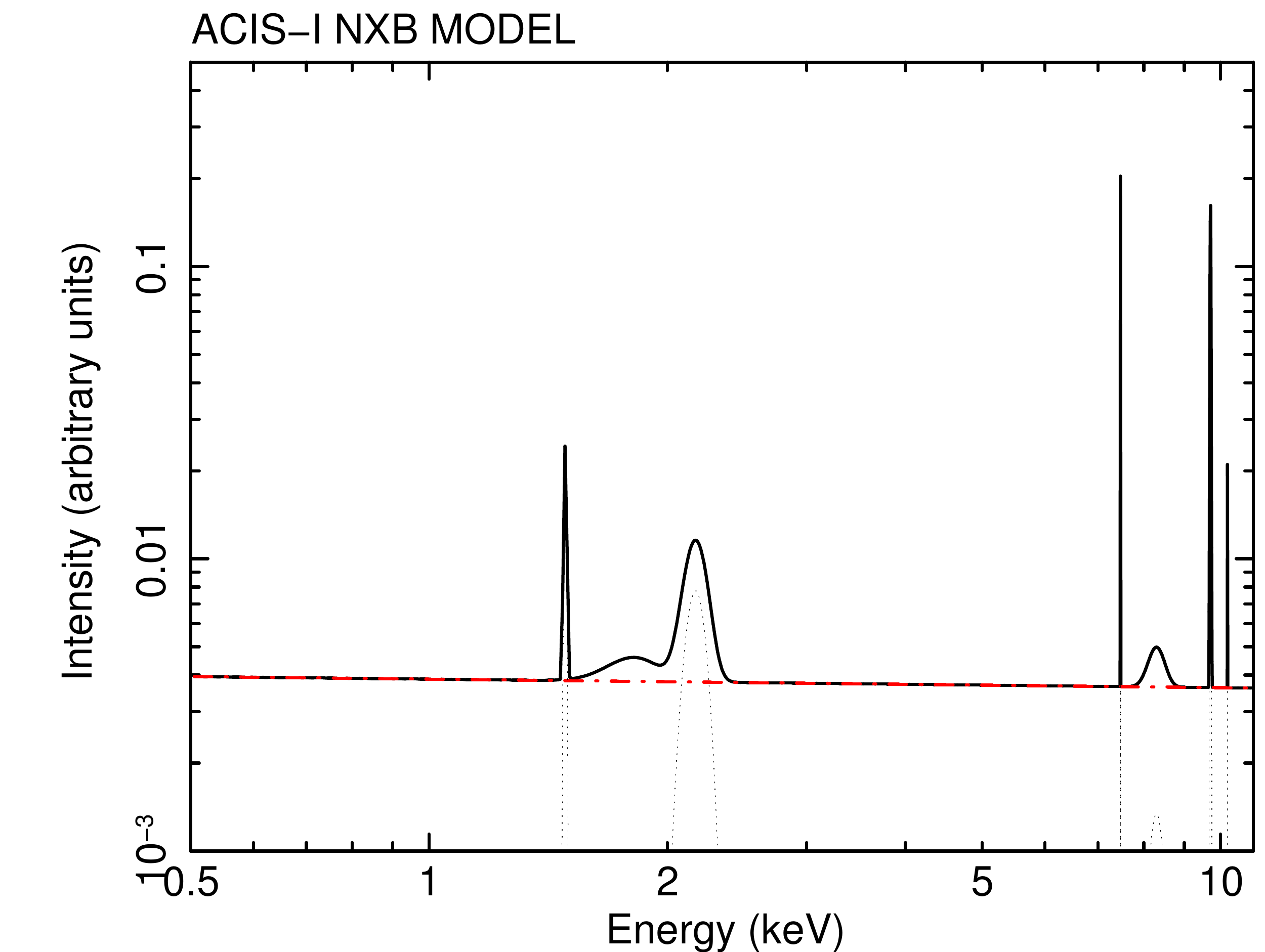}
 \includegraphics[width=.73\hsize,angle=-90]{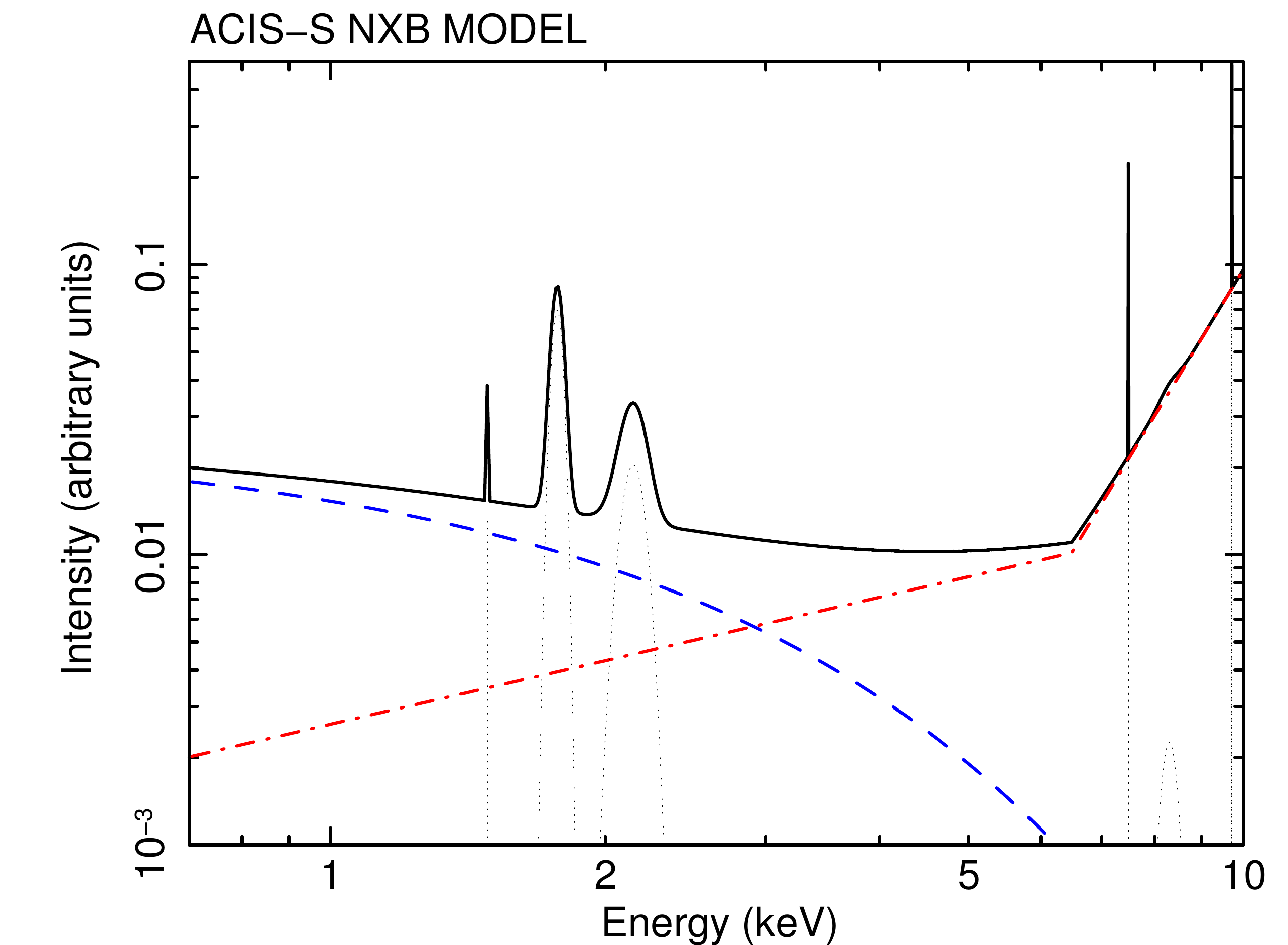}
 \caption{\acisi\ (\textit{top}) and \aciss\ (\textit{bottom}) NXB models. Gaussian lines are reported in dotted black, the exponential decay in dashed blue (not visible for \acisi) and the (broken) power-law in dot-dashed red.}
 \label{fig:nxb_plot}
\end{figure}

The NXB spectrum is different for the \acisi\ and \aciss\ detectors\footnote{http://cxc.cfa.harvard.edu/contrib/maxim/stowed/} (Eq.~\ref{eq:bkg_model}) and its continuum part can be rewritten as

\begin{equation}\label{eq:nxb_acisi}
  C(E) = K_1 e^{-A_1 E} + K_2 E^{-A_2} 
\end{equation}

\noindent
for \acisi\ and as

\begin{eqnarray}\label{eq:nxb_aciss}
  C(E) = K_1 e^{-A_1 E} + \biggl\{
  \begin{array}{llr}
   K_2 E^{-\Gamma_1} & \mbox{if} & E\leq E_b  \\
   K_2 E_b^{{\Gamma_2-\Gamma_1}} \bigl(\frac{E}{1\:{\rm keV}}\bigr)^{-\Gamma_2} & \mbox{if} & E > E_b \\
 \end{array}
\end{eqnarray}

\noindent
for \aciss, where the parameters $K$ represent the normalizations, $A$ are dimensionless factors and $E_b$ is the break point for the energy of the broken power-law described by the two photon indexes $\Gamma$. \\
\indent
The \acisi\ NXB was investigated by \citet{bartalucci14}, they performed a detailed analysis of the stowed \vfaint\ \acisi\ event files to create an analytical model of background. We adopted the values reported in Tab.~1 of \citet{bartalucci14} for the parameters of Eq.~\ref{eq:nxb_acisi}. To model the \aciss\ NXB we used a similar approach and extracted spectra from the S3 chip (used for the imaging of the target) of the stowed \aciss\ event files taken in the closer epoch to the observation. Our best-fitting values are reported in Tab.~\ref{tab:nxb_fit} for both \faint\ and \vfaint\ observing mode. The NXB models for \acisi\ and \aciss\ are shown in Fig.~\ref{fig:nxb_plot}. \\
\indent
Once that the shape of the NXB has been modeled on the stowed files, the total (astrophysical+instrumental) background of Eq.~\ref{eq:bgk_x} was fitted in a peripheral region of the target observation, where the cluster emission is almost negligible, letting the normalizations free to vary and then it was rescaled in the region of interest. In the fitting of \aciss\ spectral regions we found that for $E>E_b$ the spectrum can not be described by a single value of $\Gamma_2$ that, for this reason, was also let free to vary in the fits. This possibly indicates a spatial variation with respect to the chip coordinates of the second photon index of the broken power-law. A deeper investigation of the \aciss\ NXB (as the one performed by \citealt{bartalucci14} for \acisi) would be of great interest and certainly desirable; this is, however, beyond the scope of this paper.

\begin{table}
 \centering
 \caption{Best-fit parameters of Eq.~\ref{eq:nxb_aciss}. The normalization values $K$ are given in \xspec\ units, the break energy $E_b$ is in \kev.}
 \label{tab:nxb_fit}
  \begin{tabular}{lrr} 
  \hline
  Parameter & \multicolumn{2}{c}{\aciss} \\
            & \faint\ & \vfaint\ \\
  \hline
  $K_1$ & 0.0257 & 0.0167 \\
  $A_1$ & 0.52 & 0.40 \\
  $K_2$ & 0.0025 & 0.0028 \\
  $\Gamma_1$ & -0.73 & -0.52 \\
  $\Gamma_2$ & -5.19 & -5.57 \\
  $E_b$ & 6.48 & 6.42 \\
  \hline
  \end{tabular}
\end{table}

\section{Goodness of fits}\label{app:errors}

\begin{figure}
 \centering
 \includegraphics[width=.23\textwidth]{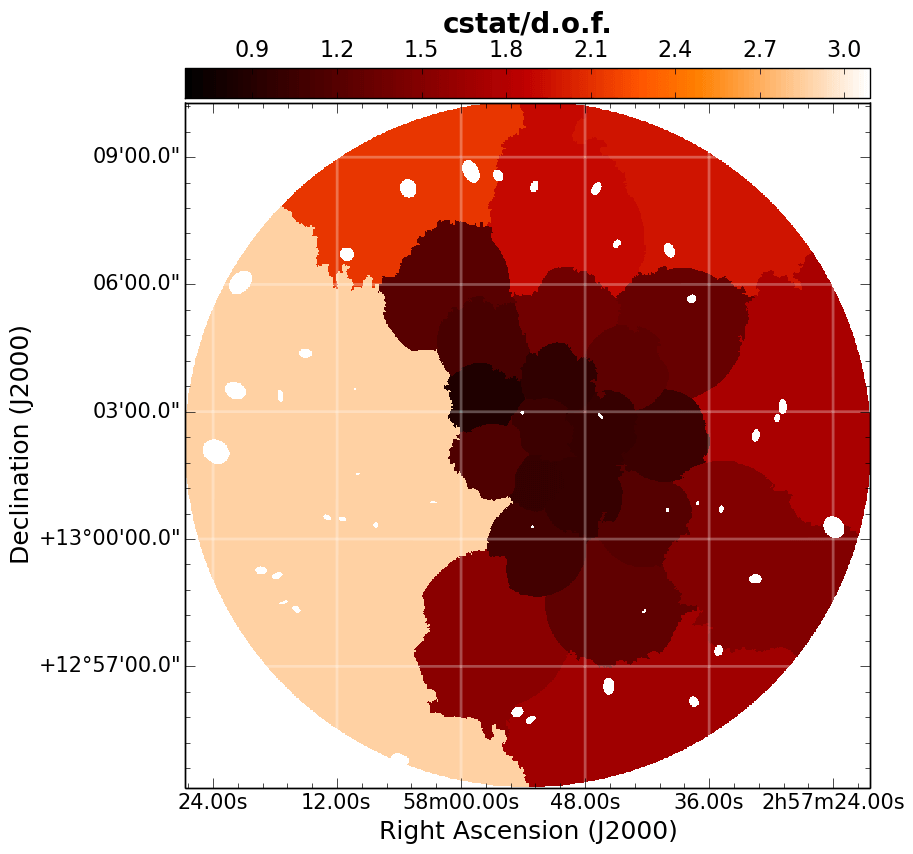}
 \includegraphics[width=.23\textwidth]{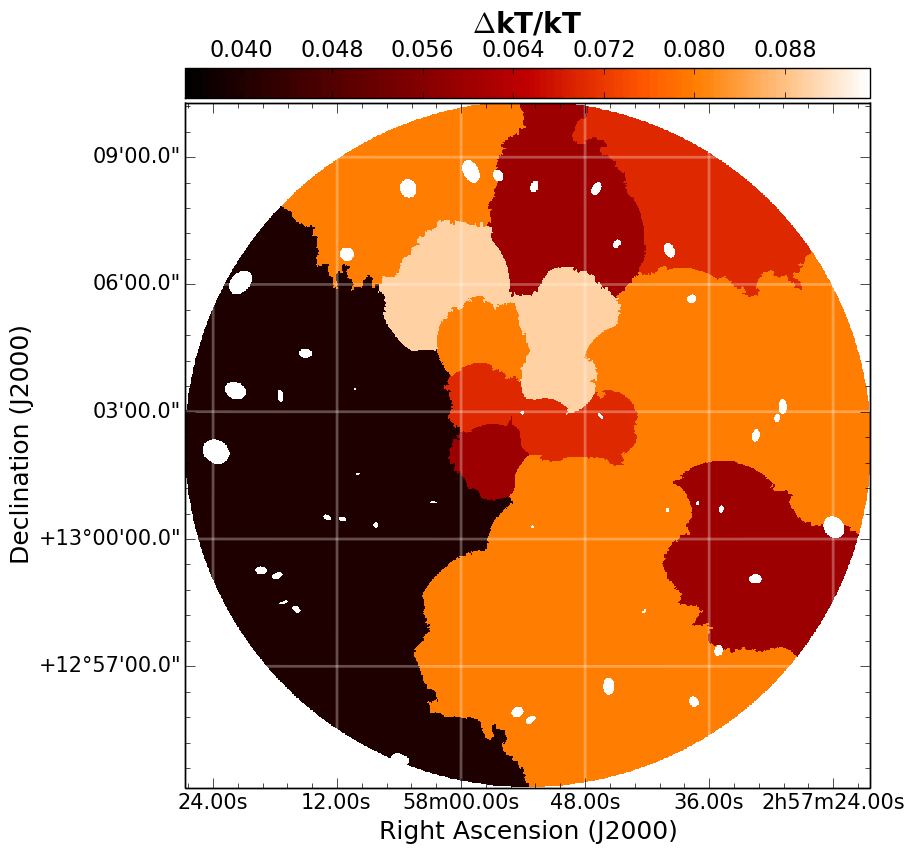}
 \caption{Values of \cstatdof\ (\textit{left}) and error map on $kT$ (\textit{right}) for A399 (\cf\ Fig.~\ref{fig:a399}).}
 \label{fig:a399_errors}
\end{figure}

\begin{figure}
 \centering
 \includegraphics[width=.23\textwidth]{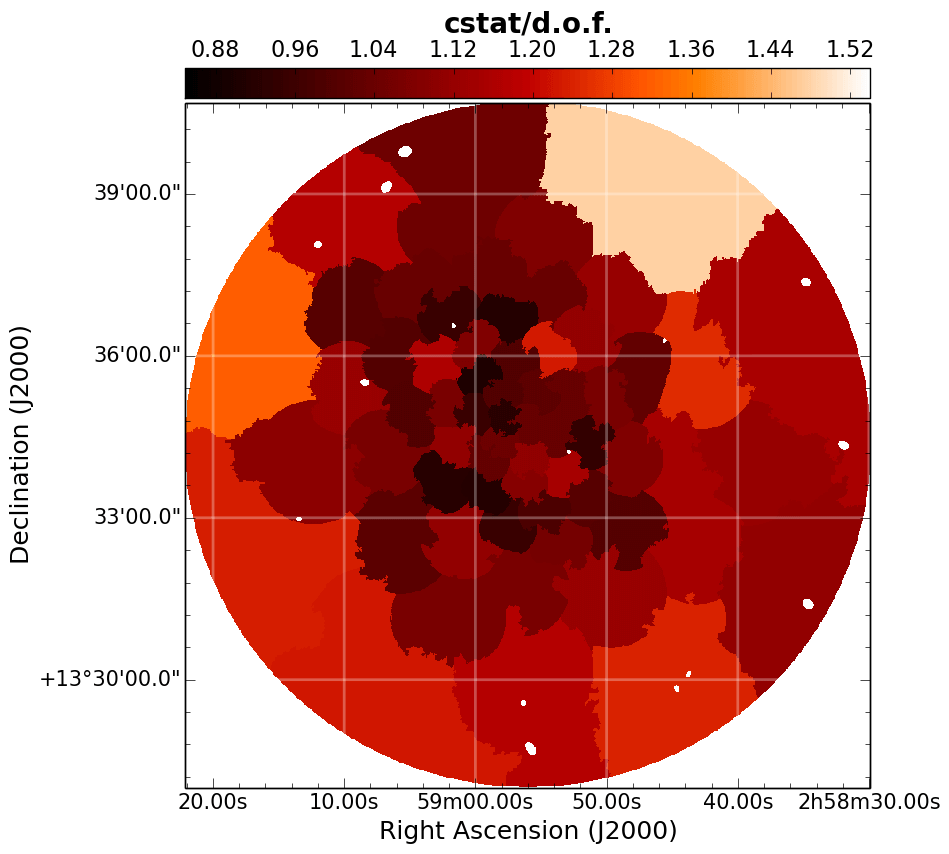}
 \includegraphics[width=.23\textwidth]{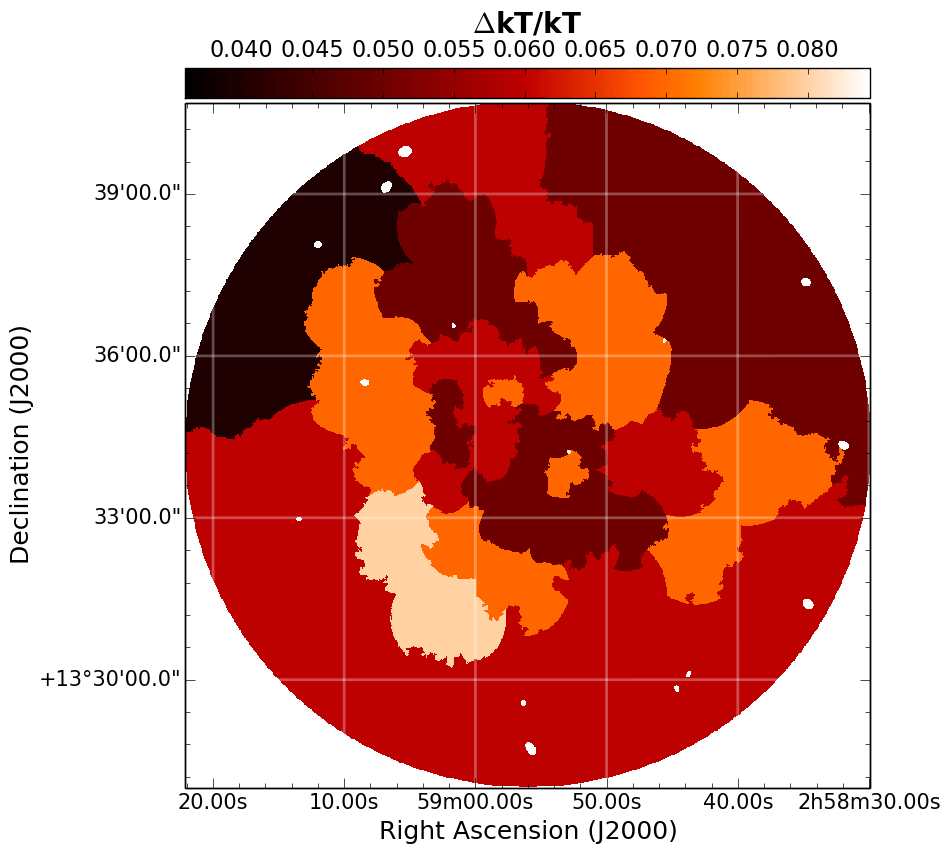}
 \caption{The same as Fig.~\ref{fig:a399_errors} but for A401 (\cf\ Fig.~\ref{fig:a401}).}
 \label{fig:a401_errors}
\end{figure}

\begin{figure}
 \centering
 \includegraphics[width=.23\textwidth]{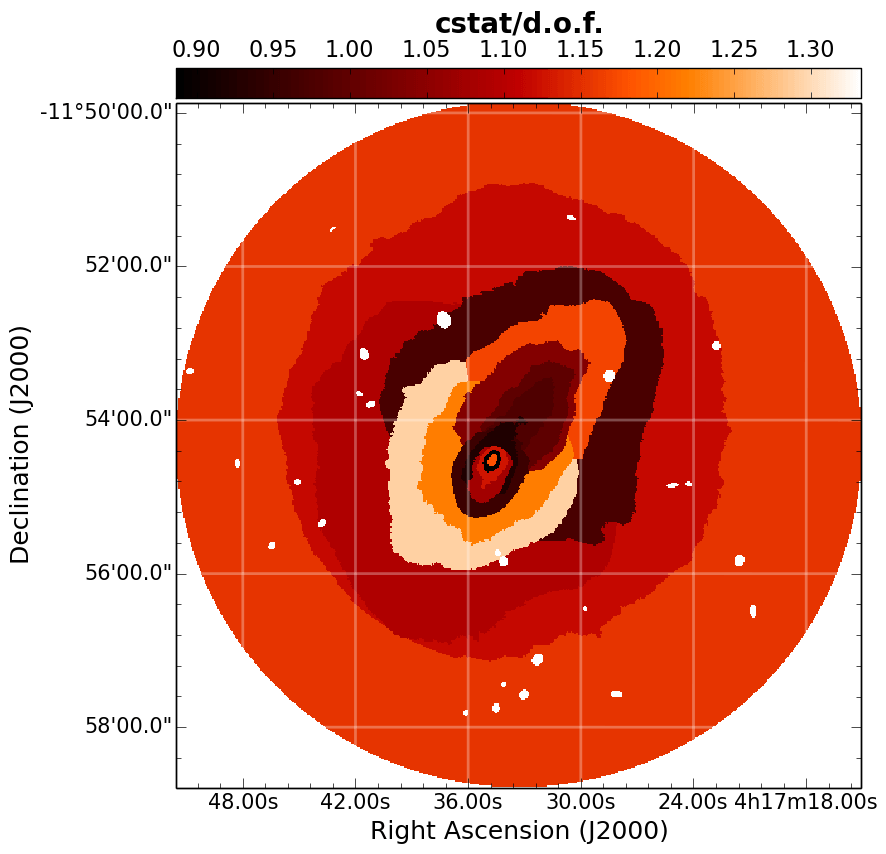}
 \includegraphics[width=.23\textwidth]{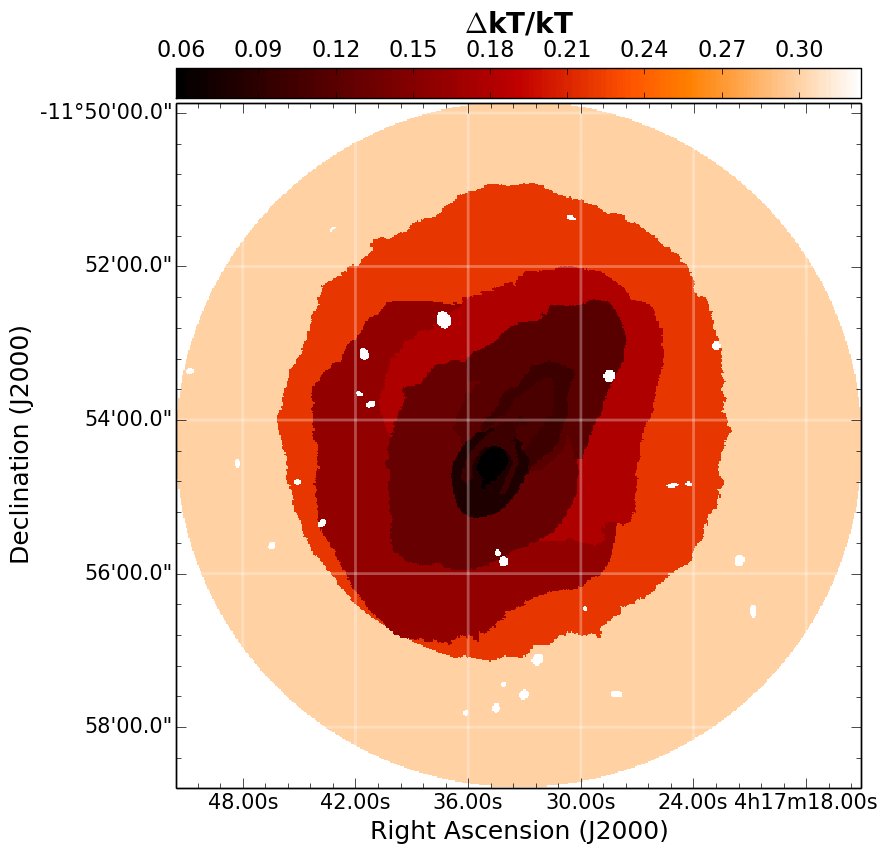}
 \caption{The same as Fig.~\ref{fig:a399_errors} but for MACSJ0417 (\cf\ Fig.~\ref{fig:macsj0417}).}
 \label{fig:macsj0417_errors}
\end{figure}

\begin{figure}
 \centering
 \includegraphics[width=.23\textwidth]{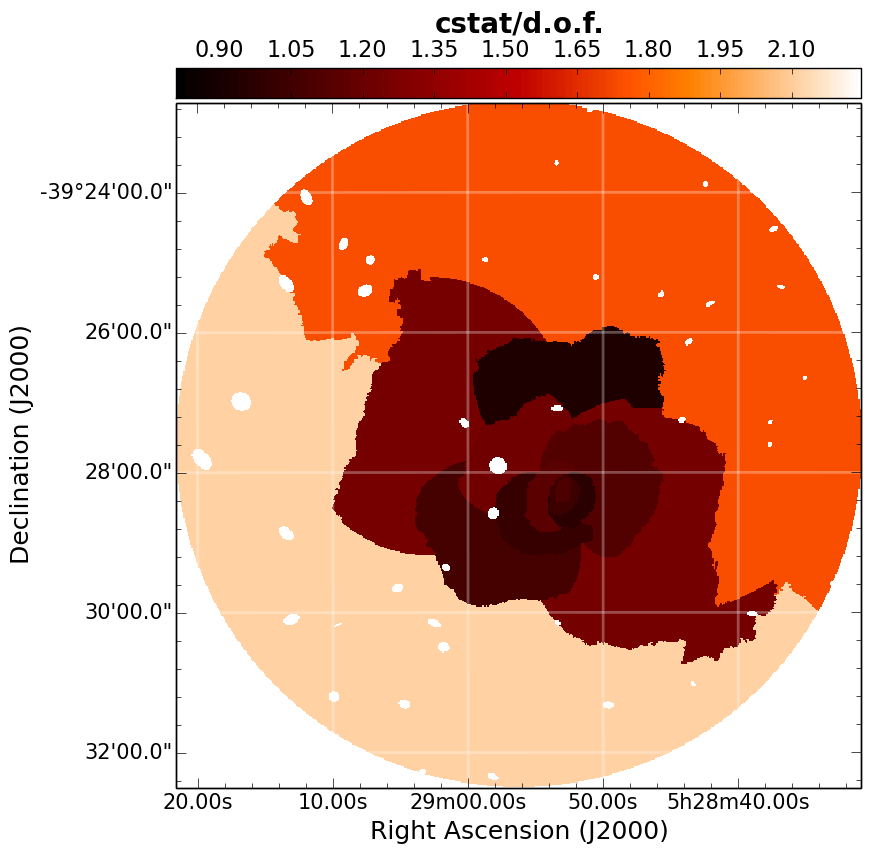}
 \includegraphics[width=.23\textwidth]{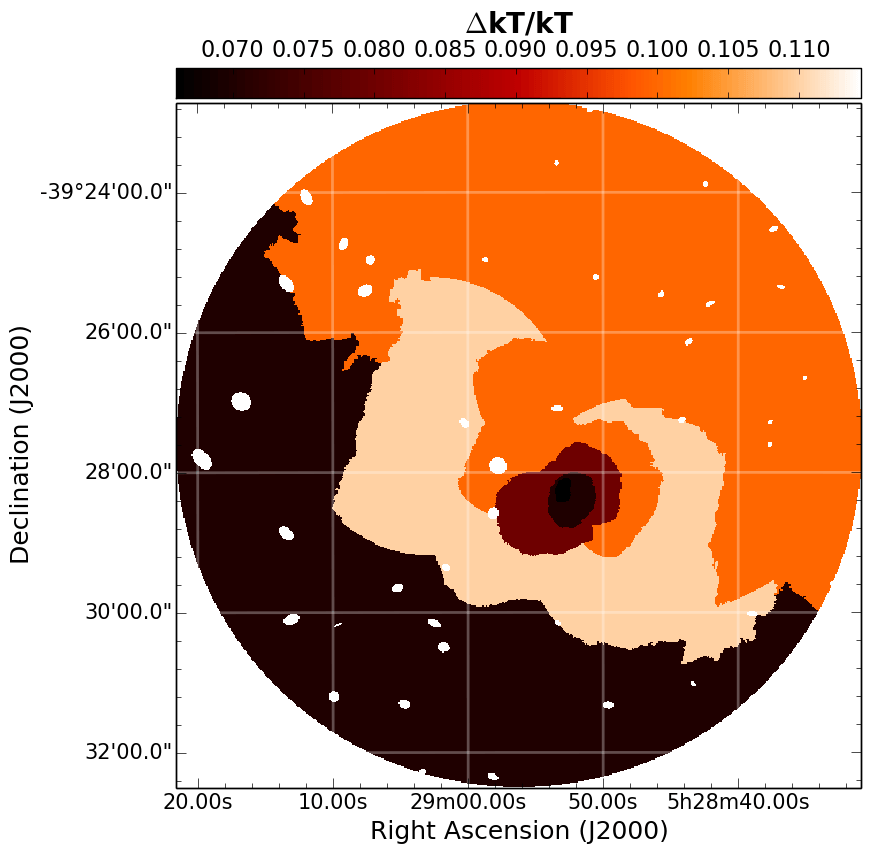}
 \caption{The same as Fig.~\ref{fig:a399_errors} but for RXCJ0528 (\cf\ Fig.~\ref{fig:rxcj0528}).}
 \label{fig:rxcj0528_errors}
\end{figure}

\begin{figure}
 \centering
 \includegraphics[width=.23\textwidth]{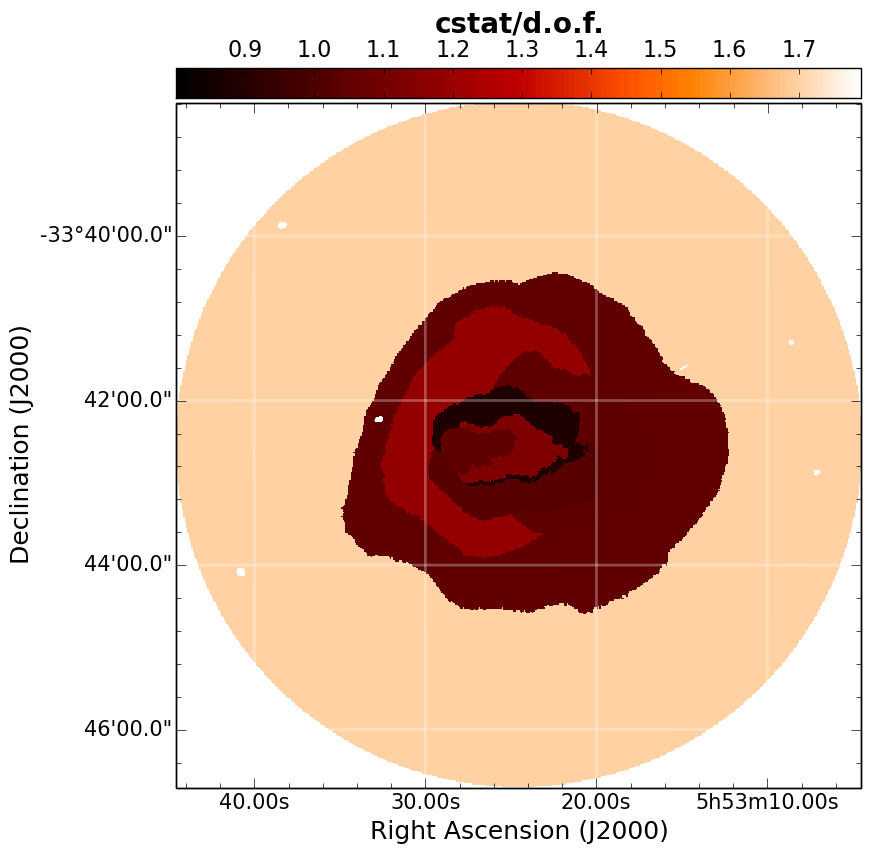}
 \includegraphics[width=.23\textwidth]{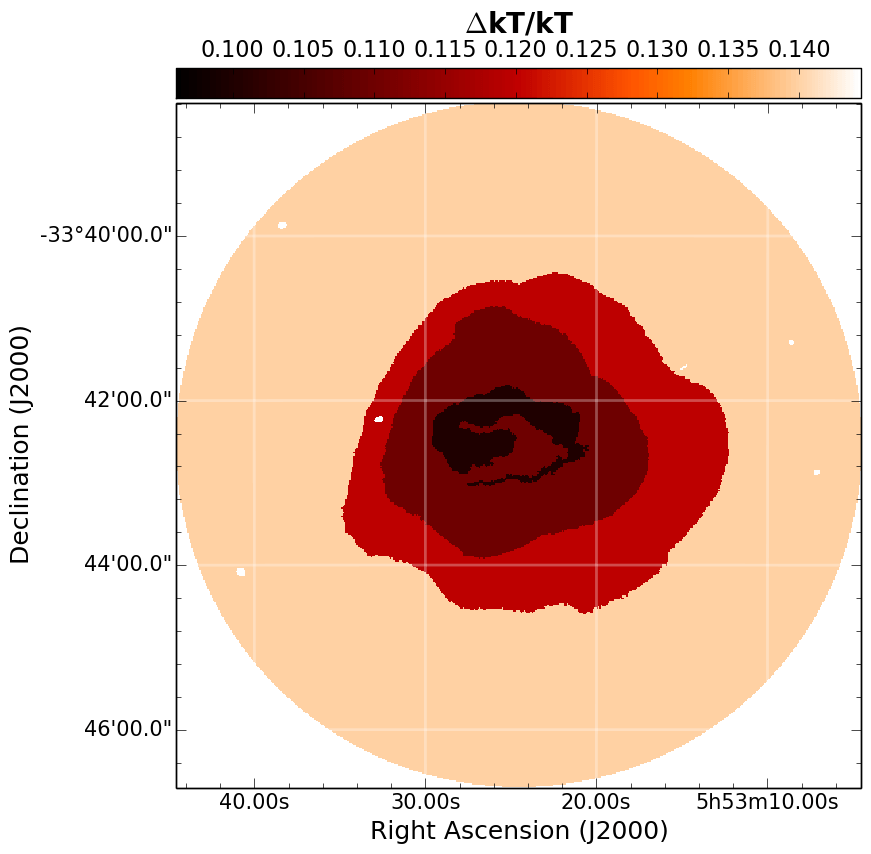}
 \caption{The same as Fig.~\ref{fig:a399_errors} but for MACSJ0553 (\cf\ Fig.~\ref{fig:macsj0553}).}
 \label{fig:macsj0553_errors}
\end{figure}

\begin{figure}
 \centering
 \includegraphics[width=.23\textwidth]{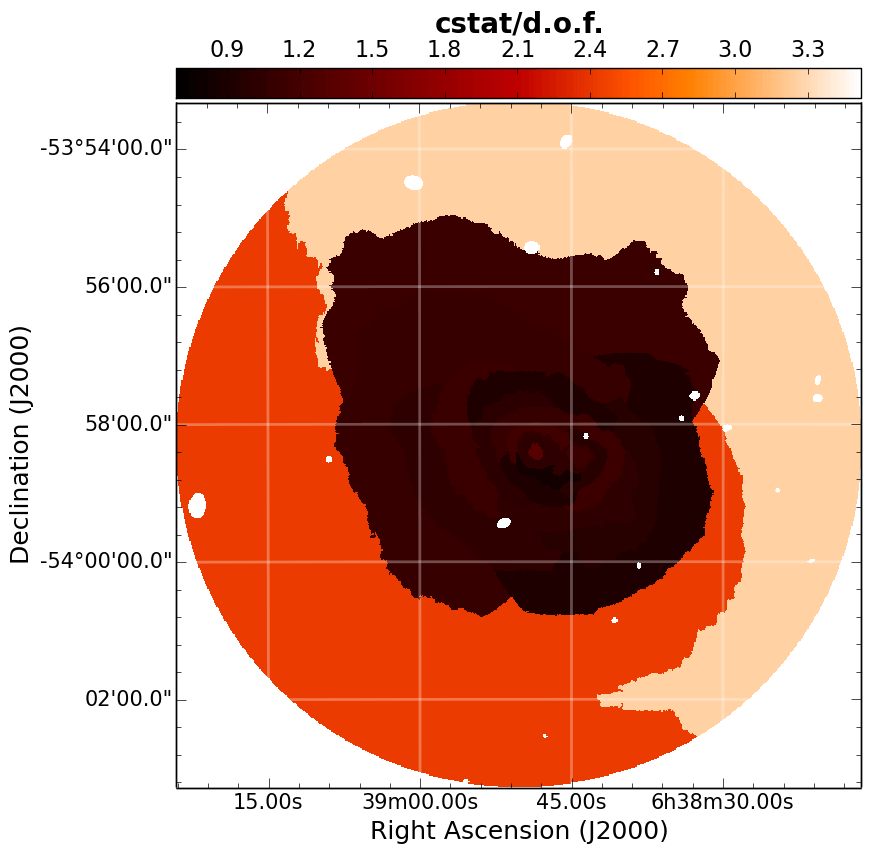}
 \includegraphics[width=.23\textwidth]{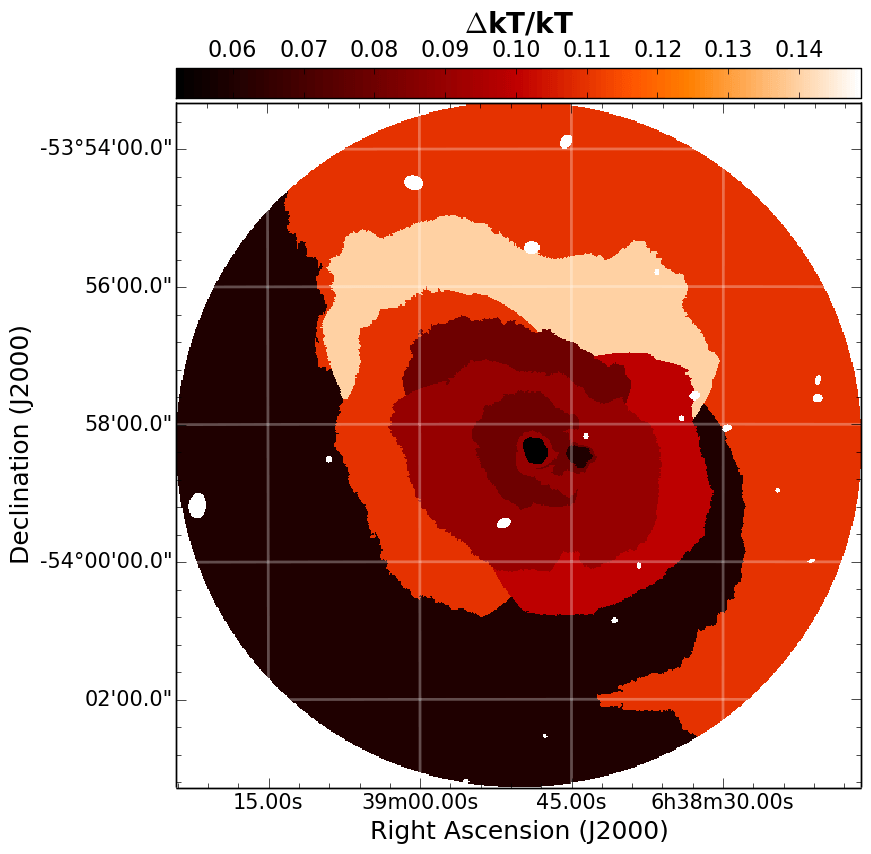}
 \caption{The same as Fig.~\ref{fig:a399_errors} but for AS592 (\cf\ Fig.~\ref{fig:as592}).}
 \label{fig:as592_errors}
\end{figure}

\begin{figure}
 \centering
 \includegraphics[width=.23\textwidth]{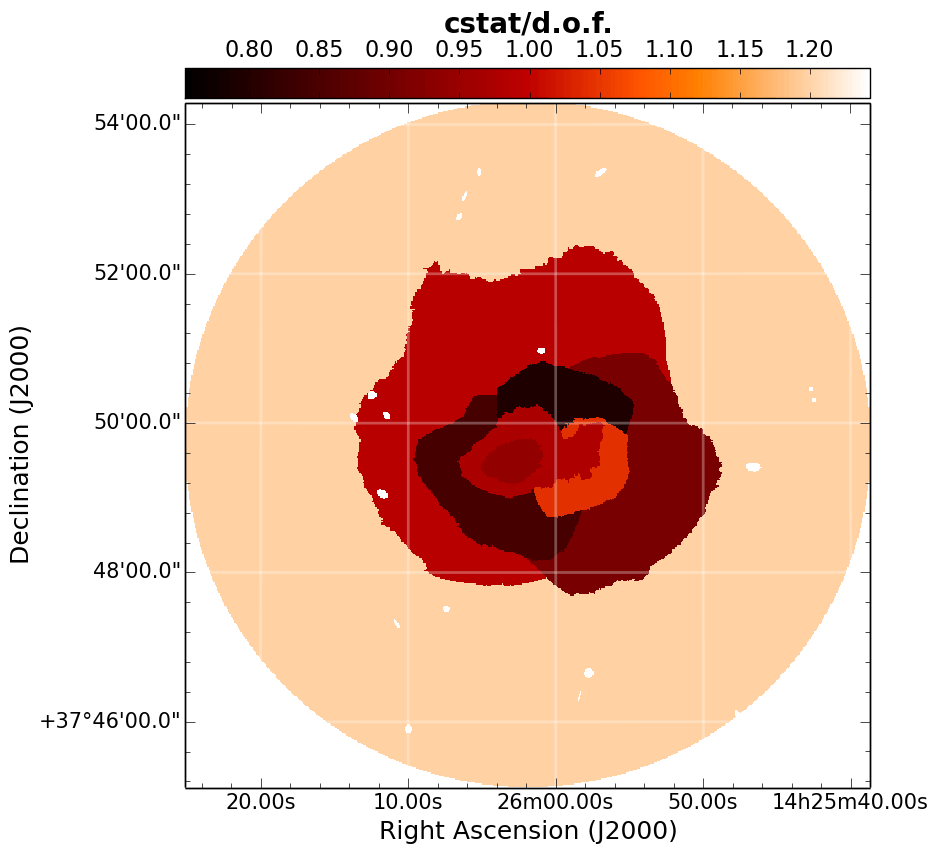}
 \includegraphics[width=.23\textwidth]{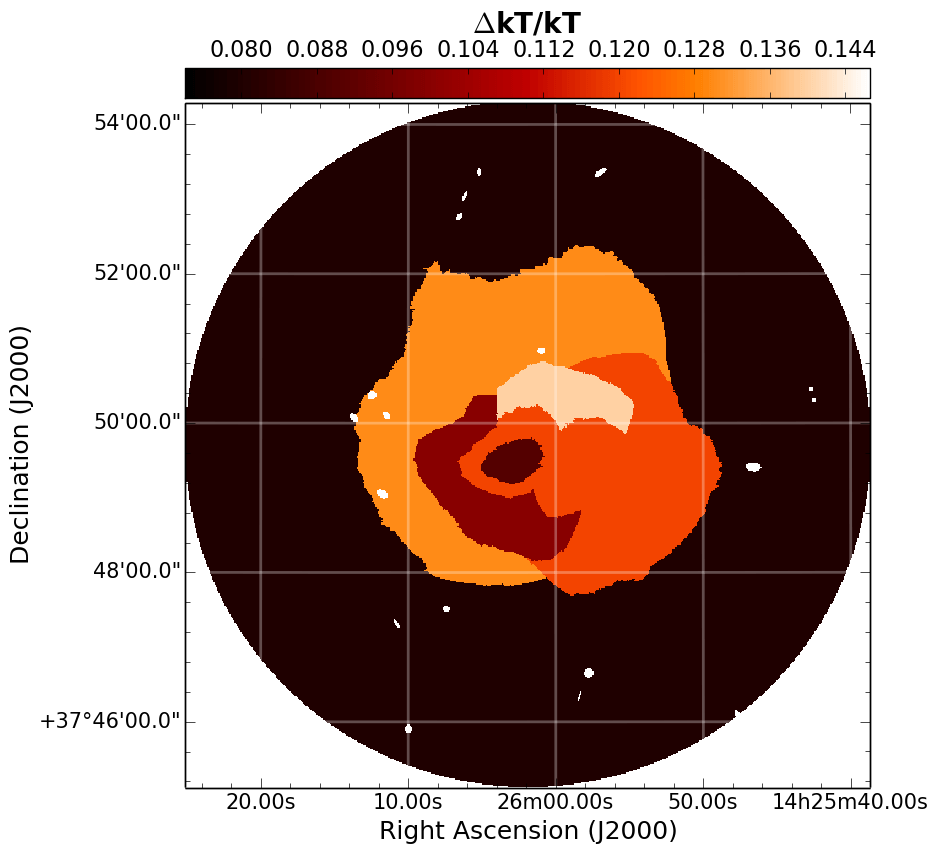}
 \caption{The same as Fig.~\ref{fig:a399_errors} but for A1914 (\cf\ Fig.~\ref{fig:a1914}).}
 \label{fig:a1914_errors}
\end{figure}

\begin{figure}
 \centering
 \includegraphics[width=.23\textwidth]{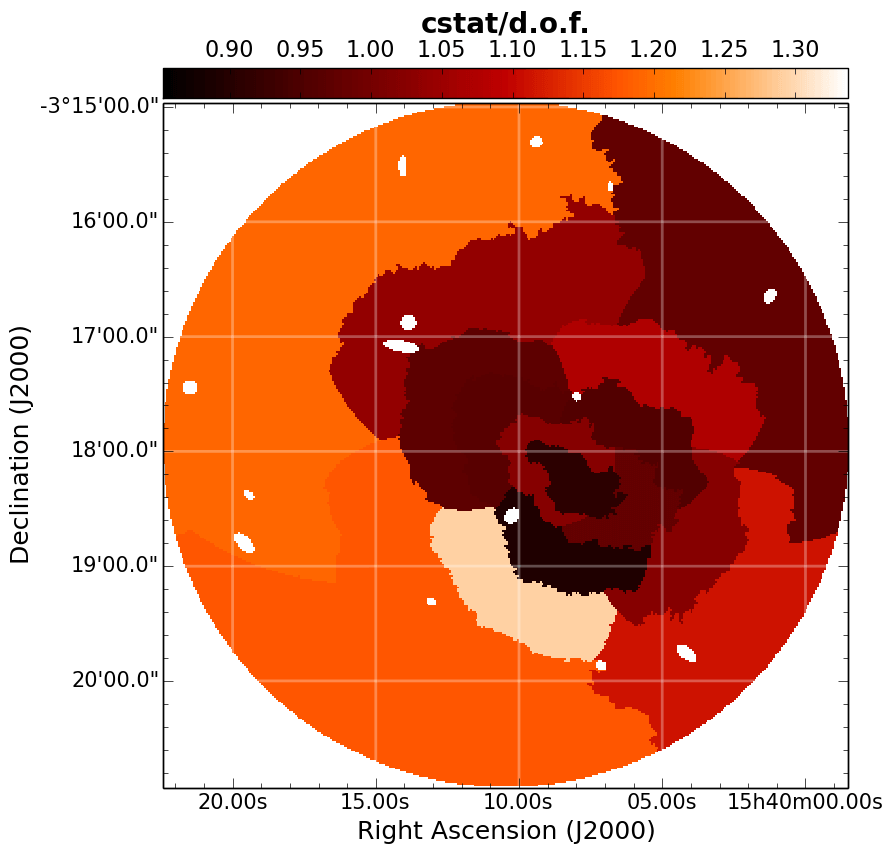}
 \includegraphics[width=.23\textwidth]{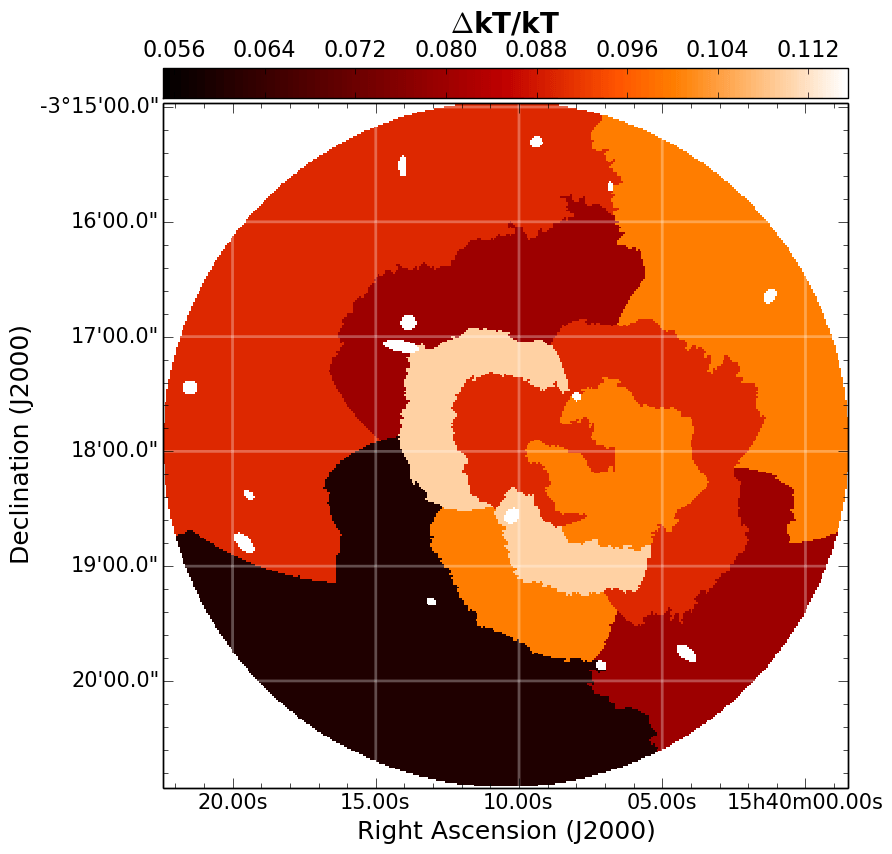}
 \caption{The same as Fig.~\ref{fig:a399_errors} but for A2104 (\cf\ Fig.~\ref{fig:a2104}).}
 \label{fig:a2104_errors}
\end{figure}

\begin{figure}
 \centering
 \includegraphics[width=.23\textwidth]{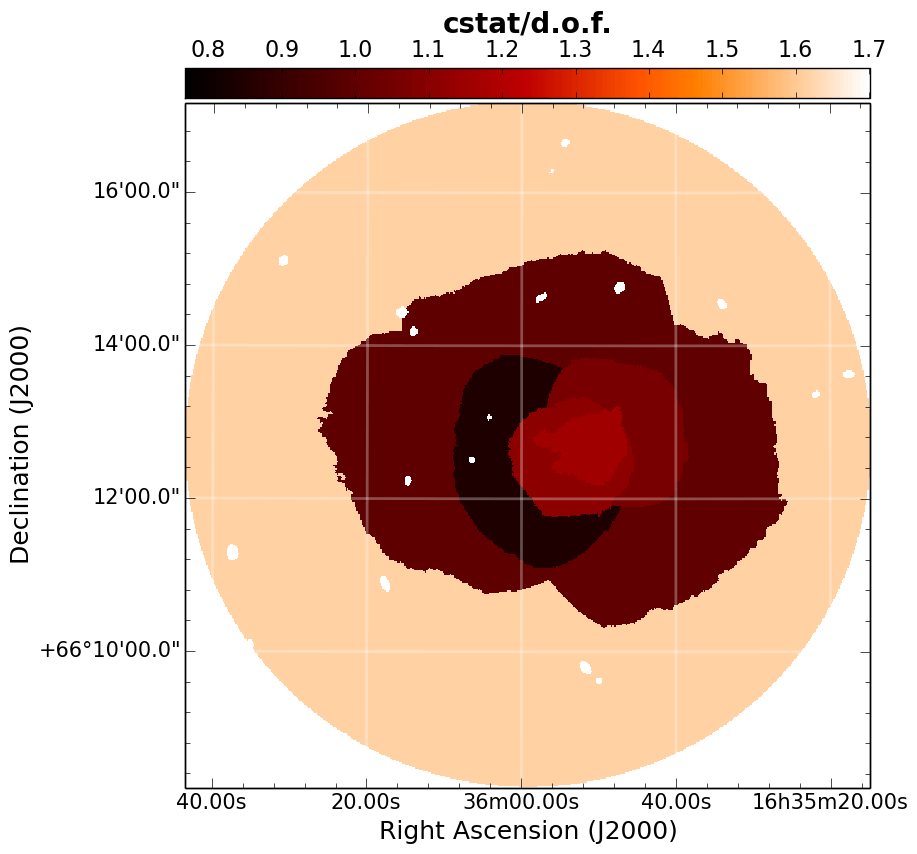}
 \includegraphics[width=.23\textwidth]{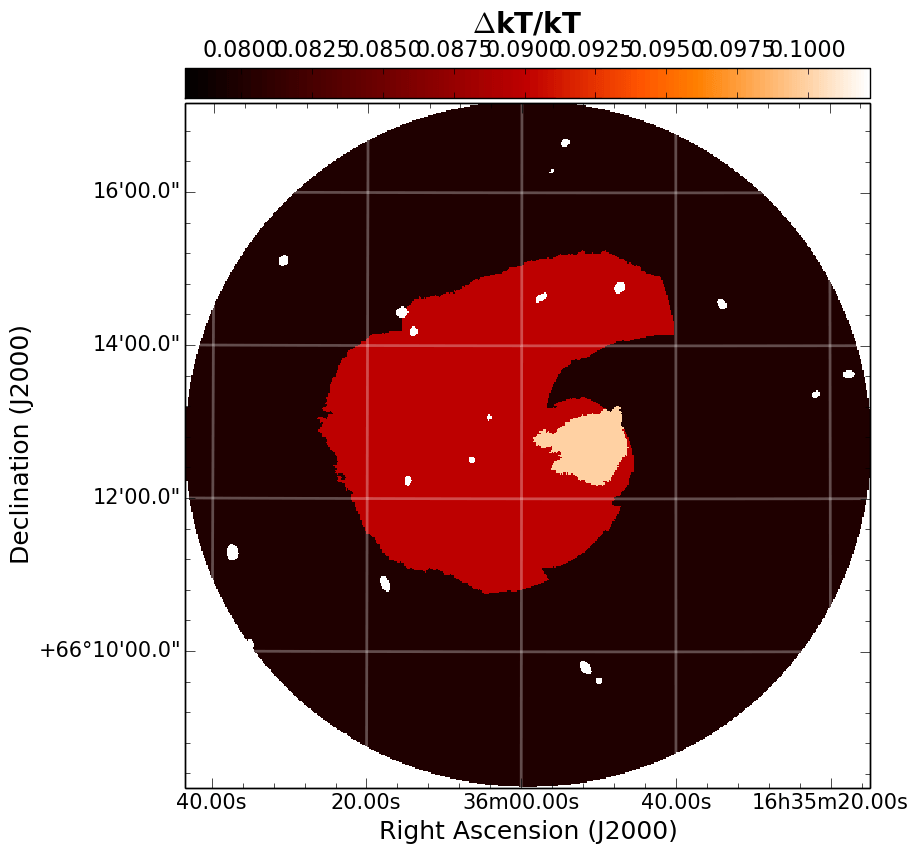}
 \caption{The same as Fig.~\ref{fig:a399_errors} but for A2218 (\cf\ Fig.~\ref{fig:a2218}).}
 \label{fig:a2218_errors}
\end{figure}

\begin{figure}
 \centering
 \includegraphics[width=.23\textwidth]{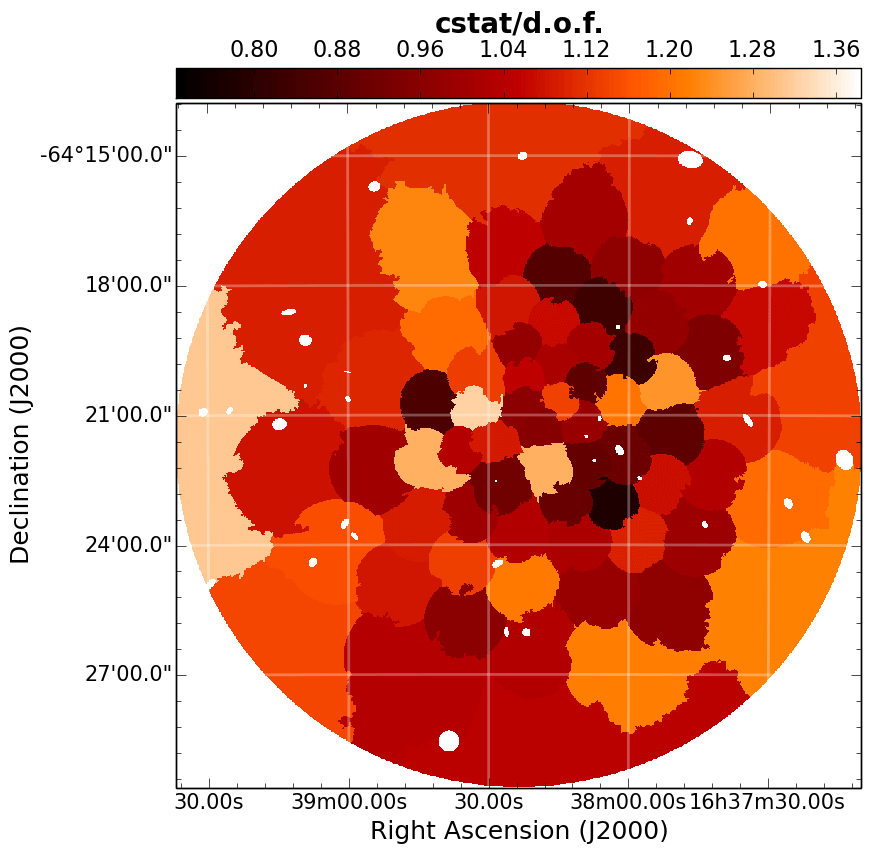}
 \includegraphics[width=.23\textwidth]{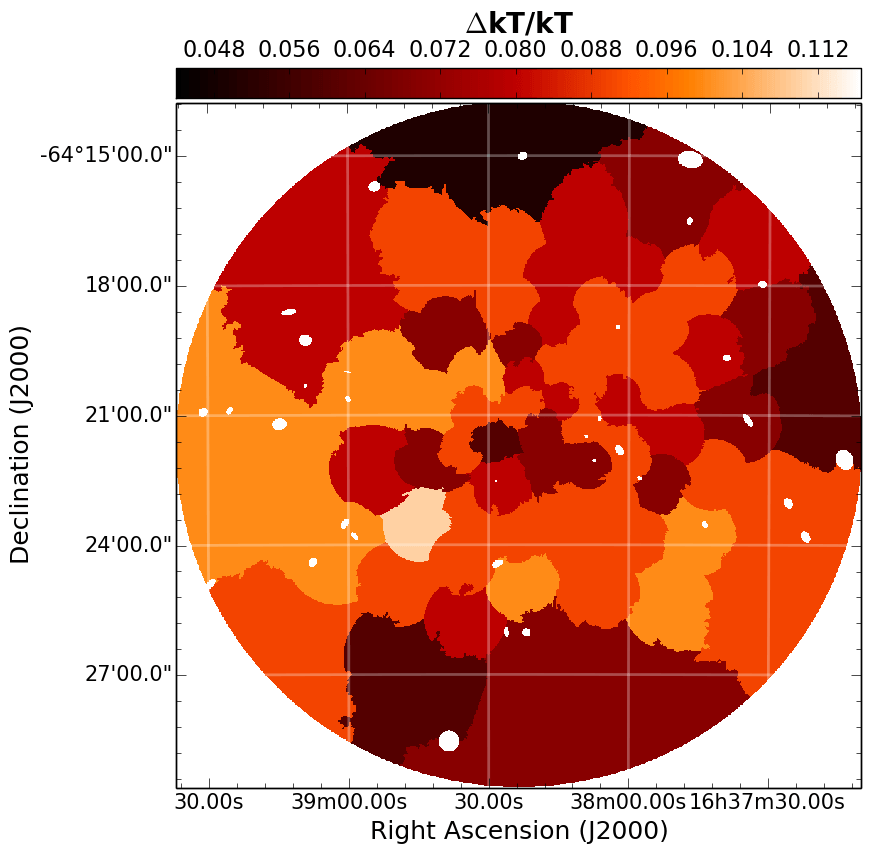}
 \caption{The same as Fig.~\ref{fig:a399_errors} but for Triangulum Australis (\cf\ Fig.~\ref{fig:triangulum}).}
 \label{fig:triangulum_errors}
\end{figure}

\begin{figure}
 \centering
 \includegraphics[width=.23\textwidth]{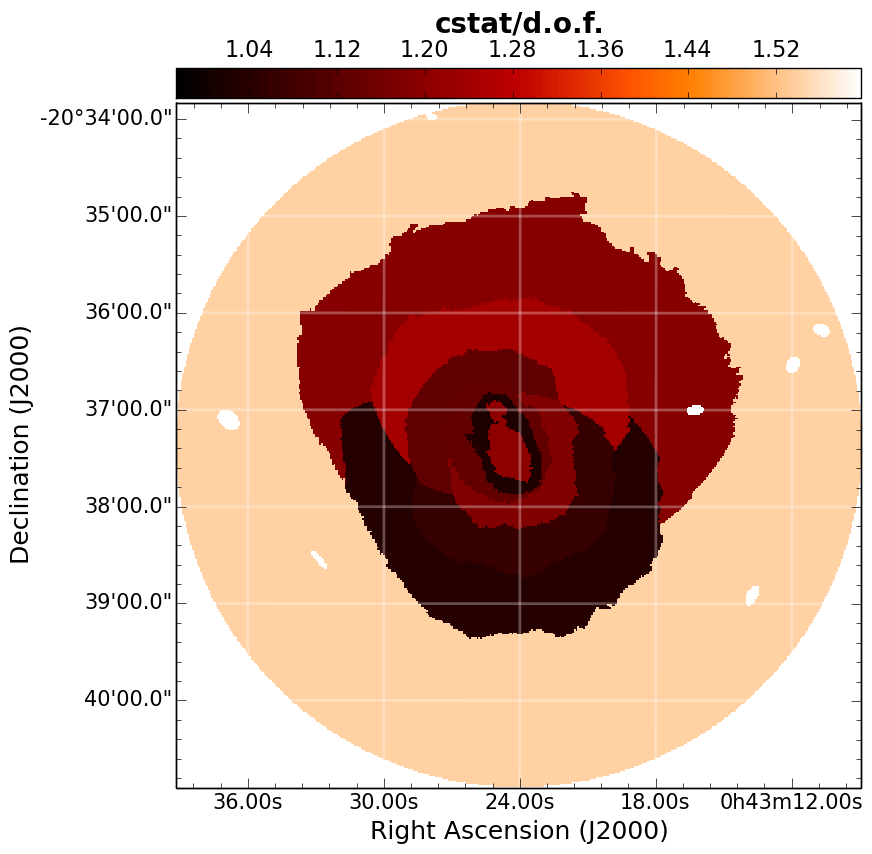}
 \includegraphics[width=.23\textwidth]{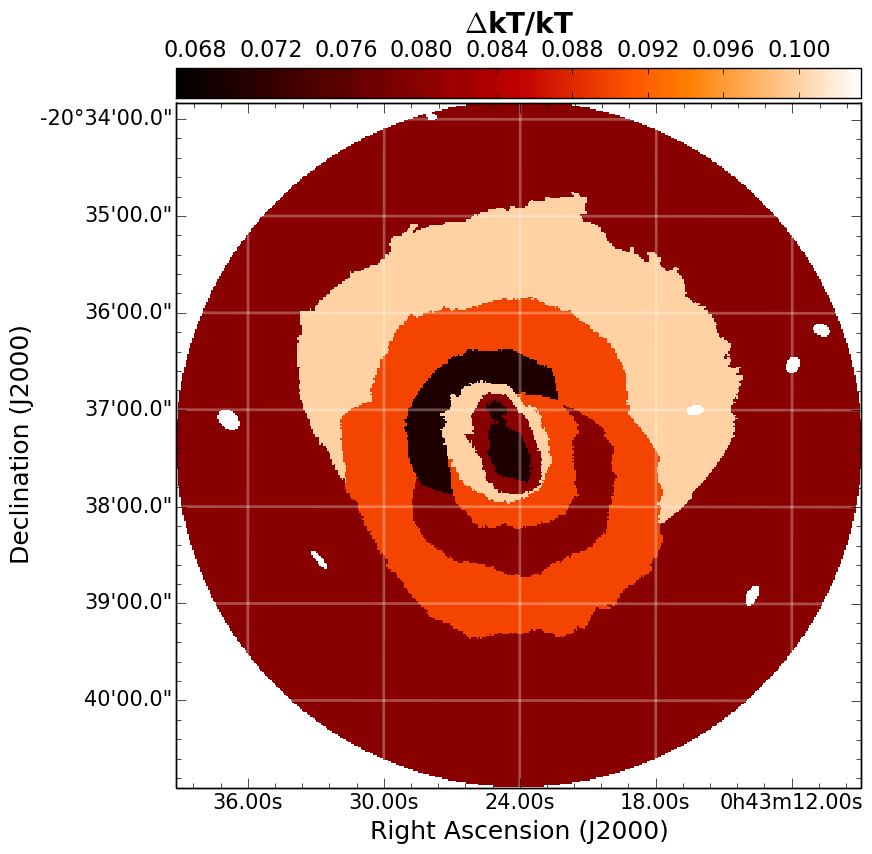}
 \caption{The same as Fig.~\ref{fig:a399_errors} but for A2813 (\cf\ Fig.~\ref{fig:a2813}).}
 \label{fig:a2813_errors}
\end{figure}

\begin{figure}
 \centering
 \includegraphics[width=.23\textwidth]{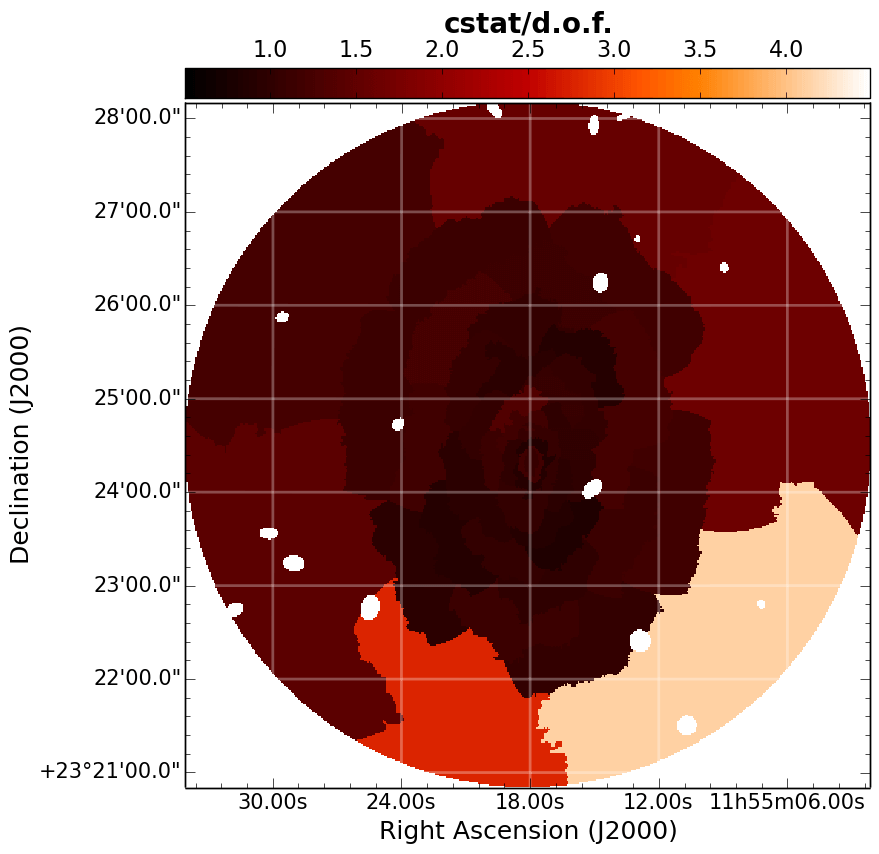}
 \includegraphics[width=.23\textwidth]{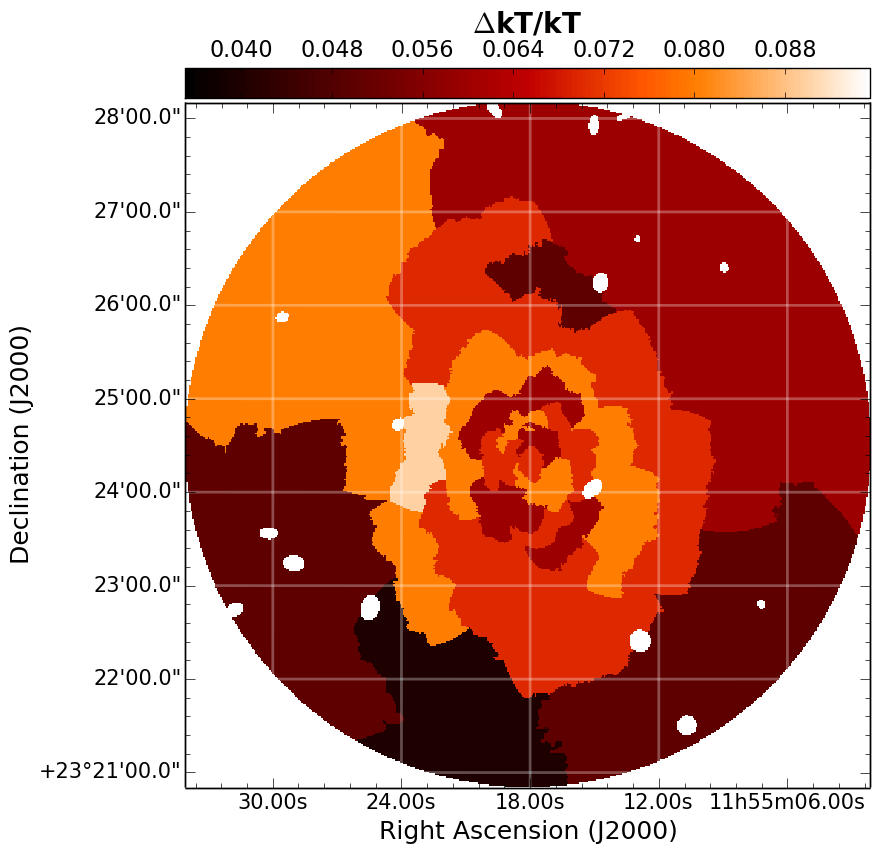}
 \caption{The same as Fig.~\ref{fig:a399_errors} but for A1413 (\cf\ Fig.~\ref{fig:a1413}).}
 \label{fig:a1413_errors}
\end{figure}

\begin{figure}
 \centering
 \includegraphics[width=.23\textwidth]{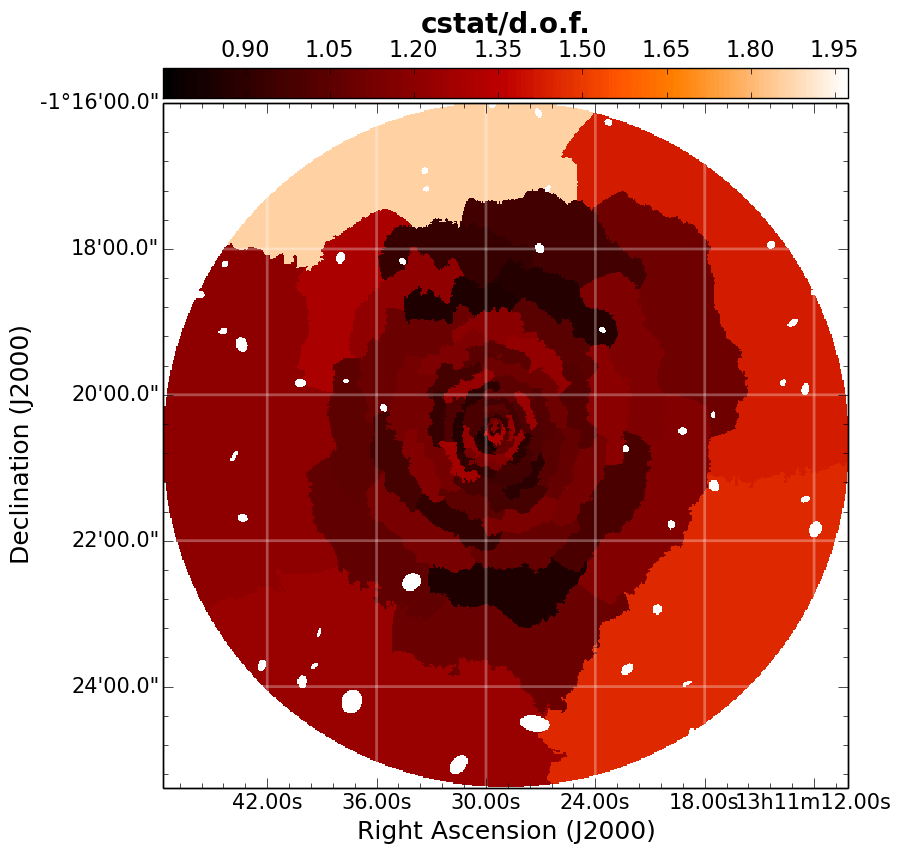}
 \includegraphics[width=.23\textwidth]{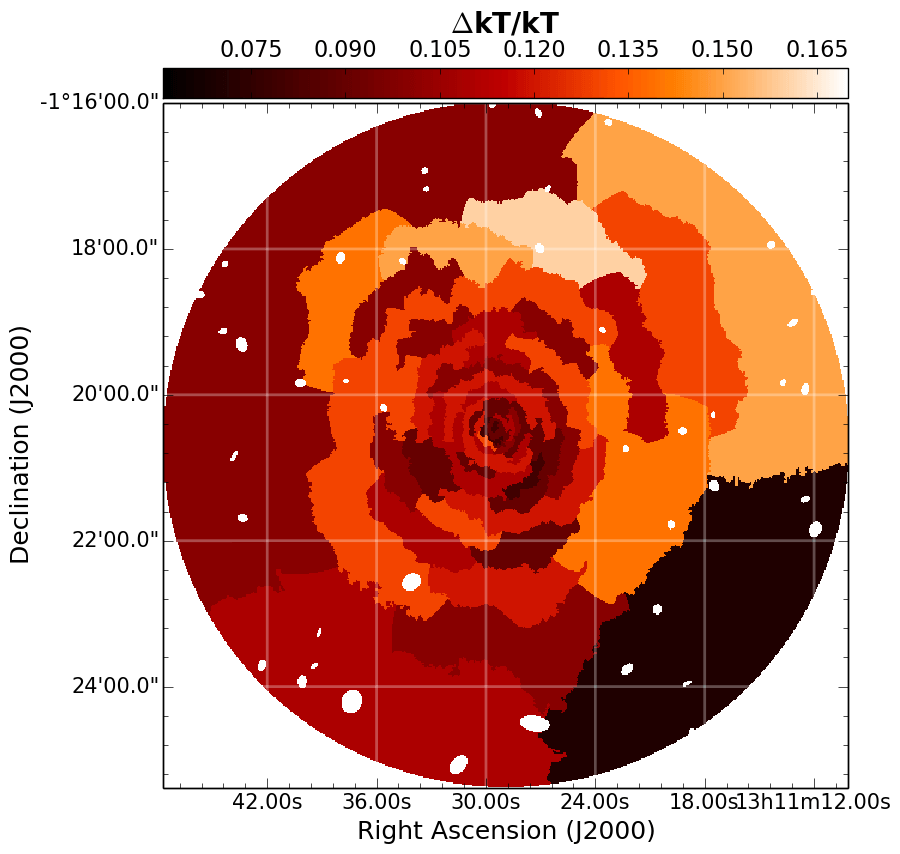}
 \caption{The same as Fig.~\ref{fig:a399_errors} but for A1689 (\cf\ Fig.~\ref{fig:a1689}).}
 \label{fig:a1689_errors}
\end{figure}

\begin{figure}
 \centering
 \includegraphics[width=.23\textwidth]{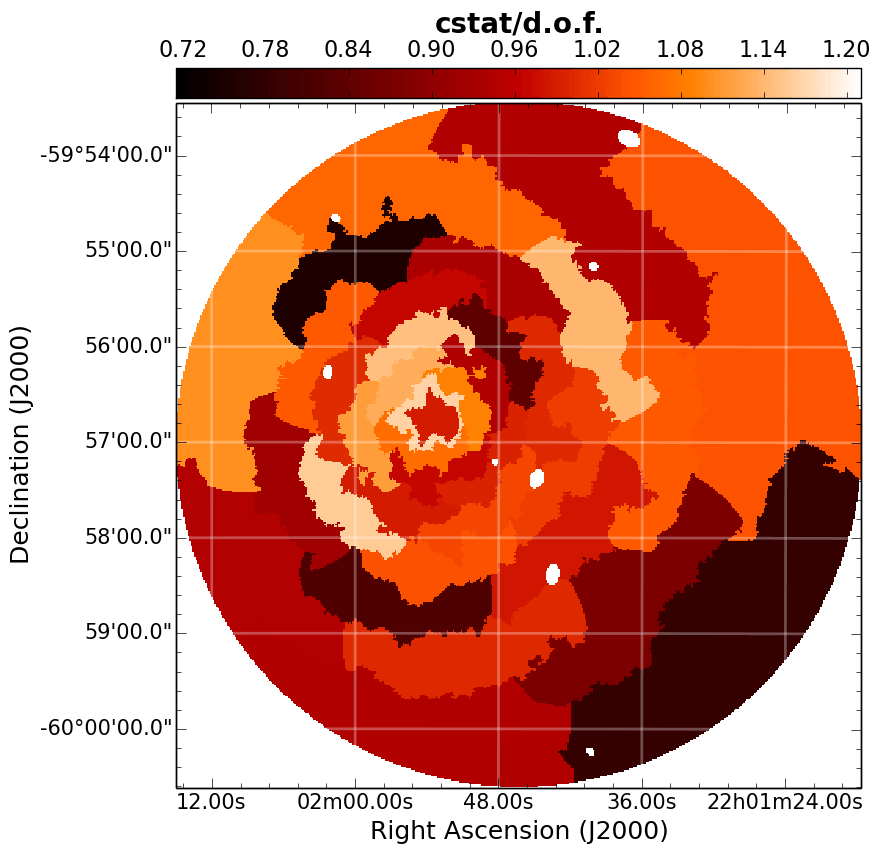}
 \includegraphics[width=.23\textwidth]{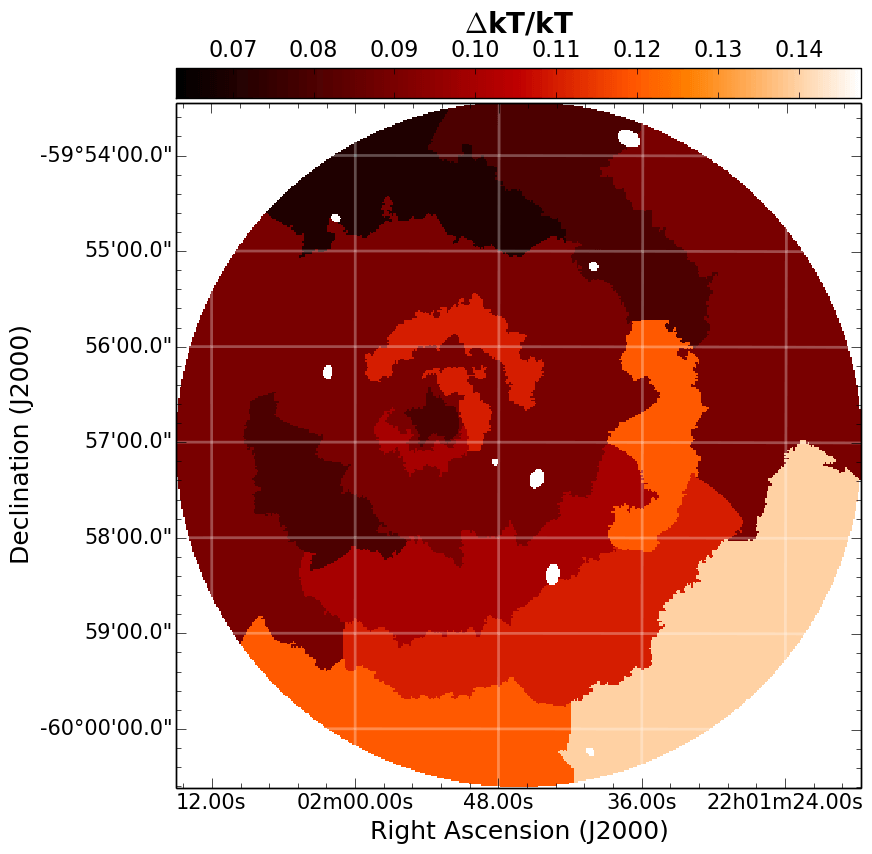}
 \caption{The same as Fig.~\ref{fig:a399_errors} but for A3827 (\cf\ Fig.~\ref{fig:a3827}).}
 \label{fig:a3827_errors}
\end{figure}

In Fig.~\cref{fig:a2813_errors,fig:a399_errors,fig:a401_errors,fig:macsj0417_errors,fig:rxcj0528_errors,fig:macsj0553_errors,fig:as592_errors,fig:a1413_errors,fig:a1689_errors,fig:a1914_errors,fig:a2104_errors,fig:a2218_errors,fig:triangulum_errors,fig:a3827_errors} we report the \cstatdof\ and the fractional error on the determination of the temperature for each spectral region shown in Fig.~\cref{fig:a2813,fig:a399,fig:a401,fig:macsj0417,fig:rxcj0528,fig:macsj0553,fig:as592,fig:a1413,fig:a1689,fig:a1914,fig:a2104,fig:a2218,fig:triangulum,fig:a3827}. Pressure and entropy uncertainties are dominated by the errors on the temperature (see Eq.~\ref{eq:pseudo-pressure}, \ref{eq:pseudo-entropy}) as the errors on the emission measure are only at a level of a few percent. We decided to not report uncertainties on these two thermodynamical quantities to reduce the number of images in the paper. The \chisqdof\ of the broken power-law fits is close to unity in the majority of SB profiles.

\section{Null results}\label{app:null}

\begin{figure}
 \centering
 \includegraphics[width=.23\textwidth]{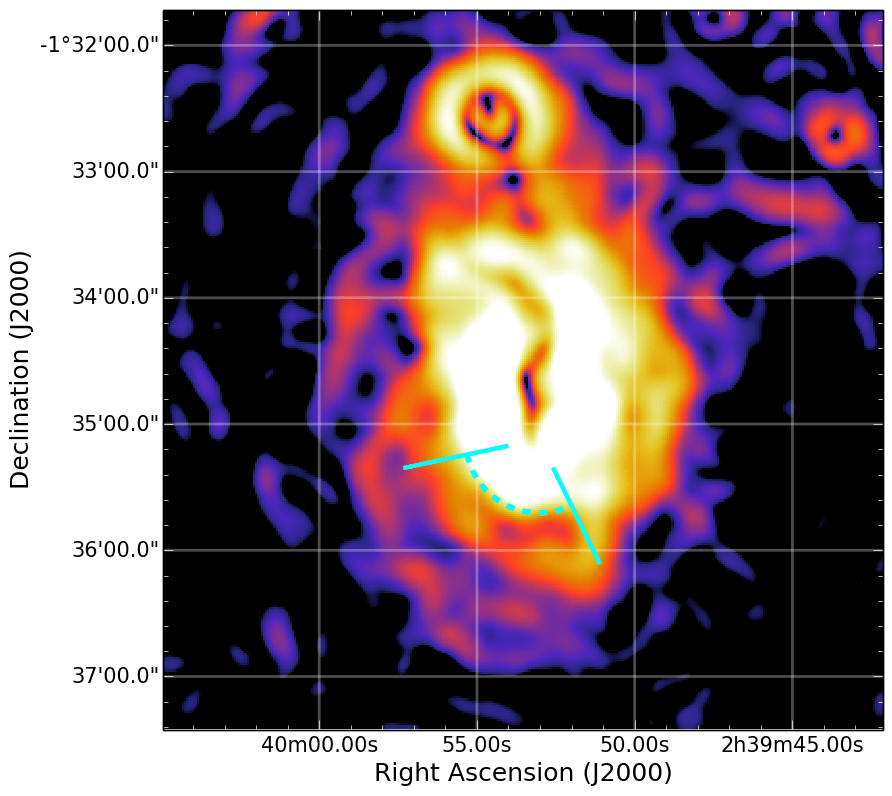}
 \includegraphics[width=.24\textwidth]{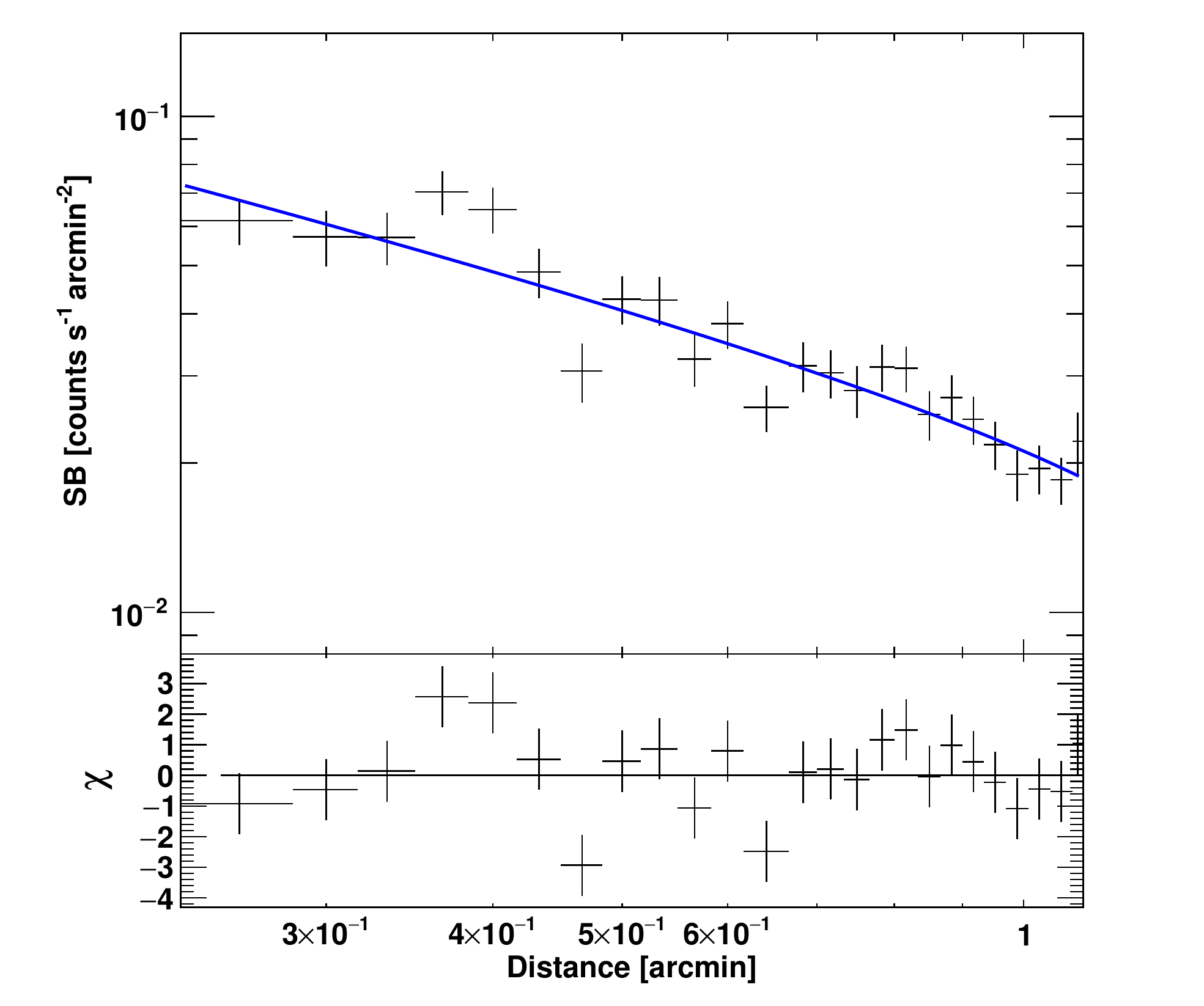}
 \caption{GGM filtered image of A370 (the same as Fig.~\ref{fig:a370}c) marked with the region used to extract the SB profile (\textit{left}) and corresponding fit (\textit{right}). The putative edge is at $r\sim0.6\arcmin$.}
 \label{fig:a370_noedge}
\end{figure}

\begin{figure}
 \centering
 \includegraphics[width=.23\textwidth]{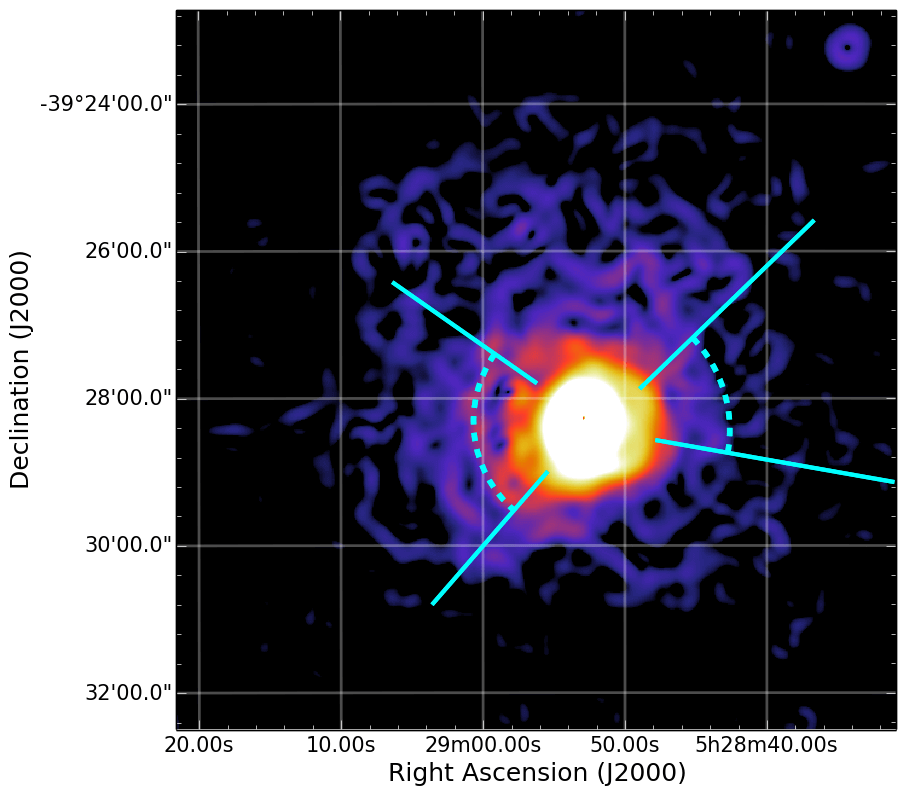}
 \includegraphics[width=.24\textwidth]{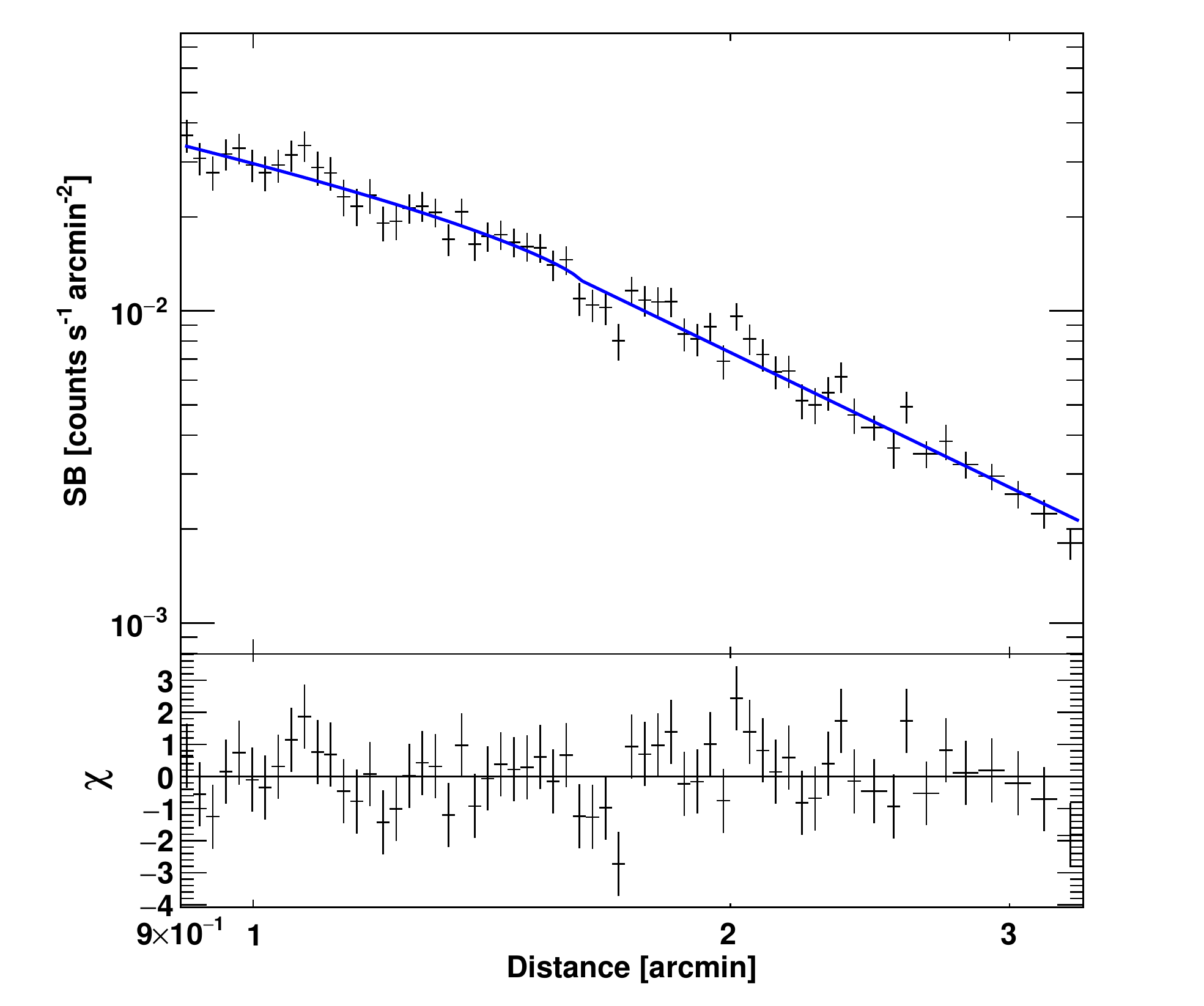}
 \includegraphics[width=.24\textwidth]{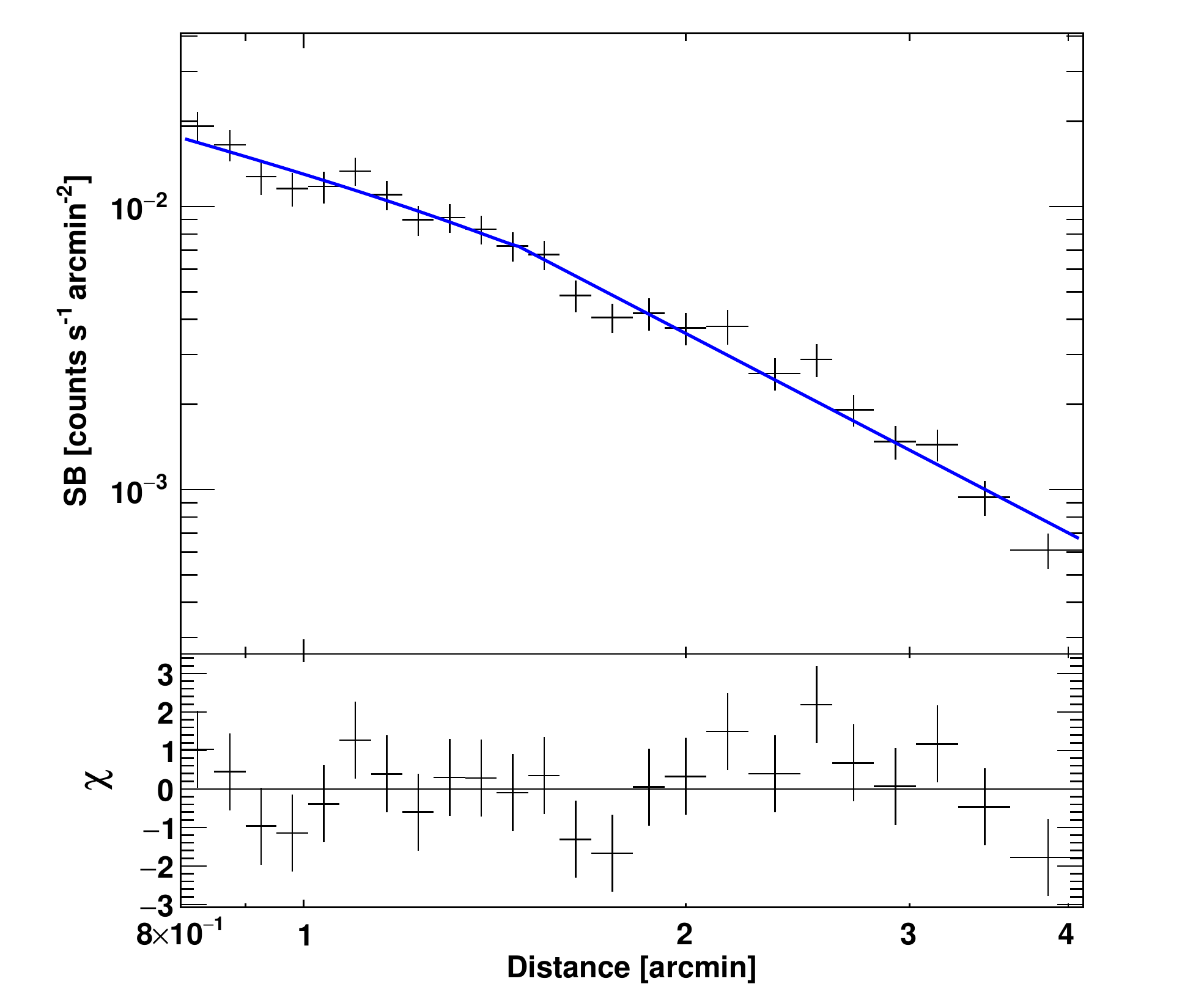}
 \caption{GGM filtered image of RXCJ0528 (the same as Fig.~\ref{fig:rxcj0528}c) marked with the regions used to extract the SB profiles (\textit{left}) and corresponding fits. The putative edge in the E sector is at $r\sim1.6\arcmin$ (\textit{right}) whereas in the W sector is at $r\sim1.8\arcmin$ (\textit{bottom}).}
 \label{fig:rxcj0528_noedge}
\end{figure}

\begin{figure}
 \centering
 \includegraphics[width=.23\textwidth]{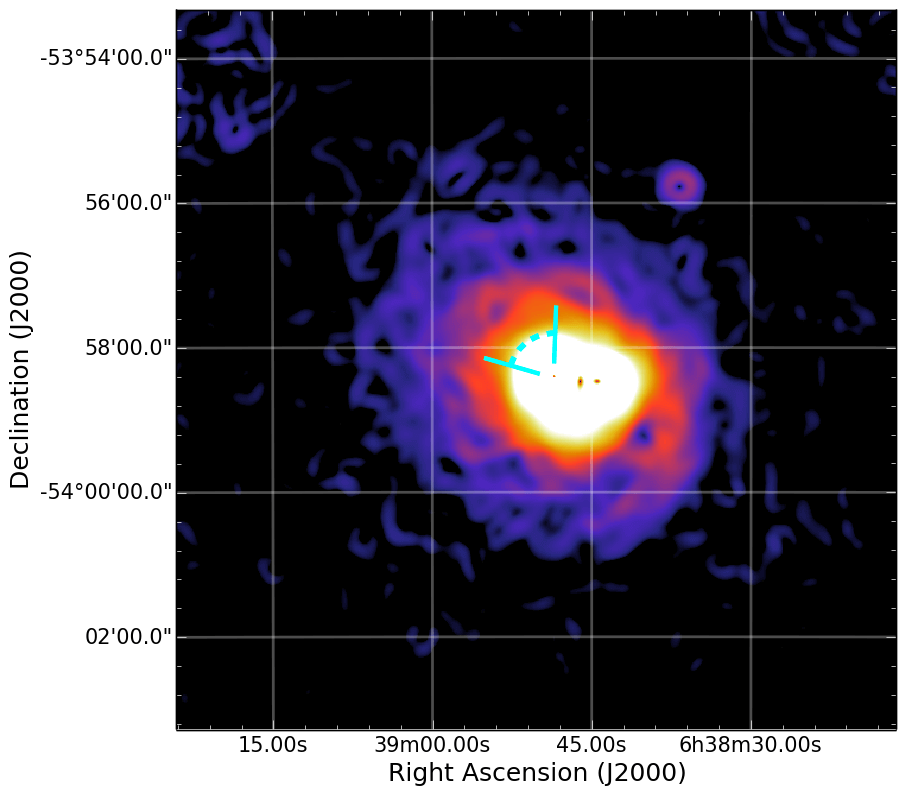}
 \includegraphics[width=.24\textwidth]{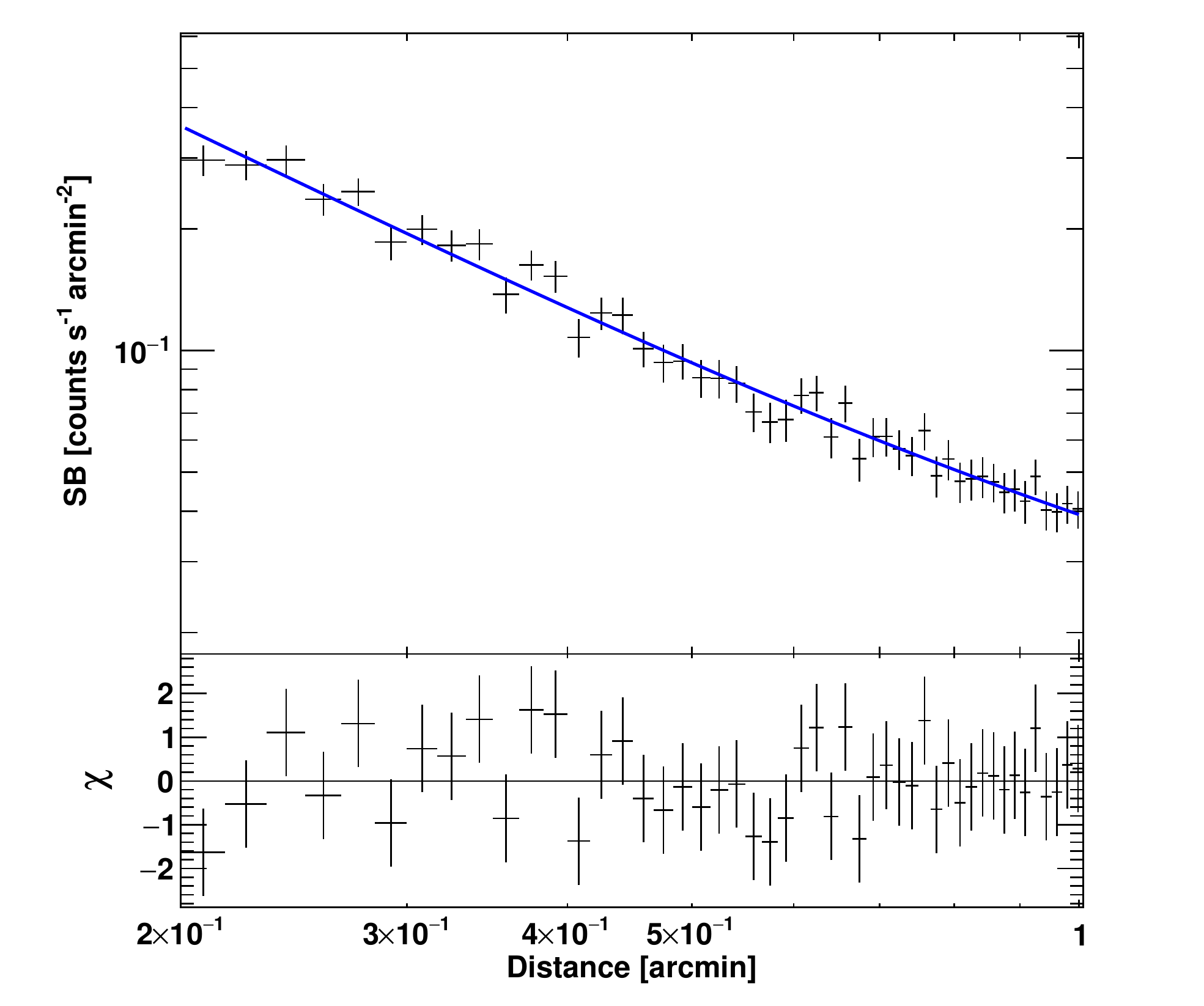}
 \caption{GGM filtered image of AS592 (the same as Fig.~\ref{fig:as592}c) marked with the region used to extract the SB profile (\textit{left}) and corresponding fit (\textit{right}). The putative edge is at $r\sim0.6\arcmin$}
 \label{fig:as592_noedge}
\end{figure}

\begin{figure}
 \centering
 \includegraphics[width=.23\textwidth]{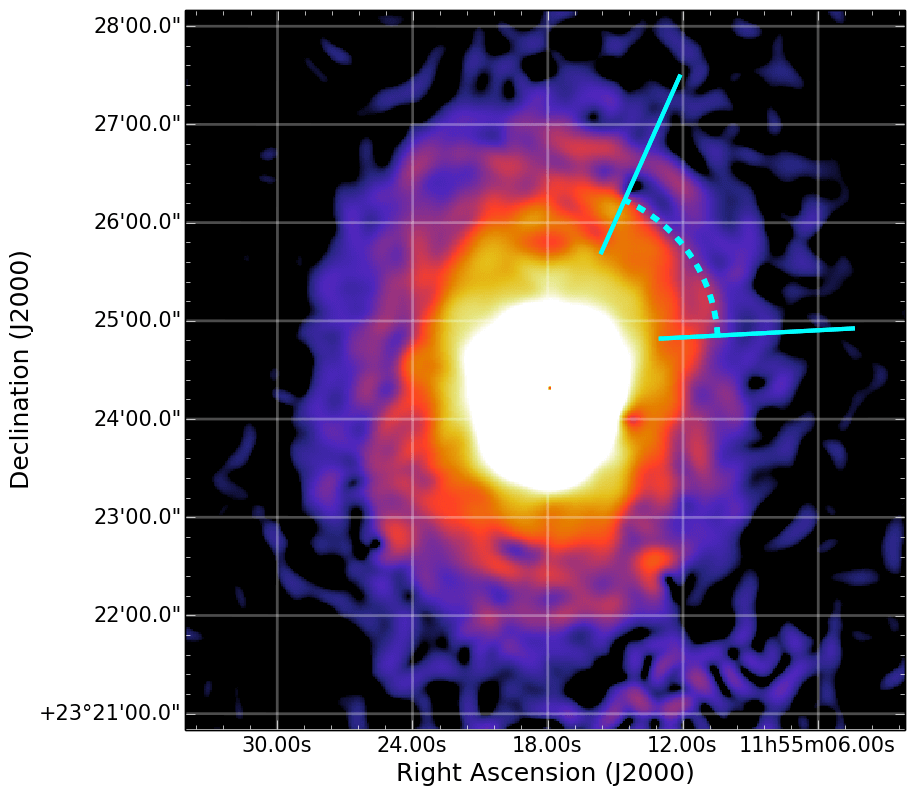}
 \includegraphics[width=.24\textwidth]{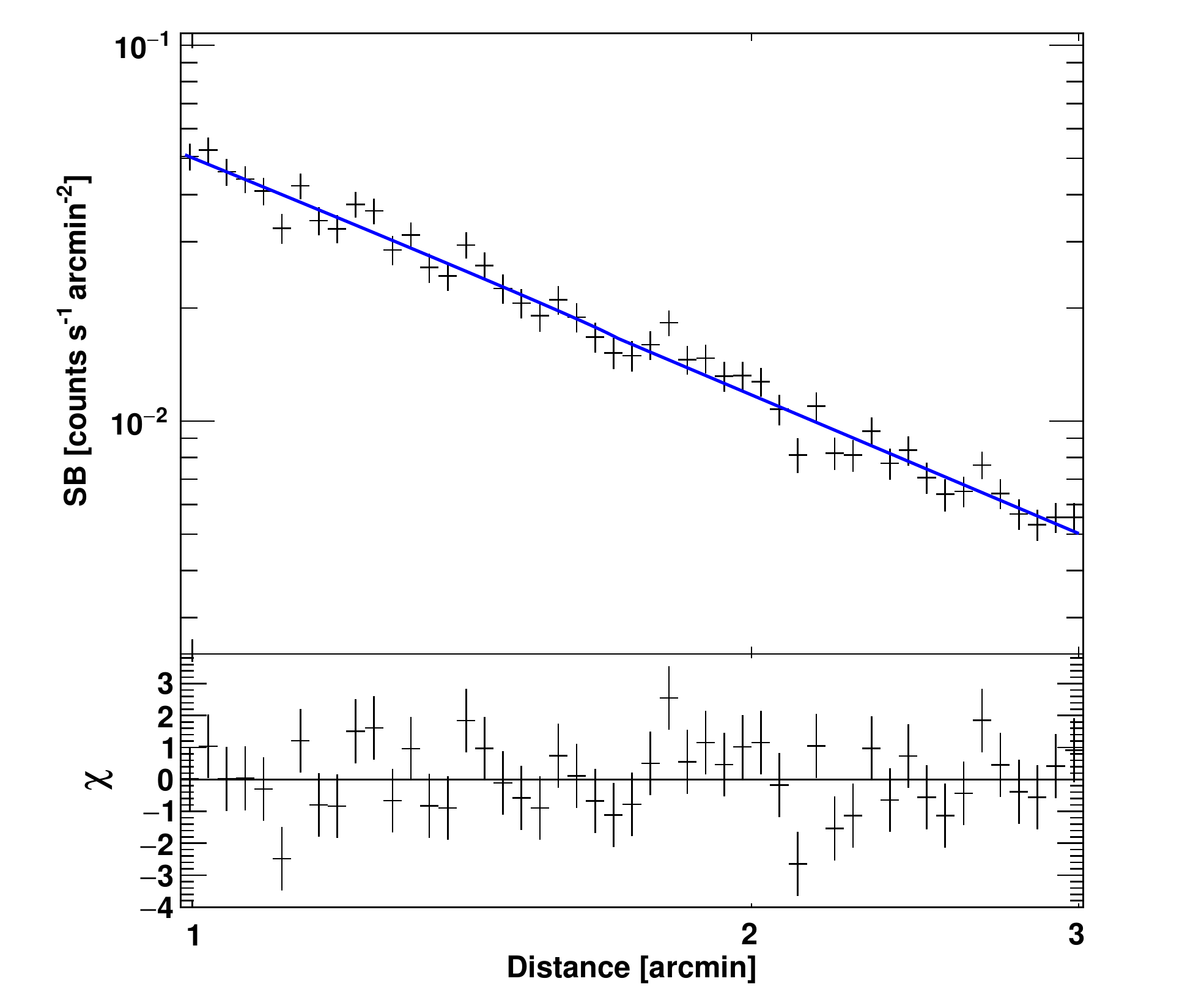}
 \caption{GGM filtered image of A1413 (the same as Fig.~\ref{fig:a1413}c) marked with the region used to extract the SB profile (\textit{left}) and corresponding fit (\textit{right}). The putative edge is at $r\sim1.6\arcmin$}
 \label{fig:a1413_noedge}
\end{figure}

\begin{figure}
 \centering
 \includegraphics[width=.23\textwidth]{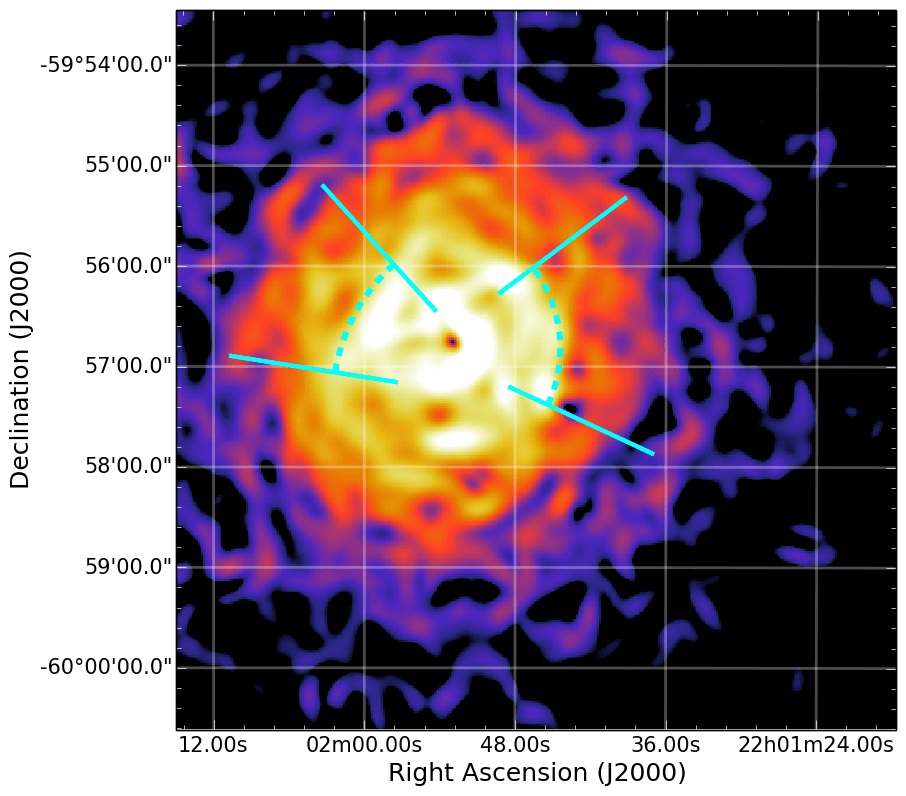}
 \includegraphics[width=.24\textwidth]{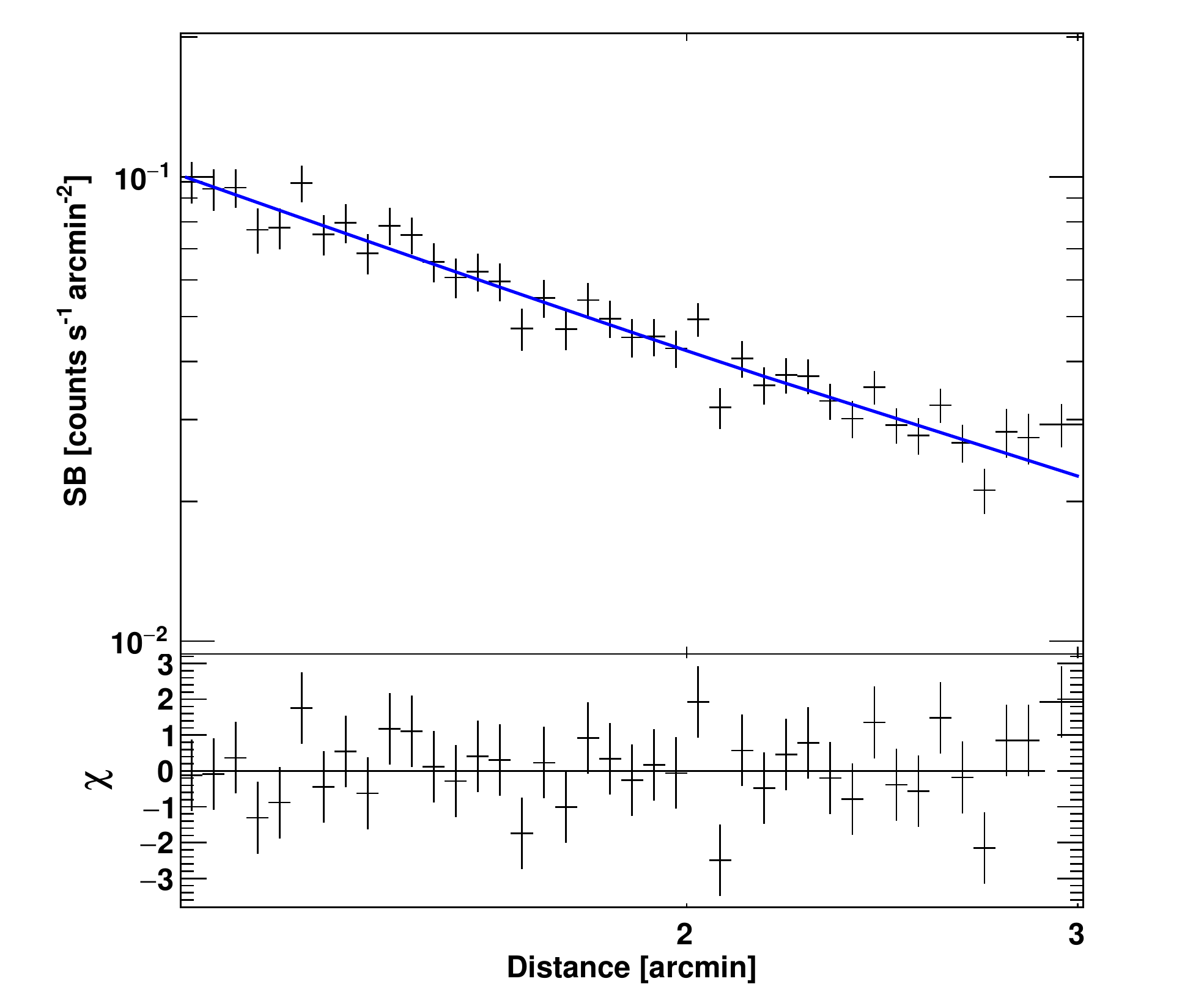}
 \includegraphics[width=.24\textwidth]{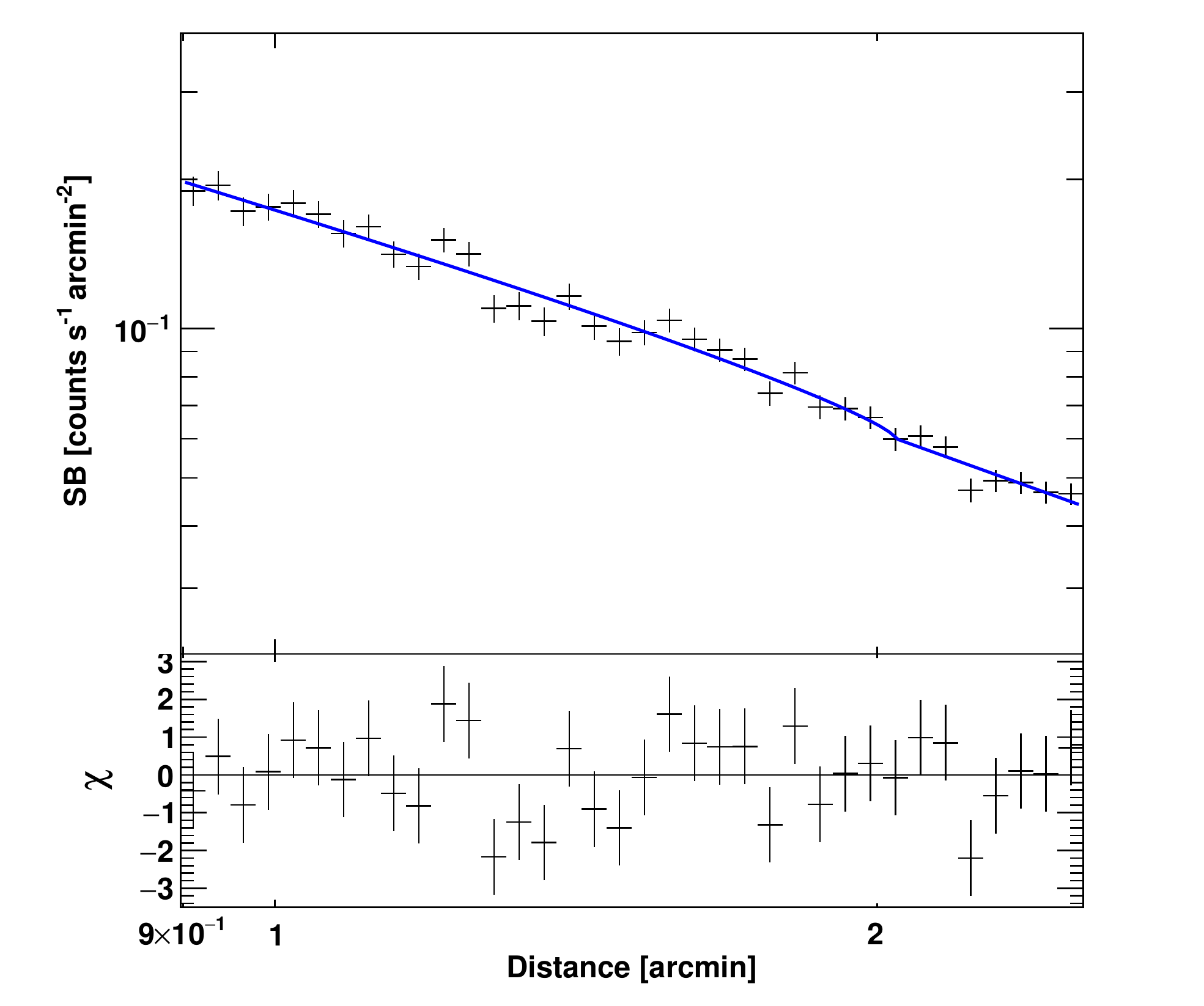}
 \caption{GGM filtered image of A3827 (the same as Fig.~\ref{fig:a3827}c) marked with the regions used to extract the SB profiles (\textit{left}) and corresponding fits. The putative edge in the E sector is at $r\sim1.8\arcmin$ (\textit{right}) whereas in the W sector is at $r\sim1.3\arcmin$ (\textit{bottom}).}
 \label{fig:a3827_noedge}
\end{figure}

Here we report seven cases where the presence of a SB gradient was suggested by the GGM filters but the fitting of the SB profile did not evidence any sharp edge. In Fig.~\cref{fig:a370_noedge,fig:rxcj0528_noedge,fig:as592_noedge,fig:a1413_noedge,fig:a3827_noedge} we show the GGM images with $\sigma=8$ pixels with overlaid the sectors used to extract the SB profiles of the candidate edges together with the corresponding broken power-law model (Eq.~\ref{eq:bknpow}) fits.

\clearpage

% Don't change these lines
\bsp	% typesetting comment
\label{lastpage}
\end{document}